\documentclass[aps,prd,twocolumn,superscriptaddress,preprintnumbers,nofootinbib]{revtex4-2}
\usepackage{amsmath}
\usepackage{amssymb}
\usepackage{natbib}
\usepackage{graphicx}
\usepackage{epsf}
\usepackage{subfigure}
\usepackage{colortbl}
\usepackage[dvipsnames, table]{xcolor}
\usepackage{threeparttable}
\usepackage{dcolumn}
\usepackage{comment}
\usepackage{epsfig}
\usepackage{xspace}
\usepackage{multirow}
\usepackage{makecell}
\usepackage{mathtools}
\usepackage{appendix}
\usepackage{booktabs}
\usepackage[utf8]{inputenc}
\usepackage[TS1,T1]{fontenc}
\usepackage{diagbox}
\usepackage[normalem]{ulem}
\usepackage[
colorlinks=True,
allcolors=blue,
linktocpage=true,
]{hyperref}
\usepackage[capitalise]{cleveref}
\usepackage{latexsym}
\usepackage{array}
\usepackage{orcidlink}
\usepackage{listings}
\usepackage{float}
\newcolumntype{C}[1]{>{\centering\let\newline\\\arraybackslash\hspace{0pt}}m{#1}}
\renewcommand{\arraystretch}{1.2}

\date{\today}

\DeclareGraphicsExtensions{.jpg,.pdf,.png,.eps,.ps}

\newcommand{\T}{\text{T}}
\newcommand{\E}{\text{E}}

\renewcommand{\P}{\text{P}}
\newcommand{\Q}{\text{Q}}
\newcommand{\U}{\text{U}}
\newcommand{\X}{\text{X}}
\newcommand{\Y}{\text{Y}}
\newcommand*{\TT}{\text{TT}}
\newcommand*{\EE}{\text{EE}}
\newcommand*{\TE}{\text{TE}}

\newcommand*{\ET}{\text{ET}}
\newcommand*{\EETE}{\text{TE/EE}}
\newcommand*{\TTTEEE}{\text{\T\&\E{}}}

\newcommand*{\XX}{\text{XX}}

\newcommand*{\XY}{\text{XY}}
\newcommand*{\ZV}{\text{ZV}}

\newcommand*{\YV}{\text{YV}}

\newcommand*{\XZ}{\text{XZ}}

\newcommand*{\pp}{\ensuremath{\phi\phi}}
\newcommand*{\PP}{\pp}
\newcommand*{\lcdm}{$\Lambda$CDM}
\newcommand*{\planck}{\textit{Planck}}
\newcommand*{\plik}{\texttt{Plik}}
\newcommand*{\camspec}{\texttt{Camspec}}

\newcommand{\WMAP}{\textsc{WMAP}}

\newcommand{\DESI}{\textsc{DESI}}
\newcommand{\ACT}{\textsc{ACT}} 
\newcommand{\ACTDR}{\textsc{ACT\,DR6}} 
\newcommand{\shoes}{\textsc{SH0ES}}
\newcommand{\ground}{\textsc{SPT+ACT}}
\newcommand{\cmball}{\textsc{CMB-SPA}}

\newcommand{\LCDM}{\lcdm}
\newcommand{\candl}{\texttt{candl}}
\newcommand{\sptlite}{\texttt{SPTlite}}
\newcommand{\OLE}{\texttt{OL\'E}}
\newcommand*{\ellmin}{\ensuremath{\ell_{\rm min}}}
\newcommand*{\ellmax}{\ensuremath{\ell_{\rm max}}}
\newcommand{\RNum}[1]{\uppercase\expandafter{\romannumeral #1\relax}}

\newcommand{\healpix}{\texttt{HEALPix}}
\newcommand*{\polspice}{\texttt{Polspice}}
\newcommand*{\agora}{\textsc{Agora}}
\newcommand*{\fisherlens}{\texttt{FisherLens}}

\newcommand{\RR}[1]{{ #1}}
\newcommand{\replace}[2]{{}{\RR{#2}}}

\def\sqdeg{{\rm deg}^{2}}
\def\degsq{\sqdeg}
\def\uk{\ensuremath{\mu \mathrm{K}}}
\def\uksq{\ensuremath{\mu \mathrm{K}^2}}
\def\ukarcmin{\uk{\rm -arcmin}}

\newcommand*{\ghz}{\,\text{GHz}}
\def\eV{\text{eV}}

\newcommand{\ie}{i.e.}

\newcommand{\SPT}{\textsc{SPT}}
\newcommand{\muse}{\textsc{MUSE}}
\newcommand{\mainfield}{SPT-3G Main\xspace}
\newcommand{\summerfield}{SPT-3G Summer}

\newcommand{\extfield}{SPT-3G Wide}
\newcommand{\allspt}{Ext-10k}
\def\sptnew{SPT-3G\xspace}
\def\sptbp{\text{SPT-3G D1 }\TTTEEE{}\xspace}
\def\sptlr{\text{SPT-3G D1}\xspace}

\def\quickmock{\texttt{Quickmock}\xspace}
\def\fullmock{\texttt{Fullmock}\xspace}
\def\master{\texttt{MASTER}\xspace}
\def\cork{\texttt{cork}\xspace}
\def\sptlite{\texttt{SPT-lite}}
\def\camb{\texttt{CAMB}\xspace}
\def\class{\texttt{CLASS}\xspace}
\def\cosmopower{\texttt{CosmoPower}\xspace}
\def\cobaya{\texttt{Cobaya}\xspace}
\def\montepython{\texttt{MontePython}\xspace}

\newcommand{\VEV}[1]{\ensuremath{\left\langle#1\right\rangle_{\rm sims}}}
\newcommand{\Nside}{\ensuremath{N_{\rm side}}}
\newcommand{\Nmc}{\ensuremath{N_{\rm MC}}}
\newcommand{\Ndof}{\ensuremath{N_{\rm dof}}}
\newcommand{\ff}{\text{f}}
\newcommand{\uu}{\text{u}}

\newcommand{\Wt}{\text{Wt}}
\newcommand{\W}{\text{W}}
\newcommand{\mmmask}{\text{MMmask}}
\newcommand{\nommmask}{\text{no\  MMmask}}
\newcommand{\TtoP}{\text{T-to-P}\xspace}

\newcommand{\QM}{\text{QM}}
\newcommand{\FM}{\text{FM}}
\newcommand{\commenter}[1]{{}}

\newcommand{\thetaMC}{\ensuremath{\theta_{\rm MC}}}
\newcommand{\thetastar}{\ensuremath{\theta_{\rm s}^\star}}
\newcommand{\ombh}{\ensuremath{\Omega_{\rm b} h^2}}
\newcommand{\omch}{\ensuremath{\Omega_{\rm c} h^2}}
\newcommand{\omegam}{\ensuremath{\Omega_{\rm m}}}
\newcommand{\As}{\ensuremath{A_{\mathrm{s}}}}
\newcommand{\logA}{\ensuremath{\log(10^{10}\,\As{})}}
\newcommand{\Hubble}{\ensuremath{H_0}}
\newcommand{\taureio}{\ensuremath{\tau_\mathrm{reio}}}
\newcommand{\ns}{\ensuremath{n_\mathrm{s}}}
\newcommand{\Age}{\ensuremath{\mathrm{Age}}}
\newcommand{\oml}{\ensuremath{\Omega_{\Lambda}}}
\newcommand{\clamp}{\ensuremath{10^9\,A_\mathrm{s} e^{-2\taureio}}}

\newcommand{\omm}{\omegam}
\newcommand{\ommh}{\ensuremath{\omega_\mathrm{m}}}
\newcommand{\rdrag}{\ensuremath{r_\mathrm{d}}}
\newcommand{\sigmaeight}{\ensuremath{\sigma_8}}
\newcommand{\ommrdsq}{\ensuremath{\omega_\mathrm{m} \rdrag^{2}}}
\newcommand{\seight}[1]{\ensuremath{S_8(#1)}}
\newcommand{\neff}{\ensuremath{N_{\rm eff}}}
\newcommand*{\Yp}{\ensuremath{Y_{\mathrm{P}}}}
\newcommand*{\mnu}{\ensuremath{\Sigma m_{\mathrm{\nu}}}}

\newcommand*{\curv}{\ensuremath{\Omega_{\rm k}}}
\newcommand{\hrd}{\ensuremath{h\rdrag}}

\newcommand{\wo}{\ensuremath{w_0}}
\newcommand{\wa}{\ensuremath{w_a}}
\newcommand{\wowa}{\wo\wa}
\newcommand{\wperp}{\ensuremath{w_\perp}}
\newcommand{\alm}{\ensuremath{a_{\ell m}}}
\newcommand{\betapol}{\ensuremath{\beta_{\rm pol}}}
\newcommand{\Tcal}{\ensuremath{A_{\mathrm{cal}}}}
\newcommand{\Ecal}{\ensuremath{\Epol_{\mathrm{cal}}}}
\newcommand*{\Epol}{\ensuremath{E}}
\newcommand{\Atwopt}{\ensuremath{A_{\rm 2pt}}}
\newcommand{\Arecon}{\ensuremath{A_{\rm recon}}}
\newcommand{\Alens}{\ensuremath{A_{\rm lens}}}

\newcommand{\Hubbleunit}{[\text{km/s/Mpc}]}
\newcommand*{\kmsmpc}{\ensuremath{\mathrm{km\,s^{-1}\,Mpc^{-1}}}}

\newcommand{\Ageunit}{[\text{Gyr}]}
\newcommand{\rdragunit}{[\text{Mpc}]}
\newcommand{\rdragunittxt}{\text{Mpc}}
\newcommand{\GHz}{\ensuremath{\mathrm{GHz}}}

\newcommand{\omegab}{\ensuremath{\ombh}}
\newcommand{\omegal}{\ensuremath{\oml}}
\newcommand{\omegac}{\ensuremath{\omch}}

\newcommand*{\Neff}{\neff}
\newcommand*{\yp}{\Yp}
\newcommand{\Omegak}{\ensuremath{\curv}}

\newcommand{\rd}{\rdrag}

\newcommand{\alens}{\Alens}

\begin{document}

\title{\sptlr: CMB temperature and polarization power spectra and cosmology from 2019 and 2020 observations of the SPT-3G Main field}

\affiliation{Sorbonne Universit\'e, CNRS, UMR 7095, Institut d'Astrophysique de Paris, 98 bis bd Arago, 75014 Paris, France}
\affiliation{High-Energy Physics Division, Argonne National Laboratory, 9700 South Cass Avenue, Lemont, IL, 60439, USA}
\affiliation{Department of Physics, University of Chicago, 5640 South Ellis Avenue, Chicago, IL, 60637, USA}
\affiliation{Kavli Institute for Cosmological Physics, University of Chicago, 5640 South Ellis Avenue, Chicago, IL, 60637, USA}
\affiliation{Kavli Institute for Particle Astrophysics and Cosmology, Stanford University, 452 Lomita Mall, Stanford, CA, 94305, USA}
\affiliation{Department of Physics, Stanford University, 382 Via Pueblo Mall, Stanford, CA, 94305, USA}
\affiliation{Department of Physics \& Astronomy, University of California, One Shields Avenue, Davis, CA 95616, USA}
\affiliation{Department of Physics, University of California, Berkeley, CA, 94720, USA}
\affiliation{Department of Astronomy and Astrophysics, University of Chicago, 5640 South Ellis Avenue, Chicago, IL, 60637, USA}
\affiliation{Center for AstroPhysical Surveys, National Center for Supercomputing Applications, Urbana, IL, 61801, USA}
\affiliation{Fermi National Accelerator Laboratory, MS209, P.O. Box 500, Batavia, IL, 60510, USA}
\affiliation{School of Physics, University of Melbourne, Parkville, VIC 3010, Australia}
\affiliation{SLAC National Accelerator Laboratory, 2575 Sand Hill Road, Menlo Park, CA, 94025, USA}
\affiliation{Enrico Fermi Institute, University of Chicago, 5640 South Ellis Avenue, Chicago, IL, 60637, USA}
\affiliation{National Taiwan University, No. 1, Sec. 4, Roosevelt Road, Taipei 106319, Taiwan}
\affiliation{Universit\'e Paris-Saclay, Universit\'e Paris Cit\'e, CEA, CNRS, AIM, 91191, Gif-sur-Yvette, France}
\affiliation{Department of Astronomy, University of Illinois Urbana-Champaign, 1002 West Green Street, Urbana, IL, 61801, USA}
\affiliation{High Energy Accelerator Research Organization (KEK), Tsukuba, Ibaraki 305-0801, Japan}
\affiliation{Department of Physics and McGill Space Institute, McGill University, 3600 Rue University, Montreal, Quebec H3A 2T8, Canada}
\affiliation{Canadian Institute for Advanced Research, CIFAR Program in Gravity and the Extreme Universe, Toronto, ON, M5G 1Z8, Canada}
\affiliation{Joseph Henry Laboratories of Physics, Jadwin Hall, Princeton University, Princeton, NJ 08544, USA}
\affiliation{Department of Astrophysical and Planetary Sciences, University of Colorado, Boulder, CO, 80309, USA}
\affiliation{Department of Physics, University of Illinois Urbana-Champaign, 1110 West Green Street, Urbana, IL, 61801, USA}
\affiliation{Department of Physics and Astronomy, University of California, Los Angeles, CA, 90095, USA}
\affiliation{Department of Physics and Astronomy, Michigan State University, East Lansing, MI 48824, USA}
\affiliation{Department of Physics and Astronomy, Northwestern University, 633 Clark St, Evanston, IL, 60208, USA}
\affiliation{CASA, Department of Astrophysical and Planetary Sciences, University of Colorado, Boulder, CO, 80309, USA }
\affiliation{Department of Physics, University of Colorado, Boulder, CO, 80309, USA}
\affiliation{Department of Physics, Case Western Reserve University, Cleveland, OH, 44106, USA}
\affiliation{Dunlap Institute for Astronomy \& Astrophysics, University of Toronto, 50 St. George Street, Toronto, ON, M5S 3H4, Canada}
\affiliation{David A. Dunlap Department of Astronomy \& Astrophysics, University of Toronto, 50 St. George Street, Toronto, ON, M5S 3H4, Canada}
\affiliation{NSF-Simons AI Institute for the Sky (SkAI), 172 E. Chestnut St., Chicago, IL 60611, USA}
\affiliation{Center for Astrophysics \textbar{} Harvard \& Smithsonian, 60 Garden Street, Cambridge, MA, 02138, USA}
\author{E.~Camphuis\,\orcidlink{0000-0003-3483-8461}}
\email[Corresponding author: ]{etienne.camphuis@iap.fr}
\affiliation{Sorbonne Universit\'e, CNRS, UMR 7095, Institut d'Astrophysique de Paris, 98 bis bd Arago, 75014 Paris, France}
\author{W.~Quan}
\affiliation{High-Energy Physics Division, Argonne National Laboratory, 9700 South Cass Avenue, Lemont, IL, 60439, USA}
\affiliation{Department of Physics, University of Chicago, 5640 South Ellis Avenue, Chicago, IL, 60637, USA}
\affiliation{Kavli Institute for Cosmological Physics, University of Chicago, 5640 South Ellis Avenue, Chicago, IL, 60637, USA}
\author{L.~Balkenhol\,\orcidlink{0000-0001-6899-1873}}
\affiliation{Sorbonne Universit\'e, CNRS, UMR 7095, Institut d'Astrophysique de Paris, 98 bis bd Arago, 75014 Paris, France}
\author{A.~R.~Khalife\,\orcidlink{0000-0002-8388-4950}}
\affiliation{Sorbonne Universit\'e, CNRS, UMR 7095, Institut d'Astrophysique de Paris, 98 bis bd Arago, 75014 Paris, France}
\author{F.~Ge}
\affiliation{Kavli Institute for Particle Astrophysics and Cosmology, Stanford University, 452 Lomita Mall, Stanford, CA, 94305, USA}
\affiliation{Department of Physics, Stanford University, 382 Via Pueblo Mall, Stanford, CA, 94305, USA}
\affiliation{Department of Physics \& Astronomy, University of California, One Shields Avenue, Davis, CA 95616, USA}
\author{F.~Guidi\,\orcidlink{0000-0001-7593-3962}}
\affiliation{Sorbonne Universit\'e, CNRS, UMR 7095, Institut d'Astrophysique de Paris, 98 bis bd Arago, 75014 Paris, France}
\author{N.~Huang\,\orcidlink{0000-0003-3595-0359}}
\affiliation{Department of Physics, University of California, Berkeley, CA, 94720, USA}
\author{G.~P.~Lynch\,\orcidlink{0009-0004-3143-1708}}
\affiliation{Department of Physics \& Astronomy, University of California, One Shields Avenue, Davis, CA 95616, USA}
\author{Y.~Omori}
\affiliation{Department of Astronomy and Astrophysics, University of Chicago, 5640 South Ellis Avenue, Chicago, IL, 60637, USA}
\affiliation{Kavli Institute for Cosmological Physics, University of Chicago, 5640 South Ellis Avenue, Chicago, IL, 60637, USA}
\author{C.~Trendafilova}
\affiliation{Center for AstroPhysical Surveys, National Center for Supercomputing Applications, Urbana, IL, 61801, USA}
\author{A.~J.~Anderson\,\orcidlink{0000-0002-4435-4623}}
\affiliation{Fermi National Accelerator Laboratory, MS209, P.O. Box 500, Batavia, IL, 60510, USA}
\affiliation{Kavli Institute for Cosmological Physics, University of Chicago, 5640 South Ellis Avenue, Chicago, IL, 60637, USA}
\affiliation{Department of Astronomy and Astrophysics, University of Chicago, 5640 South Ellis Avenue, Chicago, IL, 60637, USA}
\author{B.~Ansarinejad}
\affiliation{School of Physics, University of Melbourne, Parkville, VIC 3010, Australia}
\author{M.~Archipley\,\orcidlink{0000-0002-0517-9842}}
\affiliation{Department of Astronomy and Astrophysics, University of Chicago, 5640 South Ellis Avenue, Chicago, IL, 60637, USA}
\affiliation{Kavli Institute for Cosmological Physics, University of Chicago, 5640 South Ellis Avenue, Chicago, IL, 60637, USA}
\author{P.~S.~Barry\,\orcidlink{0000-0001-9103-9354}}
\affiliation{School of Physics and Astronomy, Cardiff University, Cardiff, CF24 3AA, UK}
\author{K.~Benabed}
\affiliation{Sorbonne Universit\'e, CNRS, UMR 7095, Institut d'Astrophysique de Paris, 98 bis bd Arago, 75014 Paris, France}
\author{A.~N.~Bender\,\orcidlink{0000-0001-5868-0748}}
\affiliation{High-Energy Physics Division, Argonne National Laboratory, 9700 South Cass Avenue, Lemont, IL, 60439, USA}
\affiliation{Kavli Institute for Cosmological Physics, University of Chicago, 5640 South Ellis Avenue, Chicago, IL, 60637, USA}
\affiliation{Department of Astronomy and Astrophysics, University of Chicago, 5640 South Ellis Avenue, Chicago, IL, 60637, USA}
\author{B.~A.~Benson\,\orcidlink{0000-0002-5108-6823}}
\affiliation{Fermi National Accelerator Laboratory, MS209, P.O. Box 500, Batavia, IL, 60510, USA}
\affiliation{Kavli Institute for Cosmological Physics, University of Chicago, 5640 South Ellis Avenue, Chicago, IL, 60637, USA}
\affiliation{Department of Astronomy and Astrophysics, University of Chicago, 5640 South Ellis Avenue, Chicago, IL, 60637, USA}
\author{F.~Bianchini\,\orcidlink{0000-0003-4847-3483}}
\affiliation{Kavli Institute for Particle Astrophysics and Cosmology, Stanford University, 452 Lomita Mall, Stanford, CA, 94305, USA}
\affiliation{Department of Physics, Stanford University, 382 Via Pueblo Mall, Stanford, CA, 94305, USA}
\affiliation{SLAC National Accelerator Laboratory, 2575 Sand Hill Road, Menlo Park, CA, 94025, USA}
\author{L.~E.~Bleem\,\orcidlink{0000-0001-7665-5079}}
\affiliation{High-Energy Physics Division, Argonne National Laboratory, 9700 South Cass Avenue, Lemont, IL, 60439, USA}
\affiliation{Kavli Institute for Cosmological Physics, University of Chicago, 5640 South Ellis Avenue, Chicago, IL, 60637, USA}
\affiliation{Department of Astronomy and Astrophysics, University of Chicago, 5640 South Ellis Avenue, Chicago, IL, 60637, USA}
\author{F.~R.~Bouchet\,\orcidlink{0000-0002-8051-2924}}
\affiliation{Sorbonne Universit\'e, CNRS, UMR 7095, Institut d'Astrophysique de Paris, 98 bis bd Arago, 75014 Paris, France}
\author{L.~Bryant}
\affiliation{Enrico Fermi Institute, University of Chicago, 5640 South Ellis Avenue, Chicago, IL, 60637, USA}
\author{M.~G.~Campitiello}
\affiliation{High-Energy Physics Division, Argonne National Laboratory, 9700 South Cass Avenue, Lemont, IL, 60439, USA}
\author{J.~E.~Carlstrom\,\orcidlink{0000-0002-2044-7665}}
\affiliation{Kavli Institute for Cosmological Physics, University of Chicago, 5640 South Ellis Avenue, Chicago, IL, 60637, USA}
\affiliation{Enrico Fermi Institute, University of Chicago, 5640 South Ellis Avenue, Chicago, IL, 60637, USA}
\affiliation{Department of Physics, University of Chicago, 5640 South Ellis Avenue, Chicago, IL, 60637, USA}
\affiliation{High-Energy Physics Division, Argonne National Laboratory, 9700 South Cass Avenue, Lemont, IL, 60439, USA}
\affiliation{Department of Astronomy and Astrophysics, University of Chicago, 5640 South Ellis Avenue, Chicago, IL, 60637, USA}
\author{C.~L.~Chang}
\affiliation{High-Energy Physics Division, Argonne National Laboratory, 9700 South Cass Avenue, Lemont, IL, 60439, USA}
\affiliation{Kavli Institute for Cosmological Physics, University of Chicago, 5640 South Ellis Avenue, Chicago, IL, 60637, USA}
\affiliation{Department of Astronomy and Astrophysics, University of Chicago, 5640 South Ellis Avenue, Chicago, IL, 60637, USA}
\author{P.~Chaubal}
\affiliation{School of Physics, University of Melbourne, Parkville, VIC 3010, Australia}
\author{P.~M.~Chichura\,\orcidlink{0000-0002-5397-9035}}
\affiliation{Department of Physics, University of Chicago, 5640 South Ellis Avenue, Chicago, IL, 60637, USA}
\affiliation{Kavli Institute for Cosmological Physics, University of Chicago, 5640 South Ellis Avenue, Chicago, IL, 60637, USA}
\author{A.~Chokshi}
\affiliation{Department of Astronomy and Astrophysics, University of Chicago, 5640 South Ellis Avenue, Chicago, IL, 60637, USA}
\author{T.-L.~Chou\,\orcidlink{0000-0002-3091-8790}}
\affiliation{Department of Astronomy and Astrophysics, University of Chicago, 5640 South Ellis Avenue, Chicago, IL, 60637, USA}
\affiliation{Kavli Institute for Cosmological Physics, University of Chicago, 5640 South Ellis Avenue, Chicago, IL, 60637, USA}
\affiliation{National Taiwan University, No. 1, Sec. 4, Roosevelt Road, Taipei 106319, Taiwan}
\author{A.~Coerver}
\affiliation{Department of Physics, University of California, Berkeley, CA, 94720, USA}
\author{T.~M.~Crawford\,\orcidlink{0000-0001-9000-5013}}
\affiliation{Department of Astronomy and Astrophysics, University of Chicago, 5640 South Ellis Avenue, Chicago, IL, 60637, USA}
\affiliation{Kavli Institute for Cosmological Physics, University of Chicago, 5640 South Ellis Avenue, Chicago, IL, 60637, USA}
\author{C.~Daley\,\orcidlink{0000-0002-3760-2086}}
\affiliation{Universit\'e Paris-Saclay, Universit\'e Paris Cit\'e, CEA, CNRS, AIM, 91191, Gif-sur-Yvette, France}
\affiliation{Department of Astronomy, University of Illinois Urbana-Champaign, 1002 West Green Street, Urbana, IL, 61801, USA}
\author{T.~de~Haan}
\affiliation{High Energy Accelerator Research Organization (KEK), Tsukuba, Ibaraki 305-0801, Japan}
\author{K.~R.~Dibert}
\affiliation{Department of Astronomy and Astrophysics, University of Chicago, 5640 South Ellis Avenue, Chicago, IL, 60637, USA}
\affiliation{Kavli Institute for Cosmological Physics, University of Chicago, 5640 South Ellis Avenue, Chicago, IL, 60637, USA}
\author{M.~A.~Dobbs}
\affiliation{Department of Physics and McGill Space Institute, McGill University, 3600 Rue University, Montreal, Quebec H3A 2T8, Canada}
\affiliation{Canadian Institute for Advanced Research, CIFAR Program in Gravity and the Extreme Universe, Toronto, ON, M5G 1Z8, Canada}
\author{M.~Doohan}
\affiliation{School of Physics, University of Melbourne, Parkville, VIC 3010, Australia}
\author{A.~Doussot}
\affiliation{Sorbonne Universit\'e, CNRS, UMR 7095, Institut d'Astrophysique de Paris, 98 bis bd Arago, 75014 Paris, France}
\author{D.~Dutcher\,\orcidlink{0000-0002-9962-2058}}
\affiliation{Joseph Henry Laboratories of Physics, Jadwin Hall, Princeton University, Princeton, NJ 08544, USA}
\author{W.~Everett}
\affiliation{Department of Astrophysical and Planetary Sciences, University of Colorado, Boulder, CO, 80309, USA}
\author{C.~Feng}
\affiliation{Department of Physics, University of Illinois Urbana-Champaign, 1110 West Green Street, Urbana, IL, 61801, USA}
\author{K.~R.~Ferguson\,\orcidlink{0000-0002-4928-8813}}
\affiliation{Department of Physics and Astronomy, University of California, Los Angeles, CA, 90095, USA}
\affiliation{Department of Physics and Astronomy, Michigan State University, East Lansing, MI 48824, USA}
\author{K.~Fichman}
\affiliation{Department of Physics, University of Chicago, 5640 South Ellis Avenue, Chicago, IL, 60637, USA}
\affiliation{Kavli Institute for Cosmological Physics, University of Chicago, 5640 South Ellis Avenue, Chicago, IL, 60637, USA}
\author{A.~Foster\,\orcidlink{0000-0002-7145-1824}}
\affiliation{Joseph Henry Laboratories of Physics, Jadwin Hall, Princeton University, Princeton, NJ 08544, USA}
\author{S.~Galli}
\affiliation{Sorbonne Universit\'e, CNRS, UMR 7095, Institut d'Astrophysique de Paris, 98 bis bd Arago, 75014 Paris, France}
\author{A.~E.~Gambrel}
\affiliation{Kavli Institute for Cosmological Physics, University of Chicago, 5640 South Ellis Avenue, Chicago, IL, 60637, USA}
\author{R.~W.~Gardner}
\affiliation{Enrico Fermi Institute, University of Chicago, 5640 South Ellis Avenue, Chicago, IL, 60637, USA}
\author{N.~Goeckner-Wald}
\affiliation{Department of Physics, Stanford University, 382 Via Pueblo Mall, Stanford, CA, 94305, USA}
\affiliation{Kavli Institute for Particle Astrophysics and Cosmology, Stanford University, 452 Lomita Mall, Stanford, CA, 94305, USA}
\author{R.~Gualtieri\,\orcidlink{0000-0003-4245-2315}}
\affiliation{High-Energy Physics Division, Argonne National Laboratory, 9700 South Cass Avenue, Lemont, IL, 60439, USA}
\affiliation{Department of Physics and Astronomy, Northwestern University, 633 Clark St, Evanston, IL, 60208, USA}
\author{S.~Guns}
\affiliation{Department of Physics, University of California, Berkeley, CA, 94720, USA}
\author{N.~W.~Halverson}
\affiliation{CASA, Department of Astrophysical and Planetary Sciences, University of Colorado, Boulder, CO, 80309, USA }
\affiliation{Department of Physics, University of Colorado, Boulder, CO, 80309, USA}
\author{E.~Hivon\,\orcidlink{0000-0003-1880-2733}}
\affiliation{Sorbonne Universit\'e, CNRS, UMR 7095, Institut d'Astrophysique de Paris, 98 bis bd Arago, 75014 Paris, France}
\author{G.~P.~Holder\,\orcidlink{0000-0002-0463-6394}}
\affiliation{Department of Physics, University of Illinois Urbana-Champaign, 1110 West Green Street, Urbana, IL, 61801, USA}
\author{W.~L.~Holzapfel}
\affiliation{Department of Physics, University of California, Berkeley, CA, 94720, USA}
\author{J.~C.~Hood}
\affiliation{Kavli Institute for Cosmological Physics, University of Chicago, 5640 South Ellis Avenue, Chicago, IL, 60637, USA}
\author{A.~Hryciuk}
\affiliation{Department of Physics, University of Chicago, 5640 South Ellis Avenue, Chicago, IL, 60637, USA}
\affiliation{Kavli Institute for Cosmological Physics, University of Chicago, 5640 South Ellis Avenue, Chicago, IL, 60637, USA}
\author{F.~K\'eruzor\'e}
\affiliation{High-Energy Physics Division, Argonne National Laboratory, 9700 South Cass Avenue, Lemont, IL, 60439, USA}
\author{L.~Knox}
\affiliation{Department of Physics \& Astronomy, University of California, One Shields Avenue, Davis, CA 95616, USA}
\author{M.~Korman}
\affiliation{Department of Physics, Case Western Reserve University, Cleveland, OH, 44106, USA}
\author{K.~Kornoelje}
\affiliation{Department of Astronomy and Astrophysics, University of Chicago, 5640 South Ellis Avenue, Chicago, IL, 60637, USA}
\affiliation{Kavli Institute for Cosmological Physics, University of Chicago, 5640 South Ellis Avenue, Chicago, IL, 60637, USA}
\affiliation{High-Energy Physics Division, Argonne National Laboratory, 9700 South Cass Avenue, Lemont, IL, 60439, USA}
\author{C.-L.~Kuo}
\affiliation{Kavli Institute for Particle Astrophysics and Cosmology, Stanford University, 452 Lomita Mall, Stanford, CA, 94305, USA}
\affiliation{Department of Physics, Stanford University, 382 Via Pueblo Mall, Stanford, CA, 94305, USA}
\affiliation{SLAC National Accelerator Laboratory, 2575 Sand Hill Road, Menlo Park, CA, 94025, USA}
\author{K.~Levy}
\affiliation{School of Physics, University of Melbourne, Parkville, VIC 3010, Australia}
\author{A.~E.~Lowitz\,\orcidlink{0000-0002-4747-4276}}
\affiliation{Kavli Institute for Cosmological Physics, University of Chicago, 5640 South Ellis Avenue, Chicago, IL, 60637, USA}
\author{C.~Lu}
\affiliation{Department of Physics, University of Illinois Urbana-Champaign, 1110 West Green Street, Urbana, IL, 61801, USA}
\author{A.~Maniyar}
\affiliation{Kavli Institute for Particle Astrophysics and Cosmology, Stanford University, 452 Lomita Mall, Stanford, CA, 94305, USA}
\affiliation{Department of Physics, Stanford University, 382 Via Pueblo Mall, Stanford, CA, 94305, USA}
\affiliation{SLAC National Accelerator Laboratory, 2575 Sand Hill Road, Menlo Park, CA, 94025, USA}
\author{E.~S.~Martsen}
\affiliation{Department of Astronomy and Astrophysics, University of Chicago, 5640 South Ellis Avenue, Chicago, IL, 60637, USA}
\affiliation{Kavli Institute for Cosmological Physics, University of Chicago, 5640 South Ellis Avenue, Chicago, IL, 60637, USA}
\author{F.~Menanteau}
\affiliation{Department of Astronomy, University of Illinois Urbana-Champaign, 1002 West Green Street, Urbana, IL, 61801, USA}
\affiliation{Center for AstroPhysical Surveys, National Center for Supercomputing Applications, Urbana, IL, 61801, USA}
\author{M.~Millea\,\orcidlink{0000-0001-7317-0551}}
\affiliation{Department of Physics, University of California, Berkeley, CA, 94720, USA}
\author{J.~Montgomery}
\affiliation{Department of Physics and McGill Space Institute, McGill University, 3600 Rue University, Montreal, Quebec H3A 2T8, Canada}
\author{Y.~Nakato}
\affiliation{Department of Physics, Stanford University, 382 Via Pueblo Mall, Stanford, CA, 94305, USA}
\author{T.~Natoli}
\affiliation{Kavli Institute for Cosmological Physics, University of Chicago, 5640 South Ellis Avenue, Chicago, IL, 60637, USA}
\author{G.~I.~Noble\,\orcidlink{0000-0002-5254-243X}}
\affiliation{Dunlap Institute for Astronomy \& Astrophysics, University of Toronto, 50 St. George Street, Toronto, ON, M5S 3H4, Canada}
\affiliation{David A. Dunlap Department of Astronomy \& Astrophysics, University of Toronto, 50 St. George Street, Toronto, ON, M5S 3H4, Canada}
\author{A.~Ouellette}
\affiliation{Department of Physics, University of Illinois Urbana-Champaign, 1110 West Green Street, Urbana, IL, 61801, USA}
\author{Z.~Pan\,\orcidlink{0000-0002-6164-9861}}
\affiliation{High-Energy Physics Division, Argonne National Laboratory, 9700 South Cass Avenue, Lemont, IL, 60439, USA}
\affiliation{Kavli Institute for Cosmological Physics, University of Chicago, 5640 South Ellis Avenue, Chicago, IL, 60637, USA}
\affiliation{Department of Physics, University of Chicago, 5640 South Ellis Avenue, Chicago, IL, 60637, USA}
\author{P.~Paschos}
\affiliation{Enrico Fermi Institute, University of Chicago, 5640 South Ellis Avenue, Chicago, IL, 60637, USA}
\author{K.~A.~Phadke\,\orcidlink{0000-0001-7946-557X}}
\affiliation{Department of Astronomy, University of Illinois Urbana-Champaign, 1002 West Green Street, Urbana, IL, 61801, USA}
\affiliation{Center for AstroPhysical Surveys, National Center for Supercomputing Applications, Urbana, IL, 61801, USA}
\affiliation{NSF-Simons AI Institute for the Sky (SkAI), 172 E. Chestnut St., Chicago, IL 60611, USA}
\author{A.~W.~Pollak}
\affiliation{Department of Astronomy and Astrophysics, University of Chicago, 5640 South Ellis Avenue, Chicago, IL, 60637, USA}
\author{K.~Prabhu}
\affiliation{Department of Physics \& Astronomy, University of California, One Shields Avenue, Davis, CA 95616, USA}
\author{S.~Raghunathan\,\orcidlink{0000-0003-1405-378X}}
\affiliation{Center for AstroPhysical Surveys, National Center for Supercomputing Applications, Urbana, IL, 61801, USA}
\author{M.~Rahimi}
\affiliation{School of Physics, University of Melbourne, Parkville, VIC 3010, Australia}
\author{A.~Rahlin\,\orcidlink{0000-0003-3953-1776}}
\affiliation{Department of Astronomy and Astrophysics, University of Chicago, 5640 South Ellis Avenue, Chicago, IL, 60637, USA}
\affiliation{Kavli Institute for Cosmological Physics, University of Chicago, 5640 South Ellis Avenue, Chicago, IL, 60637, USA}
\author{C.~L.~Reichardt\,\orcidlink{0000-0003-2226-9169}}
\affiliation{School of Physics, University of Melbourne, Parkville, VIC 3010, Australia}
\author{M.~Rouble}
\affiliation{Department of Physics and McGill Space Institute, McGill University, 3600 Rue University, Montreal, Quebec H3A 2T8, Canada}
\author{J.~E.~Ruhl}
\affiliation{Department of Physics, Case Western Reserve University, Cleveland, OH, 44106, USA}
\author{E.~Schiappucci}
\affiliation{School of Physics, University of Melbourne, Parkville, VIC 3010, Australia}
\author{A.~Simpson}
\affiliation{Department of Astronomy and Astrophysics, University of Chicago, 5640 South Ellis Avenue, Chicago, IL, 60637, USA}
\affiliation{Kavli Institute for Cosmological Physics, University of Chicago, 5640 South Ellis Avenue, Chicago, IL, 60637, USA}
\author{J.~A.~Sobrin\,\orcidlink{0000-0001-6155-5315}}
\affiliation{Fermi National Accelerator Laboratory, MS209, P.O. Box 500, Batavia, IL, 60510, USA}
\affiliation{Kavli Institute for Cosmological Physics, University of Chicago, 5640 South Ellis Avenue, Chicago, IL, 60637, USA}
\author{A.~A.~Stark}
\affiliation{Center for Astrophysics \textbar{} Harvard \& Smithsonian, 60 Garden Street, Cambridge, MA, 02138, USA}
\author{J.~Stephen}
\affiliation{Enrico Fermi Institute, University of Chicago, 5640 South Ellis Avenue, Chicago, IL, 60637, USA}
\author{C.~Tandoi}
\affiliation{Department of Astronomy, University of Illinois Urbana-Champaign, 1002 West Green Street, Urbana, IL, 61801, USA}
\author{B.~Thorne}
\affiliation{Department of Physics \& Astronomy, University of California, One Shields Avenue, Davis, CA 95616, USA}
\author{C.~Umilta\,\orcidlink{0000-0002-6805-6188}}
\affiliation{Department of Physics, University of Illinois Urbana-Champaign, 1110 West Green Street, Urbana, IL, 61801, USA}
\author{J.~D.~Vieira\,\orcidlink{0000-0001-7192-3871}}
\affiliation{Department of Astronomy, University of Illinois Urbana-Champaign, 1002 West Green Street, Urbana, IL, 61801, USA}
\affiliation{Department of Physics, University of Illinois Urbana-Champaign, 1110 West Green Street, Urbana, IL, 61801, USA}
\affiliation{Center for AstroPhysical Surveys, National Center for Supercomputing Applications, Urbana, IL, 61801, USA}
\author{A.~Vitrier\,\orcidlink{0009-0009-3168-092X}}
\affiliation{Sorbonne Universit\'e, CNRS, UMR 7095, Institut d'Astrophysique de Paris, 98 bis bd Arago, 75014 Paris, France}
\author{Y.~Wan}
\affiliation{Department of Astronomy, University of Illinois Urbana-Champaign, 1002 West Green Street, Urbana, IL, 61801, USA}
\affiliation{Center for AstroPhysical Surveys, National Center for Supercomputing Applications, Urbana, IL, 61801, USA}
\author{N.~Whitehorn\,\orcidlink{0000-0002-3157-0407}}
\affiliation{Department of Physics and Astronomy, Michigan State University, East Lansing, MI 48824, USA}
\author{W.~L.~K.~Wu\,\orcidlink{0000-0001-5411-6920}}
\affiliation{Kavli Institute for Particle Astrophysics and Cosmology, Stanford University, 452 Lomita Mall, Stanford, CA, 94305, USA}
\affiliation{SLAC National Accelerator Laboratory, 2575 Sand Hill Road, Menlo Park, CA, 94025, USA}
\author{M.~R.~Young}
\affiliation{Fermi National Accelerator Laboratory, MS209, P.O. Box 500, Batavia, IL, 60510, USA}
\affiliation{Kavli Institute for Cosmological Physics, University of Chicago, 5640 South Ellis Avenue, Chicago, IL, 60637, USA}
\author{J.~A.~Zebrowski}
\affiliation{Kavli Institute for Cosmological Physics, University of Chicago, 5640 South Ellis Avenue, Chicago, IL, 60637, USA}
\affiliation{Department of Astronomy and Astrophysics, University of Chicago, 5640 South Ellis Avenue, Chicago, IL, 60637, USA}
\affiliation{Fermi National Accelerator Laboratory, MS209, P.O. Box 500, Batavia, IL, 60510, USA}
\collaboration{SPT-3G Collaboration}
\noaffiliation
  
\begin{abstract}
	We present measurements of the temperature and E-mode polarization angular power spectra of the cosmic microwave background (CMB) from observations of 4\% of the sky with SPT-3G, the current camera on the South Pole Telescope (SPT). 
    The maps used in this analysis are the deepest used in a CMB TT/TE/EE analysis to date.
	The maps and resulting power spectra have been validated through blind and unblind tests. 
	The measurements of the lensed EE and TE spectra are the most precise to date at $\ell=1800$-$4000$ and $\ell=2200$-$4000$, respectively. 
	Combining our TT/TE/EE spectra with previously published SPT-3G CMB lensing results,
    we find parameters for the standard \lcdm\ model consistent with \planck\ and ACT DR6 with comparable constraining power.
    We report a Hubble constant of $H_0=66.66\pm0.60\,\kmsmpc$ from SPT-3G alone, $6.2\,\sigma$ away from local measurements from SH0ES. 
	For the first time, combined ground-based (SPT+ACT) CMB primary and lensing data have reached \planck's constraining power on some parameters, a milestone for CMB cosmology. 
	The combination of these three CMB experiments yields the tightest CMB constraints to date, with $H_0=\replace{67.24\pm0.35}{67.19\pm0.38}\,\kmsmpc$, and the amplitude of clustering $\sigma_8=0.8137\pm\replace{0.0038}{0.0037}$. 
	CMB data alone show no evidence for physics beyond \lcdm; however, we observe a $2.8\,\sigma$ difference in \lcdm\ between CMB and baryon acoustic oscillation (BAO) results from DESI-DR2, which is relaxed in extended models.
	The combination of CMB and BAO yields $2$-$3\, \sigma$ shifts from \lcdm\ in the curvature of the universe, the amplitude of CMB lensing, or the dark energy equation of state. 
	It also drives mild preferences for models that address the Hubble tension through modified recombination or variations in the electron mass in a non-flat universe. 
	This work highlights the growing power of ground-based CMB experiments and lays a foundation for further cosmological analyses with SPT-3G.
\end{abstract}


\maketitle
\tableofcontents

\crefformat{subsection}{\S#2#1#3}
\Crefformat{subsection}{\S#2#1#3}
\crefrangeformat{subsection}{\S#3#1#4--#5#2#6}
\Crefrangeformat{subsection}{\S#3#1#4--#5#2#6}
\crefmultiformat{subsection}{\S#2#1#3}{, \S#2#1#3}{, \S#2#1#3}{, \S#2#1#3}
\Crefmultiformat{subsection}{\S#2#1#3}{, \S#2#1#3}{, \S#2#1#3}{, \S#2#1#3}
\crefformat{subsubsection}{\S#2#1#3}
\Crefformat{subsubsection}{\S#2#1#3}
\crefrangeformat{subsubsection}{\S#3#1#4--#5#2#6}
\Crefrangeformat{subsubsection}{\S#3#1#4--#5#2#6}
\crefmultiformat{subsubsection}{\S#2#1#3}{, \S#2#1#3}{, \S#2#1#3}{, \S#2#1#3}
\Crefmultiformat{subsubsection}{\S#2#1#3}{, \S#2#1#3}{, \S#2#1#3}{, \S#2#1#3}


\section{Introduction}
\label{sec:intro}

One of the main driving forces behind the phenomenal progress of cosmology in the past thirty years has been observations of the cosmic microwave background (CMB).
CMB measurements from the \planck{} satellite have confirmed the $\Lambda$ Cold Dark Matter (\LCDM{}) model as the standard model of cosmology, constraining \LCDM{} parameters at the percent or sub-percent level~\citep{planck18-6}.  
At the same time, results from \planck{} show a number of inconsistencies when compared with other probes, which could hint at cracks developing in this heretofore very successful framework. 
The most notable of these is the Hubble tension, i.e., the discrepancy between the expansion rate of the universe inferred from early-universe probes, such as the CMB and baryon acoustic oscillations (BAO), and the value measured directly using distance-ladder measurements, such as the Cepheid-calibrated Type Ia supernova measurements from the \shoes{} collaboration \citep{breuval_small_2024}.
Moreover, recent BAO results from the second data release from the Dark Energy Spectroscopic Instrument (DESI DR2,~\citep{desi25,desi25b}) suggest an evolving dark energy equation of state, a suggestion that is reinforced by a discrepancy in \LCDM{} parameter values preferred by BAO, CMB, and uncalibrated SNe Ia observations.
Furthermore, classical questions in cosmology, such as the nature of dark matter and dark energy, remain unanswered.

While the \planck\ constraints have dominated the CMB field since the first \planck\ data release in 2013, improved measurements of CMB anisotropies, particularly at small scales and in polarization, promise to bring additional and complementary information~\citep{galli14}. 
CMB anisotropies in polarization are less contaminated by extragalactic foregrounds than in temperature, allowing the extraction of cosmological information from smaller angular scales. 
Furthermore, polarization measurements are less contaminated by atmospheric fluctuations, making them easier to achieve with ground-based experiments.
Recently, the Atacama Cosmology Telescope (ACT) collaboration published its newest data release (DR6), showing constraints on cosmology which are consistent with \planck\  with an almost equivalent constraining power on many cosmological parameters~\citep{naess_atacama_2025,louis25, calabrese25}.

In this paper, we present results from 2019 and 2020 observations with SPT-3G, the current camera mounted on the 10-meter South Pole Telescope (SPT).
The SPT~\citep{carlstrom11} is located at the Amundsen-Scott South Pole Station, one of the premier sites on Earth for CMB research, and was designed specifically for low-noise, high-resolution observations of the CMB. 
SPT observations began in 2007 with the SPT-SZ camera, which was sensitive to total intensity in three bands~\citep{story13}, and SPT capabilities were expanded to polarization with the SPTpol camera in 2012~\citep{henning12}.
SPT-3G is the third-generation SPT camera, installed in 2016-2017. It features $\sim\,16\,000$ polarization-sensitive detectors (over 10 times more detectors than SPTpol or SPT-SZ) configured to observe at 95, 150, and 220\ghz{}~\citep{sobrin18}.

The first cosmological results from SPT-3G~\citep{dutcher21,balkenhol21,balkenhol23,pan23} were derived from observations taken in 2018 of the $1500\,\sqdeg$ \mainfield{} field over four months (half a normal SPT ``winter'' season) and using half of the focal plane.
We refer to these results as SPT-3G 2018. 
Results were found to be consistent with the \lcdm\ model and in agreement with other CMB experiments, with error bars on cosmological parameters such as the Hubble constant a factor of three larger than those from \planck.

Starting from austral summer 2018-2019, the whole focal plane was operational, and SPT-3G has been used to observe the Main field every winter season since (except 2024, during which the SPT-3G Wide Survey was conducted) and with consistently high observing efficiency (typically $60 \%$ of all time spent scanning the CMB field). 
We refer to the observations of the \mainfield\ field taken during the austral winters in 2019 and 2020 as \sptnew{} D1 observations, and first cosmological results based on them were published in~\citep{ge24}. 
There, a Bayesian map-based approach was used to infer the unlensed EE power spectra and the CMB lensing power spectrum using polarization alone. 
We refer to these results as \muse{}. 
Despite being based on polarization alone, this work provided competitive constraints on cosmological parameters---for example, the Hubble constant was constrained with precision within a factor of 1.5 of \planck.
Recently,~\citep{zebrowski25} used the same observations to present a measurement of the B-mode polarization power spectrum on large angular scales ($\ell \in [32, 502]$). 
Enabled by new techniques for mitigating polarized atmospheric emission~\citep{coerver24}, the analysis of the resulting data places a 95\% confidence upper limit on the tensor-to-scalar ratio, $r<0.25$, which is the second-best ground-based B-mode constraint after the one from BICEP/Keck~\citep{bicepkeck21c}.
Finally,~\citep{raghunathan24} combined these same observations with SPTpol and {\it Herschel}-SPIRE~\citep{pilbratt10} data on $100\,\sqdeg$ of the sky. The analysis set constraints on the duration of the epoch of reionization using the non-Gaussianity of the kinematic Sunyaev-Zel’dovich effect.

In this paper, we present \TT/\TE/\EE{} (or \T\&\E{}) power spectrum measurements and cosmological results from the same observations of the \mainfield{} field.
We call our measurement of CMB power spectra \sptbp. 
The inclusion of \TT{} and \TE{} data significantly increases the constraining power of these observations. 
In particular, when combining with the SPT-3G measurements of the CMB lensing power spectrum from~\citep{ge24}, results from SPT-3G alone have comparable (within $\sim25\%$) constraining power to that of \planck\ on some cosmological parameters, such as $\Hubble$ and the amplitude of the matter density perturbations today, $\sigmaeight$.

The maps used in this analysis are the deepest high-resolution CMB data for a \TTTEEE{} analysis, with coadded white noise levels of $3.3\,\ukarcmin$\ in temperature and $5.1\,\ukarcmin$ in polarization. 
The production and validation of the maps are described in detail in an upcoming companion paper (W. Quan et al., in preparation, hereafter Q25).

Producing these CMB power spectrum measurements motivated a series of advances with respect to previous SPT-3G analyses to address the sensitivity of our data.
These include:
\begin{itemize}
\item A curved-sky pipeline that uses \healpix{}~\citep{gorski05} and fast spherical-harmonic tools~\citep{szapudi01,chon04}, an improvement from the flat-sky approximation used in previous SPT-3G \T\&\E{} analyses.
\item A new code to produce realistic and fast end-to-end simulations called \quickmock{}, which will be described in a future publication (E. Hivon et al., in preparation).
\item Modeling of systematic effects, including map-making filtering artifacts, quadrupolar temperature-to-polarization leakage, and polarized beams.
\item A new algorithm to inpaint point source masks with Gaussian constrained realizations, called \cork.
\item An accurate semi-analytical covariance matrix, based on the work of~\citet{camphuis_accurate_2022}.
\item A differentiable parameter estimation pipeline, using a JAX-based likelihood, called \candl{}~\citep{jax18, balkenhol_candl_2024}, and machine learning emulators, such as \OLE{}~\citep{gunther23} and \cosmopower~\citep{spuriomancini22, piras23}.
\item A CMB-only foreground-marginalized likelihood, \sptlite{}~\citep{balkenhol25}.
\end{itemize}
The analysis pipeline is kept blind to the final results until all cosmology-independent consistency tests pass a pre-determined threshold. 
We did discover effects that were not caught in the blind validation process, and we chose to implement non-trivial post-unblinding changes to address these.

Our analysis pipeline allows us to produce a robust and validated set of band powers from the SPT data. 
These are the most precise determinations of the lensed  \EE\ spectrum at $\ell=1800$-$4000$, and of the lensed \TE\ spectrum at $\ell=2200$-$4000$, with comparable uncertainties to \ACTDR\ at $\ell=1800$-$2200$ in \TE.  
The \ACTDR{} data set is more sensitive at large angular scales, owing to its broader sky coverage. 
In contrast, SPT-3G achieves better performance at small angular scales because of its lower noise levels.
We use the \sptbp binned power spectrum measurements, or band powers, described in this paper, together with the results of the measurement of CMB lensing from \muse{}, to set constraints on cosmology. 
The results presented in this paper, as well as the combination of data products associated with them, is referred to as \sptlr.

This paper represents a milestone for CMB cosmology in many respects, and we summarize our key findings in \cref{{sec:summary}}.
For the first time, a combination of ground-based experiments, \sptlr and \ACTDR{}, reaches \planck's precision on some cosmological parameters, such as the Hubble constant. 
This is the beginning of a new era for CMB cosmology, in which our knowledge of the universe will increasingly be driven by experiments from the ground.
Furthermore, while we find that the results from SPT-3G, \planck, and \ACTDR{}  are in excellent agreement and consistent with \lcdm, we report borderline statistically significant differences with the BAO results from DESI DR2. 
The results from the combination of SPT-3G, \planck{}, and \ACTDR{} are discrepant with those of DESI DR2 at the $2.8\,\sigma$ level in the \lcdm\ model. 
This is alleviated in extended models of cosmology, and in some cases we find deviations from \lcdm\ at the level of $2$-$3\,\sigma$ when CMB and BAO data are combined.

The rest of this work is structured as follows.
In \cref{sec:maps} we provide a summary of the processing of raw telescope data into CMB maps.
In \cref{sec:power_spec} we present our power spectrum pipeline and the measured band powers.
In \cref{sec:likelihood} we present our model for the data and detail our cosmological inference procedure (the likelihood).
In \cref{sec:validation} we present the validation of our data products through an extensive suite of consistency and robustness tests. We highlight that our model accurately describes the data to a high degree of precision.
In \cref{sec:pars} we present the cosmological analysis before closing with concluding remarks in \cref{sec:conclusion}.
The data and likelihood code used in this paper are publicly available.\footnote{\url{https://pole.uchicago.edu/public/data/camphuis25/}}


\section{Summary of results}
\label{sec:summary}

\begin{figure*}[ht]
	\includegraphics[width=\textwidth]{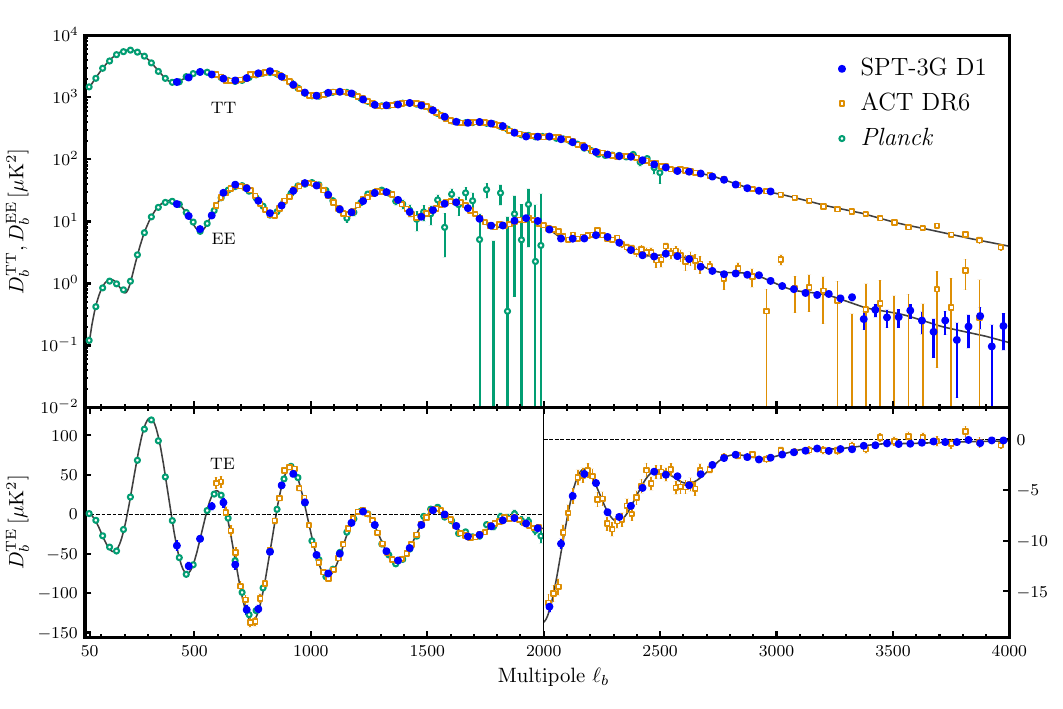}
	\caption{\TT, \TE, and \EE\ band powers from SPT-3G D1 (blue dots), \ACTDR{} (orange empty squares) and \planck{} PR3 (green empty dots).
	Band powers from each experiment are foreground- and nuisance-parameter cleaned combinations of all auto- and cross-frequency spectra.
    We also show the best-fit \lcdm{} model to \sptbp (solid line). \emph{Top}: \TT{} and \EE{} band powers on a logarithmic scale. 
	SPT \TT{} band powers are estimated in the multipole range $\ell=400$ to $3000$, 
	while the range for \TE{} and \EE{} band powers is $\ell=400$ to $4000$, see \cref{sec:bandpowers} for details.
	\emph{Bottom}: \TE{} band powers in linear scale, with a zoomed-in view of the $\ell>2000$ region where ground-based experiments dominate the measurement.
	These data sets demonstrate excellent agreement with each other, and the \sptbp data provide the tightest measurement of the lensed \EE{} and \TE{} band powers at $\ell=1800$-$4000$ and $\ell=2200$-$4000$, respectively.}
	\label{fig:experiments}
\end{figure*}
\begin{table*}[htbp]
    \begin{tabular}{l | l l l l l l} \hline
        \textbf{Parameter}               & \textbf{\textit{Planck}} \quad & \textbf{\sptlr} \quad & \textbf{ \ACTDR{}} \quad & \textbf{\ground{}} \quad & \textbf{SPT+\textit{Planck}} \quad & \textbf{\cmball{}} \quad \\ \hline
        \hline
        \emph{Sampled}                                                                                                                                                                                                     \\
        ${10^{4}\thetastar}$             & $ 104.184\pm 0.029$            & $ 104.171\pm 0.060$       & $ 104.157\pm 0.030$      & $ 104.158\pm 0.025$       & $ 104.176\pm 0.026$                & $\replace{104.162\pm 0.023}{104.161\pm 0.023}$    \\
        ${100\,\ombh}$                   & $ 2.238\pm 0.014$              & $ 2.221\pm 0.020$         & $ 2.257\pm 0.016$        & $ 2.247\pm 0.013$         & $ 2.230\pm 0.011$                  & $\replace{2.2381\pm 0.0093}{2.2398\pm 0.0095}$    \\
        ${100\,\omch}$                   & $ 11.98\pm 0.11$               & $ 12.14\pm 0.16$          & $ 12.26\pm 0.17$         & $ 12.22\pm 0.12$          & $ 12.050\pm 0.089$                 & $\replace{12.009\pm 0.086}{12.028\pm 0.094}$     \\
        ${\ns}$                          & $ 0.9657\pm 0.0040$            & $ 0.951\pm 0.011$         & $ 0.9682\pm 0.0069$      & $ 0.9671\pm 0.0058$       & $ 0.9636\pm 0.0035$                & $\replace{0.9684\pm 0.0030}{0.9679\pm 0.0033}$    \\
        ${\logA}$                        & $ 3.042\pm 0.011$              & $ 3.054\pm 0.015$         & $ 3.038\pm 0.012$        & $ 3.042\pm 0.011$         & $ 3.046\pm 0.010$                  & $\replace{3.0479\pm 0.0099}{3.047\pm 0.010}$    \\
        ${\taureio}$                     & $ 0.0535\pm 0.0056$            & $ 0.0506\pm 0.0059$       & $ 0.0513\pm 0.0060$      & $ 0.0514\pm 0.0059$       & $ 0.0538\pm 0.0054$                & $\replace{0.0559\pm 0.0055}{0.0549\pm 0.0055}$    \\
        \hline
        \emph{Derived}                                                                                                                                                                                                     \\
        ${\Hubble\,{\text \Hubbleunit}}$ & $ 67.41\pm 0.49$               & $ 66.66\pm 0.60$          & $ 66.51\pm 0.64$         & $ 66.59\pm 0.46$          & $ 67.07\pm 0.38$                   & $\replace{67.24\pm 0.35}{67.19\pm 0.38}$       \\
        ${\Age\,{\text{\Ageunit}}}$      & $ 13.797\pm 0.022$             & $ 13.826\pm 0.027$        & $ 13.797\pm 0.021$       & $ 13.805\pm 0.016$        & $ 13.812\pm 0.017$                 & $\replace{13.805\pm 0.014}{13.805\pm 0.014}$     \\
        ${\clamp}$                       & $ 1.883\pm 0.010$              & $ 1.915\pm 0.021$         & $ 1.884\pm 0.013$        & $ 1.889\pm 0.011$         & $ 1.8890\pm 0.0092$                & $\replace{1.8843\pm 0.0060}{1.8871\pm 0.0079}$    \\
        ${\oml}$                         & $ 0.6854\pm 0.0067$            & $ 0.6753\pm 0.0091$       & $ 0.670\pm 0.010$        & $ 0.6722\pm 0.0072$       & $ 0.6810\pm 0.0054$                & $\replace{0.6833\pm 0.0051}{0.6824\pm 0.0055}$    \\
        ${\omm}$                         & $ 0.3145\pm 0.0067$            & $ 0.3246\pm 0.0091$       & $ 0.330\pm 0.010$        & $ 0.3277\pm 0.0072$       & $ 0.3189\pm 0.0054$                & $\replace{0.3166\pm 0.0051}{0.3175\pm 0.0055}$    \\
        ${\rdrag\,{\text \rdragunit}}$   & $ 147.13\pm 0.25$              & $ 146.92\pm 0.47$         & $ 146.20\pm 0.46$        & $ 146.43\pm 0.34$         & $ 147.06\pm 0.23$                  & $\replace{147.07\pm 0.22}{147.00\pm 0.24}$      \\
        ${\sigmaeight}$                  & $ 0.8099\pm 0.0051$            & $ 0.8158\pm 0.0058$       & $ 0.8171\pm 0.0055$      & $ 0.8169\pm 0.0042$       & $ 0.8132\pm 0.0042$                & $\replace{0.8137\pm0.0038}{0.8137\pm 0.0037}$     \\
        \hline
    \end{tabular}
    \caption{\lcdm{} parameter constraints from different CMB experiments. We report mean values and 68\% confidence intervals. Data sets are described in \cref{tab:dataset}. 
        All data sets include \TTTEEE{} measurements, lensing reconstruction and a prior on  $\taureio$ from \planck\ PR4~\citep{planck20-57}.}
    \label{tab:lcdm}
\end{table*}

\begin{figure*}
	\includegraphics[width=\textwidth]{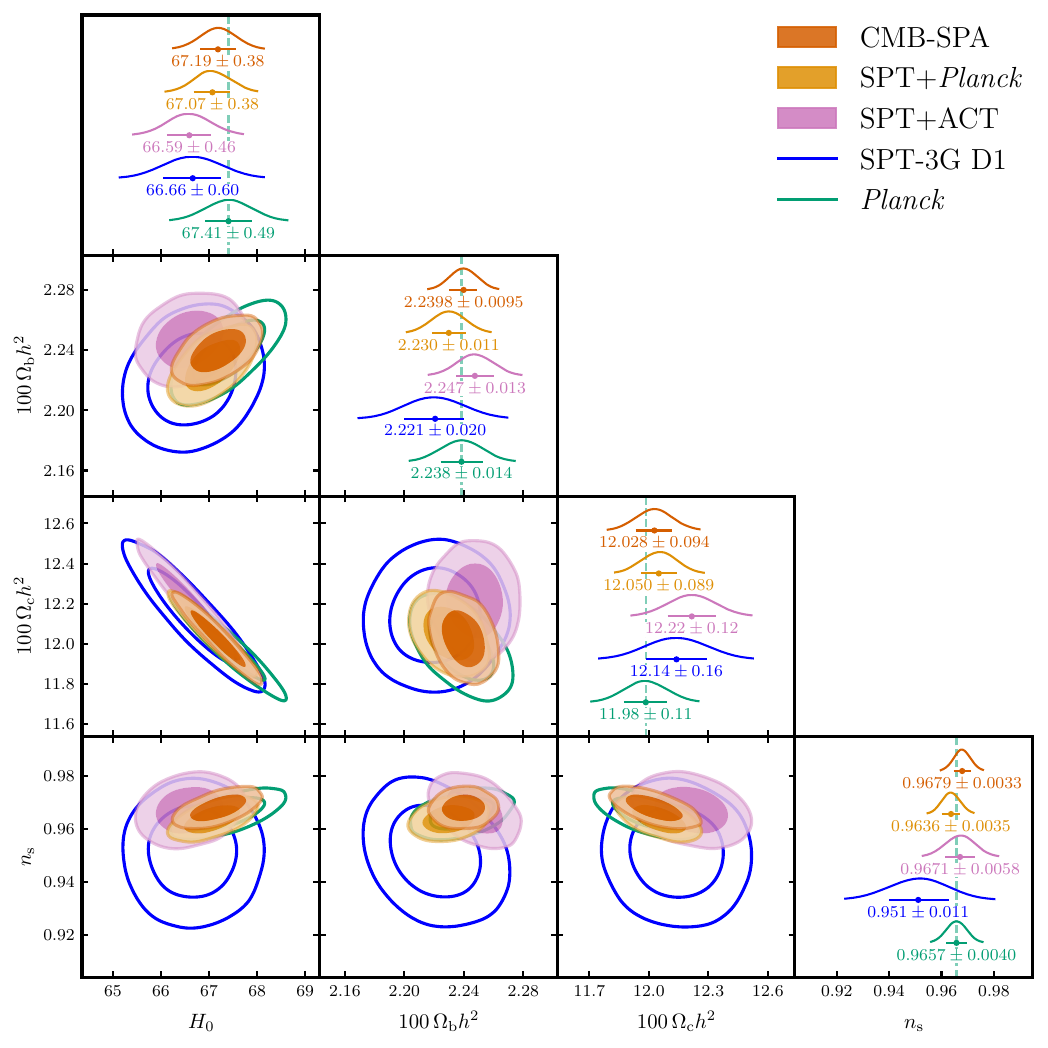}
	\caption{Summary of CMB (primary and lensing) \lcdm{} cosmological constraints. 
	The data sets used are described in \cref{tab:dataset}, with \ground{} being the combination of (ground-based) \SPT{} and \ACTDR{}, data, and \cmball{} being the combination of \ground{} with \planck{}. 
	The diagonal panels display the 1D posterior distributions of the parameters, with corresponding error bars. 
	The off-diagonal panels show the 2D 68\% and 95\% confidence intervals. 
	}
	\label{fig:lcdm_main}
\end{figure*}

In this section, we summarize the key results of the paper. Specifically, we present the band powers estimated from SPT-3G measurements of the temperature and polarization anisotropies of the CMB and the cosmological parameter constraints that these band power measurements enable. 
The results are based on observations of the \mainfield{} field, a region that covers roughly 4\% of the sky, taken during the austral winter seasons of 2019 and 2020. The maps used in this analysis are the deepest CMB data used for a \TTTEEE\ analysis to date. 
The minimum-variance band powers obtained from these maps (by combining band powers from all combinations of SPT-3G frequency maps) are shown in \cref{fig:experiments}. 
These data provide the tightest measurements of the lensed \EE{} and \TE{} power spectra at angular multipoles $\ell=1800$-$4000$ and $\ell=2200$-$4000$, respectively, while being comparable to \ACTDR\ at $\ell=1800$-$2200$ in \TE. 
We use these band powers, in combination with the lensing power spectrum from~\citep{ge24}, to set constraints on cosmology. Here we highlight our key findings.

\begin{enumerate}

	\item The SPT-3G data align well with the \lcdm{} model when considered independently of other datasets, providing strong confirmation on angular scales not accessible to \planck{}~\citep{planck18-1}. 
	The \TT{}, \TE{}, and \EE{} channels independently are well fit by \lcdm{} and yield consistent results. Our strongest cosmological constraints come from the \TE\ channel. 
	The \lcdm{} parameter results are reported in \cref{tab:lcdm}. 
	They are consistent with the results of \planck\footnote{We use the combination of \planck{} \TTTEEE\ from Public Release 3 (PR3)~\citep{planck18-5} with the \planck\ $\phi\phi$ CMB lensing~\citep{carron22} estimated from PR4 maps~\citep{planck20-57}. 
	To constrain reionization, instead of using the PR3 large scale \EE\ polarization, we use a prior on $\taureio$ from the analysis of PR4 maps~\citep{planck20-57}.} and \ACTDR{}\footnote{We use the combination of \ACTDR\ \TTTEEE\ band powers~\citep{louis25} and CMB lensing $\phi\phi$~\citep{qu24} with a prior on $\taureio$ from the analysis of PR4 maps~\citep{planck20-57}.}, as shown in \cref{fig:lcdm_main}.
	
	\item The  constraining power of SPT-3G is comparable to that of \planck{} on some \lcdm{} cosmological parameters. 
	For the Hubble constant, we find $H_0=66.66 \pm 0.60\,\kmsmpc$, in excellent agreement with \planck\ and \ACTDR\ and $6.2\,\sigma$ away from the local measurements of \shoes{}~\citep{breuval_small_2024}. The SPT-3G data alone is thus able to confirm the Hubble tension with a high level of statistical significance.

	\item For the first time, a combination of CMB ground-based experiments, \sptlr and \ACTDR, which we refer to as \ground{}, reaches \planck's constraining power on some \lcdm{} and extended model parameters. 
	This is a milestone for CMB cosmology. With \ground, we obtain $\Hubble{} = 66.59\pm 0.46\,\kmsmpc$. 
	We report the strongest CMB constraints to date by combining SPT, ACT, and \planck{} together into \cmball, resulting in a constraint of $\Hubble{} = \replace{67.24 \pm 0.35}{67.19 \pm 0.38}\,\kmsmpc$, a $6.4\,\sigma$ discrepancy from \shoes{}. 
	We also highlight that \cmball\ constrains the scalar spectral index to $\ns=\replace{0.9684\pm 0.0030}{0.9679\pm0.0033}$, a $\replace{10.5}{9.9}\,\sigma$ difference from a scale-invariant primordial power spectrum with $\ns = 1$.
		  
	\item The amplitude of matter density perturbations today, $\sigmaeight$, and of the matter density, $\omm$, from  SPT-3G are in excellent agreement with the findings of other CMB experiments. 
	Many large scale structure probes now provide results consistent with those of the CMB on these parameters, including the latest cosmic shear analysis of the Kilo-Degree Survey (KiDS)~\citep{wright25}, the 3$\times$2 point analysis of the Dark Energy Survey (DES)~\citep{abbott22a}, the CMB lensing analysis of~\citep{qu25}, and the galaxy cluster statistics analysis of~\citep{bocquet24}.

	\item We investigate the amplitude of CMB lensing implied from its effect on the primary CMB power spectra and find a value consistent with the \lcdm{} prediction, $\alens = 1.016^{+0.048}_{-0.054}$ from \ground{} \TTTEEE\ data, a result that differs at $\sim 2\,\sigma$ from the mild anomaly in the \planck{} data~\citep{planck18-6}.

	\item We report a borderline statistically significant differences between CMB data and BAO data from DESI DR2 in \lcdm{}, at the level of 2.8$\,\sigma$ in the \omm{}-\hrd{} plane\footnote{\rd{} is the sound horizon at the drag epoch and $h\equiv\Hubble/100\,\kmsmpc$.} when \SPT{}, \ACTDR{}, and \planck{} are combined. 

	\item While the CMB data alone do not prefer any extended model over \lcdm, the discrepancy between the CMB experiments and DESI is alleviated in some extended models of cosmology. The combination of CMB and BAO yields $2$-$3\,\sigma$ deviations from the standard model of cosmology.
\end{enumerate}


\section{Sky maps}
\label{sec:maps}

There are several steps between observations of the microwave sky and cosmological analysis of the CMB. 
In this section, we discuss the steps to produce CMB maps from the raw observations.
The observations themselves and the processing of the data are described in great detail in~Q25; in this work we highlight characteristics of the data that we need to take into account in our power spectrum modeling. 

\subsection{Data and map-making approach}
\label{sec:map-making}
The maps used in this work were constructed using SPT-3G data from the austral winter observing seasons of 2019 and 2020. 
The $1500\,\degsq$ SPT-3G Main field footprint spans declinations from $-42$ to $-70$ degrees and right ascensions from 20h40m0s to 3h20m0s, see \cref{fig:footprints}.
In the same figure, we also display the footprints of the \summerfield{} and \extfield{} fields. 
The combination of these \sptnew{} fields will constitute the total \allspt{} field, 
which will probe 25\% of the sky with low noise and high resolution (see~\citep{prabhu24}). 
Additionally, \cref{fig:footprints} includes survey masks from other CMB experiments, such as ACT and \planck{}, which are used in the cosmological analysis of this work in combination with the \mainfield{} field data.

The SPT-3G Main field that this work is based on is split into four subfields in declination (equivalent to elevation at the South Pole) to avoid detector nonlinearity from large changes in airmass without retuning the detectors.
The data are acquired in the form of timestreams for each detector following an observing strategy that raster scans each subfield many times.
We call one two-hour raster scan of each subfield an {observation}, and there were roughly 3000 total observations of the four subfields over the two years.

Following previous SPT analyses (see for example~\citep{dutcher21}), we use the filter-and-bin approach of map-making~\citep{hivon02}. 
The timestreams are first low-pass filtered 
(to remove high frequency noise and reduce aliasing) and high-pass filtered 
(to remove low frequency sources of noise).
The low-pass filtering is performed in Fourier space, while the high-pass filtering is achieved through fitting the individual detector timestreams to a set of low-order polynomials and low-frequency sinusoids.
The filtered timestreams are combined and projected into sky maps of the temperature (\T{}) and linear polarization Stokes parameters (\Q{} and \U{}) in each frequency band for each individual observation of a \mainfield{} subfield.
Alongside the sky maps, we also build a weight map, which is a measure of the inverse variance of the observation map in each pixel. 
Some details of the data set and timestream processing are discussed in Section II of \citep{ge24}, and additional details are discussed in Q25. 
We note that \citep{ge24} and this work are based on the same data and timestream processing but use maps with different pixelization schemes (see below).

In the following paragraphs, we highlight two important differences between the maps used here and those used in previous SPT analyses.

In previous SPT \T\&\E{} power spectrum analyses, we pixelized the maps on a two-dimensional grid using a projection of the sphere, and we used two-dimensional Fourier transforms to substitute for spherical harmonic transforms (SHTs) under the flat-sky approximation. 
In particular, \citep{ge24} used a map in the 
Lambert azimuthal equal-area projection (ZEA) with $0.56'$ pixels.
In this new analysis (and the BB analysis in~\citep{zebrowski25}), we pixelize the maps using \healpix{}\footnote{\url{https://healpix.sourceforge.io/}}~\citep{gorski05}. 
We choose an \Nside{} parameter of 8192, which is equivalent to $0.4'$ resolution. 
With the size of the \mainfield field and the sensitivity levels achieved in this work, the non-idealities caused by the flat-sky approximation are no longer negligible or trivial to account for. 
In particular, we found that projection effects were causing excess correlation between power spectrum bins at high $\ell$, in a manner that is difficult to compute analytically. 
For this reason, the current analysis is performed with full SHTs, and we take advantage of the many tools that have been developed for fast estimation of SHTs and resulting power spectra from \healpix{} maps.

The filtering scheme is broadly similar to previous work. 
In particular, as a result of the high-pass filtering, information at multipoles below 300 along the scan direction is absent from our maps, which results in a power spectrum bias that we evaluate and discuss in \cref{sec:tf}. 
An important difference involves a newly identified systematic effect from timestream filtering, one that only appears at the level of sensitivity of the data in this work.
As in most previous SPT analyses, we do not include timestream samples near bright emissive sources or strong Sunyaev-Zel'dovich-effect decrements from galaxy clusters when estimating the amplitudes of the polynomials and sinusoids to remove.
This avoids creating extended features along the scan direction near strong sources, which are often referred to as ``filtering wings.''\footnote{An illustration of those can be found in Fig. 3 of \citep{archipley_2025}. \RR{Discussion of the effect of filtering wings on the power spectrum can be found in \citep{rvhm-fsdc}.}}
We refer to this procedure of avoiding strong sources in the timestream filtering as ``map-making masking'' in the rest of this work.
We mask all sources above $6\,$mJy at 150\ghz{} and any galaxy cluster detected at more than $10\,\sigma$ in a separate set of maps optimized for cluster detection, for a total of 2655 objects.
While eliminating filtering wings is advantageous, in this work we detect small and previously undiagnosed side effects of this masking process. We refer to these as ``filtering artifacts' which we describe in detail---along with our method for mitigating and accounting for them in our power spectrum estimation pipeline---in \cref{sec:filtering-artifacts}.

\subsection{Coadds}
\label{sec:coadd}
We produce different weighted averages of single observation maps, in order to produce the different data products that are the baseline of the power spectrum pipeline. 
These are: the full depth coadd, the bundles, and the noise realizations or sign-flips. For each of these, we use the {observation} weight maps to weight the observation maps in the coadd.

We call the maps produced by the inverse-variance weighting of all of the observation maps at each frequency the full-depth coadds. 
Those maps are shown in~Q25. 
We use them in the construction of the noise maps described below, as well as for inpainting emissive sources and galaxy clusters (\cref{sec:inpainting}).

The main product that is used in the power spectrum estimation are the {bundles}. 
We randomly group all of our observation maps at each frequency into 30 bundles, ensuring that the maps formed by the inverse-variance-weighted coaddition of all the maps of each bundle have a similar signal-to-noise. 
All bundles have similar noise levels but no noise correlation between them. 
We use the cross-spectra of these bundle maps to form the noise-unbiased spectra that are used in the rest of this analysis. 
This strategy is similar to that used in other CMB power spectrum analyses, which use various data split strategies to produce noise-unbiased power spectra (\cite{planck18-5,dutcher21,louis25}). 
In contrast to most other CMB experiments, the highly redundant SPT observing strategy allows us, as in previous SPT analyses, to use a higher number of splits in order to allow for a more precise estimation of the noise covariance (see~\citep{lueker10,dutcher21}).
While the depth of the 2019-2020 Main field data would allow us to produce more than 30 bundles to improve the  noise covariance estimate, we find that this number is a good compromise between accuracy and efficiency as the computing cost of increasing the number of bundles grows quadratically.

Finally, we call {noise realizations}, or {sign-flips}, the noise estimates formed by the difference between random selections of the observations:
\begin{equation}
	N^i \equiv w_C^{-1} \sum_o \epsilon^i_o w_o (M_o - C),
\end{equation}
where $\epsilon^i_o \in \{-1, 1\}$, $M_o$ and $w_o$ are the map and weight map for observation $o$, $C$ is the {full-depth coadd}, $w_C$ is the weight map of the {full-depth coadd}, and $N^i$ is the $i$th noise realization. 
We form 500 of these noise realizations which are used in our simulations described in \cref{sec:simulations}. 
More details on the noise realization procedure are given in~Q25.

\subsection{Masks and point sources}
\label{sec:mask}

\begin{figure*}
	\includegraphics[width=\textwidth]{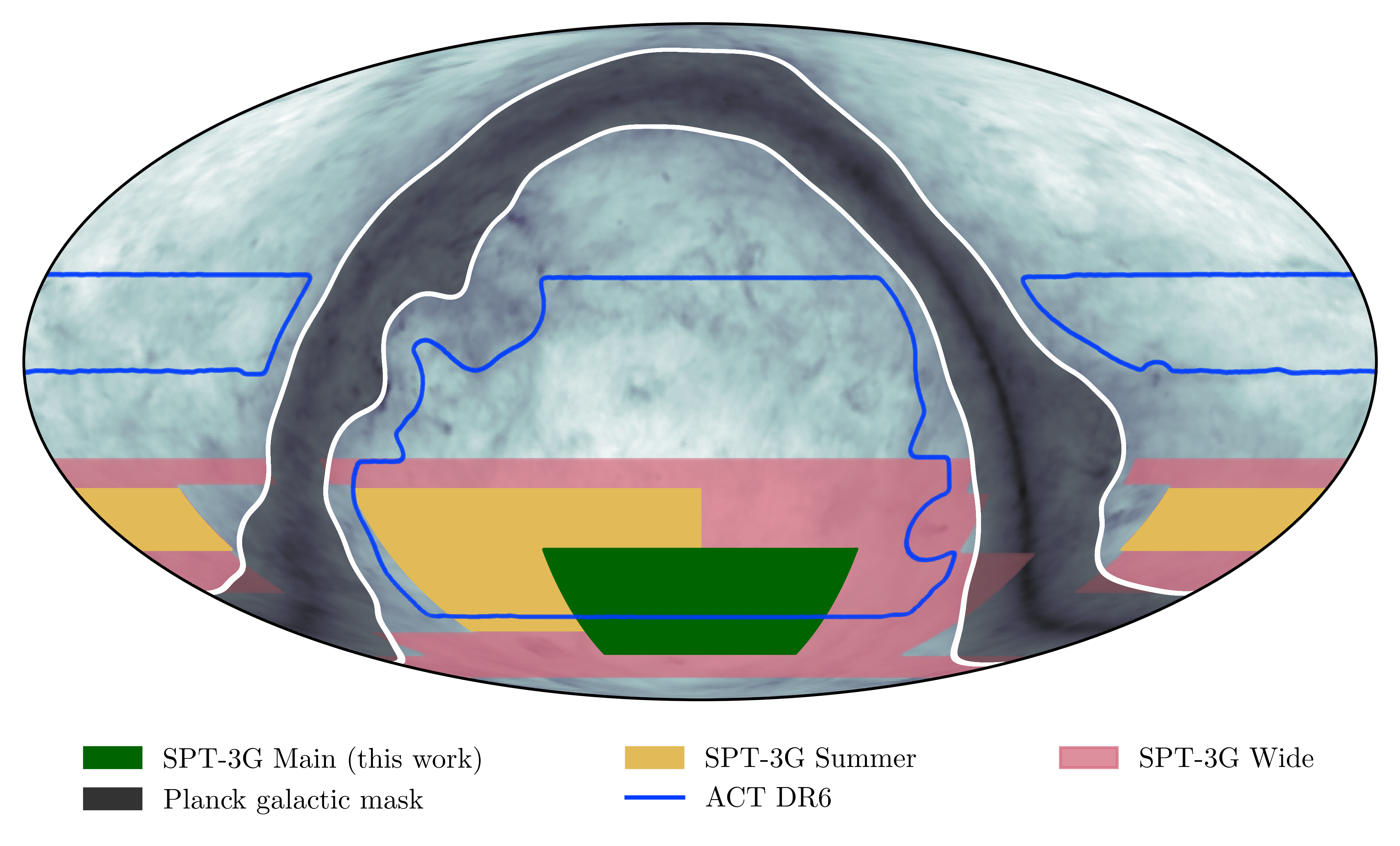}
	\caption{
	Footprints of the \sptnew surveys---Main (green filled area), Summer (orange filled area), and Wide (red filled area)---shown alongside the \ACTDR{} survey mask (blue solid line).
	We show Galactic dust as measured by \planck{} in the background~\citep{planck15-10} and indicate the most contaminated region with the \planck{} GAL080 mask\footnote{\url{https://pla.esac.esa.int/}} shown in gray shading with a white border.
	This work is based on observations of the \mainfield field.
	The combination of the \mainfield{}, \summerfield{}, and \extfield{} fields will constitute the \allspt{} field, which will probe 25\% of the sky with low noise and high resolution.
	}
	\label{fig:footprints}
\end{figure*}

For power spectrum analysis of the \mainfield{} field, we use two different masks: a {sky mask} and a {point source mask}.
The \mainfield{} field binary sky mask is obtained by applying a threshold to the weight map, excluding regions where the weights are below 10\% of the median across the map.
The resulting footprint is shown in \cref{fig:footprints}. 

For power spectrum estimation, we further apodize the \mainfield field mask using a $0.4^\circ$ Gaussian taper.
To produce the binary point source mask, we mask the 2655 emissive sources and clusters which have already been masked during map-making, using a radius where the signal-to-noise of beam-convolved sources falls to 1 (the same radius used in masking these objects during map-making), ranging from 2 to 15 arcmin.
The point source mask removes 0.06\% of the sky.
We apodize this point source mask using a Gaussian taper with $\sigma=0.03^\circ$. 
An important distinction from previous SPT analyses (and CMB analyses more generally) is that, 
when estimating signal power spectra from the maps, we do not use this point source mask; 
instead we replace the pixels where the sources are localized with a Gaussian constrained realization of the CMB informed by the rest of the map, in a process generally known as inpainting. 
The reason for this choice is the effect of the large number of point-source holes on the power spectrum covariance (\cref{sec:inpainting}).

The total apodized mask is the product of the apodized border mask and the apodized point source mask. 
This mask is used for computing the noise band powers, as our analysis and map inpainting rely on accurate knowledge of the noise within this region. 
While the sum of the weights in the apodized border mask corresponds to 4.0\% of the sky, the sum of the weights in the total apodized mask is reduced to 3.8\% as a result of point source masking and apodization.

\subsection{Calibration and cleaning}
\label{sec:calibration}
As discussed in~Q25, the signal and noise maps at each frequency need to be recalibrated and cleaned to account for inaccuracies in the gain, polarization efficiency, and polarization angle of each detector. 
This process consists of four different operations, applied in the order listed below.
\begin{enumerate}
	\item \emph{Gain calibration}: We multiply all the T, Q, and U coadds by a common scaling factor to correct for the overall miscalibration of detector gains.
	\item \emph{Leakage from differential gain calibration}: We subtract small copies of a T coadd from the corresponding Q and U coadds to correct for the monopole temperature-to-polarization leakage due to differential gain miscalibration between detectors with different polarization angles.
	\item \emph{Polarization angle calibration}: 
	We apply small rotations to the \Q{} and \U{} coadds to correct for the overall miscalibration of the detector polarization angles.
	The calibration angle is determined by nulling the correlation between E and B modes in each frequency channel.
	\item \emph{Polarization efficiency calibration}: We multiply the Q and U coadds by a common scaling factor to correct for the overall miscalibration of the detector polarization efficiencies.
\end{enumerate}

Note that in the case of gain and polarization efficiency calibration, the correction is done in two steps.
First, we perform an external recalibration of the SPT-3G $150\ghz$ maps by cross-correlating them with the \planck{} PR3 $143\ghz$ maps within the \mainfield{} field mask, using the multipole range from 800 to 1200 for gain recalibration and from 400 to 1300 for polarization efficiency.
To account for the filtering of the SPT map, the \planck{} map is mock observed through our data processing pipeline, in the same manner as the sky simulations used for transfer function modeling (see \cref{sec:simulations}). 
We note that this comparison uses only a small portion ($\sim$4\%) of the total \planck{} PR3 $143\ghz$ map and thus does not compromise our blindness to the \planck{} data in the cosmological analysis.
After the external recalibration of the SPT-3G $150\ghz$ map, we perform an internal recalibration of the SPT-3G 95 and $220\ghz$ maps by comparing them to the recalibrated $150\ghz$ map. 
The four steps summarized above are what \citep{ge24} refers to as the alternative systematics estimates and discusses in some detail in its Appendix B. 
Additional information on these steps is presented in Q25.

We propagate the gain calibration and polarization efficiency calibration uncertainties in the likelihood analysis as described in \cref{sec:data_model}. Note that in most cases (except when using the polarization data alone), we ignore the external polarization efficiency priors and let this calibration be determined from the data within a given cosmological model. 
Similarly, we only use the internal gain calibration priors and internal polarization efficiency priors in specific validation tests and only explore the residual uncertainty of the internal recalibration under flat priors.

The polarization angle calibration and differential gain calibration do not correct for all of the sources of polarization leakage. We discuss in \cref{sec:quadrupolar} how we model and correct for quadrupolar leakage induced by beam ellipticity correlated with detector polarization angle.

\subsection{Map-level null tests}
\label{sec:null}
In addition to mitigating biases from known systematic effects, we search for potential unmodeled systematic errors in our maps through null tests, \RR{described in Q25}. In a null test, we split the full set of individual-observation maps into two halves in a way that maximizes sensitivity 
to the suspected systematic error. We then subtract one half of the data from the other and check whether the difference is consistent with expectation spectra calculated from simulations to within noise fluctuations \RR{using a $\chi^2$ test}.
The tests performed for this analysis include splits of the data based on the following: scan direction, detector wafer, date of observation, moon position, sun position, and observation azimuth. 
\RR{In total, we perform 54 independent null tests, yielding a threshold of $0.05/54=0.00093$ for a single test to fail.
This is due to the different spectra and frequency combinations among the six null splits.}

The initial scan direction and wafer null tests failed our predetermined threshold for consistency with noise, and we traced both failures to excess power localized in spherical harmonic space.
We choose to remove this excess power by applying a harmonic-space mask, defined as
\begin{equation}
	\label{eq:notch}
	W^{\rm notch}_{\ell m} = \begin{cases}
		0 & \text{if } \ell \in [500, 680] \text{ and } m \in [350, 425], \\
		1 & \text{otherwise}.
	\end{cases}
\end{equation}
This masking is applied to the \alm{} coefficients of the data maps, as described in \cref{sec:filtering-artifacts}, and we account for it in the transfer function and covariance matrix estimations. 

After applying this correction, all the \TE{} and \EE{} null spectra are consistent with expectations (see Section IV.B.1 of \citep{ge24} for those spectra and more information on the null tests). 
\RR{While most of the \TT{} null spectra are also consistent with expectations, which can be non-zero due to instrumental effects such as time constants, we find that s}everal \TT{} null spectra formally fail our pre-determined threshold. 
\RR{This is the case for azimuth test at 95\ghz{} and 150\ghz{}, and for the year test at 95\ghz{}.}
\RR{However, the detected contaminations have amplitudes that are a smaller than $1\%$ of the sample variance error bars on the signal band powers, such that if we add $10^{-4}$ times the sample variance to the null spectrum covariance, all tests pass (see Q25).} 
\replace{We consider this acceptable, as the potential systematic errors are too small to affect the cosmological analysis.}{As those tests are designed to maximize the sensitivity to systematic errors, we interpret these failures as indications of small unmodeled systematic errors in the \TT{} data. We verified that these are too small to affect the cosmological analysis, and consider this acceptable.}

\subsection{Simulations}
\label{sec:simulations}

In this analysis, we rely on simulations to propagate the effect of the data processing pipeline to our power-spectrum estimation and to validate our pipeline. When possible, we use analytical models to compute data products, which we validate with simulations.

All simulations are generated based on the \planck{}-\lcdm{} cosmology, with added Gaussian realizations of extragalactic foreground components based on \agora{} simulations~\citep{omori22} which have been calibrated to match the mid to high-$\ell$ portion of the measured spectra from data.
Note that \agora{} does not contain the reionization kSZ signal.
We do not include Galactic foregrounds.
We discuss in \cref{sec:filtering} how we process these simulated skies to reproduce the effect of the data processing.

We also generate some simulations with an alternative underlying cosmological model to test our pipeline, which we describe in more detail in \cref{sec:altcosmo}.

In most cases, we only simulate the signal part of the data. When needed (for inpainting in \cref{sec:inpainting} or for the validation of our pipeline in \cref{sec:pipesims}), we add a simulation of the noise contribution using the noise realizations (see \cref{sec:coadd}).

As an improvement over previous SPT analyses, we have developed a tool for fast modeling of the effects of the data processing on the power spectrum signal and covariance. This results in two types of simulations used in this analysis:
\begin{enumerate}
	\item \fullmock{}: a set of 500 full-sky simulations that we use to calibrate the transfer function and validate the pipeline. This simulation set is similar to the one used in previous SPT analyses and described in Section~IV.D.1 of~\citep{dutcher21} and in~Q25. In this simulation set, the input skies are projected into individual-detector timestreams using the exact pointing and weights stored in the observation data files. These timestreams are then processed into maps with the same map-making pipeline used for the data, including the map-making masking procedure described in \cref{sec:map-making} even though there are no point sources in the simulations.
	\item \quickmock{}: a set of \replace{2000}{4000} fast simulations that we use to compute the covariance matrix and the transfer function. The \quickmock{} simulations use a lighter and faster simulation pipeline that implements several approximations to the scanning strategy to reduce computation while preserving accuracy (E. Hivon et al., in preparation). We typically achieve percent-level accuracy or better with \quickmock{}. 
\end{enumerate}
Furthermore, inspired by the \texttt{CarPool} approach of~\citep{chartier_carpool_2021}, we use common CMB input skies for both \fullmock{} and \quickmock{} and can correct for any discrepancy between the two methods to obtain higher accuracy while limiting the impact of the Monte Carlo (MC) noise. We discuss in \cref{app:transferfunction} a particular application of this method for the calibration of the power spectrum transfer function model and the propagation of the residual MC variance.


\section{Power Spectrum}
\label{sec:power_spec}

The data vector that we use in our likelihood analysis is formed by combining all of the available cross-bundle power spectra for each pair of frequency bands.
To go from the maps described in \cref{sec:maps} to the auto- and cross-frequency band powers requires a series of steps, which we summarize here and then describe in detail in the following sections.

First, the bundle maps are inpainted at the location of bright point sources and massive galaxy clusters with a process described in detail in \cref{sec:inpainting}. 
The bundle maps are then multiplied by the apodized sky mask described in \cref{sec:mask}, and bundle cross-power spectra are estimated. 
For this operation, as well as for all power spectrum estimations throughout this work, we use \polspice{}\footnote{\url{http://www2.iap.fr/users/hivon/software/PolSpice/}}~\citep{szapudi01,chon04}, a pseudo-power-spectrum framework similar to the \master{} algorithm~\citep{hivon02} which corrects for mode coupling induced by the mask. 
As discussed in~\citep{camphuis_accurate_2022}, this method regularizes the inversion of the \master{} matrix by apodizing the correlation function at the scales poorly explored in the mask.\footnote{We use a Gaussian taper with $\sigma^{\texttt{PolSpice}}_{\rm apo}=30^\circ$.} 
This regularization needs to be accounted for in the data model.

Taking into account the weights designed to mitigate excess power causing null test failures (see \cref{sec:null}) and filtering artifacts,
the timestream filtering (see \cref{sec:filtering}), 
the suppression or reweighting of spherical harmonic modes from the beam (see \cref{sec:beams}),  
the effect of pixelization, 
the residuals from the inpainting procedure (see \cref{sec:inpainting}),
and the \polspice{} regularization,
we form the following data model, which states that, on average, the measured estimation of the \XY{} ($\X{},\Y{} \in \{\T{}, \E{}\}$) power spectrum between two frequencies $\mu, \nu$ ($\mu, \nu \in \{95, 150, 220\}$\;GHz) can be related to the underlying signal by:
\begin{widetext}
	\begin{align}
		\label{eq:simsmodel}
		{\hat{C}^{\XY;\mu\nu}_{\ell}} = \sum_{\ell'} K_{\ell\ell'}^\XY \left[
			F_{\ell'}^{\XY;\mu\nu}
			P_{\ell'}^2 B_{\ell'}^{\mu} B_{\ell'}^{\nu}  C_{\ell'}^{\XY;\mu\nu;\rm signal} + A_{\ell'}^{\XY;\mu\nu} + I_{\ell'}^{\XY;\mu\nu}
			\right],
	\end{align}
\end{widetext}
where $C_{\ell'}^{\XY;\mu\nu;\rm signal}$ is the underlying signal power spectrum,  $B_{\ell'}^{\mu}$ is the instrument beam, $P_{\ell}$ is the \healpix{} pixel window function, and $K_{\ell\ell'}^\XY$ is the residual kernel from the \polspice{} regularization. 
We describe in \cref{sec:filtering} how the timestream filtering and the filtering artifacts around the  masked point sources can be modeled by a multiplicative ($F_{\ell'}^{\XY;\mu\nu}$) and an additive ($A_{\ell'}^{\XY;\mu\nu}$) bias, respectively.
We show in \cref{sec:inpainting} that the inpainting residual $I_{\ell'}^{\XY;\mu\nu}$ can be treated as an additive correction. 

In the following sections, we describe how we estimate each part of this data model necessary to debias our final band powers (see \cref{sec:bandpowers}). 
At the end of this section, after accounting for all processing steps and the estimation of the noise power spectra (see \cref{sec:noise}), we describe the computation of the covariance matrix for the final band powers (see \cref{sec:covariance}).
The final band powers and covariance matrix are used in the next section to form our primary CMB likelihood (\cref{sec:likelihood}).

\subsection{Filter modeling}
\label{sec:filtering}

During map-making, we apply a high-pass filter to the timestreams to remove large-scale noise and a low-pass filter to prevent aliasing, as described in \cref{sec:map-making}.
These operations are performed at the level of the individual-detector time-ordered data and are by nature anisotropic and inhomogeneous. 
For the specific case of the SPT, operating at the geographical South Pole, the filtering affects spherical harmonic modes primarily as a function of $m$ (assuming the map is in equatorial coordinates). However, as the filter cutoffs are defined in terms of absolute angle on the sky, not angle in right ascension, the effective cutoff in $m$ varies with declination.
This makes any attempt at a fully analytical estimation of the filtering challenging. 

For this reason, in this work (as in past SPT analyses) we use MC simulations to estimate the effect of the filtering at the power spectrum level.
We expand significantly on the approach from previous analyses to reach the level of precision required by the sensitivity of the new data set. 

As before, we assume that a transfer function description is sufficient to capture the effect of the filtering on the signal power spectrum. In an extension of previous work, we allow for a different filtering correction at the covariance level (beyond the effect of mode loss), under the assumption that the effect is faithfully captured by a diagonal rescaling of the matrix. In estimating the effects on both the power spectrum and the covariance, we use the combination of \quickmock{} and \fullmock{} simulations 
to achieve the required precision on the calibration of the filter model.

We assume that, on average, the effects of our filtering (including artifacts) on an input signal at the power spectrum level can be approximated by the combination of a multiplicative and an additive bias
\begin{equation}
	\label{eq:filtering}
	\VEV{C_\ell^{\ff\W;\XY;\mu\nu}} = F_\ell^{\XY;\mu\nu}\VEV{C_\ell^{\uu;\XY;\mu\nu}} + A_\ell^{\XY;\mu\nu},
\end{equation}
where $C_\ell^{\uu,\XY;\mu\nu}$ is the power spectrum of a simulation in the absence of filtering, $C_\ell^{\ff\W,\XY;\mu\nu}$ is that same spectrum in the presence of filtering, $F_\ell^{\XY;\mu\nu}$ is the {transfer function}, $A_\ell^{\XY;\mu\nu}$ represents the residual {filtering artifacts} after the $\ell,m$ weighting described in the next section, and all averages $\VEV{\ldots}$ are over a set of MC simulations. All spectra are computed using the point source mask and corrected for mask effects with \polspice{}. We further added a $\W$ marker to $C_\ell^{\ff\W,\XY;\mu\nu}$ to indicate that we have applied an \alm{} filter when computing the power spectrum and need to take its effect into account along with the filter.
We discuss these two biases in the following sections.

\subsubsection{Filtering artifacts}
\label{sec:filtering-artifacts}

One of the key features of our maps is the presence of a large number of point sources due to the high resolution of the instrument and the low noise level in the maps.
As discussed in \cref{sec:map-making}, 
if the brightest point sources are included in the parametric fit used to high-pass filter the data, the resulting map has large scan-direction features near the locations of those sources, which we referred to as {filtering wings} in \cref{sec:maps}. 
These features make it difficult to mask or inpaint over the sources in subsequent analyses. 
Removing the point source region from the parametric fit, a procedure we refer to as map-making masking, solves this issue, but at a price. 

By coupling with the effective holes in the maps at the location of point sources, the filtering now introduces smaller spurious features in the map near the location of these sources; we refer to these as {filtering artifacts}.
This effect leaks a fraction of the large-scale modes along the scan direction of the map that are targeted by the high-pass filter to small scales (also along the scan direction). 
The amplitude of these new features is on the order of the CMB signal that we are missing in the fit. 
This is much smaller than the filtering wings that have been avoided, the amplitude of which is on the order of 10\% of the peak point-source amplitude;
thus, the choice to mask point sources is a good trade-off for our data analysis.
Further details are provided in \cref{app:filtering-artifacts-map}. 
A map-level illustration is shown in \cref{fig:artifacts_map}.

The filtering artifacts, which depend on the location and size of the masked regions, can be described by an additive bias to the measured power spectrum. 
We perform an initial estimate of the bias from filtering artifacts using the difference of simulation pairs with and without map-making masking.
This first estimate is performed to identify the angular scales at which the filtering artifacts are significant and to determine the range of multipoles that we need to target with our filtering and weighting, but is not used in the final analysis.

The artifacts are found to be well-localized in pseudo-\alm{} space, specifically at $m \lesssim 200$ (which is not surprising, as this is the range of modes targeted by the filter). 
We use this fact to mitigate the artifacts through a weighting of the pseudo-\alm{} coefficients.
Again, assuming that the effect is on average well described by an additive effect, we decompose the signal \alm{} as a sum of filtered CMB (corresponding to our \fullmock{} simulations without map-making masking) and artifacts.
Then, we can write the Wiener filter
\begin{align}
	\label{eq:wieneralm}
	W^{\rm wiener}_{\ell m} & \equiv \VEV{\frac{|a^{\ff;\nommmask}_{\ell m}|^2}{|a^{\ff}_{\ell m}|^2}},  \\
	\ &= \VEV{
		\frac{|a^{\ff;\nommmask}_{\ell m}|^2}
			   {|a^{\ff;\nommmask}_{\ell m} + a^{\rm artifacts}_{\ell m}|^2}
			   },
\end{align}
where
$a^{\ff;\nommmask}_{\ell m}$ is the pseudo-\alm{} of a \fullmock{} simulation without map-making masking and $a^{\ff}_{\ell m}$ is the pseudo-\alm{} of the same simulation with the standard map-making masking procedure. We perform the MC estimation on 110 pairs of \fullmock{} simulations and estimate the pseudo-\alm{} using the point source mask.

We combine this new filter with the notch filter, see \cref{eq:notch}, such that
\begin{equation}
	\label{eq:totalweightlm}
	W^{\rm total}_{\ell m} \equiv W^{\rm notch}_{\ell m}W^{\rm wiener}_{\ell m}.
\end{equation}
We apply this weighting, $W^{\rm total}_{\ell m}$, to mitigate both the excess power contributing to initial null test failures and the filtering artifacts in the power spectrum estimation. This procedure reduces the amplitude of the filtering artifacts by a factor of $\sim 10$.
Of course, this weighting also affects the measured power spectrum and covariance, and we include this contribution when accounting for the filtering multiplicative effect in the next section.

\begin{figure}
	\includegraphics[width=\columnwidth]{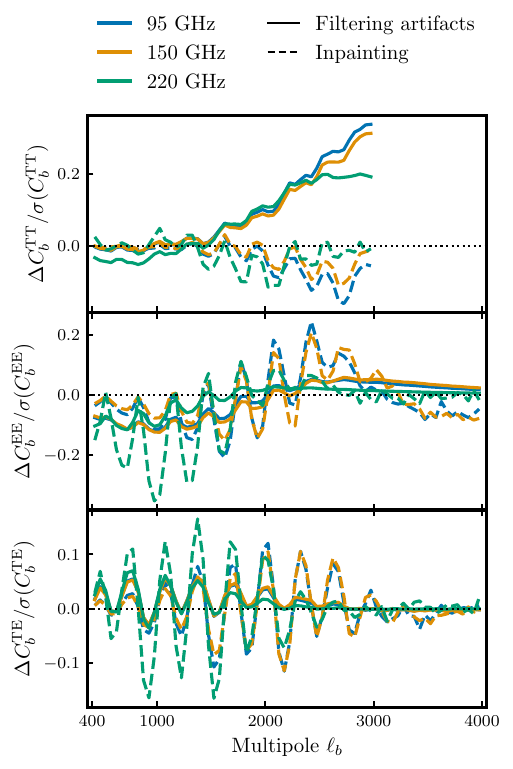}
	\caption{Filtering artifacts (solid lines) and inpainting biases (dashed lines) at $95\,\ghz$ (blue), $150\,\ghz$ (orange), and $220\,\ghz$ (green). 
	We remove both effects from the power spectrum estimates using a template subtraction approach. 
	The filtering artifacts are the result of masking bright sources in the high-pass filter applied to the maps and the inpainting biases are due to the inpainting of the masked point sources.
	The filtering artifacts are negligible on large angular scales but reach up to $0.3\,\sigma$ at $\ell=3000$. 
	The inpainting biases are the largest in \EE{} but always remain below $0.3\,\sigma$.}
	\label{fig:specbiases}
\end{figure}

After application of the Wiener filter, despite the strong reduction in the filtering artifact power, there is still a detectable residual effect on the measured power spectra. 
We account for this residual bias with a template subtraction approach at the power spectrum level.
The shape of the template for the contribution of the residual artifacts to the measured power spectra, which we denote as $A_\ell$ in \cref{eq:simsmodel}, is determined by the large-scale power in the map, the position and size of holes in the maps, and the timestream filtering strategy and the Wiener filter. This final $A_\ell$  is estimated using
\begin{equation}
	\label{eq:filtering_artifacts_notch}
	A_\ell = \VEV{C_\ell^{\ff\Wt;{\rm i}} - C_\ell^{\ff\Wt;{\rm i};\nommmask}},
\end{equation}
with similar notation to \cref{eq:wieneralm}, and using the marker ``$\Wt$'' to note that we are now using the total filter in \cref{eq:totalweightlm} and the marker ``i'' to note that the maps have been inpainted, as detailed later in \cref{sec:inpainting}.
We show the templates for the various spectra and band combinations, along with the inpainting bias templates discussed in \cref{sec:inpainting}, in \cref{fig:specbiases}.
The template for residual filtering artifacts in \TE{} and \EE{} is $<0.1\,\sigma$ at all scales; for \TT{}, it is negligible on large angular scales but reaches $0.3\,\sigma$ at $\ell=3000$.
While we use these templates to clean simulations and data, we also verify that leaving them uncleaned would result in negligible ($<0.1\,\sigma$) biases on cosmological parameters; 
we thus conclude that our power spectrum pipeline is robust to our modeling of the filtering artifacts.

\subsubsection{Transfer functions}
\label{sec:tf}

As discussed at the beginning of this section, the combined effects of the filtering, the \alm{} weighting, inpainting, and the mask can result in multiplicative and additive biases and mix power between multipoles. 
We also noted that in our modeling of the power spectra we assume that effects from the mask can be dealt with separately from the rest. 
We further assume that the multiplicative bias from the filtering and \alm{} weighting can be treated as a one-dimensional function of $\ell$,  i.e. a {transfer function}.

Similarily to the filtering artifact estimation performed above, we rely on simulations to estimate the transfer function. We define the transfer function in terms of an average ratio of the (weighted) power spectrum of filtered simulations without map-making masking $C_\ell^{\ff\Wt,\XX;\mu\nu;\nommmask}$  and the (unweighted) power spectrum of unfiltered simulations $C_\ell^{\uu;\XX;\mu\nu}$:
\begin{align}
	F^{\XX;\mu\nu}_\ell          & \equiv \frac{ \VEV{C_\ell^{\ff\Wt,\XX;\mu\nu;\nommmask}} } { \VEV{C_\ell^{\uu;\XX;\mu\nu} }}.
	\label{eq:Fl_def}
\end{align}
In the case of \TE{}, the denominator of \cref{eq:Fl_def} can vanish and, therefore, we define the corresponding transfer function as the geometric mean of the auto-spectrum transfer functions
\begin{align}
	F^{\TE;\mu\nu}_\ell & \equiv \left(F^{\TT;\mu\mu}_\ell \times F^{\EE;\nu\nu}_\ell\right)^{1/2}.
\end{align}

\begin{figure}
	\includegraphics[width=\columnwidth]{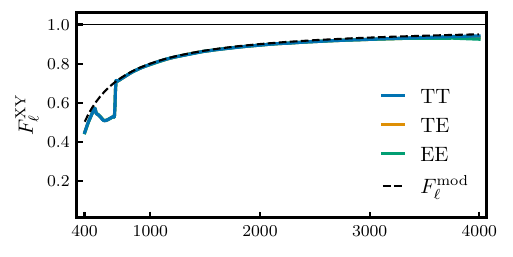}
	\caption{One-dimensional transfer function for the \sptbp data at $150\times150\ghz$\replace{.}{, for the \TT{} (blue), \TE{} (orange), and \EE{} (green) spectra.}
	The transfer function is computed using \quickmock simulations not masking point sources during map-making and corrected for the measured discrepancy between \quickmock and \fullmock simulations. 
	\replace{\emph{Upper panel:} The transfer function for the \TT{} (blue), \TE{} (orange), and \EE{} (green) spectra. The lines overlap for most of the multipole range and are just distinguishable above $\ell=3000$.}{}
	Except for masking due to the notch filter in \cref{eq:notch}, the transfer function is close to a simple model $F^{\rm mod}_\ell = 1 - 200/\ell$.
	\replace{\emph{Lower panel:} Ratio of the \TE{} and \EE{} transfer functions to the \TT{} transfer function.}{}	
	}
	\label{fig:transfer_function}
\end{figure}

We show the $150\times150$\ghz\ transfer functions in \cref{fig:transfer_function}. (The transfer functions at other frequency combinations look similar.)
The effect of our filters is important: a strong suppression at the lower range of multipoles owing to the high-pass filter is evident, as is a sharp cut induced by the notch filter, see \cref{eq:notch}. 
While the high-pass is a sharp cut in the Fourier conjugate to scan-direction angle, the dependence of the effective $m$ cutoff on declination in the map translates this into a softer cutoff as a function of $m$.
Effectively averaging over $m$ at each multipole $\ell$ results in the high-pass filter transfer function being quite broad in one-dimensional $\ell$ space.
We find that, up to the effect of the notch filter, the transfer function is close to a simple model $F^{\rm mod}_\ell = 1 - 200/\ell$. 
This can be understood by looking at Eq. (B11) of~\citep{hivon02} and considering that the \SPT{} scanning strategy consists of parallel scans.
At small angular scales, the transfer function approaches unity. 
\replace{The \EE{} transfer function differs from the \TT{} transfer function due to residual leakage from large to small scales.
This leakage is due to the cut-off, it is unrelated with filtering artifacts, and it is relatively more significant in \EE{} than in \TT{}, owing to the presence of small-scale foregrounds in \TT{} and their absence in \EE{}.}{
The \TE{} and \EE{} transfer functions shown in \cref{fig:transfer_function} differ very slightly from the TT transfer function at high multipoles due to a \polspice{} configuration error discovered after initial paper submission. 
We have confirmed that the difference in best-fit parameter values between using the original and corrected \TE{} and \EE{} transfer functions is less than $0.03\,\sigma$, and we continue to use the originally calculated versions for consistency.
}

We achieve excellent precision for this MC  estimation, as can be seen in the lower panel of \cref{fig:transfer_function}, thanks to our joint use of the \replace{2000}{4000} \quickmock{} simulations and the use of the 500 \fullmock{} simulations to calibrate them. We discuss in \cref{app:transferfunction} how we use both simulations to achieve the best possible precision and how we propagate residual errors from the MC estimate.

The filtering, masking, and reweighting of the \alm{} also affect the covariance of the power spectra in a non-trivial way, potentially modifying the off-diagonal structure of the covariance. 
Similarly to the case of the power spectrum, we assume that the mask and filtering effects decouple and we discuss how we deal with the former in \cref{sec:covariance}. 
For filtering and \alm{} reweighting, see \cref{eq:totalweightlm}, we again assume that the multiplicative effect is dominant compared to the coupling of multipoles and approximate the effect of the filters by a diagonal rescaling (see \cref{eq:Hrescale} in the covariance section).  
We estimate this rescaling from our simulations using the same method we used for the transfer function: 
\begin{align}
	 & H^{\XY\mu\nu;\X'\Y'\alpha\beta}_{\ell}  \equiv\label{eq:Hl_def} \\& \frac{\VEV{\delta C_{\ell}^{\ff\Wt;\nommmask;\XY\mu\nu}\ \delta C_{\ell}^{\ff\Wt;\nommmask;\X'\Y'\alpha\beta} }} %
	{\VEV{ \delta C_{\ell}^{\uu;\XY\mu\nu}\ \delta C_{\ell}^{\uu;\X'\Y'\alpha\beta,} }},\nonumber
\end{align}
where  
\begin{equation}
	\delta C_{\ell}^{\XY;\mu\nu} \equiv C_{\ell}^{\XY;\mu\nu} - \VEV{C_{\ell}^{\XY;\mu\nu}}
    \label{eq:delta_Cl}
\end{equation}  
is the discrepancy between a random realization of the power spectrum and its MC average.
A similar method was used in \citep{atkins_atacama_2024} in the case of inhomogeneous survey depth.

The power spectrum transfer function $F^{\XY;\mu\nu}_{\ell}$ and the multiplicative correction $H^{\XY,\X'\Y';\mu\nu}_{\ell}$ are, respectively, quadratic and quartic in mode filtering, and one would therefore expect $H^{\XY,\X'\Y';\mu\nu}_{\ell} \sim (F^{\XY;\mu\nu}_{\ell})^2.$
However, since the high pass filtering implemented here is by nature mostly binary, meaning that the frequency modes along the scan direction are either left unchanged or totally removed, we expect the effects of the filtering on the mean and variance of the power spectrum to be similar.
Indeed, using the large number of \quickmock simulations, we found that, at the $\sim 10\%$ level, 
\begin{equation}
	H_{\ell}^{\XY,\XY} \simeq F_{\ell}^{\XY}.
	\label{eq:Hl_Fl}
\end{equation} 
This was confirmed in \fullmock simulations, with a larger scatter. We show how we integrate this correction in our estimation of the covariance matrix in  \cref{sec:covariance}.

\subsection{Beams}
\label{sec:beams}

The beam, or point-spread function, describes the angular dependence of the instrument response to a point source. 
For microwave instruments (for which atmospheric seeing is not a significant contribution), the beam is primarily determined by diffraction in the optical system but can also include effects such as finite detector response time.
The effect of the beam on sky signal can be represented as a convolution in real space. 
The effective beam is well described by a single transfer function in harmonic space, represented by the $B_\ell$ terms in \cref{eq:simsmodel}.
The SPT-3G beams are described in Appendix 4 of \citep{ge24}, and a more complete characterization will be given in an upcoming work (N. Huang et al., in preparation, hereafter H25).  
Here, we only review the aspects of the beam modeling that are relevant to our analysis.

\subsubsection{Temperature beams}

\begin{figure*}
	\includegraphics[width=\textwidth]{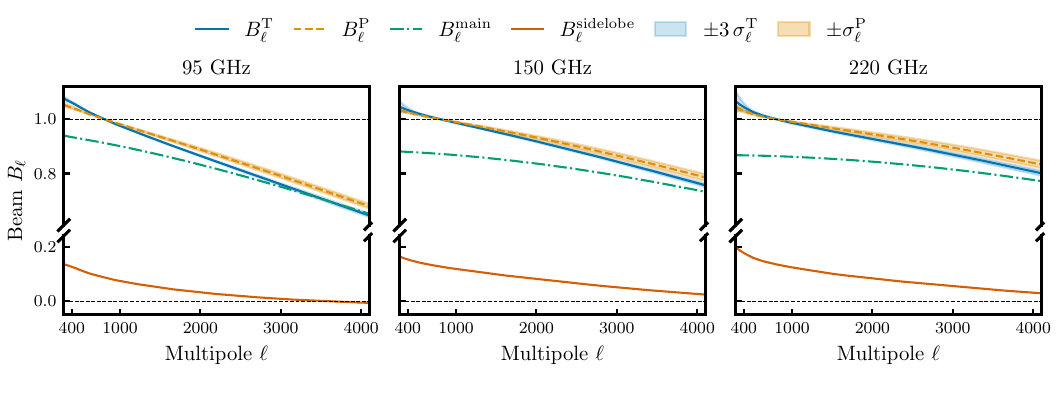}
	\caption{
		Harmonic-space beam functions for each frequency band.
		The solid blue lines show the measured temperature beams, with shaded regions indicating three times the measurement uncertainty.
		Note that the beam uncertainty is correlated across both angular scales and frequency bands.
		The orange dashed lines represent the polarized beams, which are modeled as the sum of the main beam and a fractional sidelobe contribution, as described in \cref{eq:polbeams}.
		The sidelobe beams (full red line) are obtained by subtracting the main beam (green dash-dotted line) from the temperature beam, following \cref{eq:sidelobe}.
		We display the best-fit polarized beams using the parameters from \cref{tab:nuisance-priors}, as determined by fitting the \sptbp{} data.
		The orange shaded regions indicate the uncertainty on the polarized beams, which is propagated from the uncertainty in the sidelobe polarization fraction. 
		This uncertainty is correlated across both angular scales and frequency bands.
		Both the temperature and polarized beams are normalized to unity at $\ell=800$.
	}
	\label{fig:beams}
\end{figure*}

We measure the angular response in total intensity to an unpolarized source, or the temperature beam, using the combination of bright active galactic nuclei (AGN) located in the Main field and dedicated observations of Saturn.
While bright AGN allow us to characterize the main beam close to the peak of the response, they are too faint to map the telescope beam far from the center, i.e. the beam sidelobes.
Meanwhile, planet observations are useful for mapping the sidelobes, but planets are sufficiently bright to saturate some detectors and thus cannot be used for the main beam.
We stitch together observations of Saturn and individual AGN into a composite real-space beam map.
A harmonic decomposition of this map gives us the temperature version of the $B_\ell^\mu$ term in \cref{eq:simsmodel}, which we notate as  
$B_\ell^{\T;\mu}$. We normalize the beam to unity at $\ell=800$ to decorrelate the beam shape from the \planck{} based calibration, see \cref{sec:calibration}.
One-dimensional harmonic-space temperature beams $B_\ell^{\T;\mu}$ are plotted in \cref{fig:beams}.

We quantify the uncertainty on $B_\ell^{\T;\mu}$ related to noise, systematic effects, and analysis choices (such as the radius at which the Saturn and AGN beams are stitched) in a beam covariance matrix, which covers all angular scales and all observational frequencies. 
We use it to propagate the beam uncertainty to the band powers in the signal part of our data model as shown in \cref{eq:datamodel}.
The uncertainty in the temperature beam represents at most $0.2\%$ of the beam at $\ell=3000$ for all frequencies.
We display the uncertainty on the temperature beams in \cref{fig:beams}.
More details on the uncertainty estimation are given in H25.

\subsubsection{Quadrupolar beam leakage}
\label{sec:quadrupolar}

\begin{figure}
	\includegraphics[width=\columnwidth]{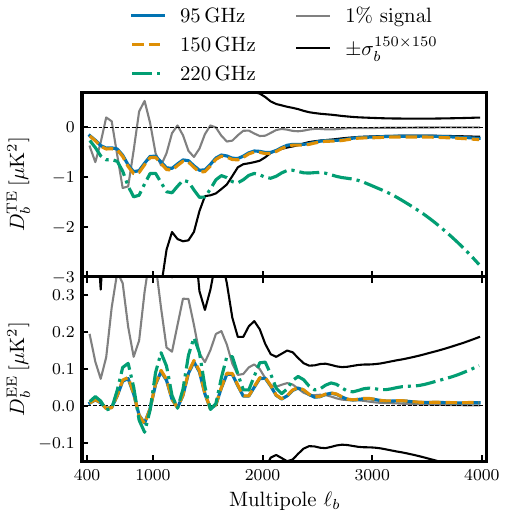}
	\caption{Quadrupolar leakage templates for the \TE{} (top) and \EE{} (bottom) spectra. 
	The colored lines indicate the leakage spectra, see \cref{eq:t2p}, for the three frequency bands. 
	The solid gray line represents 1\% of the best-fit CMB power spectrum, while the solid black lines represent the $150\times150\ghz$ error bars. 
	The \TE{} $150\times150\ghz$ leakage reaches $1\,\sigma$ at small scales. 
	The contamination is notably larger at $220\,\mathrm{GHz}$, primarily due to the higher foreground power in \TT{} leaking into \TE{} and \EE{}, however, the associated frequency channel error bars are also larger.
	}

	\label{fig:leak}
\end{figure}

The ``leakage beam'' describes the response in polarization to an unpolarized source.
The monopole leakage from temperature to polarization caused by gain differences between detectors has been removed from the maps in a previous analysis step, see \cref{sec:calibration}.
However, the maps also contain a significant contribution from higher-order quadrupolar leakage sourced by the differential beam ellipticities of the detectors~\citep{hivon_quickpol_2017,hu03,shimon_cmb_2008}.
This effect is detected and measured by analyzing \Q{} and \U{} maps at the location of bright sources in temperature, which we present in \cref{app:t2p}.
We propagate the measured map-level contamination to band powers using an analytical model derived from \cref{eq:t2p_beams}, which we confirm with simulations.
We model quadrupolar leakage at the band power level as
\begin{equation}
	\label{eq:t2p}
	\begin{split}
		C_\ell^{\TE;\mu\nu;\text{leak}} = & \epsilon^{\nu}_2\sigma^{2}_\nu \ell^2 C_\ell^{\TT;\mu\nu},                                   \\
		C_\ell^{\EE;\mu\nu;\text{leak}} = & \epsilon^{\mu}_2 \sigma^{2}_\mu \ell^2 C_\ell^{\TE;\mu\nu}
		+ \epsilon^{\nu}_2 \sigma^{2}_\nu \ell^2 C_\ell^{\ET;\mu\nu}                                                                     \\
		\                                 & + \epsilon^{\mu}_2 \epsilon^{\nu}_2 \sigma^{2}_\mu\sigma^{2}_\nu \ell^4 C_\ell^{\TT;\mu\nu},
	\end{split}
\end{equation}
where $\epsilon_2^\mu$ is the amplitude of the quadrupolar leakage and $\sigma_\mu$ is the size of the leakage beam in the $\mu$ band.
From the map-level study described in \cref{app:t2p} we deduce the amplitudes of the quadrupolar leakage:
\begin{align}
	& \epsilon_2^{\rm 95} = -(0.65 \pm 0.11)/100, \\ 
	& \epsilon_2^{\rm 150} = -(1.2 \pm 0.21)/100, \\ 
	& \epsilon_2^{\rm 220} = -(2.3 \pm 0.66)/100,  
\end{align}
which are, respectively, 6, 6, and 3.5$\,\sigma$ detections of the quadrupolar leakage. 
In \cref{fig:leak}, we display the leakage template, which corresponds to contamination amplitudes of $\sim1\%$ of the signal on large scales in \TE{} and, with respect to error bars, affects mostly the highest range of multipoles of this analysis, between $\ell=3000$ and $4000$. 
We find that the leakage is largest at $220\ghz$ due to the higher foreground power in \TT{} leaking into \TE{} and \EE{}.
The leakage is smaller, and less significant, in \EE{}. 

We note in \cref{app:post-unblind} that the quadrupolar leakage correction
was implemented only after unblinding (for details of the blinding procedure, please refer to \cref{sec:blinding}). 
This is one of the reasons why the band powers are not debiased for this effect, which  is instead treated at the likelihood level. 
The fit to the map-level measurement is used as a prior for the beam leakage parameters. 

\subsubsection{Polarized beams}
\label{sec:polbeams}

We define the polarized beam as the angular response of the instrument in the linear-polarization Stokes parameters to a 100\% linearly polarized source.
In principle, this can be different from the temperature beam. 
The main beam, which is formed by optical paths that are well-controlled and pass through all optical elements as designed, 
is expected to be uniformly and highly polarized, but the polarization of the sidelobes, which include radiation scattered and reflected from non-ideal optical elements and the diffraction pattern formed by the gaps and surface imperfections in the telescope mirrors, can be more complex.
We do not have sufficiently bright polarized sources in the survey region to directly measure the polarized beam on the relevant angular scales. 
For this reason, we allow for depolarization of the beam sidelobes relative to the main beam and marginalize over the fraction of sidelobe polarization at the likelihood level.
This additional freedom slightly degrades the constraining power of the data set. 
In \cref{app:polbeams} we show that this polarized beam model is preferred by the data. 
We emphasize that the evidence for sidelobe depolarization comes from requiring internal consistency between frequency bands in the \EE\ data. 
Its detection does not require assuming any particular cosmological model.

We model the polarized beam as the sum of the main beam $B^{\rm main;\mu}$ and the sidelobe contribution $B^{\rm sidelobe;\mu}_\ell$ modulated by a scale-invariant polarization fraction $\betapol$:
\begin{equation}
	B^{\rm P;\mu}_\ell = \frac{B^{\rm main;\mu}_\ell + \betapol^{\mu} B^{\rm sidelobe;\mu}_\ell}
	{B^{\rm main;\mu}_{800} + \betapol^{\mu} B^{\rm sidelobe;\mu}_{800}}.
	\label{eq:polbeams}
\end{equation}
When $\beta=1$ we recover the temperature beam, 
and if the sidelobes are depolarized we expect to recover $\beta<1$.
The main beam $B^{\rm main;\mu}$ is calculated analytically from our knowledge 
of the optics, as described in Appendix 4 of~\citep{ge24}, and shown in \cref{fig:beams}.
The sidelobe contribution is taken as the difference between the
measured temperature beam and the calculated main beam,
\begin{equation}
	B^{\rm sidelobe;\mu}_\ell \equiv B^{\rm T;\mu}_\ell - B^{\rm main;\mu}_\ell.
	\label{eq:sidelobe}
\end{equation}
We normalize the harmonic space polarized beam to unity at $\ell=800$ to preserve 
the relative polarization efficiency priors obtained by comparison with 
\planck{} independent of changes to the assumed beam shape. 
As discussed in \cref{sec:calibration}, the external polarization efficiency 
calibration is only needed when using polarization data alone.

Similarly to \cref{sec:quadrupolar}, the baseline polarized beam model
was implemented only after unblinding  and we similarly fit for it at the likelihood level, see \cref{eq:datamodel}.
We find strong support for this model with cosmology-independent methods,
see~\citep{ge24} and the \sptlite{} discussion in \cref{app:polbeams}.
In the context of \lcdm{}, the data strongly support the polarized beam model 
with a $5\,\sigma$ preference for $(\betapol^{95},\betapol^{150},\betapol^{220})\neq(1,1,1)$
from SPT data alone. 
Posterior values for the $\betapol$ parameters are given in \cref{tab:nuisance-priors}.
In \cref{fig:beams}, we show the best-fit polarized beams and their propagated uncertainties, which are derived from the uncertainty in the sidelobe polarization fraction $\betapol$.
At small angular scales ($\ell > 2000$), these beams are nearly identical to the temperature beam, except for the effect of normalization at $\ell=800$.
At large angular scales ($\ell < 2000$), however, the sidelobe contribution alters the polarized beam shape relative to the temperature beam.

\subsection{Inpainting}
\label{sec:inpainting}

\begin{figure}
	\includegraphics[width=\columnwidth]{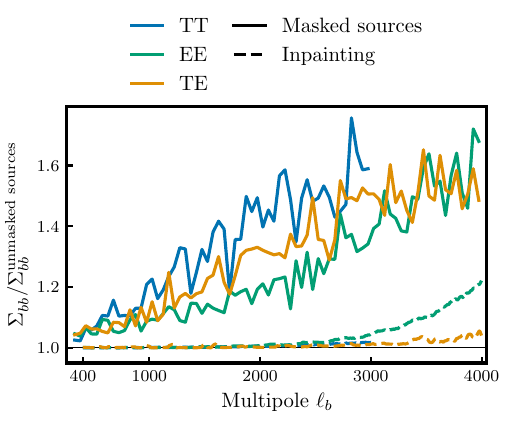}
	\caption{Ratio of the covariance matrix diagonal elements with and without point source masking. 
	The ratio is computed from 500 \fullmock simulations, for which we either mask (or not) the point sources during power spectrum estimation. 
	The source masking results in 50\% more variance at small scales than without. 
	In comparison, we show the increase in variance due to inpainting, see \cref{eq:rhoell,eq:cov_inpainting}, which is much smaller, justifying our choice to use the inpainting method.}
	\label{fig:covariance_masked_ptsrc}
\end{figure}

We mentioned in \cref{sec:map-making} that the unprecedented depth of the \mainfield{} field and the high resolution of the SPT results in many thousands of emissive sources and galaxy clusters detected at high signal-to-noise. 
We choose to mask over 2000 of them in map-making and to remove them from the map before estimating power spectra. 
Traditionally this has been accomplished by multiplying the map by a mask with (apodized) holes at the location of all sources and clusters we wish to remove.
Under the assumption of statistical isotropy of the signal, such features in the mask do not bias the power spectrum estimation, as the mask effect can be properly taken into account in the \master{} framework.
However, point-source masking strongly impacts the statistical properties of the estimator by inducing correlations in the estimated power spectrum across angular scales.
In particular, we see correlation of large-scale power to smaller scales: bigger modes across the patch look like small-scale modes when interrupted by the point source holes and can be disambiguated from the real small-scale modes only on average.
Using the \fullmock simulations, we show that we can expect point-source masking to increase the variance of the estimator on the three \TTTEEE{} channels at $\ell=3000$ by 50\%, see \cref{fig:covariance_masked_ptsrc}.

To avoid the consequences of point source masking on the statistical properties of our power spectrum estimator, we choose to fill the locations of point sources and galaxy clusters in our maps with Gaussian realizations constrained over the power in the rest of the map, a method also known as inpainting~\citep{hoffman_constrained_1991,matrixcookbook}.
This allows us to use only the border mask, which does not have any small-scale features, for power spectrum estimation, thereby reducing the mode-coupling of the estimator.
However, we need to increase the covariance of our power spectra to account for the fact that we have added fake simulated signal to the data.

Gaussian-constrained inpainting is now a well-established technique in CMB lensing analysis~\citep{benoitlevy13,bucher_filling_2012,planck13-17,omori17,raghunathan19c,pan23,omori23,qu24}. 
This work is the first use of it for primary CMB analysis, which presents different challenges.

We follow here the method presented in~\citep{benoitlevy13,bucher_filling_2012,pan23} and only use a small region around each point source to predict the CMB signal inside the region we wish to inpaint. Even with this simplification, the inpainting of the SPT-3G maps is challenging due to the large number of point sources and the high resolution of the maps.
For this work, we improved on the inpainting code called \cork{} used in the \planck{} lensing analysis~\citep{benoitlevy13}, extending it to apply to multiple frequencies and polarization data and perform efficiently in the high-resolution regime ($\Nside=8192$).

Schematically, the method proposes to create a new set of \T{}, \Q{}, and \U{} bundle maps $M^\mathrm{i}$ from the $M$ observed bundle maps such that

\begin{equation}
	\label{eq:inpainting}
	\begin{bmatrix}
		M^\mathrm{i}_\mathrm{msk} \\
		M^\mathrm{i}_\mathrm{brd} \\
		M^\mathrm{i}_\mathrm{rst}
	\end{bmatrix}
	=
	\begin{bmatrix}
		0 & W & 0 \\
		0 & 1 & 0 \\
		0 & 0 & 1
	\end{bmatrix}
	\begin{bmatrix}
		M_\mathrm{msk} \\
		M_\mathrm{brd} \\
		M_\mathrm{rst}
	\end{bmatrix}
	+
	\begin{bmatrix}
		1 & - W & 0 \\
		0 & 0   & 0 \\
		0 & 0   & 0
	\end{bmatrix}
	\begin{bmatrix}
		{M}^{\rm sim}_\mathrm{msk} \\
		{M}^{\rm sim}_\mathrm{brd} \\
		{M}^{\rm sim}_\mathrm{rst}
	\end{bmatrix},
\end{equation}
where $M_\mathrm{msk}$ is the set of pixels of the observed bundle maps inside the point source masks, $M_\mathrm{brd}$ is the set of pixels in the constraining zone, and $M_\mathrm{rst}$ is the set of pixels in the rest of the bundle maps outside of the mask and the constraining zone. 
$M^{\rm sim}$ is a set of simulated bundle maps that have the same power spectrum and noise as the observed one. 
Finally, following~\citep{benoitlevy13},  $W$ is the Wiener filter that predicts the data inside the masked region from the border region and is given by
\begin{equation}
W = \Sigma_\mathrm{msk, brd}^{\phantom{-1}} \Sigma_{\mathrm{brd,brd}}^{-1},
\end{equation}
the two covariances here being the joint pixel covariances of the bundle of maps. 
When all of the remaining data is used as a constraining region and the size of the masked region is much smaller than the total map, the power spectra of the inpainted map bundle are unbiased.

In full generality, the pixel covariance $\Sigma$ must couple pixels across bundles (each bundle sees the same sky up to noise), frequencies (each frequency band sees the same CMB up to foregrounds), and temperature and polarization (because of the cosmological \TE{} correlation). 
The resulting $W$ matrix is large, dense, and based on the inversion of the also large and full $\Sigma_{\mathrm{brd,brd}}$.
To reduce the computational complexity of the $W$ matrix and enable parallelization, we adopt several simplifying assumptions that decouple its components. 
These approximations introduce a residual inpainting-induced correlated noise, which we calibrate using simulations and correct for in the power spectrum analysis.

First, following the lensing analysis procedure, we reduce the constraining region to a small border around each point source hole of the mask. 
This simplification allows us to decouple the problem by inpainting region and parallelize over the list of regions. 
While our tests show that constraining regions of approximately degree scale are needed to leave a vanishingly small residual, a smaller constraining region corresponding to a few tens of arcminutes, which we adopt, is a sufficient trade-off between computation time at $\Nside=8192$ and amplitude of the residual.

Second, instead of inpainting all of the bundles together taking into account their correlations, we instead inpainted only the full-depth coadd (see \cref{sec:coadd}) and used it to fill the masked regions in each bundle. This correctly takes into account the correlation between bundles. 
This simplification does not increase the inpainting residual.

Third, we assume that the different frequencies are uncorrelated and inpaint each frequency separately. The assumption of uncorrelated frequencies induces biases in the cross-frequency spectra that we estimated.

After those assumptions, the problem can be reduced to a suite of parallelizable Gaussian constrained realizations for each inpainting region on a triplet of \T{}, \Q{}, and \U{} maps using small constraining regions.

We checked that a perfect knowledge of the exact power spectrum of the map was not a strong requirement for the building of the Wiener filters. 
Eventually, we used the mean of the \fullmock{} power spectra to compute the filter. This directly includes the transfer function correction. 
The noise contribution needed for the filter is obtained from the data noise spectra (discussed in \cref{sec:noise}).

The quality of the simulation used for the inpainting ($M^{\rm sim}$ in \cref{eq:inpainting}) is paramount. 
In particular, since we only account for the timestream filtering in the pixel covariances at the 1D transfer function level, we miss the anisotropic effects of the filtering. 
For this reason, to inpaint the maps, we use \fullmock{} simulations plus noise realizations to include the full filtering effect and our best noise model.

Finally, we estimate the residual inpainting bias $I_{\ell'}^{\XY;\mu\nu}$ arising from all of those approximations by inpainting the set of \fullmock{} plus sign-flip simulations with the same settings, which we define as
\begin{equation}
	I_{\ell'}^{\XY;\mu\nu} = \VEV{C_\ell^{\ff\Wt;{\rm i}} - C_\ell^{\ff\Wt}}.
\end{equation}
This bias is displayed in \cref{fig:specbiases}. 
It is found to be of similar order of magnitude to the filtering artifact biases and we remove it from the computed power spectra.

The inpainting procedure must be taken into account in the covariance of the power spectra, as a small part of the maps has been replaced with fake simulated signal. 
We must marginalize over it in the estimation of the covariance matrix of the band powers. 
To do so, following \cref{eq:inpainting}, we first note that the power spectra of the inpainted maps can be split into a data component and a random component, both Wiener-filtered,
\begin{equation}
    \label{eq:inpainting_splitting}
    C^{\rm i}_{\ell} = C_{\ell}^{W;{\rm data}} + C_{\ell}^{1-W;{\rm sim}}.
\end{equation}
Up to the inpainting bias $I_\ell$ (that we neglect in this discussion), $C^{\rm i}_{\ell}$ is by construction unbiased and its covariance $\Sigma$ is given by the calculation that we detail in \cref{sec:covariance}. If we fix the data part and explore the covariance of $C^{\rm i}_{\ell}$ under the variation of simulations, we expect that
\begin{equation}
\VEV{\delta C^{\rm i}_{\ell} \delta C^{\rm i}_{\ell'}} = \VEV{\delta C_{\ell}^{1-W;{\rm sim}}\delta C_{\ell'}^{1-W;{\rm sim}}}.
\end{equation}
The marginalization over the random fake data can be approximated by adding to $\Sigma$  an extra term corresponding to the covariance of $C_{\ell}^{1-W;{\rm sim}}$.
Our problem is reduced to the computation of the covariance of the Wiener-filtered fake data spectra $C_{\ell}^{1-W;{\rm sim}}$. 
We assume that a 1D transfer function sufficiently captures the effect at the power spectrum level, so that
\begin{align}
	\VEV{\delta C_{\ell}^{1-W}\delta C_{\ell'}^{1-W}} \approx \rho_\ell\rho_{\ell'} \VEV{\delta C_{\ell}\delta C_{\ell}}, \\
    \label{eq:rhoell}
    \text{with} \quad \rho_\ell \equiv \frac{\VEV{C_{\ell}^{1-W}}}{\VEV{C^{\rm i}_{\ell}}},
\end{align}
which we estimate from the inpainted \fullmock{} simulations. 
We show the impact of inpainting on the diagonal of the covariance matrix in \cref{fig:covariance_masked_ptsrc}. 
The inpainting covariance is significantly smaller than the masked point source covariance, as intended by the inpainting procedure. 
We employ this methodology to increase the band power covariance matrix, thereby marginalizing over the artificial simulated signal introduced into the maps, see \cref{sec:covariance}.

\subsection{Band powers}
\label{sec:bandpowers}

\begin{figure*}
	\includegraphics[width=\textwidth]{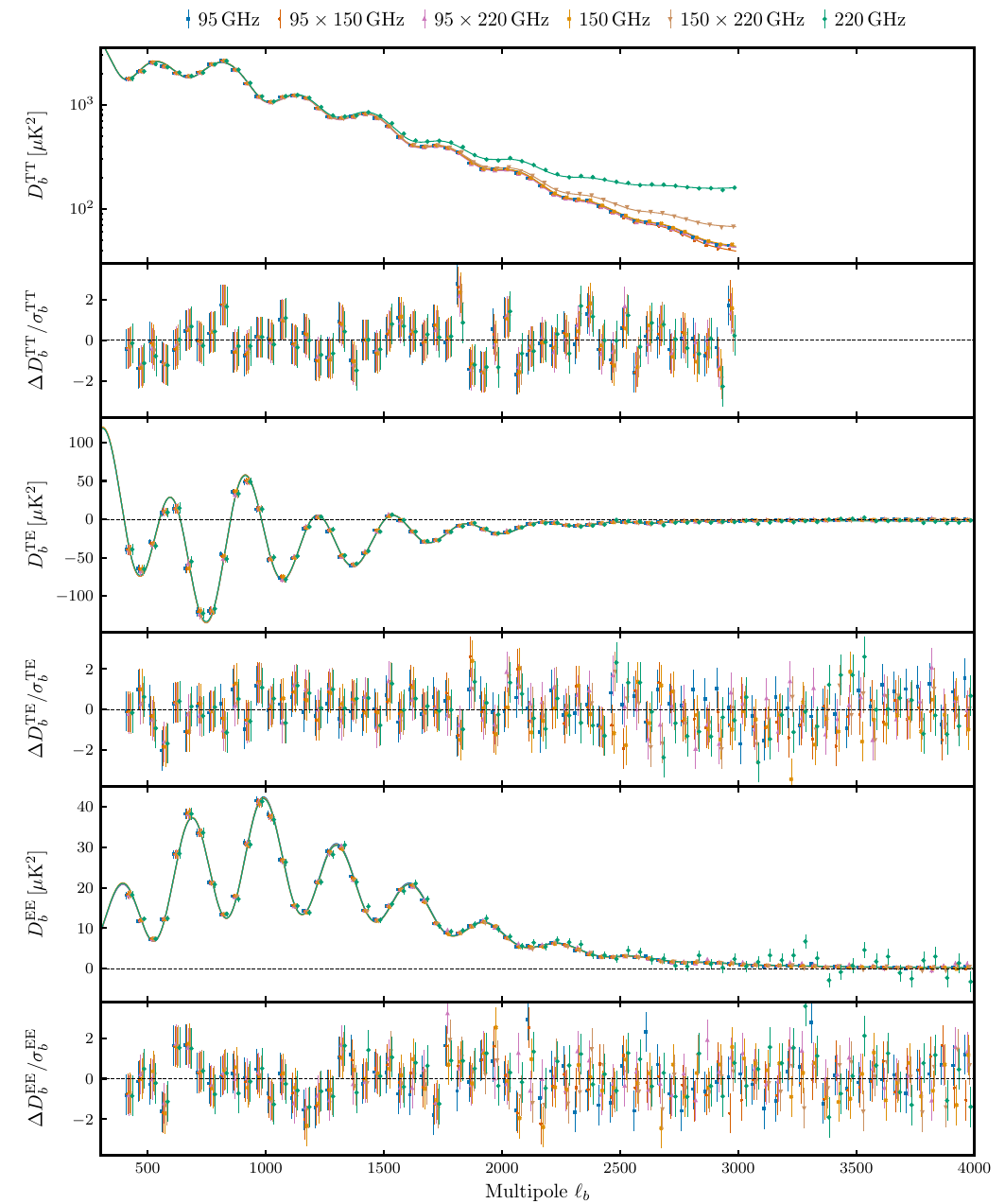}
	\caption{\textit{Large panels:} Measured \sptbp band powers for each cross-frequency spectrum, together with the best-fit theoretical prediction under the \lcdm{} cosmological model, incorporating the complete data model as described in \cref{eq:datamodel}. For clarity, the $150\times95\,\GHz{}$, $220\times150\,\GHz{}$, and $220\times95\,\GHz{}$ \TE{} band powers are omitted. The error bars represent the square roots of the diagonal elements of the covariance matrix as computed in \cref{sec:covariance}. 
	\textit{Small panels:} Corresponding residuals with respect to the best-fit prediction. The \lcdm{} model, combined with the full data model, provides an excellent fit to the measurements, with all residuals consistent with statistical expectations. The residuals further illustrate the transition between signal-dominated (correlated residuals) and noise-dominated (uncorrelated residuals) multipole ranges.}
	\label{fig:bandpowers}
\end{figure*}

After inpainting the maps, we estimate the auto- and cross-frequency pseudo-power spectra from the 95, 150, and 220\ghz{} maps on the curved sky using \polspice{} and the harmonic-space filter defined in \cref{eq:totalweightlm}.
Each spectrum is calculated as the average of the cross-power spectra calculated from different bundles $(i, j)$, defined in \cref{sec:coadd}, as
\begin{equation}
	\bar{C}^{\XY;\mu\nu}_\ell = \frac{1}{N_\mathrm{bundles} (N_\mathrm{bundles}-1)} \sum_{i\neq j} C^{\XY;\mu\nu;ij}_{\ell}.
	\label{eq:crosspower}
\end{equation}
Note that for the case where $\mu=\nu$, the normalization of the sum is still correct as the right-hand side sum double counts the spectra.
We do not use auto-frequency spectra and cross-frequency spectra calculated from the same bundle to avoid noise bias and reduce co-temporal systematics, \ie{} we never use $C_\ell^{\XY;\mu\nu;ij}$ with $i=j$.

Due to our filtering strategy, we limit ourselves to multipoles above $\ellmin\equiv400$, where the transfer function is above 0.5 (see \cref{fig:transfer_function}). 
This threshold is above the hard limit of our filtering at $\ell=300$.
This conservative choice avoids the multipoles where the filtering effect is large and where we would be limited by the MC precision of our transfer function estimate, see \cref{app:transferfunction}.

Given that the main goal of this work is measuring the CMB primary anisotropies, we cut the temperature power spectrum at $\ellmax^{\T}\equiv3000$ where the signal starts to be dominated by foreground contamination in all six cross-frequency spectra. 
The foreground contamination is much weaker in polarization, and we include multipoles up to $\ellmax^{\E}\equiv4000$ in \TE{} and \EE{}.
\RR{We have verified that extending the analysis to include multipoles up to $\ellmax^{\T}=4000$ does not lead to a significant improvement in cosmological constraints. 
The largest observed improvement is a 10\% reduction in the error bar on $\ombh$ for the \lcdm{} model when fitting to \TT{}-only data. 
As for single parameter extensions to \lcdm{}, including more multipoles can significantly improve constraints when fitting \TT{}-only data, but the overall gain remains modest when including \TE{} and \EE{} data. 
For instance, the error bar on $\neff$ and $\ns$ improves by 1\% and 3\%, respectively, when additionally including multipoles up to $\ellmax^{\T}=4000$ to the baseline analysis.} 

We debias the band powers for all the effects included in \cref{eq:simsmodel}, i.e. the multiplicative \healpix{} pixel window function, transfer function, and temperature beam, and the additive filtering artifact and inpainting residuals.
As described in \cref{sec:beams}, the quadrupolar beam leakage as well as the polarized beams are treated as parametrized systematics in the signal data model at the likelihood level. 
We bin the power spectrum estimates into $\Delta \ell = 50$ band powers using uniform weighting in $D_\ell$ with the binning operator
\begin{equation}
	Q_{b\ell} = \begin{cases}
		& 0  \text{ if } \ell \notin [50b,50(b+1)], \\
		& \frac{\ell(\ell+1)}{\sum_{50b \le \ell' <50(b+1)}\ell'(\ell'+1)} \ {\rm otherwise}.
		\label{eq:binmatrix}
	\end{cases}
\end{equation}
Binning reduces the impact of possible errors on the off-diagonal terms of the covariance matrix and reduces the residual mode mixing due to the \polspice{} regularization. \citet{camphuis_accurate_2022} show that with our mask, the covariance precision is of order 1\% at this binning size. 
\RR{As noted in \cref{sec:covariance}, this binning results in a $2.5\%$ correlation between adjacent bins, which arises from the effects of masking.}
Binning also reduces the size of the data vector, reducing the computational cost of likelihood estimation.
Binning can potentially hide features in the power spectrum and reduce constraining power, but Fig. 1 of~\citep{balkenhol24} shows that the binning choice made here leads to a negligible increase in cosmological parameter errors. This yields the final expression for the debiased binned band powers,
\begin{equation}
	\hat{C}^{\XY;\mu\nu}_{b} \equiv \sum_\ell Q_{b\ell} \frac{\bar{C}^{\XY;\mu\nu}_\ell - A^{\XY;\mu\nu}_\ell - I^{\XY;\mu\nu}_\ell}{F_{\ell}^{\XY;\mu\nu} B_{\ell}^{\T;\mu} B_{\ell}^{\T;\nu} P_{\ell}^2},
	\label{eq:bandpowers}
\end{equation}
where $\bar{C}^{\XY;\mu\nu}_\ell$ is the average of the cross-power spectra calculated from different bundles from \cref{eq:crosspower} and other terms have been defined above.

We also calculate $Q^{\XY;\mu\nu}_{b\ell}$, the band power window functions that summarize all the effects of debiasing and binning and  relate the theoretical sky signal power spectra to the data vector:
\begin{align}
  \label{eq:bpwf}
  Q^{\XY;\mu\nu}_{b\ell} = {F_{\ell}^{\XY;\mu\nu} B_{\ell}^{\T;\mu}B_{\ell}^{\T;\nu}P_{\ell}^2}\sum_{\ell'} &   \frac{Q_{b\ell'} K_{\ell'\ell}^{\XY}}{F_{\ell'}^{\XY;\mu\nu}B_{\ell'}^{\T;\mu} B_{\ell'}^{\T;\nu}P_{\ell'}^2} .
\end{align}
We use the band power window functions to build our likelihood in \cref{sec:likelihood,eq:datamodel}.

We show the final multi-frequency band powers in \cref{fig:bandpowers}, along with the best-fit predictions assuming \lcdm{} (see \cref{sec:lcdm}). Contributions from foregrounds are clearly seen at high $\ell$ in \TT{}. At this stage, we can already appreciate the excellent agreement between the different cross-spectra. We discuss the internal consistency of the data in \cref{sec:powspenulltests}.

\subsection{Noise power spectra}
\label{sec:noise}
\begin{figure*}
  \includegraphics[width=\textwidth]{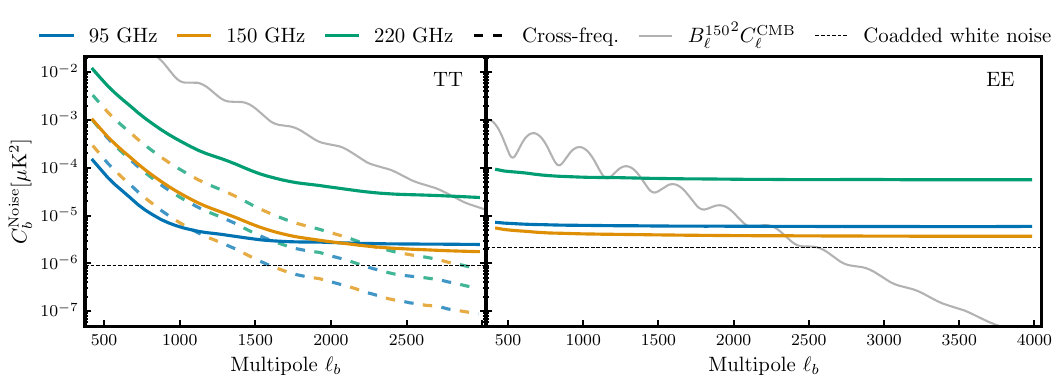}
  \caption{Noise power spectra of the maps based on SPT-3G D1 observations.
  Solid lines correspond to auto-frequency spectra ($95\,\ghz$: blue, $150\,\ghz$: orange, $220\,\ghz$: green), dashed lines to cross-frequency spectra (alternating colors according to the frequency combination: $95 \times 150 \,\ghz$: blue-orange, $95 \times 220 \,\ghz$: blue-green, $150 \times 220 \,\ghz$: orange-green). 
  The gray line shows the expected beam-convolved CMB signal for the \lcdm{} best-fit of the \sptbp data. 
  The horizontal dashed lines indicate the coadded white noise levels in temperature ($3.3\,\ukarcmin$) and \E{}-mode polarization ($5.1\,\ukarcmin$).
  We do not show the cross-spectra for \EE{} channels, as they are consistent with zero. 
  The noise spectra have been corrected for the transfer function (see \cref{sec:tf}) and the pixel window function. 
  The white noise levels in temperature and \E{}-mode polarization are 5.4, 4.4, and 16.2 $\ukarcmin$ and 8.4, 6.6, and 25.8 $\ukarcmin$ at $95$, $150$, and $220\,\ghz$, respectively. 
  The SPT-3G D1 \EE{} data is signal-dominated up to $\ell\sim2550$.}
  \label{fig:noise}
\end{figure*}

We use the noise realizations described in \cref{sec:coadd} to estimate noise power spectra. 
To estimate it, we take the average of the auto-power spectrum of  sign-flip maps over the $N_{\rm signfip}=500$ realizations, masked using the combination of the point source and border masks.
We show \TT{} and \EE{} noise power spectra in \cref{fig:noise}.

The temperature noise is dominated on large scales by contributions from the atmosphere, resulting in a characteristic red spectrum~\citep{kolmogorov_local_1941}.
At high multipoles, white instrumental noise dominates.
We report a white noise floor of 5.4, 4.4, and $16.2\,\ukarcmin$ at 95, 150, and 220\ghz{}, respectively,
and a coadded noise level of $3.3\,\ukarcmin$.
As noted in~\citep{balkenhol23}, 
the atmospheric noise is significantly correlated in temperature between different frequencies. 
At high multipoles, we expect the cross-frequency noise to vanish since the white noise of the detectors and readout is uncorrelated, 
and we confirm that the cross-frequency noise vanishes at sufficiently high $\ell$ up to the uncertainty of our estimation. 
We include the cross-frequency noise in the covariance estimation. 
We recall here that the cross-frequency noise only affects co-temporal data, i.e., bundles at different frequencies that see the same atmosphere at the same time. 
The cross-bundle based auto-frequency or cross-frequency band powers are not noise-biased. 

In polarization, the noise is dominated by white detector and readout noise and is hence nearly scale-independent.
We report a white noise level of 8.4, 6.6, and 25.8 $\ukarcmin$ at 95, 150, and 220\ghz{}, respectively, 
and a coadded noise level of $5.1\,\ukarcmin$.
These values are larger than the expected noise from temperature, $n^\P{} = \sqrt{2} n^\T{}$, as they are affected by the correction for the polarization efficiencies.
Atmospheric noise in polarization is subdominant on the angular scales probed by this analysis and hence cross-frequency noise is negligible (see~\citep{coerver24} for detailed discussion of polarized atmospheric noise). We find no evidence of significantly correlated noise between temperature and polarization. 

\subsection{Covariance matrix}
\label{sec:covariance}

In contrast with previous SPT-3G analyses which relied mostly on simulations for the power spectrum covariance estimation~\citep{dutcher21,balkenhol23}, here we use a semi-analytic procedure akin to what was used in \planck{} analyses (e.g.,~\citep{planck18-5}). 
We highlight in this section the different steps of the covariance estimation and where we differ from the \planck{} analysis.

Since the \mainfield field  covers 4\% of the sky, an analytical computation of the covariance matrix requires proper modeling of the mask coupling effect. 
The problem is also present in the \planck{} analysis, but the near-unity sky fraction observed by \planck{} enabled the use of a simpler approximation. 
In this work, we compute the band power covariance matrix based on the framework developed in~\citep{camphuis_accurate_2022}.
Following the notation in that work, we compute the covariance matrix of the biased pseudo-power spectrum $\tilde{C}_{\ell}^{\XY;\mu\nu}$ (i.e. before the \polspice{} debiasing) with
\begin{align}
	\label{eq:covariance_pseudo}
	 & \tilde{\Sigma}_{\ell\ell'}^{\XY\mu\nu;\ZV\alpha\beta} = \Xi_{\ell\ell'}^{\XY;\ZV}[W^2] \sum_{L_1,L_2} \\ &  \Big[
	\bar{C}^{\XZ\mu\alpha}_{L_1} \bar{\Theta}_{\ell\ell'}^{\XZ;\YV;L_1L_2}[W] \bar{C}^{\YV\nu\beta}_{L_2}  + ({\scriptstyle Z \leftrightarrow V}) \Big],\nonumber
\end{align}
where $\Xi[W^2]$ and $\bar{\Theta}[W]$ are coupling matrices describing the mask effect and are entirely computed from the apodized border-only mask $W$ described in \cref{sec:mask}.

Following \cref{eq:simsmodel}, the model signal is
\begin{equation}
	\label{eq:covariance_pseudo_model}
	\bar{C}_\ell^{\XZ\mu\nu} \equiv F_\ell^{\XY\mu\nu} B_\ell^{\mu} B_\ell^{\nu} P_\ell^2 \left(C_\ell^{\XY\mu\nu;\rm signal} + N_\ell^{\XY\mu\nu}\right).
\end{equation}
This model neglects the contribution of the inpainting and filtering artifact residuals to the covariance but does include the isotropic part of the filtering and \alm{} weighting, beams, and pixelization effects. 
At the end of the analysis, the signal part is evaluated at the best-fit cosmology and foreground model. 
For previous iterations of the analysis we used a best-fit \planck{} model, along with the foreground components used to generate simulated skies (\cref{sec:simulations}). 
The noise contribution is described in \cref{sec:noise} and includes the cross-frequency \TT{} noise. 
The noise maps are found to be statistically isotropic, and, contrary to some previous CMB studies~\citep{planck18-5,atkins_atacama_2024}, we do not include any anisotropic noise correction.

Similarly to the power spectrum, the covariance matrix is debiased from the mask effects. 
The debiasing is done using the  $G^{\XY}$ kernel from~\citep{camphuis_accurate_2022} that includes the mask debiasing and \polspice{} residual kernel. 
We further correct for the instrumental beam, transfer function, and factors of the pixel window function, such that
\begin{align}
	\hat{\Sigma}_{\ell\ell'}^{\XY\mu\nu;\ZV\alpha\beta}  =
	 & \frac{\sum_{L_1L_2} G^{\XY}_{\ell L_1} \tilde{\Sigma}_{L_1L_2}^{\XY\mu\nu;\ZV\alpha\beta} G^{\ZV}_{L_2\ell'}}{F_{\ell}^{\XY\mu\nu} B_{\ell}^{\mu}B_{\ell}^{\nu}P_{\ell}^2F_{\ell'}^{\ZV\alpha\beta} B_{\ell'}^{\alpha}B_{\ell'}^{\beta}P_{\ell'}^2}.
\end{align}

The data vector is formed by the averaged cross spectra of the 30 bundles. 
Instead of summing the covariance of each of the cross spectra, we follow~\citep{lueker10} by computing the covariance of the coadded map and correcting it following the methodology described in \cref{app:covcross}, so that 
\begin{align}
	\label{eq:covnoisecontribution}
	\hat{\Sigma} \rightarrow \hat{\Sigma} + \frac{1}{N_{\rm bundles}-1}\hat{\Sigma}^{{\rm noise}-{\rm only}}.
\end{align}

We discussed in \cref{sec:tf} that at the level of the covariance matrix, the anistropic contribution from the filtering and weighting can be modeled by a rescaling of the covariance, $H^{\XY;\mu\nu}_{\ell}$ estimated from simulations, so that

\begin{equation}
	\hat{\Sigma}_{\ell\ell'}^{\XY\mu\nu;\ZV\alpha\beta} \to \frac{1}{\sqrt{H^{\XY\mu\nu}_\ell H^{\ZV\alpha\beta}_{\ell'}}}\hat{\Sigma}_{\ell\ell'}^{\XY\mu\nu;\ZV\alpha\beta}.
	\label{eq:Hrescale}
\end{equation}
In order to avoid large-to-small scale couplings in the covariance, we have filled the map at the location of bright point sources and massive galaxy clusters with Gaussian constrained realizations of the CMB. Following the discussion in \cref{sec:inpainting}, we correct the covariance with
\begin{equation}
    \hat{\Sigma}_{\ell\ell'}^{\XY\mu\nu;\ZV\alpha\beta} \to \hat{\Sigma}_{\ell\ell'}^{\XY\mu\nu;\ZV\alpha\beta}(1 + \rho^{\XY;\mu\nu}_\ell \rho^{\ZV;\alpha\beta}_{\ell'})
    \label{eq:cov_inpainting}
\end{equation}
where $\rho_\ell$ is the inpainting ratio defined in \cref{eq:rhoell}. 
We compare the impact of the inpainting correction on the covariance matrix to the effect of point source masking in \cref{fig:covariance_masked_ptsrc}, and we show that the increase in variance from inpainting is much smaller than what would have been incurred from masking.

At this stage, we do not include other data processing effects in the covariance. 
Beam uncertainties, quadrupolar beam leakages, and polarized beam corrections are explored along with the CMB and foreground signal in the likelihood and, in contrast to the approach in \ACTDR{}~\citep{louis25}, we do not marginalize over them in the covariance.
We note that we marginalize over some foreground contributions in the covariance, as described in \cref{sec:foreground}, in order to speed up our likelihood evaluation. 

We account for the lensing contribution to the covariance by adding to the final covariance the lensing checkerboard term described in~\citep{benoitlevy12}, computed using \fisherlens\footnote{\url{https://github.com/ctrendafilova/FisherLens}}~\citep{hotinli_benefits_2022}. 
As in previous SPT analyses~\citep{dutcher21,balkenhol23} and, in contrast with~\citep{louis25}, we do not include the contribution of super-sample lensing modes in the covariance, but marginalize over its effect in the likelihood.
Also, since we limit the \TT{} analysis to $\ell\le3000$, we ignore the non-Gaussian contributions to the covariance from foregrounds.
Finally, the resulting matrix is binned as we did for the band powers with the  $Q_{b\ell}$ matrix defined in \cref{eq:binmatrix}
\begin{align}
	\Sigma^{\XY\ZV}_{bb'} = \sum_{\ell,\ell'} Q_{b\ell} \hat{\Sigma}_{\ell\ell'}^{\XY\ZV} Q_{b'\ell'}.
	\label{eq:covariance}
\end{align}
\RR{The resulting binned covariance matrix exhibits a correlation of approximately 2.5\% between adjacent bins. This value is determined prior to any foreground marginalization, as described in \cref{sec:foreground}.}

The covariance matrix is highly conditioned because of the low noise levels, the high resolution, and the large number of band powers. 
Slight inconsistencies between the different frequencies in the data vector that are not modeled in the approximated covariance, and can be much smaller than the cosmological constraining power, have a large effect on the goodness-of-fit of any cosmological model solution. 
While this is not an issue when performing band-power difference tests discussed in the validation section of this article (\cref{sec:difference}), we fail our more stringent conditional tests (\cref{sec:conditional}) because the associated error bars are much smaller than the sample variance error bars.
We acknowledge this limitation of our covariance approximation and regularize the matrix by minimally increasing its diagonal.
This slightly decorrelates the different cross-frequency spectra.
Specifically, we add a small fraction of the sample variance to the diagonal
\begin{equation}
	\Sigma^{\XY\XY} \to \Sigma^{\XY\XY} + \alpha_{\XY}^2 \text{diag}\left(\Sigma^{\XY\XY}\right)
	\label{eq:fudge}
\end{equation}
where diag is an operator extracting only the diagonal of the matrix.
The regularization factor is chosen to be $\alpha_{\TT} = 0.1\%$, $\alpha_{\TE} = \alpha_{\EE} = 1\%$, so that the precision of our tests (consistency tests or goodness-of-fit) cannot be sensitive to effects which are below 1\% of the sample variance error bars. 
This factor is similar to that used in the map null tests, see~Q25. 
We measured on \fullmock{} simulations that the regularization decreases the average $\chi^2$ by 60 points. 
We checked that the  $\chi^2$ nevertheless follows a $\chi^2$ distribution with the expected number of degrees of freedom, allowing us to easily interpret it.
Finally, we verified that, as expected, the regularization step has no measurable impact on cosmological results.

We describe the validation of the covariance matrix in \cref{sec:covmatval} and show that it is consistent with the \fullmock{} covariance matrix.

\begin{figure}
	\includegraphics[width=\columnwidth]{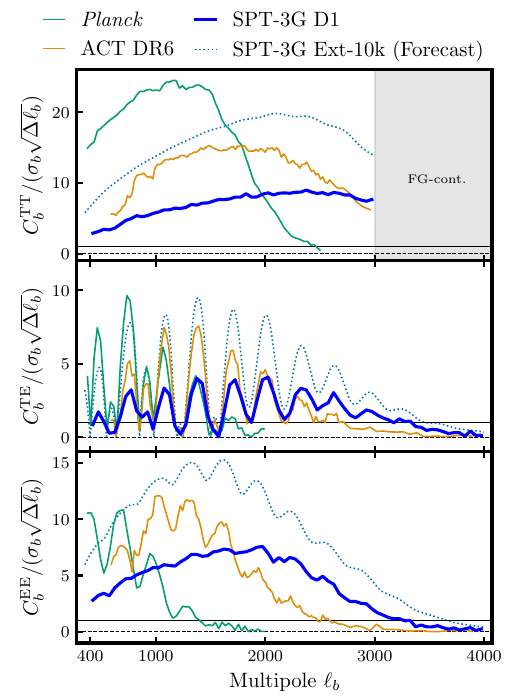}
	\caption{Signal-to-noise ratio (SNR) of the minimum-variance band powers for \sptbp (blue), \planck{} (green), and \ACTDR{} (orange). 
	At multipoles $\ell > 1800$, \sptbp achieves the highest sensitivity in \EE{}. 
	For \TE{}, the signal-to-noise ratio is similar to that of \ACTDR{} between $\ell=1800$ and $\ell=2200$ and exceeds it at higher multipoles.
	In dotted blue, we show the forecasted signal-to-noise for the complete SPT-3G \allspt{} data, including all fields (see \cref{fig:footprints} and~\citep{prabhu24}).
	In that line, we included the effect of filtering assuming the simple model of the transfer function described in \cref{fig:transfer_function} and the relation of \cref{eq:Hl_Fl}.
	We do not display the \TT{} SNR above $\ellmax^{\T}=3000$, as the minimum-variance band powers become correlated due to foreground contamination and as these band powers are not measured in this work.
	The black dashed line indicates SNR$=0$, while the solid black line indicates SNR$=1$.
	}
	\label{fig:snr}
\end{figure}

The resulting covariance matrix can be used to investigate the contribution of each of the frequency channels to the minimum-variance combination of the
band powers, as discussed in \cref{app:mixingmatrix}. We build the minimum-variance band powers using the covariance matrix, as described in
\cref{eq:mixingmatrix}, and the associated minimum-variance covariance matrix. Applying the same steps to \planck{} and \ACTDR{} data, we show the
resulting signal-to-noise ratio in \cref{fig:snr}.
We note that the bin width factor is divided out in this plot, so that the SNR for all experiments is shown with an effective bin width of $\Delta \ell = 1$.
The \sptbp data are the most sensitive in \EE{} and \TE{} at $\ell= 1800$-$4000$ and $\ell= 2200$-$4000$\footnote{At higher
	multipoles, the tightest constraints were published in~\citep{chou25}.},
respectively. In the multipole range $\ell= 1800$-$2200$ in \TE{}, the constraining power of SPT-3G is comparable to that of \ACTDR.
The signal-to-noise per multipole is greater than one at
$\ell \lesssim 3300$ in both \TE\ and \EE.

Uncertainties in the \sptbp{} band powers are dominated over much of the $\ell$ range by the large sample variance resulting from the small
sky fraction used in this analysis. 
For example, the signal contribution to the covariance is larger than the noise contribution up to $\ell \simeq 2600$ in \EE, and over the full range reported here in \TT. 
This makes the signal-to-noise per bin 
at large scales lower than that for \planck{} and
\ACTDR{}, which used larger sky areas. 
This limitation will be largely overcome by the addition of the SPT-3G Summer and Wide fields, with observations obtained on a total of $\sim 10\,000\,\sqdeg$, 25\% of the sky. 
The forecasts for the full SPT-3G \allspt{} survey shown in \cref{fig:footprints}, including the effect of filtering as described in \cref{eq:Hl_Fl}, are shown as blue dotted lines in the plot.

\section{Likelihood}
\label{sec:likelihood}

The temperature and polarization band powers described in \cref{sec:bandpowers} and shown in \cref{fig:experiments} and \cref{fig:bandpowers} are one of the two major results of this paper. 
The other is the constraints on cosmological parameters that these unprecedentedly sensitive power spectrum measurements enable. 
To go from one to the other, we need a framework in which to compare the measured band powers to cosmological models; we describe that framework in this section. 
The four primary components to this framework are: 
(1) the likelihood function quantifying the probability of obtaining the measured band powers given an underlying model;  
(2) the cosmological model and the procedure to produce model band powers from it; 
(3) the model of the instrument, data analysis, and foreground contamination that transforms the theory band powers into a data vector we can compare directly to the measured data; 
and (4) the method for exploring posterior distributions of the cosmological and data-model (nuisance) parameters. 
We discuss all of these in detail below.

\subsection{Likelihood function}

On the angular scales probed in this analysis, even given our limited sky coverage, each band power bin averages over a sufficiently large number of independent modes that the distribution of band powers can be well approximated as Gaussian~\citep{gerbino20}.
Under this assumption, the formal likelihood $\cal{L}$, i.e. the probability of the data given the model, is
\begin{equation}
	\begin{split}
		&-\ln{\cal L}(\hat{C} | C^{\mathrm{model}}(\theta)) \propto \\ & \qquad \frac{1}{2} \left[\hat{C}_b - C^{\rm model}_b(\theta)\right] \Sigma_{bb'}^{-1}
		\left[\hat{C}_{b'} - C^{\rm model}_{b'}(\theta)\right],
	\end{split}
	\label{eq:basic-likelihood}
\end{equation}
where $\hat{C}_b$ represents the binned power spectrum estimates (band powers) after all debiasing steps, defined in \cref{eq:simsmodel} and \cref{eq:bandpowers}; 
$C^{\rm model}_b$ are the model band powers, including CMB and foreground contributions, and transformed into a quantity comparable to the data vector as described below; 
$\Sigma$ is the band power covariance matrix described in \cref{sec:covariance};  and summation over repeated indices is implied. The full data vector $\hat{C}_b$ consists of 21 auto- and cross-frequency \TT{}, \TE{}, and \EE{} spectra (including \TE{} and \ET{} separately for cross-frequency pairs) for a total of 1392 elements.

The full set of model parameters $\theta$ consist of {cosmological parameters} $\psi$ (six \lcdm{} parameters plus any extension parameters), 
which are handed to Boltzmann solver routines or emulators to create theory CMB power spectra (see \cref{sec:theory_codes}), 
and 43 {nuisance parameters} $\phi$, which are used along with the band power window functions (\cref{sec:bandpowers}) to transform the theory CMB spectra into $C^{\rm model}_b$, the model band powers appropriate for comparing to our data band powers. 
This transformation, which we refer to as the data model, includes both additive and linear corrections to the theory spectra, accounting for binning and residual mask effects, astrophysical foregrounds, 
and any residual systematic biases not fully removed in the processes described in \cref{sec:power_spec}.
The data model is performed within the likelihood implementation described in \cref{sec:candl_tricks}.

In the following sections, we discuss the data model in detail, 
including implementation details and efforts to speed up the computation. 
We also introduce the different codes we relied on in this work to evaluate the CMB power spectra and briefly describe their relative merits.

\subsection{Data model}
\label{sec:data_model}

The theory CMB spectra are computed externally to our likelihood implementation, while the transformation of these spectra into band powers that can be compared with the data (including the addition of foreground spectra) occurs within our likelihood code. 
With this in mind, we write the transformation, which we refer to as the data model for our likelihood, as follows:
\begin{widetext}
	\begin{equation}
		\label{eq:datamodel}
		C^{\rm model}(C^{\rm CMB}(\psi), \phi) = 
		\mathbb{A}_{\rm cal}
		\cdot
		\mathbb{\Epol}_{\rm cal}
		\cdot
		\mathbb{Q}
		\cdot
		\mathbb{B}^{\rm P} \left(\betapol\right)
		\cdot
		\mathbb{B}^\Delta \left(\beta_{i}\right)
		\cdot
		\mathbb{L}\left(\epsilon_{2}\right)
		\cdot
		\left[
			\mathbb{A} \cdot \mathbb{S}\left(\kappa\right) \cdot C^{\rm CMB}(\psi)
			+ C^{\rm fg}\left(\phi^{\rm fg}\right)
			\right],
	\end{equation}
\end{widetext}
where the nuisance parameters $\phi$ have been split into foreground parameters $\phi^{\rm fg}$ and individual systematic parameters;
double line symbols (such as $\mathbb{Q}$) denote linear operators that can depend on nuisance parameters;
and indices indicating spectrum types (i.e. \TT{}, \EE{}, \TE{}), frequency pairs, or band power bins have been suppressed for clarity.
The different ingredients of this long equation are, from left to right:
\begin{itemize}
	\setlength{\itemsep}{2pt}
	\item $\mathbb{A}_{\rm cal}$: calibration factors; one external calibration $A_{\rm cal}^{\rm ext}$, and the relative calibration factors compared to the 150\ghz{} channel, $A^{\rm rel;95}_{\rm cal}$ and $A^{\rm rel;220}_{\rm cal}$
	\item $\mathbb{\Epol}_{\rm cal}$: polarization efficiencies, affecting only the \TE{}, \ET{}, and \EE{} spectra.
	They are similarly separated into an external parameter $\Ecal$, and relative polarization efficiencies compared to the 150\ghz{} channel, $\Ecal^{\rm rel;95}$ and $\Ecal^{\rm rel;220}$.
	\item $\mathbb{Q}$: the band power window functions defined in \cref{eq:bpwf}, accounting for the binning and residual \polspice{} mixing.
	\item $\mathbb{B}^{\rm P}$: the polarized beam correction (\cref{sec:polbeams}) defined in \cref{eq:polbeams}.
	\item $\mathbb{B}^\Delta$:  the beam error modes operator which we use to propagate the error on the effective beam measurement (\cref{sec:beams}), defined in \cref{eq:beammodes}.
	\item $\mathbb{L}$: the quadrupolar beam leakage (\cref{sec:quadrupolar}) defined in \cref{eq:t2p}.
	\item $\mathbb{A}$: the aberration defined in \cref{eq:aberration}.
	\item $\mathbb{S}\left(\kappa\right)$: the super-sample lensing correction as defined in \cref{eq:ssl}.
	\item $C^{\rm fg}(\phi^{\rm fg})$: the additive foreground contamination model. It does not depend on the CMB signal, and we describe its content in \cref{sec:foreground}.
\end{itemize}
We detail the different linear and additive corrections in the following two sections: \cref{sec:sys} for the linear operators and \cref{sec:foreground} for the additive foreground model. \cref{tab:nuisance-priors} and \cref{tab:foreground-priors} list all of the nuisance parameters and, when relevant, the priors we use.
We note that we do not include the uncertainty associated with filtering artifacts (see \cref{sec:filtering-artifacts}) and inpainting (see \cref{sec:inpainting}) in the likelihood. 
We have tested that the impact of these effects on the cosmological parameters we constrain is less than $0.1\,\sigma$.

Finally, we note that after unblinding, the foreground model was updated and the \TtoP leakage corrections and polarized beam modeling beam were added. We only report the final model here and discuss the changes in detail in \cref{app:post-unblind}.
 
\subsubsection{Linear corrections}
\label{sec:sys}

Most of the linear corrections arise from the non-idealities of the instrument and data processing and depend on parameters summarized in \cref{tab:nuisance-priors}.
Two of them, $\mathbb{S}$ and $\mathbb{A}$, are of astrophysical origin.

The super-sample lensing $\mathbb{S}$ is the distortion of the CMB on our field from weak gravitational lensing due to modes larger than the survey area.
Following~\citep{balkenhol23} and~\citep{manzotti14}, we treat the overall mean convergence in the survey field $\kappa$ as a free parameter and include its effect on power spectra as:
\begin{equation}
	\label{eq:ssl}
	\left[\mathbb{S}\left(\kappa\right) \cdot C\right]_\ell \equiv C_\ell - \frac{\partial\ell^2 C_\ell}{\partial \ln{\ell}} \frac{\kappa}{\ell^2}.
\end{equation}
We use a prior on $\kappa$ of $\mathcal{N}(0, 0.00045)$ based on~\citep{dutcher21}.\footnote{Note that as defined above $\kappa>0$ corresponds to a de-magnification effect.}

The aberration $\mathbb{A}$ accounts for the distortion due to the motion of the Earth with respect to the rest frame of the CMB~\citep{jeong14}. 
The \mainfield{} field covers a small fraction of the sky and we do not explore the forward/backward symmetry of the local dipole, in which case a first-order approximation to the aberration is sufficient:
\begin{equation}
	\label{eq:aberration}
	\left[\mathbb{A} \cdot C\right]_\ell \equiv C_\ell - \beta_{\rm ab}  \ell  \frac{\partial C_\ell}{\partial \ell}.
\end{equation}
The correction depends on the dipole and mean survey angle relative to the direction of the dipole, the product of which~\citep{balkenhol23} computed for the \mainfield{} survey as $\beta_{\rm ab} = -0.0004826$.\footnote{We note that we only apply the aberration correction to the CMB, while~\citep{louis25} applies it to both CMB and foregrounds.}

The group of nuisance operators, $\mathbb{L}$, $\mathbb{B}^\Delta$, and $\mathbb{B}^{\rm P}$ all describe the limitations of our beam modeling and debiasing. 
The first of the operators, $\mathbb{L}$, encodes the quadrupolar temperature-to-polarization leakage described in \cref{sec:quadrupolar} and \cref{eq:t2p}. 
We describe in \cref{app:t2p} how we propagate the leakage measured on point source maps to the priors on the parameters of $\mathbb{L}$.

Like the analyses of~\citep{planck13-7, planck13-15,henning18}, we treat uncertainty on the temperature beam at the parameter level (as opposed to including it in the power spectrum covariance, as in~\citep{planck18-5, balkenhol23, louis25}).
We first estimate a beam covariance matrix by varying the AGN and Saturn observations included in the stitched beam, as well as parameters that explore our analysis choices and instrument systematics (see H25 for details).
We then extract from that matrix nine eigenmodes, discarding eigenvalues smaller than 1\% of the largest eigenvalue, and include the amplitudes of these eigenmodes as free parameters in the likelihood, in order to fit for potential residual systematic errors and marginalize over the associated uncertainty.
We introduce the beam error modes operator $\mathbb{B}^\Delta$, which modifies power spectra as:
\begin{align}
	\label{eq:beammodes}
	\mathbb{B}^\Delta \cdot C \equiv  \left(1+ \sum_{i=1}^{9} \beta_i B^{\Delta,\mu}_i\right)\left(1+ \sum_{i=1}^{9} \beta_i B^{\Delta,\nu}_i\right)C^{\mu\nu},
\end{align}
where $B^\Delta_i$ is the $i$th beam error mode and $\beta_i$ is the amplitude of that mode. We include nine error modes and thus add nine free $\beta_i$ parameters to the model. 
The diagonalized beam covariance matrix spans the full frequency and spectrum (\TT{}, \TE{}, \EE{}) space, so that the $B^\Delta_i$ error modes span this full space and include any correlations of beam error between frequency pairs and spectra. 
The modes are normalized by their eigenvalues, so we use uncorrelated standard normal priors for each of them. We show in \cref{tab:nuisance-priors} that in our standard cosmological fits all $\beta_i$ are compatible with zero. 
Because some of the constraints in the standard fits appear prior-dominated, we also checked that after widening the priors by a factor of 10 the constraints on all $\beta_i$ remain compatible with zero.

Finally, the $\mathbb{B}^{\rm P}$ operator accounts for the difference between the beams in temperature and polarization. Using the notation in \cref{sec:polbeams}, this operator is defined through the relations:
\begin{align}
	\mathbb{B}^{\rm P} \cdot C^{\EE;\mu\nu} &\equiv  \frac{B^{\T;\mu} B^{\T;\nu}}{B^{\P;\mu}\left(\betapol^\mu\right)B^{\P;\nu}\left(\betapol^\nu\right)}C^{\EE;\mu\nu},\\
	\mathbb{B}^{\rm P} \cdot C^{\TE;\mu\nu} &\equiv  \frac{B^{\T;\mu}}{B^{\P;\mu}\left(\betapol^\mu\right)}C^{\TE;\mu\nu}.
\end{align}
As discussed in \cref{sec:polbeams}, when $\betapol^\mu = 1$ the polarized beam at frequency $\mu$ is identical to the temperature beam, and when $\betapol^\mu = 0$ the polarized beam at frequency $\mu$ is equal to the model of the main temperature beam only (no sidelobes). 
In the likelihood we use uniform priors between 0 and 1 on all $\betapol^\mu$, and, as shown in \cref{app:nuisance_pars,tab:nuisance-priors}, $\betapol^\mu < 1$ is clearly preferred by our data in all frequency bands. 

The last two operators $\mathbb{A}_{\rm cal}$ and $\mathbb{\Epol}_{\rm cal}$ recalibrate the absolute gain and polarization efficiency, respectively, and propagate our uncertainty on the calibration of our maps (\cref{sec:calibration}) to the constraints on cosmology. 
In most cases, we only use an informative prior from our external calibration of the 150\ghz{} map on the \planck{} map, $A_{\rm cal}^{\rm ext}$ (\cref{sec:calibration}), and let the internal calibration parameters vary in flat priors as reported in \cref{tab:nuisance-priors}. 
The same is true for the internal polarization efficiency estimates. 
The external polarization efficiency is jointly fitted with all other cosmological plus nuisance parameters assuming a cosmological model~\citep{galli21}. 
We adopt a flat prior for the external polarization efficiency parameter, except when reporting results from \TE{} or \EE{} spectra only. 
In this case, certain cosmological parameters are fully degenerate with polarization efficiency and we break the degeneracy with our \planck-based prior.

\subsubsection{Foreground model}
\label{sec:foreground}

As has been documented in many results over the last 25 years, signals from the CMB are contaminated by many different astrophysical foregrounds and this contamination must be taken into account in cosmological modeling. 
Which foregrounds are necessary to model depends on many factors, including which area of sky and multipole range are targeted, what frequency bands are used, and whether the temperature or polarization power spectra (or both) are used (see, e.g.,~\citep{planck15-10} and~\citep{millea12} for reviews).

In this analysis, we model foregrounds in a manner similar to~\citep{balkenhol23}. We include Galactic (Milky Way) and extragalactic contributions in the model. 
For the Galactic model we include dust emission in both temperature and polarization, but we neglect synchrotron emission as it is expected to be negligible compared to Galactic dust at the frequency bands used in this work~\citep{planck18-4}.
Galactic dust emission affects the \TT{}, \TE{}, and \EE{} spectra.
Our extragalactic foreground modeling includes synchrotron and quasi-thermal dust emission from background galaxies and the thermal and kinematic Sunyaev-Zel'dovich effects (tSZ and kSZ), all in temperature only. 
Using improved estimates of the polarized point source contribution from~\citep{chou25} results in a prediction of negligible contribution from this component.

The  computation of each of the foreground components follows many published works and we defer the details of it to \cref{app:foreground_functions}. 
We show the total foreground contribution to the band powers in \cref{fig:foregrounds}.
We discuss here some specific choices made in the foreground modeling that are different than~\citep{balkenhol23}. 
The priors we place on foreground parameters, as well as the posterior constraints on those parameters in the context of the \lcdm{} model, are listed in \cref{tab:foreground-priors}.
We stress that the foreground model described here is intended to be a flexible phenomenological model that captures the main features of the foreground contamination in our data.
It is based on priors and templates validated within the baseline multipole range of this analysis.
It is expected to be inaccurate outside of this range.
Variations on this model have little impact on cosmological parameters, as explored in \cref{app:foreground_functions}.

We adopt the same Galactic dust model as in~\citep{balkenhol23}, featuring a modified black-body spectral dependence and a power law spatial dependence. 
The \mainfield{} field was designed to be far from the Galactic plane and the contribution from Galactic dust is predicted to be small compared to the CMB anisotropy in both temperature and polarization.
 The priors on Galactic dust in \cref{tab:foreground-priors} were derived from power spectra of \planck{} maps on the \mainfield{} field, with a color correction to account for the SPT-3G bandpasses.

We model the contribution in temperature from Poisson-distributed unresolved radio sources and dusty star-forming galaxies with a fixed Poisson power law ($D_\ell \propto \ell^2$) and a free amplitude parameter for each cross-frequency combination.
We set uninformative uniform priors $(0, 200) \ [\uksq]$ on the amplitude parameters, as they should be well determined by the data.

We model the clustering term of the dusty star-forming galaxy distribution (hereafter ``CIB clustering'') with an angular dependence parametrized by a power law index (similar to, e.g., ~\citep{george15,dunkley13}). 
We use the results of~\citep{mak_measurement_2017} to impose a prior on the power law index, though we widened the prior to $\alpha^{\rm CIB} \sim \mathcal{N}(0.53 \pm 0.1)$.
We do not attempt to model the spectral energy distribution (SED) of this term and instead adopt individual uncorrelated amplitudes for the $150\times150$, $150\times220$, and $220\times220\ghz{}$ cross-spectra.
We do not detect any contamination from CIB clustering in cross-spectra involving 95\ghz{} data but we nevertheless add a suitable constant contribution to the band power covariance matrix by propagating the prior from the 2018 analysis to the updated data model, see \cref{tab:foreground-priors}.

Finally, we model tSZ and kSZ signals using templates derived from the \agora{}~\citep{omori22} simulations.\footnote{The templates are derived from the version of \agora{} tSZ/kSZ maps with AGN heating temperature $10^8\,{\rm K}$. 
When deriving these templates, a mask is applied with the same emissive source flux cut of 6\,mJy at 150\ghz{} as used in this analysis.} 
We allow the amplitudes of these templates to freely float in the likelihood, with priors adapted from the 2018 analysis and based on~\citep{reichardt21}.
The tSZ amplitude parameter is defined at the \planck{} reference frequency of 143\ghz{} and the SED of this term is the standard tSZ frequency dependence relative to primary CMB fluctuations (see for example~\citep{shaw10}), taking into account color corrections due to our bandpasses and ignoring any relativistic corrections.
We marginalize over the tSZ-CIB cross-correlation by adding a constant contribution to the band power covariance matrix computed from the prior on tSZ-CIB cross-correlation used in~\citep{balkenhol23} (see \cref{app:foreground_functions,tab:foreground-priors}).
The kSZ SED is constant in CMB units.

\subsection{Building a robust likelihood with \candl{}}
\label{sec:candl_tricks}

We implement the likelihood code in \candl{}~\citep{balkenhol24}\footnote{\url{https://github.com/Lbalkenhol/candl}},
a python-based CMB likelihood library with \texttt{JAX} support.
Crucially, \texttt{JAX} exposes the code to an automatic differentiation algorithm, which allows for the easy and fast computation of accurate gradients.
By combining \candl{} with a differentiable theory code, we can then build a fully differentiable pipeline from cosmological and nuisance parameters, $\theta$, through to the likelihood value, $\cal{L}$.
This allows us to trivially evaluate the functions $\cal L(\theta)$, $\partial{ \cal L}/\partial \theta|_{\theta}$, and $\partial^2 {\cal L}/\partial\theta_i\partial\theta_j |_{\theta}$, which opens up a plethora of applications~\citep[see e.g.][]{campagne23, balkenhol24}.

In this analysis, we couple our \candl{} likelihood to a differentiable model, \cosmopower{} (described in \cref{sec:theory_codes}), to greatly increase our ability to test the robustness of our analysis pipeline.
Specifically, this unlocks two key tests:
\begin{enumerate}
	\item Shortcutting MCMC analyses. We perform a gradient-descent minimization using the truncated Newton-Raphson algorithm implemented in \texttt{scipy}~\citep{nash84, nocedal06, scipy} and then approximate the parameter posterior distributions as Gaussian by evaluating the Hessian at the best-fit point to obtain the Fisher matrix~\citep{heavens_generalisations_2016}.
	\item Translating biases in band powers to biases in cosmological parameters. By performing a Taylor expansion of the likelihood around the best-fit point one can show that, to first order, parameter biases $\delta\theta$ from band power biases $\delta D_\ell$ are given by $\delta\theta = F^{-1} \frac{\partial D}{\partial \theta} \Sigma^{-1} \delta D_\ell$~\citep{kable20}\footnote{We find higher-order contributions to be negligible.}, where $F$ is the Fisher matrix and $\Sigma$ is the band power covariance matrix (\cref{sec:covariance}).
\end{enumerate}
Both types of analyses can be performed in less than a minute, even for our high-dimensional multi-frequency likelihood.
This allowed us to propagate any change to the likelihood—whether to the data vector (e.g., the residual bias from the inpainting procedure), the band power covariance matrix (e.g., the lensing checkerboard), or the data model (e.g., the instrumental beam models)-to parameter constraints with negligible computational cost.
Still, since the above methods rely on certain approximations of the likelihood, they were used for testing purposes only; the final cosmological results presented in \cref{sec:pars} are calculated via traditional MCMC analyses.

\subsection{CMB-only likelihood}
\label{sec:lite_likelihood}

We follow the procedure of~\citep{balkenhol25} to construct a CMB-only, \emph{lite}, likelihood.
The underlying framework was first introduced by~\citep{dunkley13}.
In this procedure we extract the best-fit CMB band powers and covariance from the combination of cross-frequency spectra of the same \TT{}, \TE{}, or \EE{} channel, while marginalizing over nuisance parameters.
These products are then used to construct a simple Gaussian likelihood following the functional form of \cref{eq:basic-likelihood} that can be used to explore cosmological models.
The advantage of this approach is that it provides a fast and interpretable cosmological likelihood that compresses the information from different frequencies and marginalizes over foreground contamination and systematic effects.
For details on the framework we refer the reader to~\citep{dunkley13} as well as to previous applications of this framework~\citep{calabrese13, planck15-11, choi20, prince24, balkenhol25, louis25}.

Using the new approach put forward by~\citep{balkenhol25}, we exploit the differentiability of our multi-frequency likelihood implementation to perform the data compression quickly and accurately.
As in~\citep{balkenhol25}, we retain a global temperature and a global polarization calibration parameter (\Tcal{}, \Ecal{}) to minimize bin-to-bin correlations of the CMB-only band powers and account for the effect of aberration in the \emph{lite} likelihood.
The compression reduces the length of the data vector from 1392 to 196 and the number of operations in the data model from 18 to four (aberration, calibration of \TT{}, \TE{}, and \EE{} spectra).
This leads to a speed-up of a factor of 50 in the evaluation of the likelihood.
Together with the reduction of the number of nuisance parameters from 43 to two, this greatly speeds up MCMC analyses.
We refer to our compressed likelihood as \sptlite{}.

Results from \sptlite{} and the multi-frequency likelihood are consistent; in \lcdm{}, the central values of cosmological parameter posteriors shift by $\lesssim 0.1\,\sigma$ and their widths match to within $10\%$; this is compatible with MC noise.
The \emph{lite} likelihood is made publicly available alongside the multi-frequency likelihood on the SPT website.\footnote{\url{https://pole.uchicago.edu/public/data/camphuis25/}\\\url{https://github.com/SouthPoleTelescope/spt_candl_data}}
Note that the compressed likelihood has been constructed for the complete \TTTEEE{} data set; for constraints from individual spectra the multifrequency likelihood should be used.
We provide further details on the construction of the \emph{lite} likelihood and its performance in \cref{app:cmblite}.

\subsection{Theory codes}
\label{sec:theory_codes}

A key component of the likelihood is the step of computing predictions of the CMB power spectra for a given set of cosmological parameters.
The baseline cosmological model we use is the standard flat cold dark matter model with a constant dark energy component, \lcdm{}. The six parameters we use to parameterize \lcdm{} are: the physical density of baryons and dark matter, $\ombh$ and $\omch$, the amplitude and spectral tilt of initial scalar perturbations, $\As$ and $\ns$, the optical depth to reionization $\taureio$, and either the angular size of the sound horizon at recombination $\thetastar$ or the expansion speed of the universe today $\Hubble$. In \cref{sec:pars}, we also explore several extended models and we define the extension parameters in that section.
We assume one massive neutrino with $m_\nu=0.06\,\eV$, unless otherwise stated.
Definitions of all cosmological parameters appearing in this manuscript are provided in \cref{app:par_defs,tab:par_def}.

We generally use the Boltzmann solvers \camb~\citep{lewis11b} and \class~\citep{blas11} to compute the CMB power spectrum expected for a given cosmological model.
We use the same accuracy settings as the \ACTDR{} results~\citep{louis25}.
\RR{For recombination computations, we use \texttt{HyRec}~\citep{alihamound11,lee20}.\footnote{\texttt{CosmoRec}~\citep{chluba10} is also available and yields similar results.}}
While these codes are accurate, they are slow; for some purposes, we use two different interpolation methods to replace the full Boltzmann computations by quicker, approximated ones: \cosmopower{} and \OLE{}. We briefly introduce these two codes below and specify when we use them.

We use the models for the neural-network based \cosmopower{} emulator~\citep{spuriomancini22, piras23, jense25} developed in~\citep{balkenhol23} at the following times: during the development of the likelihood (to enable the methods described in \cref{sec:candl_tricks}), in the test of the CMB-only likelihood in \lcdm{} (\cref{sec:lite_likelihood,app:cmblite}), for the results presented in \cref{sec:cosmoparametertests}, and when testing for scale coherence in \cref{sec:lcdm_integrity}.
The emulator is trained on high-precision \camb spectra and covers \lcdm{}, as well as \alens{} and \neff{} extensions.\footnote{The metadata for the particular \cosmopower{} emulators we used is based on a training set with a higher accuracy than the one released with~\citep{spuriomancini22}. They are publicly available at \url{https://github.com/alessiospuriomancini/cosmopower/tree/main/cosmopower/trained_models/SPT_high_accuracy}}
The use of \cosmopower{} emulators over full Boltzmann solvers has two key advantages:
\begin{enumerate}
	\item once trained, the emulator runs $10^5$ times faster than \camb or \class~\citep{spuriomancini22}, and 
	\item it is differentiable, which facilitates several robustness tests (as discussed in \cref{sec:candl_tricks}).
\end{enumerate}
However, our use of \cosmopower{} is limited by the availability of models that have been trained on high-precision reference spectra and cover the necessary multipole ranges for primary CMB and CMB lensing power spectra.
One downside of the emulator compared to full Boltzmann solvers is also it does not return the full set of derived parameters of interest.
Still, the list of publicly available models grows and derived parameters can also be emulated or may be calculated cheaply during post-processing using the emulator itself~\citep{bolliet23, jense25}.

It is difficult and expensive to train \cosmopower{} emulators on all possible cosmological extensions.
We thus use a second code, \OLE{}, to compute the CMB power spectra for cosmological models not covered by \cosmopower{}.
\OLE{} is an online learning emulator framework; integrated within the MCMC exploration codes \cobaya and \montepython, it trains a Gaussian-process emulator model while exploring the posterior distribution of the parameters~\citep{guenther25}.
After a short training period, the emulator is used during the MCMC and regularly validated and retrained to improve its accuracy and the efficiency of the parameter exploration.
Inference with \OLE{} is more flexible than \cosmopower{}, as the training takes place on the fly, exclusively where the posterior distribution has most of its mass.
However, this also limits the speed gains compared to \cosmopower{} as the \OLE{} emulator covers a smaller area of parameter space, focusing only on the region where high likelihood values are possible, relying on slower computation for the low probability region.
During this work we encountered compatibility issues with the use of \muse{} \PP{} and \ACTDR{} \PP{} lensing likelihoods (see data set definitions in \cref{tab:dataset}) in \OLE{}
which limited our ability to exploit this promising tool at the time.
Still, we were able to use \OLE{} throughout \cref{sec:pars} for analyses without lensing data and to quickly obtain good proposal matrices for the other cases.
We refer the reader to~\citep{guenther25} for details on \OLE{}.

Finally, when using \class{}, we use \texttt{HaloFit}~\citep{HaloFit,Takahashi:2012em} to compute dark-matter only non-linear corrections, while we use \texttt{HMcode2020}~\citep{HMcode,HMcode2020} when using \camb{}. 
To assess the impact on our results, we first compared the primary and lensing CMB spectra for most of the models considered in \cref{sec:cosmo_data} using \texttt{HaloFit} or \texttt{HMcode2020}.
Note that the current version of \texttt{HaloFit} included in \class incorporates corrections due to massive neutrinos or the CPL parametrization of dark energy~\citep{HaloFit_Mnu,HaloFit_w0wa}.
We found the relative differences between the two to be well below experimental uncertainties at all relevant scales. 
Moreover, as an additional test, we did runs using \texttt{HMcode2020} 
for $\mnu$ with data sets \SPT+\ACT+\WMAP+\DESI{}~\citep{bennett13} 
and $\neff$ with data sets \ground{} + \DESI{} and compared them to equivalent runs with \texttt{HaloFit}. 
We find no deviations in parameter constraints beyond the $0.2\,\sigma$ level between the two codes. 
However, we should caution that true non-linear corrections, especially for non-classical extensions of \lcdm{}, 
might require performing N-body simulations, 
which is beyond the scope of this work (see~\citep{trendafilova_end_2025} for the impact of non-linear corrections on cosmological analysis).


\section{Validation of the analysis pipeline}
\label{sec:validation}

In this section, we discuss the process for verifying that the model formulated in \cref{sec:likelihood} accurately describes the measured data from \cref{sec:power_spec}.

\subsection{Blinding}
\label{sec:blinding}

Building on~\citep{balkenhol23}, we constructed and validated our analysis pipeline ``blind,'' i.e. restricting ourselves from looking at certain results until a series of robustness tests have been passed.
This methodology was designed to mitigate confirmation bias and, more concretely, to avoid stopping investigations early once results align with expectations or unconsciously modifying the analysis to achieve desired results.

During the blind stage of the analysis, we did not allow ourselves to compare the measured band powers to those from any other experiment, including previous SPT analyses\footnote{While the absolute calibration in temperature and polarization is obtained by comparing to \planck\ data (see \cref{sec:calibration}), these comparisons are performed only over the \mainfield{} field mask; at no point before unblinding did we compare SPT-3G band powers to the full-sky \planck\ power spectra.} or theoretical predictions. 
When deriving cosmological parameters for internal consistency checks or assessing the impact of analysis choices, the mean values were either hidden or systematically offset by an unknown amount. 
Finally, we did not perform any comparison that depends on the cosmological model, such as the cosmological parameter consistency test between \TT{}, \TE{}, and \EE{} data, during the blind period.

These restrictions were lifted once the pre-defined tests detailed below were successfully passed. 
We committed to publishing the obtained results and documenting any post-unblinding modifications to the pipeline.
For each test, a pre-defined passing criterion was established based on an associated probability-to-exceed (PTE), see \cref{app:pte}. 
The PTE threshold was set to $0.05/N$, where $N$ represents the total number of independent tests performed and incorporates the correction for the look-elsewhere effect~\citep{dunn61}.

We performed the following consistency tests before unblinding: (1) null tests at the map level (see \cref{sec:null} and~Q25), (2) differences between  frequencies at the power spectrum level (\cref{sec:difference}), (3) conditional frequency tests at the power spectrum level\footnote{Note that these two tests are slightly cosmology-dependent, since they require subtracting or correcting for any frequency-dependent contribution from each spectrum in order for them to be compared. 
From a practical point of view, this entails calculating a best-fit model assuming \lcdm{} and using the inferred nuisance (foreground plus instrumental) parameters to correct the spectra before comparison. 
The relatively limited correlations between nuisance and cosmological parameters allow us to proceed in this way.} 
(\cref{sec:conditional}), and (4) differences of \lcdm{} parameters obtained from different frequencies (\cref{sec:cosmoparametertests}). 
Finally, we ensured that our pipeline is unbiased and robust to differences in cosmological parameters by performing all our consistency tests on mock observations, both for our fiducial cosmology and alternate models.

While this analysis was still in the blind stage, the \muse{} pipeline underlying~\citep{ge24} was mature enough to unblind those results.
The \muse{} analysis uses exclusively polarization data to estimate the unlensed \EE{} power spectrum and lensing power spectrum \PP{}, while the pipeline presented here also includes temperature information and estimates lensed \TTTEEE{} power spectra. For these reasons, we do not expect identical results from the two analyses.
Initial results on \lcdm{} parameters from \muse{} were communicated across the collaboration on April 1st, 2024, after which key people working on this analysis were shut off from any further discussion of \muse{} results.
None of the \muse{} findings and comparisons conducted after unblinding and changes to the \muse{} pipeline were communicated to the analysis group for this work until they were ready to unblind as well.

We unblinded on September 9, 2024, after passing all pre-defined tests. 
Subsequently, we identified two previously unaccounted-for systematic effects in our data---quadrupolar temperature-to-polarization leakage (\cref{sec:quadrupolar}) and depolarization of beam sidelobes (\cref{sec:polbeams})---and we made several other minor updates. 
The quadrupolar leakage is seen most clearly in stacked maps of bright point sources, while the beam sidelobe depolarization manifests in part as a subtle inconsistency between the \EE{} band powers from different frequencies, just below our blinding threshold.
Although our data passed the pre-defined consistency tests, 
these systematics resulted in failures of the consistency tests in the $\ell \in [3000, 4000]$ range for \TE{} and \EE{}. 
These multipole ranges had initially been excluded from the analysis and were only reintroduced after unblinding.
While these corrections were made post-unblinding, and discovered at least in part through inconsistencies in cosmological fits, we emphasize that there is strong cosmology-independent evidence for them, as detailed in \cref{app:post-unblind}.
We are confident that the final pipeline provides a more accurate description of the data.
Prior to these corrections, there were significant inconsistencies among the \lcdm{} cosmological results derived from the individual \TT{}, \TE{}, and \EE{} spectra, resulting in unreliable combined \TTTEEE{} constraints and a poor overall fit, with the $\chi^2$ statistic exceeding expectations.
Consistent with our blinding protocol, we did not perform any tests strongly dependent on the cosmological model—such as cross-checks between \TT{}, \TE{}, and \EE{}—before unblinding.
Addressing these systematics substantially improved the agreement between temperature and polarization data within \lcdm{}, as reflected in the improved fit quality.
We compare cosmological constraints before and after the post-unblinding improvements in \cref{fig:unblinding}.

In the rest of this section we describe our suite of robustness tests, performed with the baseline likelihood, which includes post-unblinding corrections.

\subsection{Power-spectrum level tests}
\label{sec:powspenulltests}

We perform two types of consistency tests at the power spectrum level:
(1) we assess the difference between power spectra estimated using different frequency pairs (hereafter ``frequency spectra'') and (2) we compare a given frequency spectrum to the prediction for that spectrum conditioned on all other spectra.
These tests have similarly been used in \planck{}~\citep{planck18-5} and SPT-3G 2018 analyses~\citep{balkenhol23}.
As discussed in \cref{sec:covariance}, these tests are highly sensitive, with uncertainty margins significantly smaller than the band power error bars across most of the multipole range considered.
This sensitivity is achieved by canceling common fluctuations, notably the sample variance, between spectra and enables stringent assessments of frequency-dependent systematic effects, such as residual foreground contamination, inter-frequency calibration, and beam effects.
Notably, these tests are insensitive to systematic effects which impact all observation bands identically.
While the covariance regularization described in \cref{sec:covariance} make our tests insensitive to inconsistencies that are below 1\% of the level of sample variance uncertainty, possible features of this size are irrelevant for our cosmological analysis.

\begin{figure*}
	\includegraphics[width=\textwidth]{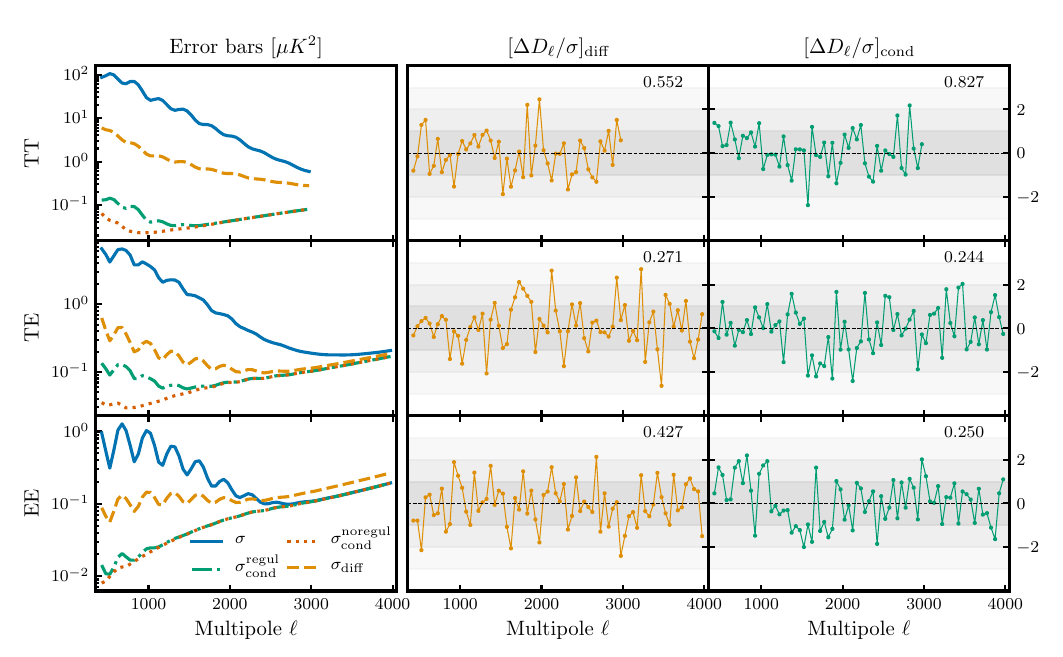}
	\caption{Example of power spectrum consistency tests.
	The left column shows the error bars for the two $150\times150$\ghz{} band-power tests (difference test: dashed orange, conditional test: dash-dotted green) and compares them to the uncertainty of the power spectrum measurement (solid blue).
	We also show the error bars of the conditional test without covariance regularization, labeled `noregul', for reference (dotted red).
	The central column displays the difference test for ($150\times 150-150\times95$\ghz), while the right column shows the conditional prediction test for the $150\times150$\ghz{} spectrum, both in relative units, i.e. deviations with respect to the expected uncertainty from the first column.
	PTEs are indicated in the top right corner of each panel.
	All tests across all spectra and frequencies pass, indicating the exceptional consistency of the measured data. The rest of the tests are shown in \cref{app:consistency} \cref{fig:TTdifference,fig:TEdifference,fig:EEdifference,fig:freq_cond}.
	}
	\label{fig:errorbars}
\end{figure*}

As an example, in \cref{fig:errorbars} we compare the error bars for the $150\times150$\ghz{} band powers with those associated with the difference test ($150\times 150-150\times95$\ghz) and the conditional prediction test for the $150\times150$\ghz{} spectrum.
Uncertainties are shown both with (`regul') and without (`noregul') covariance regularization.
Notably, at large angular scales the conditional error bar with covariance regularization is two orders of magnitude smaller than the band-power error bars.
Without covariance regularization, this reduction would reach three orders of magnitude for \TT{} at large scales, highlighting the exceptional sensitivity of these tests.
While the conditional test is more stringent than the difference test, the latter is easier to interpret, which is why we perform both.
Both tests operate on CMB-only spectra and we remove the best-fit \lcdm{} foreground and systematics contamination from our measured data.
As such, there is a small model dependence to these tests.

\subsubsection{Frequency difference}
\label{sec:difference}

The first test we perform is the difference between pairs of frequency spectra.
We correct each frequency spectrum for the contribution of foregrounds and systematic effects as described above, such that the spectra contain only the common CMB signal.
We then calculate the difference of a pair of spectra as:
\begin{equation}
	\label{eq:difftestspec}
	\Delta = \hat{C}^{\mu\nu}-\hat{C}^{\alpha\beta}
\end{equation}
and the covariance of the difference $\Sigma^{\Delta}$ as:
\begin{align}
	\label{eq:difftestcov}
	\Sigma^{\Delta} = \Sigma^{\mu\nu;\mu\nu}+\Sigma^{\alpha\beta;\alpha\beta} - \Sigma^{\mu\nu;\alpha\beta} - \Sigma^{\alpha\beta;\mu\nu}
\end{align}
where $\Sigma^{\mu\nu;\mu\nu}$ and $\Sigma^{\alpha\beta;\alpha\beta}$ are the blocks of the covariance matrix for the two power spectra and $\Sigma^{\mu\nu;\alpha\beta}$ is the cross-covariance between the two.
For each test, the number of degrees of freedom is equal to the number of bins in the difference spectrum, \ie{} 52 for \TT{} and 72 for \TE{} and \EE{}, as we cut \TT{} data at $\ellmax^\T=3000$, see \cref{sec:bandpowers}.
In total, there are five independent tests for \TT{} and \EE{} and eight for \TE{} spectra, such that the corresponding PTE threshold to pass is  $0.05/18 = 0.0028$.

We list the PTEs of all tests and show the difference spectra in \cref{app:frequencydiff,fig:TTdifference,fig:TEdifference,fig:EEdifference}.
All tests pass, with the smallest PTE being 0.067, and no striking features that may indicate significant residual foreground contamination, calibration offsets, or beam mismodeling.
In fact, all of the difference tests also pass even when not applying the covariance regularization (see \cref{eq:fudge}), signaling good consistency across frequencies.

\subsubsection{Conditional spectra}
\label{sec:conditional}

The second way to test inter-frequency consistency is to calculate the difference between a frequency spectrum and its conditional prediction obtained from all the other frequency spectra~\citep{planck15-11,planck18-5,balkenhol23}.
The conditional prediction is obtained by decomposing, for each frequency combination $\mu\nu$, the data vector in two blocks
\begin{equation}
	{\hat C} = \left[ {\hat C}^{\mu\nu}, {\hat C}^{\overline{\mu\nu}}\right]
\end{equation}
where ${\hat C}^{\mu\nu}$ is the frequency spectrum being considered and ${\hat C}^{\overline{\mu\nu}}$ is the vector containing all the other frequency spectra. We similarly decompose the model vector, $C$, and the covariance matrix, $\Sigma$.
For a Gaussian likelihood, the conditional prediction is
\begin{equation}
	{C}^{\mu\nu|{\rm cond}} = {C}^{\mu\nu}+ \Sigma^{\mu\nu;{\overline{\mu\nu}}}{\Sigma^{{\overline{\mu\nu}};{\overline{\mu\nu}}}}^{-1}\left({\hat{C}}^{\overline{\mu\nu}} - {C}^{\overline{\mu\nu}} \right),
	\label{eq:condcoadd}
\end{equation}
with an associated covariance of
\begin{equation}
	\Sigma^{\mu\nu;\mu\nu|{\rm cond}} =\Sigma^{\mu\nu;\mu\nu} - \Sigma^{\mu\nu;{\overline{\mu\nu}}}\left(\Sigma^{{\overline{\mu\nu}};{\overline{\mu\nu}}}\right)^{-1} \Sigma^{{\overline{\mu\nu}};\mu\nu}.
	\label{eq:condcov}
\end{equation}
We then take the difference between the measured data and the conditional prediction, i.e. $\Delta={\hat C}^{\mu\nu}-{C}^{\mu\nu|{\rm cond}}$ and calculate a $\chi^2$ test statistic using the covariance above and the relevant block of the band power covariance matrix.
Similar to before, for each test the number of degrees of freedom is equal to the number of bins in the conditional spectrum and the total number of independent tests is the same as for the difference spectrum test.

We show the conditional test residuals and report their associated PTEs in \cref{fig:freq_cond}.
All PTEs lie above the pre-defined threshold.
Remembering the small error budget of this test shown in \cref{fig:errorbars}, this is a strong sign for internal consistency.
Given that this test, as well as the difference test, passes, we conclude that the differences between frequencies in our data are well-described by our data model and the band power covariance matrix.

\subsection{Testing the pipeline on simulations}
\label{sec:pipesims}

We validate the analysis pipeline on \fullmock{} simulations, treating them in the same manner as we do for the data.
This validation is done to ensure that the covariance matrix is accurate and to confirm that the pipeline is unbiased and insensitive to the fiducial cosmology.

\subsubsection{Covariance matrix validation}
\label{sec:covmatval}

We begin by validating the analytic covariance matrix described in \cref{sec:covariance}.
We do so individually for the three constituents of the covariance---the sample variance, the chance correlation, and the noise variance terms---as well as for the complete matrix.
First, we validate the sample variance term of our analytical covariance against $500$ \fullmock{} simulations.
When not masking point-sources in the simulations, we find that the covariance of the signal-only power spectra matches our analytical estimate.
When point sources are masked, we recover the expected deviations discussed in \cref{sec:inpainting}.

Second, we validate the noise variance term.
We feed only the noise term of \cref{eq:covariance_pseudo_model} into \cref{eq:covariance_pseudo} and compare this with the variance of the noise power spectra measured on the $500$ noise realizations.
We report agreement between the two estimates of the noise variance term.
This test validates not only our covariance matrix framework, but also the computation of the four-point transfer function introduced in \cref{eq:Hl_Fl}, as we compare our purely analytical prediction to an entirely data-based estimator.

Third, we validate the chance correlation term.
We use our data bundle cross-spectra in the estimator presented in~\citep{lueker10} to obtain the combined chance correlation and noise variance contribution to the covariance.
We compare the resulting matrix to our analytical covariance after subtracting the sample variance contribution from the latter.
We report good agreement given the expected precision for the number of available bundles.

\begin{figure}
	\includegraphics[width=\columnwidth]{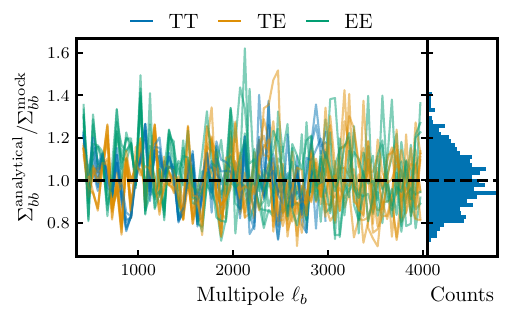}
	\caption{\emph{Left}: Ratio of the analytical covariance matrix diagonal to the covariance matrix diagonal based on end-to-end simulations for all frequency combinations.
	\emph{Right}: Histogram of values of the ratio. We find excellent agreement between the two estimates of the covariance matrix.}
	\label{fig:covariance_validation}
\end{figure}

Finally, to obtain an estimate of the complete covariance matrix, we use end-to-end simulations, created by combining simulated, inpainted \fullmock{} maps with noise realizations.
From $100$ effectively independent realizations constructed this way, we calculate the variance of the power spectra of these maps.
The diagonal of the resulting matrix agrees well with our analytical calculation, given the number of realizations, as shown in \cref{fig:covariance_validation}.
There is no apparent trend with multipole number; this is worth noting as the different constituents of the covariance dominate at different angular scales.
This agreement indicates that our analytic framework provides a good description of the true covariance of our data vector.
Because our analysis uses a curved-sky pipeline, we expect off-diagonal terms arising from sky masking to be negligible, as discussed in~\citep{camphuis_accurate_2022}. 
We have verified this expectation for the pure signal component using the full set of 500 \fullmock{} simulations. 
Off-diagonal contributions from lensing and marginalized foreground components are included in our analysis.
We conclude that the analytical framework presented here accurately describes the covariance of the data, in the case of the full matrix as well as for the sample variance, chance correlation, and noise variance terms individually.

\subsubsection{Validation of the full pipeline}

\begin{figure}
	\includegraphics[width=\columnwidth]{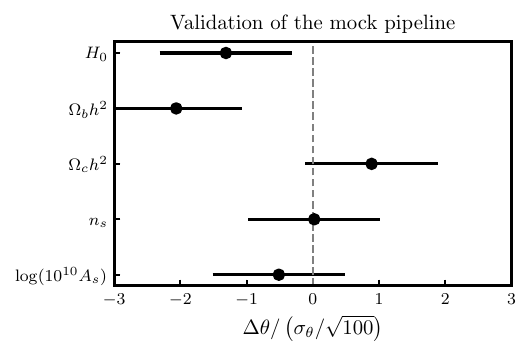}
	\caption{Validation of the pipeline on \fullmock{} simulations.
	We show the offset between the input cosmological parameter values and the product of the posteriors obtained by analyzing the $100$ \fullmock{} simulations in units of the expected uncertainty $\sigma = \sigma_\theta / \sqrt{100}$, where $\sigma_\theta$ is the expected data uncertainty on parameter $\theta$.
	We report a PTE of 0.25 across the full parameter space.}
	\label{fig:mock_validation}
\end{figure}

We run our full analysis pipeline on \fullmock{} simulations to check that our methodology is unbiased. 
This operation also validates that the scatter is accurately modeled by the covariance matrix.
First, we take simulated signal-only maps, add sign-flip noise maps, and inpaint at the location of masked sources as we do for the data.
Second, we run the band power estimation pipeline on these combined maps using the same procedure as for the data.
Third, we use our cosmological likelihood from \cref{sec:likelihood} to obtain \lcdm{} parameter constraints for each set of band powers from step two. 
For those runs, we set polarized beam, quadrupolar beam leakage, and dust-related parameters to their default values (listed in \cref{tab:foreground-priors,tab:nuisance-priors}), since the simulations do not include these effects.
We perform these steps for a total of $100$ realizations and compare the parameter constraints from simulations to the input values.
We find that the mean of the posterior distributions is consistent with the input values and that the scatter of the posteriors is well described by the expected covariance matrix. 
We show the results of this test \replace{, for which the PTE is 0.25,}{} in \cref{fig:mock_validation}.
\RR{The PTE is computed over the full five-dimensional parameter space, accounting for correlations among parameters. It is found to be 0.25, above the pre-defined threshold of 0.05.}

\subsubsection{Alternate cosmology test}
\label{sec:altcosmo}

The simulation pipeline, used to evaluate the transfer function and the additive biases and to validate the covariance matrix, is based on a fiducial cosmological model.
To verify the robustness of our results with respect to this choice, we perform a test on ten mock observations based on a different cosmology.
We do so by using the same simulation pipeline, but changing the input cosmological parameters, setting \Hubble{} to a value consistent with the \shoes{} measurement in~\citep{riess22}, and changing other parameters to yield a $5\,\sigma$ discrepancy across the five constrained \lcdm{} parameters compared to the fiducial values.
We then run our regular pipeline on these alternate cosmology simulations and check that we recover the correct input cosmology within the expected uncertainties.

\begin{figure}
	\includegraphics[width=\columnwidth]{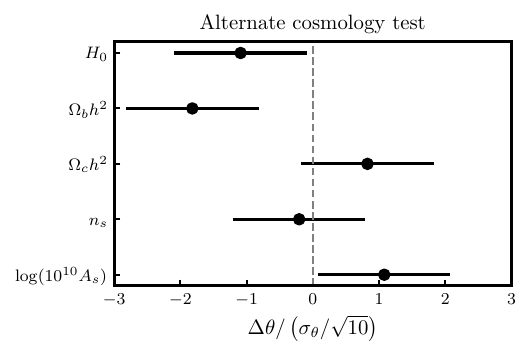}
	\caption{Validation of the pipeline on 10 \fullmock simulations with an alternate cosmology.
	We show the offset between the input cosmological parameter values and the product of the posteriors obtained by analyzing the ten alternate cosmology simulations in units of the expected uncertainty $\sigma = \sigma_\theta / \sqrt{10}$, where $\sigma_\theta$ is the expected data uncertainty on parameter $\theta$.
	All individual shifts are below $2\,\sigma$ and we report a PTE of 0.30 across the full parameter space, demonstrating that our analysis pipeline is largely insensitive to the chosen fiducial cosmological model.}
	\label{fig:alternate_cosmology}
\end{figure}

We show the averaged parameter constraints from the analyzed alternate cosmology simulations in \cref{fig:alternate_cosmology}.
The shift we observe is consistent with zero at $1.0\,\sigma$ over the five dimensional parameter space and the overall PTE of this test is 0.30 \RR{(threshold 0.05)}.
We conclude that our analysis pipeline is largely insensitive to the fiducial cosmological model.

\subsection{Parameter-level tests}
\label{sec:cosmoparametertests}

We now assess the consistency of the \LCDM{} cosmological parameters obtained from the spectra of data at different frequencies. 
This test is explicitly performed in \LCDM{} and therefore has some model dependence.
However, this is limited to the assumption that \lcdm{} provides an acceptable description of the common signal across frequencies.
This allows us to further probe any potential inconsistencies across observational frequencies and short-comings of our data model.
Since we are explicitly only interested in the differences between cosmological parameters, and not their absolute values, we were able to perform this test during the blind stage of the analysis.
We emphasize that our consistency checks were performed only within the \TTTEEE{} constraints or within the individual \TT{}, \TE{}, and \EE{} constraints, but not across these channels. Cross-channel consistency tests inherently depend on the assumed cosmological model and were therefore not conducted during the blind analysis stage.

We predict the expected correlation between parameters obtained from different frequency spectra using the framework developed in~\citep{kable20}.
The covariance of the parameters derived from different spectra is given by
\begin{equation}
	\begin{split}
		             & \text{cov} \left( \bar{\theta}^{\mu\nu}, \bar{\theta}^{\alpha\beta}\right) = (M^{\mu\nu})^T \Sigma^{\mu\nu;\alpha\beta} M^{\alpha\beta}, \label{eq:kable} \\
		\text{where} & \quad M^{\mu\nu} \equiv (\Sigma^{\mu\nu;\mu\nu})^{-1} \frac{\partial D^{\mu\nu}_\ell}{\partial \theta} (F^{\mu\nu;\mu\nu})^{-1},
	\end{split}
\end{equation}
where $F$ is the Fisher matrix~\citep{heavens_generalisations_2016}.
Our differentiable likelihood gives us easy access to the derivatives $\partial D^{\mu\nu}_\ell / \partial \theta$ (as demonstrated by~\citep{balkenhol24}), which we evaluate at the best-fit point in the full-frequency likelihood.
We use parameter covariances from MCMC analyses (see \cref{sec:pars} for details) to obtain the relevant Fisher matrices.
This allows us to compare parameter constraints from the individual frequency-spectrum likelihoods to the full-frequency likelihood and among themselves. There are five independent tests for each \TT{}, \TE{}, and \EE{} channel, and the PTE threshold to pass is $0.05/15 = 0.0033$.

When analyzing subsets of the data, it is necessary to impose additional priors on the calibration and polarization efficiency parameters (see \cref{sec:calibration}), as these parameters are not sufficiently constrained by the data alone. The specific priors adopted in these cases are summarized in \cref{tab:nuisance-priors}.

\begin{figure*}
	\centering
	\includegraphics[width=\textwidth]{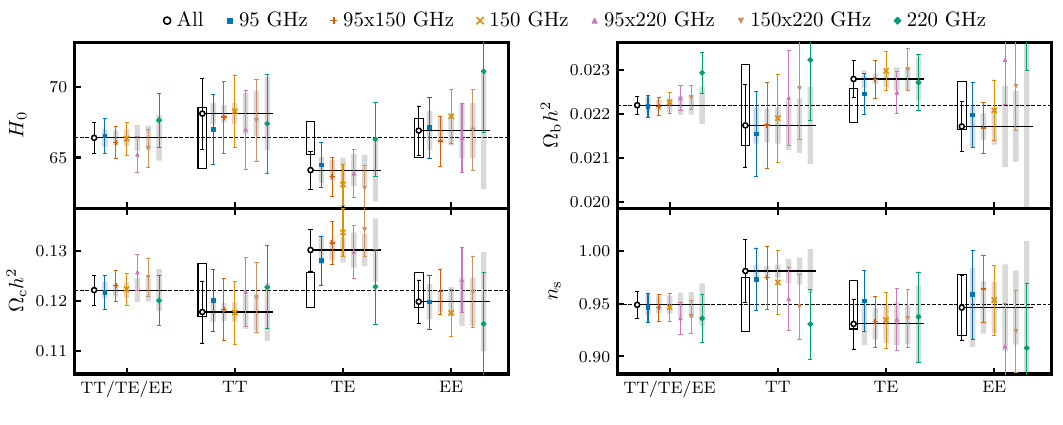}
	\caption{Cosmological parameters obtained from single frequency-spectrum likelihoods ($95\times 95\,\ghz$: blue square, $95\times 150\,\ghz$: orange cross, $150\times 150\,\ghz$: yellow x, $95\times 220\,\ghz$: pink triangle, $150\times 220\,\ghz$: brown upside-down triangle, $220\times 220\,\ghz$: green diamond) and the full-frequency likelihoods (open black circles) for combined \TTTEEE{} and individual \TT{}, \TE{}, and \EE{} fits.
	The whiskers show the $1\,\sigma$ error bars obtained by the given subset, while the shaded area shows the size of the expected $1\,\sigma$ fluctuation between the \TT{}, \TE{}, and \EE{} likelihoods and the subset, calculated via \cref{eq:kable}.
	The open boxes similarly show the size of expected fluctuation of the full-frequency \TT{}, \TE{}, and \EE{} likelihoods from the complete likelihood.
	Although we display parameters on an absolute scale for readability, we only looked at the difference before unblinding.
	Similarly, the comparison between temperature and polarization results was carried out after unblinding (see \cref{sec:lcdm_integrity}).
	There appears to be a frequency-dependent trend in temperature constraints on $\ombh$ and $\ns$, which is likely influenced by the foreground model.
	However this trend is not statistically significant and we report excellent consistency of our data across frequencies and spectra.
	Similarly, the excursion of the \TE{} multi-frequency constraints is not statistically significant, as indicated in \cref{tab:pte_ttvstevsee}.}
	\label{fig:parameter-difference-test}
\end{figure*}

In \cref{fig:parameter-difference-test} we show the results for the parameters \Hubble{}, \omegab{}, \omegac{}, and \ns{}.
The PTEs for inter-frequency comparisons are reported in \cref{tab:interfrequency_ptes} and pass the associated threshold, indicating once again the excellent internal consistency of our data. 
The ordering of the frequency spectra in the plot is designed to highlight potential foreground mismodeling. 
The rising (falling) trend for \omegab{} (\ns{}) on the \TT{} channel might be indicative of a foreground contribution, but the PTEs are still consistent with the null hypothesis.
Note that during the blind stage of the analysis, the absolute parameter values were obscured and no comparison between temperature and polarization data was carried out. 
We performed the latter after unblinding, as discussed in \cref{sec:lcdm_integrity}, but report it in the figure for completeness.

\begin{table*}
\centering
\begin{tabular}{|{c}|*{6}{c}|c|c|*{6}{c}|}
\cline{1-7}
\cline{9-15}
 \diagbox{T\&E}{TT} &$95$&$95\times 150$&$150$&$95\times 220$&$150\times 220$&$220$& \qquad & \diagbox{TE}{EE}  &$95$&$95\times 150$&$150$&$95\times 220$&$150\times 220$&$220$ \\
\cline{1-7}
\cline{9-15}
$95$ & \diagbox{\,}{} & 0.93 & 1.00 & 0.48 & 0.79 & 0.49 &  & $95$ & \diagbox{\,}{} & 0.07 & 0.05 & 0.92 & 0.11 & 0.82 \\
\cline{1-7}
\cline{9-15}
$95\times 150$ & 0.87 & \diagbox{\,}{} & 0.97 & 0.83 & 0.91 & 0.40 &  & $95\times 150$ & 0.99 & \diagbox{\,}{} & 0.82 & 0.65 & 0.89 & 0.55 \\
\cline{1-7}
\cline{9-15}
$150$ & 0.86 & 0.82 & \diagbox{\,}{} & 0.82 & 0.84 & 0.55 &  & $150$ & 0.99 & 0.70 & \diagbox{\,}{} & 0.28 & 0.99 & 0.33 \\
\cline{1-7}
\cline{9-15}
$95\times 220$ & 0.62 & 0.36 & 0.51 & \diagbox{\,}{} & 0.99 & 0.29 &  & $95\times 220$ & 0.55 & 0.89 & 0.66 & \diagbox{\,}{} & 0.05 & 0.54 \\
\cline{1-7}
\cline{9-15}
$150\times 220$ & 0.36 & 0.19 & 0.08 & 0.74 & \diagbox{\,}{} & 0.33 &  & $150\times 220$ & 0.99 & 1.00 & 0.89 & 0.93 & \diagbox{\,}{} & 0.14 \\
\cline{1-7}
\cline{9-15}
$220$ & 0.25 & 0.16 & 0.17 & 0.39 & 0.60 & \diagbox{\,}{} &  & $220$ & 0.60 & 0.74 & 0.88 & 0.73 & 0.95 & \diagbox{\,}{} \\
\cline{1-7}
\cline{9-15}
\end{tabular}
\caption{Inter-frequency PTEs for the different frequency combinations. 
The PTEs are computed within each channel combination, and for comparison of single frequencies only.
For example, the consistency between parameters derived from \TT{} 95\ghz{} and all other \TT{} frequencies is presented in the first row of the first section of the table; the PTE of the comparison of \TT{} 95\ghz{} constraints with the \TT{} 95$\times$150\ghz{} constraints is 0.93.
The left panel displays the PTEs for the \TTTEEE{} and \TT{} data. The right panel presents the PTEs for the \TE{} and \EE{} data. 
The comparison is performed on the 5 \lcdm{} parameters excluding $\tau$. 
We obtained the cosmological parameter covariance matrix according to \cref{eq:kable}.}
\label{tab:interfrequency_ptes}
\end{table*}

Overall, we find that the frequency channels are consistent across the full multipole range considered. 
The pipeline is validated to be unbiased on simulations and the analytic covariance matrix is shown to provide an accurate description of the data. 
We also confirm that the analysis is largely insensitive to the choice of fiducial cosmological model. 
Finally, the \lcdm{} constraints derived from individual frequency spectra are consistent with each other and with those from the full-frequency likelihood, as illustrated in \cref{fig:parameter-difference-test}.
We conclude that the analysis pipeline is robust and suitable to be used for cosmological inference.


\section{Cosmological analysis}
\label{sec:pars}

Having demonstrated the internal consistency of our data and verified our ability to model it, we now use it for cosmological inference.
This section is organized as follows.
After introducing all the data sets that are used (\cref{sec:cosmo_data}), we first report constraints using only CMB data (\cref{sec:cosmo_cmb}).
We assess the consistency of the \sptbp{} data with \lcdm{} (\cref{sec:lcdm_integrity}) and with other CMB data, and then report results for the standard model for SPT alone and in combination with other CMB data (\cref{sec:lcdm}).
We test whether the signature of gravitational lensing in CMB data is consistent with the \lcdm{} prediction (\cref{sec:alens}) and afterwards report constraints on new light particles (\cref{sec:neff}) and reconstruct the recombination history (\cref{sec:modrec}).
Following this, we consider joint constraints from CMB and BAO data.
We first evaluate the consistency of the relevant data sets in \lcdm{} (\cref{sec:CMB+BAO_lcdm}) before reporting constraints on extended cosmological models (\cref{sec:CMB+BAO}).
We revisit models from \cref{sec:alens}-\cref{sec:modrec} in \cref{sec:alens_BAO}-\cref{sec:modrec_BAO}, now adding BAO data, and report results on additional models not considered before (\cref{sec:curv}-\cref{sec:w0wa}).

For the analyses presented here, we explore parameter posteriors via a Markov Chain Monte Carlo (MCMC) approach using \cobaya{}~\citep{torrado21}.
We consider chains with a Gelman-Rubin statistic of $R-1\sim 0.02$ to be converged unless otherwise specified.
Although power spectrum emulators (\cref{sec:theory_codes}) were indispensable during this analysis, all of the final results reported below were generated using traditional Boltzmann solvers\footnote{We used \class\ v3.2.3~\citep{blas11} and \camb\ v1.5.8~\citep{lewis11b}} unless otherwise stated.
Almost all of the SPT results in this section are run with the \sptlite{} likelihood (\cref{sec:lite_likelihood}). We use the full-frequency likelihood only when analyzing \SPT{} \TT{}, \EE{}, or \TE{} alone, or when showing \sptlr{} only results in \lcdm{}.

\subsection{Additional data sets}
\label{sec:cosmo_data}
\begin{table*}
	\centering
	\begin{tabular}{l|l}
	\toprule
	\textbf{Name} & \textbf{Data Set} \\
	\midrule
	{$\taureio$ prior} & $\taureio \sim \mathcal{N}(0.051, 0.006)$~\citep{planck20-57}, unless specified otherwise \\
	\midrule
	{\sptbp \qquad} & This work, i.e.  \TTTEEE{} band powers from SPT-3G D1 observations  \qquad \\ 
	{\muse{} \PP{}} & \PP{} band powers from SPT-3G D1 observations~\citep{ge24} \\
	{\sptlr} & \sptbp + \muse{} \PP{} \\
	\midrule
	{\planck{} T\&E} & \planck\ 2018 PR3 \plik\ high-$\ell$  \TTTEEE{} + low-$\ell$ \TT{}~\citep{planck18-1}\\
	{\planck{} \PP{}} & \planck{} NPIPE PR4 \PP{}~\citep{carron22}\\
	{\planck{}} & {\planck{} T\&E} + {\planck{} \PP{}} \\
	\midrule
	{\SPT{}+\planck} & \sptlr + \planck{} \\
	\midrule
	{\ACTDR{} \T\&\E{}} &  \ACTDR{} \TTTEEE{}~\citep{naess_atacama_2025,louis25,calabrese25} \\ 
	{\ACTDR{} \PP{}} & \ACTDR{} \PP{}~\citep{qu24,madhavacheril24}\\
	{\ACTDR{}} & \ACTDR{} \T\&\E{} + {\ACTDR{} \PP{}}\\
	\midrule
	{P-ACT \T\&\E{}} & \planck{} + \ACTDR{} combined \TTTEEE{} likelihood~\citep{louis25} \\
	{P-ACT \PP{}} & \planck{} + \ACTDR{} combined \PP{} likelihood~\citep{carron22} \\
	{P-ACT} & P-ACT \T\&\E{} + P-ACT \PP{} \\
	\midrule
	{\ground{} \T\&\E{}} \qquad \qquad & \sptbp + \ACTDR{} \T\&\E{} \\
	{\ground{} \PP{}} & \muse{} \PP{} + \ACTDR{} \PP{} \\
	{\ground} & \sptlr{} + \ACTDR{} \\ 
	\midrule
	{\cmball{} \T\&\E{}} & P-ACT \T\&\E{} + \SPT{} \T\&\E{}\\
	{\cmball{} \PP{}} &  P-ACT \PP{} + \muse{} \PP{}~\citep{qu25}\\
	{\cmball} & \cmball{} \T\&\E{} + \cmball{} \PP{}\\
	\midrule
	{DESI} & DESI DR2 BAO data~\citep{desi25}\\
	{SDSS} & SDSS BAO data~\citep{beutler11,ross15,alam17,alam21}\\
	{\SPT{}+\DESI{}} & \sptlr + DESI DR2 BAO data \\
	\bottomrule
	\end{tabular}
	\caption{Summary of data sets used in the analysis.}
	\label{tab:dataset}
\end{table*}

We present a list of data sets used in this work in \cref{tab:dataset}.
The CMB power spectrum measurements presented in this work do not cover large-scale E-mode polarization anisotropies and hence cannot constrain with high precision the optical depth to reionization, $\taureio$.
We impose a Gaussian prior on $\taureio \sim \mathcal{N}(0.051, 0.006)$ based on~\citep{planck20-57} unless stated otherwise.
We combine our primary CMB data with the SPT-3G gravitational lensing potential reconstruction presented in~\citep{ge24}.
We expect a vanishing correlation between the lensing data and the lensed CMB band powers presented in this work~\citep{trendafilova23} and we, therefore, combine the data sets at the likelihood level after verifying their consistency (see \cref{sec:lcdm}).

We compare and combine our results with CMB data from \planck{} and \ACT.
For \planck{}, we choose to use the primary CMB data from the PR3 release~\citep{planck18-5} and the lensing data from PR4~\citep{carron22}, which is more constraining than the PR3 lensing data. This is the same combination of \planck{} data adopted in~\citep{ge24}.
We take advantage of the \texttt{python} implementation of the \planck{} \texttt{clik} likelihood, \texttt{clipy}\footnote{\url{https://github.com/benabed/clipy}}~\citep{planck18-5}.
For ACT, we use the DR6 lensing and primary CMB data~\citep{qu24,madhavacheril24,naess_atacama_2025,louis25,calabrese25}.
When needed, we rerun the ACT analyses imposing our baseline $\taureio$ prior to allow for a consistent comparison. We use the \texttt{\ACT-lite} likelihood for all of the results in this section, except when running \TT{} data alone, for which we use the \ACTDR{} multi-frequency likelihood.
We combine the \SPT{} and \ACTDR{} data under the assumption that the two data sets are uncorrelated and call this combination \ground{}.
We also combine \planck{}, \SPT{}, and \ACTDR{} and refer to this as \cmball{}. When doing this, in order to minimize the correlations between the \planck{} and \ACTDR{} data sets, we use the multipole cuts of the P-ACT combination of~\citep{louis25}.
Note in particular that P-ACT uses only \ACTDR{} TT data at $\ell>1000$, excising the \planck{} \TT{} high-$\ell$ data;
this choice is relevant in \cref{sec:CMB+BAO_lcdm} and \cref{sec:CMB+BAO} as the high-$\ell$ \TT{} data of \ACTDR{} and \planck{} prefer slightly different cosmologies.
We assume no correlation between \SPT{} and other data sets, justified by the small overlap in the sky regions, see \cref{fig:footprints}.\footnote{As \planck{} data are based on full sky observations, the fraction of the \mainfield footprint within the \planck{} mask is minimal.
For ACT, the common area observed constitutes about $10\%$ of the ACT DR6 mask.}
We neglect correlations between lensing measured by \ACTDR{} and \SPT, as these (small) correlations were shown to have a negligible effect on cosmological parameters in~\citep{qu25}.

We further contextualize our results using non-CMB data.
Here, we draw on the latest BAO measurements from DESI DR2 data~\citep{desi25}.
We also consider the SDSS BAO data~\citep{beutler11,ross15,alam17,alam21} and SNe Ia data from the Pantheon+ data set~\citep{brout_pantheon_2022}. 

\subsection{Constraints from CMB data}
\label{sec:cosmo_cmb}
\RR{In the following, we present constraints from CMB data alone. First, we verify the consistency of the \sptbp{} data with the \lcdm{} model (\cref{sec:lcdm_integrity}). 
We then report results for SPT alone and in combination with other CMB data (\cref{sec:lcdm}). 
Next, we explore extensions to the \lcdm{} model, beginning with an assessment of the internal consistency of CMB data with respect to the gravitational lensing signal amplitude (\cref{sec:alens}). 
We subsequently report constraints on new light particles (\cref{sec:neff}) and reconstruct the recombination history (\cref{sec:modrec}). 
We focus on these three extensions because they are particularly well suited to CMB data alone and have been the subject of recent interest in the literature. 
A companion paper~\citep{khalife2025spt3gd1axionearly} presents constraints on the axion-dark-energy model using the same data sets.}

\subsubsection{Consistency of the \lcdm{} model with SPT data}
\label{sec:lcdm_integrity}

We find that the \lcdm{} model provides a good description of the \sptbp spectra.\footnote{We note that all \lcdm{} results presented in this section are based on the multi-frequency \sptbp likelihood and are obtained using \camb{} as the Boltzmann solver.}
For our full data we report a best-fit $\chi^2$ value of 1359, which corresponds to a PTE of 0.52.\footnote{Following the methodology of~\citep{raveri19} (see Eq. 29 therein) we count 1362 effective degrees of freedom for the multi-frequency likelihood in \lcdm{}.}
We present the best-fit \lcdm{} model for the band powers in \cref{fig:bandpowers} together with the measured band powers, and display the residuals between the data and the model.
\lcdm{} also fits well each spectrum individually; analyzing only \TT{}, \TE{}, or \EE{} spectra we report $\chi^2$ (PTE) values of 267 (0.84), 631 (0.52), and 429 (0.38) (for 291, 633, and 421 effective degrees of freedom), respectively.
While this is a general affirmation that \lcdm{} is a good description of the SPT data, we verify the consistency of model predictions between temperature and polarization as well as across angular scales in more detail.

\begin{figure}
	\includegraphics[width=\columnwidth]{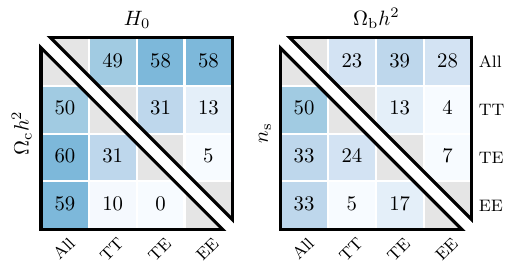}
	\caption{Correlation matrix between constraints from the individual and joint \TT{}, \TE{}, and \EE{} likelihoods.
	We show, as an example, the correlations for the parameter constraints on $\Hubble, \omch, \ns$, and $\ombh$.
	For instance, the determination of \Hubble{} from \TT{} is correlated at $31\%$ with the measurement of \Hubble{} from \TE{}.
    The correlation matrices are non-trivial and their calculation allows for interesting data and model consistency tests.}
	\label{fig:par_corr_TTvsTEvsEE}
\end{figure}

\begin{figure*}
	\includegraphics[width=\textwidth]{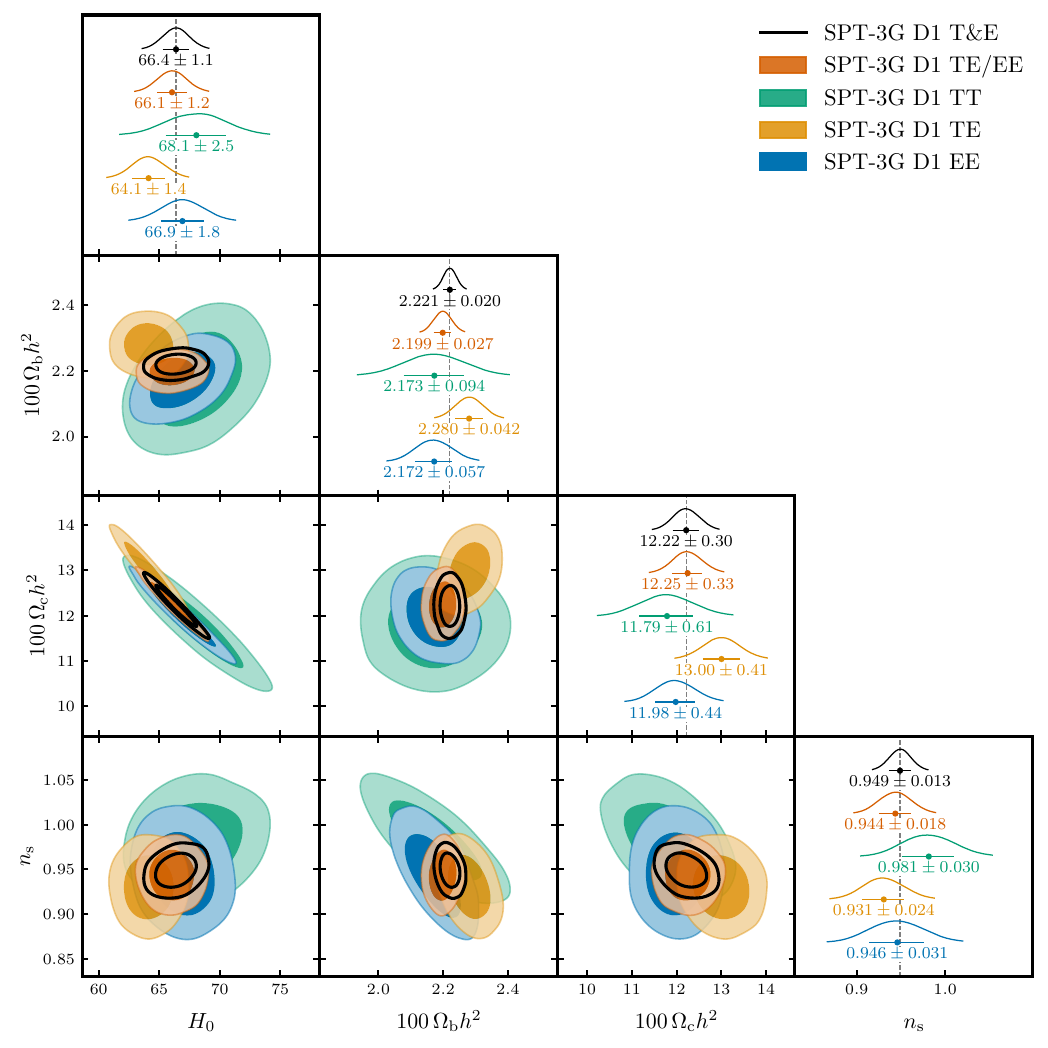}
	\caption{Comparison of \lcdm{} constraints from the \sptbp data.
	The polarization data are exceptionally constraining and, in particular, the \TE{} data holds most of the cosmological information.
	We report consistent results from temperature and polarization data in \lcdm{}, with PTEs reported in \cref{tab:pte_ttvstevsee}.
	}
	\label{fig:lcdm_TTvsTEvsEE}
\end{figure*}

First, we check the agreement of \lcdm{} parameter constraints from temperature and polarization data.
This not only allows us to assess the ability of \lcdm{} to jointly describe temperature and polarization data, but it is also a powerful test to understand frequency-coherent biases in specific channels.\footnote{This test is model dependent, which is why we did not perform it prior to unblinding.}
To evaluate consistency between spectra, we use the same framework as for inter-frequency parameter consistency in \cref{sec:cosmoparametertests}.
The calculated parameter correlation matrices are non-trivial, and we illustrate the expected correlation of constraints produced by data subsets for $\Hubble$, $\omch$, $\ns$ and $\ombh$ in \cref{fig:par_corr_TTvsTEvsEE} as an example.
We find consistency (at the $\sim 1\,\sigma$ level) between constraints from all individual and joint \TT{}, \TE{}, and \EE{} likelihoods, which we show in \cref{fig:lcdm_TTvsTEvsEE,tab:pte_ttvstevsee}. 
The precise measurements of the CMB polarization anisotropies drive our combined \TTTEEE{} result; excising the \TT{} data only slightly loosens constraints, as shown in \cref{fig:lcdm_TTvsTEvsEE}.

\begin{table}
	\begin{tabular}{c|cccc}
		\toprule
		\textbf{Spectrum}        & \textbf{All} & \textbf{\TT{}}  & \textbf{\TE{}} & \textbf{\EE{}} \\
		\midrule
		\textbf{All}          & -            & $0.4\,\sigma$ & $1.2\,\sigma$         & $0.6\,\sigma$ \\
		\textbf{\TT{}}          & 0.67         & -             & $1.0\,\sigma$         & $0.3\,\sigma$ \\
		\textbf{\TE{}} & 0.22         & 0.31          & -                     & $1.0\,\sigma$ \\
		\textbf{\EE{}}         & 0.57         & 0.78          & 0.33                  & -             \\
		\bottomrule
	\end{tabular}
	\caption{PTE table indicating levels of consistency between cosmological constraints from temperature and polarization and their combination. The lower triangle indicates the PTE values, while the upper triangle indicates the corresponding Gaussian equivalent $\sigma$ values.}
	\label{tab:pte_ttvstevsee}
\end{table}

\begin{figure}
	\includegraphics[width=\columnwidth]{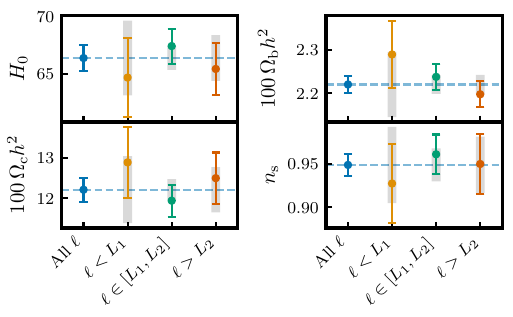}
 \begin{tabular}{l|cccc}
		\toprule
		\textbf{Range}      & All $\ell$ \  & \ $\ell<L_1$ \  & \ $\ell \in[L_1,L_2]$ \  & \ $\ell>L_2$ \ \\
		\midrule
		All $\ell$          & -             & $0.5\,\sigma$   & $0.5\,\sigma$            & $0.2\,\sigma$  \\
		$\ell<L_1$          & 0.63          & -               & $0.3\,\sigma$            & $0.2\,\sigma$  \\
		$\ell \in[L_1,L_2]$ & 0.64          & 0.80            & -                        & $0.5\,\sigma$  \\
		$\ell>L_2$          & 0.85          & 0.88            & 0.64                     & -              \\
		\bottomrule
	\end{tabular}
	\caption{\emph{Top}: Comparison of \lcdm{} parameter constraints from the low, intermediate, and high multipole ranges, with $L_1=1000$ and $L_2=2000$. The whiskers indicate the 68\% confidence intervals of the marginalized posteriors, while the gray shaded regions indicate the expected variation of the parameters assuming the full $\ell$-range following~\citep{gratton19}.
	The different subsets are effectively independent of one another and their corresponding constraints are consistent.
	This is also true for the comparison with the full likelihood, in which case we do account for the expected correlation due to the shared data (gray boxes). We conclude that our data are consistent across angular scales given \lcdm{}. \emph{Bottom}: PTE table indicating level of consistency between cosmological constraints from different $\ell$ ranges. The lower triangle indicates the PTE values, while the upper triangle indicates the corresponding Gaussian equivalent $\sigma$ values.}
	\label{fig:lcdm_ell_test}
\end{figure}

Next, we evaluate the consistency of \lcdm{} parameter constraints obtained from different angular scales.
We split our data into three multipole ranges, defined from the relative signal-to-noise of our data, see \cref{fig:snr}: low ($\ell<1000$), intermediate ($\ell \in [1000, 2000]$), and high ($\ell>2000$) and perform a cosmological analysis on each of these subsets.
We then compare the results using the same framework as in \cref{sec:cosmoparametertests}.
Again, we report no discrepancies between the subsets and show our results in \cref{fig:lcdm_ell_test}.

Together, these tests speak to the ability of the \lcdm{} model to consistently and accurately predict (1) temperature and polarization data and (2) data across a wide range of angular scales.
Moreover, they indicate that there are no substantial frequency-coherent biases in our data, assuming \lcdm{}.

\subsubsection{\lcdm{} results}
\label{sec:lcdm}

After verifying that the \lcdm{} model provides a good description of the power spectrum measurements presented in this work, we establish that \muse{} lensing and \sptbp{} data are compatible, and discuss the combined \sptlr{} results.
We then verify that \sptlr{} is consistent with other CMB data sets, before reporting joint constraints and situating these in the wider cosmological context.

From now on, we combine primary CMB data with CMB lensing reconstructions as per \cref{tab:dataset}, unless otherwise explicitly mentioned.
We first assess the consistency between the \sptbp measurements and the \muse{} lensing data.
The difference in the $\seight{0.25} \equiv \sigmaeight (\omm/0.3)^{0.25}$ measurements corresponds to a $0.5\,\sigma$ fluctuation, indicating agreement between these two probes.
We find $\seight{0.25} = 0.824 \pm 0.016$ for \muse{} and $\seight{0.25} = 0.838 \pm 0.021$ for \sptbp{}. 
Since the correlation between the lensing reconstruction and the lensed CMB data is expected to be negligible~\citep{trendafilova23}, and this has been confirmed by simulations, we combine the two data sets at the likelihood level.
Comparing our measurement of the lensed CMB power spectrum to the \muse{} reconstruction of the unlensed CMB on the other hand is not straightforward as the correlation between the two data sets remains to be accurately quantified.\footnote{The means of the marginalized one-dimensional posterior distributions of \lcdm{} parameters derived from the combination of SPT-3G Main 19/20 \EE{} and \muse{} \PP{} are within $1.1\,\sigma$ of the \muse{} $\EE{}+\PP{}$ results.
Expensive joint simulations are needed to understand whether these parameter shifts are expected, given that the two pipelines use different methodologies, survey masks, and angular scales.
This work is currently under way.}
Having demonstrated the consistency of the \sptbp{} and \muse{} lensing data, we now combine them and report constraints on cosmological parameters in \cref{tab:lcdm}.
We show marginalized posteriors for some parameters in \cref{fig:lcdm_main}.

We highlight the tight constraints that the \sptlr data place on cosmology.
Notably, for $\Hubble$, SPT data yield a $\sigma(\Hubble)=0.60\,\kmsmpc$ constraint, compared to $0.49\,\kmsmpc$ for \planck{} and $0.64\,\kmsmpc$ for \ACTDR{}.
Similarly, $100\,\omch$ is determined using SPT data with a precision of $1.3\%$ to $\sigma(100\,\omch)=0.16$; the \planck{} and \ACTDR{} data achieve constraints of $0.11$ ($0.92\%$ precision) and $0.17$ ($1.4\%$ precision) on this parameter, respectively.
For these two parameters, the SPT data benefit greatly from the degeneracy-breaking power of CMB lensing.

The constraints from the SPT data are in excellent agreement with other contemporary CMB experiments.
The contours corresponding to \planck{} overlap visibly with those for \sptlr in \cref{fig:lcdm_main}.
This agreement also holds up quantitatively; calculating the agreement over $(\Hubble, \ombh, \omch, \ns, \clamp)$ we obtain PTEs that correspond to one-dimensional Gaussian fluctuations of $0.4\,\sigma$ when comparing \sptlr to \planck{} and $1.1\,\sigma$ when comparing \sptlr{} to \ACTDR{}.\footnote{Since we derive our absolute calibration from \planck{}, we verify that the parameter-level agreement persists when excluding the combined amplitude parameter, \clamp{}, from the comparison.
In this case, the agreement between \sptlr{} and \planck{} holds steady at $0.4\,\sigma$, while the distance to \ACTDR{} slightly increases, to $1.4\,\sigma$.}
Overall, the agreement between CMB experiments at this level of precision is a remarkable achievement of the standard model of cosmology.
The three data sets contain independent information and span a wide range of angular scales for both temperature and polarization data.
This agreement motivates the combination of the data sets to further improve cosmological constraints.

While \planck{} remains the most constraining single CMB experiment, the \ground{} combination of SPT and ACT data achieves equally tight constraints on most \lcdm{} parameters, as shown in \cref{fig:lcdm_main} and quantified in \cref{tab:lcdm}.
This is a significant milestone for modern cosmology; for the first time, ground-based experiments reach \planck{}'s constraining power, most notably on the Hubble constant ($66.59\pm0.46\,\kmsmpc$ vs.~$67.41\pm 0.49\, \kmsmpc$ for \planck{}) and the amplitude of matter fluctuations parameterized by $\sigmaeight$ ($0.8169\pm 0.0042$ vs.~$0.8099\pm 0.0051$ for \planck{}).
At the same time, the \ground\ data set is consistent with the \planck{} results at $1.2\,\sigma$ (approximating them as completely independent).

We verified that the \ground\ constraints are still as good as \planck\ even if 
we replace the \planck\ 2018 high-$\ell$ \TTTEEE{} likelihood, \plik, which is the baseline for our \planck\ data set, with the \texttt{camspec} likelihood from
\citep{rosenberg22}.\footnote{The \camspec\ likelihood from \citep{rosenberg22} is based on \planck\ PR4 maps, which contain about 10\% more data than \planck\ PR3, and use a larger fraction of the sky with respect to the \plik\ PR3 likelihood.} We call the \planck\ data set using the \camspec\ likelihood \planck-\camspec. 
We find $\Hubble=67.22\pm0.44\, \kmsmpc$ and $\sigmaeight=0.8076 \pm 0.0050$ from \planck-\camspec. 
The consistency between \planck-\camspec\ and \sptlr remains excellent, at the $0.2\,\sigma$ level. 
Similarly, we find consistency between \planck-\camspec\ and \ground\ at the $1.8\,\sigma$ level.\footnote{Note that the \ACTDR\ results show a worse level of agreement with the \camspec\ PR4 likelihood compared to the \plik\ 2018 one, see \citep{louis25}.}
We conclude that the choice of \planck\ likelihood does not have a large impact on the findings reported above.

Crucially, the SPT and ACT data are highly complementary to \planck{}.
The satellite data leverage full-sky access to deliver measurements of large angular scales that are difficult to constrain from the ground and---by virtue of measuring more modes---are the best measurements in the sample-variance-dominated regime.
The \ground{} data, on the other hand, have been collected by instruments with higher resolution and significantly more detectors, thus providing exquisite measurements of the polarization of the CMB on scales where the \planck{} data is noise-dominated.
Together, these data form an incredibly rich cosmological data set.
Given the agreement demonstrated above, we combine \ground{} and \planck{} into \cmball{}.
We report constraints from this combination in \cref{fig:lcdm_main,tab:lcdm}; these are the most precise determinations of \lcdm{} parameters from CMB observations to date.

We use the square-root of the determinant of parameter covariance matrices as a measure of the allowed volume in higher dimensional spaces (i.e. $\sqrt{\det{C}}$, where $C$ is the relevant matrix);\footnote{The inverse of this is commonly referred to as the figure of merit.} ratios of this metric allow us to quantify the volume reduction SPT data enable in joint constraints.
In the five-dimensional space of $(\Hubble, \omch, \ombh, \ns, \clamp)$, adding SPT to \planck{} reduces the allowed parameter volume by a factor of $1.9$.
Similarly, when adding SPT to ACT data to form \ground{}, the allowed region shrinks by a factor of $2.8$.
These are significant improvements and speak to the constraining power of the SPT data.
Compared to the previous SPT-3G \TTTEEE{} results presented  in~\citep{balkenhol23}, the new \sptbp{} data decrease the allowed parameter volume by a factor of $2.4$.

Despite the small observed sky area this release is based on, SPT data are able to constrain the scalar spectral index to $\ns = 0.951\,\pm\,0.011$, which disfavors a scale-invariant spectrum of initial density perturbations at $4.1\,\sigma$.
While ACT and \planck{} data are more sensitive to this parameter, the SPT data make a non-negligible contribution to the combined constraints.
ACT data alone rule out $\ns = 1$ at $4.6\,\sigma$, which becomes $5.7\,\sigma$ in \ground{} when SPT data are added.
Similarly, the preference for a scale-dependent spectrum increases from $7.8\,\sigma$ for P-ACT-L~\citep{louis25} to $\replace{10.5}{9.8}\,\sigma$ for \cmball{} when SPT data are included.
In addition to shrinking the width of the $\ns$ posterior in joint constraints, SPT data also shift the central value low, which increases the evidence for $\ns < 1$.

\begin{figure}
	\includegraphics[width=\columnwidth]{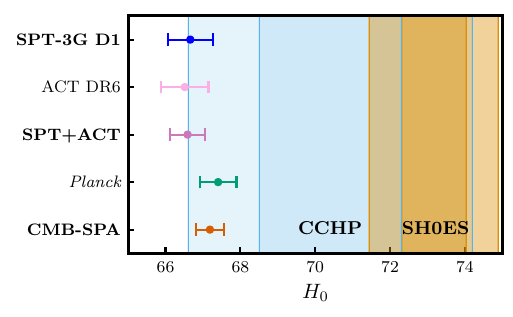}
	\caption{Status of \Hubble{} constraints. The data sets are described in \cref{tab:dataset} and the values are reported in \cref{tab:lcdm}.
	In bold, we highlight the results of this work. We show the \shoes{} and CCHP late-universe results in orange and blue, respectively.}
	\label{fig:Hubble_compilation}
\end{figure}

We situate our results in the wider cosmological landscape, beginning with constraints on the expansion rate today:
\begin{align}
	\Hubble{} &= 66.66 \pm 0.60 \, \kmsmpc \ \text{for} \  \sptlr, \\
	\Hubble{} &= 66.59\pm 0.46 \, \kmsmpc \ \text{for} \  \ground, \\
	\Hubble{} &= \replace{67.24 \pm 0.35}{67.19 \pm 0.38} \, \kmsmpc \ \text{for} \  \cmball.
\end{align}
These results are in stark contrast to the local universe determination of the expansion rate by \shoes, \ie{} $\Hubble^{\shoes{}} = 73.17 \pm 0.86\, \kmsmpc$~\citep{riess22,breuval_small_2024} as shown in \cref{fig:Hubble_compilation}. We report a $6.2\,\sigma$ tension between \shoes\ and the result from SPT data alone, and a $6.4\,\sigma$ tension between \shoes\ and \cmball{}.
\cmball{} comprises three nearly independent CMB data sets that have different weighting across angular scales and relative contributions of the \TT/\TE/\EE/$\phi\phi$ spectra\footnote{From our \EETE{} spectra alone we constrain $\Hubble{} = 66.1\,\pm\,1.2$ (see \cref{fig:lcdm_TTvsTEvsEE}), which is in $4.9\,\sigma$ tension with \shoes.} and were produced by different pipelines, all of which involve stringent consistency tests, yet they all agree and individually yield comparable offsets from \shoes.
Therefore, we are inclined to consider a systematic error in the CMB data an unlikely explanation for this discrepancy.

Local universe determinations of the Hubble constant have also undergone much scrutiny.
While the analysis of \shoes{}  is not the only result, it is currently the most statistically precise one.
We also highlight the work of the CCHP collaboration~\citep{freedman_status_2025}, who calibrate SNe Ia with Cepheids, the J region of the asymptotic giant branch, and the tip of the red giant branch.
The distances they obtain with these techniques to nearby galaxies hosting SNe Ia are consistent with the same distances found by the SH0ES team.
For the Hubble constant they report $\Hubble = 70.4 \pm 1.9\, \kmsmpc$,  which is 1.6$\,\sigma$ lower than the \shoes{} result and 2$\,\sigma$ higher than the \cmball{} constraint.
While it is possible that the difference between CMB and local data sets is caused by a real failure of \lcdm{} and not by systematics in any of the data sets, signatures of beyond-\lcdm{} models that can accommodate a high $\Hubble$ have yet to be clearly discerned in CMB data~\citep[\cref{sec:pars} in this work,][]{calabrese25}.

We now turn our attention to constraints on structure growth.
We report constraints on the amplitude of matter fluctuations, $\sigmaeight$, and on the matter density, $\omm$, of:
\begin{align}
	 &
	\left.
	\begin{array}{l}
		\sigmaeight = 0.8158\pm 0.0058, \\
		\omm = 0.3246\pm 0.0091
	\end{array}\right\} \quad \text{for} \  \sptlr,    \\
	 &
	\left.
	\begin{array}{l}
		\sigmaeight =0.8169\pm 0.0042, \\
		\omm = 0.3277\pm 0.0072
	\end{array}\right\} \quad \text{for} \  \ground, \\
	 &
	\left.
	\begin{array}{l}
		\sigmaeight = 0.8137\pm\replace{0.0038}{0.0037}, \\
		\omm = \replace{0.3166\pm 0.0051}{0.3175\pm 0.0055}
	\end{array}\right\} \quad \text{for} \  \cmball.
\end{align}
We highlight that the SPT results correspond to a $0.7\%$ determination of $\sigmaeight$, which is competitive with both \planck{} and ACT.

\begin{figure}
	\includegraphics[width=\columnwidth]{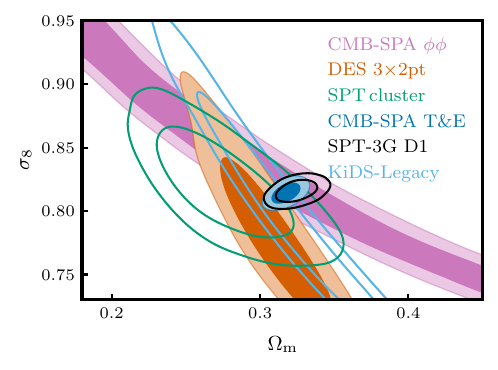}
	\caption{Constraints on structure growth in the $\omm$-$\sigmaeight$ plane using \sptlr{} (black, this work), primary \cmball{} data (blue, this work), combined SPT, ACT, \planck{} CMB lensing data (pink,~\citep{qu25}), SPT galaxy cluster statistics (green,~\citep{bocquet25}), \replace{and}{} DES-Y3 3$\times$2pt analysis with fixed neutrinos (orange,~\citep{abbott22a}), \RR{and KiDS cosmic shear (light blue,~\citep{wright25})}.
	The different data sets are sensitive to different parameterizations of $\seight{\alpha}=\sigmaeight(\omm/0.3)^{\alpha}$, where $\alpha$ corresponds to the slope in the $\omm$-$\sigmaeight$ plane.
	Yet, the four cosmological probes yield consistent results.}
	\label{fig:S8}
\end{figure}

Other cosmological probes are also able to constrain structure growth; specifically, individual and joint analyses of the following measurements allow for precise determinations: galaxy weak lensing~\citep{amon22, secco22, asgari21, li23c, li23, dalal23, jefferson25}, galaxy clustering and its combination with galaxy weak lensing (referred to as 3$\times$2 point analyses)~\citep{heymans20,abbott22a,miyatake23}, and galaxy cluster statistics~\citep{bocquet24, bocquet25}.
These probes are less sensitive to either parameter individually, but constrain the combination $\seight{\alpha}=\sigmaeight(\omm/0.3)^{\alpha}$ tightly, where $\alpha$ is typically chosen to maximize the precision in $\seight{\alpha}$ for a given experiment.
Until recently, there was moderate statistical evidence for a significantly lower $\seight{0.5}$ value from cosmic shear \citep{asgari21,li23,li23c} and 3$\times$2 point analyses \citep{des17,heymans20,miyatake23} compared to primary CMB and CMB lensing constraints.
However, many of the most recent analyses, with more data and improved modeling, as well as following comprehensive comparisons between experiments, are reporting \seight{0.5} values that differ from the CMB \lcdm{} constraints by less than $2\,\sigma$~\citep{abbott23b, wright25, stoelzner25}.
We are now in a position where constraints on $\seight{\alpha}$ from vastly different cosmological probes agree.
For example, the combined primary CMB constraint agrees at $1.\replace{8}{9}\,\sigma$ with the 3$\times$2 point analysis of DES-Y3~\citep{abbott22a} (using $\alpha=0.5$), at $0.1\,\sigma$ with the CMB lensing analysis of~\citep{qu25} (using $\alpha=0.25$), \replace{and}{} at $1.\replace{1}{2}\,\sigma$ with the galaxy cluster statistics analysis of~\citep{bocquet24} (using $\alpha=0.3$), \RR{and at $1.0\,\sigma$ with the latest KiDS cosmic shear analysis~\citep{wright25} (using $\alpha=0.58$)}, as illustrated in \cref{fig:S8}.
\replace{Moreover, the latest KiDS cosmic shear analysis agrees at $0.86\,\sigma$ with the combined primary CMB result~\citep{wright25}.}{}
This consistency is impressive, as the underlying experiments target different phenomenological signatures and span a wide range of redshifts and scales, which lead to different degeneracy directions in the $\omm$-$\sigmaeight$ plane.
Additionally, the instruments carrying out the surveys observe at a range of wavelengths and are affected by different systematics.
In this light, the consistent determination of $\seight{\alpha}$ is remarkable.

Finally, we estimate $\taureio$ excising the \planck{} PR4 based prior used in our baseline.
The amplitude of the CMB anisotropy power spectra is proportional to $A_\mathrm{s} e^{-2\taureio}$.
In the absence of large scale E-mode information, or a prior on $\taureio$, the degeneracy between $A_s$ and the amplitude suppression factor $e^{-2\taureio{}}$ can be broken via gravitational lensing, which is not affected by $\taureio{}$.
We remove the $\taureio{}$ prior from \cmball{} and report $\taureio = \replace{0.078\,\pm\,0.013}{0.076\,\pm 0.013}$, which is within $2\, \sigma$ of our \planck{}-based prior of $\mathcal{N}(0.051, 0.006)$.  We note that because CMB lensing is driving this low-E-free $\tau$ constraint, a slight preference for higher $\tau$ in \cmball{} can effectively be restated as a slight \cmball{} preference for an amplitude of lensing larger than the \lcdm\  expectations, as shown in the next section.
Moreover, this result is in excellent agreement with the corresponding \planck{} constraint on \taureio{} free of large scale \EE\ (``low-E'') data, $\taureio = 0.079\,\pm\,0.018$, but about $30\,\%$ more precise.\footnote{Compared to \texttt{base\_plikHM\_TTTEEE\_lowl\_lensing} from~\citep{planck18-6}}
Note that while the \cmball{}-based constraint is twice as wide as the \planck{} low-E result, it is as precise as the \WMAP{} final mission result of $\taureio = 0.089\,\pm\,0.014$~\citep{bennett13}.
Still, the \taureio{} constraint above is too wide to serve as an accurate cross-check of the \planck{} E-mode measurement.

\subsubsection{Lensing amplitude}
\label{sec:alens}

We now study the consistency of the signature of gravitational lensing in primary CMB and CMB lensing reconstruction data.
We stress that the models used in this section are non-physical: they are designed to facilitate further tests of the validity of the \lcdm{} description of CMB data.
As has been widely reported, analyses of \planck{} primary CMB anisotropy power spectra show a preference for an excess in the effects of lensing on the primary CMB anisotropy power spectra compared to \lcdm{} expectations at the $2$-$3\,\sigma$ level \citep{planck18-5, planck18-6}.
This preference is largely driven by an apparent excess smoothing of the \planck{} high-multipole temperature data.
We look for evidence of similar effects in the SPT data analyzed in this work, both alone and in combination with the latest ACT power spectrum measurements.
Following~\citep{ge24}, we introduce $\Atwopt$, which quantifies the effect of gravitational lensing on CMB \TTTEEE{} power spectra.
This parameter scales the amplitude of the lensing power spectrum used to predict the lensed CMB spectra from the unlensed expectation, but does not affect the prediction for the amplitude of the reconstructed CMB lensing potential ($\phi \phi$).

\begin{figure}
	\includegraphics[width=\columnwidth]{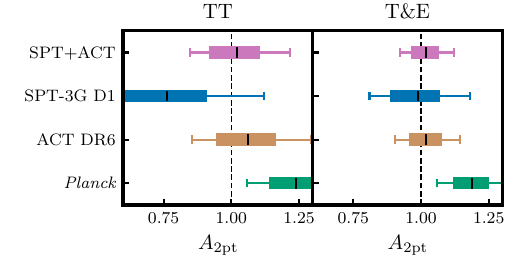}
	\begin{tabular}{l | l | l }
		\toprule
		\textbf{Data Set} & \multicolumn{1}{c|}{\TT{}} & \multicolumn{1}{c}{\T\&\E{}} \\
		\midrule
		\ground{} & $\Atwopt = 1.014 \pm 0.098$ & $\Atwopt = 1.016^{+0.048}_{-0.054}$  \\
		\sptlr & $\Atwopt = 0.76^{+0.15}_{-0.19}$ & $\Atwopt = 0.991^{+0.083}_{-0.10}$  \\
		\ACTDR{} & $\Atwopt = 1.06^{+0.10}_{-0.12}$ & $\Atwopt = 1.020\pm 0.060$  \\
		\planck{} & $\Atwopt = 1.239\pm 0.095$ & $\Atwopt = 1.185\pm 0.067$  \\
		\bottomrule
	\end{tabular}
	\caption{Constraints on $\Atwopt$, the amplitude of lensing inferred from effects on the \TTTEEE{} power spectra, from \planck{} (green), \ACTDR{} (brown), \sptlr{} (blue), and \ground (purple) temperature data (\emph{left}) and the combination of temperature and polarization data (\emph{right}), both excluding lensing \PP{} data. 
	We show the 68\% and 95\% confidence regions as the filled contours and whiskers, respectively. 
	The black bar indicates the mean value of the posterior, while the dashed line indicates the \lcdm{} expectation of $\Atwopt = 1$. 
	While the \planck{} data prefer a value of $\Atwopt > 1$, driven by the temperature data, the ground-based data sets are consistent with the \lcdm{} expectation. 
	The combination of ACT and SPT data is in excellent agreement with the \lcdm{} prediction.
	We report 68\% confidence intervals in the table below the figure.}
	\label{fig:alens_triangle}
\end{figure}

First, we consider constraints based exclusively on primary CMB temperature data.
We show the marginalized posteriors of $\Atwopt$ for different combinations of CMB temperature data in \cref{fig:alens_triangle}.
Constraints from SPT temperature data alone are relatively wide due to the limited survey area.
They are in agreement with the standard model prediction of unity at 1.6$\,\sigma$.
The combination of ground-based temperature power spectrum measurements, ACT+SPT \TT{} (see \cref{fig:alens_triangle}) is in excellent agreement with the \lcdm{} expectation and the associated posterior mean is $2.3$ standard deviations below the value preferred by the \planck{} temperature data.\footnote{We do not quote a quantitative statistic for the consistency of the \ground{} and \planck{} posterior distributions as \ground{} contains ACT data that is significantly correlated with \planck{} data.}
While this may suggest that the features in the \planck{} spectrum that cause the observed preference for $\Atwopt > 1$ may not be cosmological and rather statistical or systematic, we cannot make a definitive judgment at this point.
We note that re-analyses of \planck{} data have generally led to a reduction of this feature~\citep{rosenberg22, tristram24}.

In the right panel of \cref{fig:alens_triangle}, we show the posteriors for the \T{}\&\E{} case.
For SPT data, the bulk of the constraining power lies in polarization data and we see a considerable tightening of the $\Atwopt$ posterior.
The result is centered close to unity,\footnote{
Past SPT analyses have yielded $\lesssim 2\,\sigma$ fluctuations below one in $\Atwopt$, however we stress that there is no one-to-one comparison.
The closest case is~\citep{balkenhol23} which is based on observations of the same part of the sky.
However, the data presented here have a substantially lower noise level, which weights the constraint more towards polarization and shifts the posterior towards unity.
Significant differences in the observation fields and in the spectra used compared to~\citep{story13} and~\citep{chou25} prevent a meaningful comparison.}
which is also true for the \ground{} combination.
To further investigate the trends in the \planck{} data that project onto $\Atwopt > 1$, it would be interesting to compare the \planck{} and \ACTDR{} temperature data at $\ell>1000$.
However, this requires a quantification of the correlation of the two measurements, which is beyond the scope of this work.

We now also consider CMB lensing reconstruction data.
We follow the prescription of~\citep{ge24} and add $\Arecon$ to our model.
$\Arecon$ scales the prediction for the reconstructed CMB lensing power spectrum but has no effect on the prediction for the effects of lensing on the \T{} and \E{} power spectra. In other words, $\Atwopt$ and $\Arecon$ control independent subsets of the effects of the lensing power spectrum on CMB observables, which are often lumped together under one parameter $A_L$ or $A_{\rm lens}$ (see below).
Varying $\Atwopt$ and $\Arecon$ independently, we report
\begin{align}
	&
	\left.
	\begin{array}{l}
		\Atwopt = 0.986^{+0.078}_{-0.097} \\
		\Arecon = 0.974^{+0.081}_{-0.11}
	\end{array}\right\} \quad \text{for} \  \sptlr, \\ 
	&
	\left.
	\begin{array}{l}
		\Atwopt = 1.026\pm 0.048 \\
		\Arecon = 0.990\pm 0.050
	\end{array}\right\} \quad \text{for} \  \ground, \\
	&
	\left.
	\begin{array}{l}
		\Atwopt = \replace{1.083\pm 0.037}{1.079\pm 0.038} \\
		\Arecon = \replace{1.048\pm 0.031}{1.041\pm 0.031}
	\end{array}\right\} \quad \text{for} \  \cmball.
\end{align}
With these results, we see no evidence for inconsistent signatures of gravitational lensing between primary CMB and CMB lensing data for ground-based and large-scale satellite data.
The \cmball{} results are consistent with the \lcdm{} expectation at the $\replace{1.7}{1.6}\,\sigma$ level in the two-dimensional space of $(\Atwopt, \Arecon)$.

We now vary the amplitude of the lensing power spectrum and the signature of gravitational lensing in the primary CMB coherently, i.e. $\Atwopt = \Arecon$.
We collapse the two parameters into one, $\Alens$, and report:
\begin{align}
	\label{eq:alens_withlensing}
	\Alens &= 0.972^{+0.079}_{-0.089} \ \text{for} \  \sptlr, \\
	\Alens &= 1.011^{+0.045}_{-0.051} \ \text{for} \  \ground, \\
	\Alens &= \replace{1.057\pm 0.030}{1.052\pm 0.032} \ \text{for} \  \cmball, \label{eq:alensall}
\end{align}
These results are all consistent within $\replace{1.9}{1.6}\,\sigma$ with the standard prediction of $\Alens = 1$.
The fact that the amplitude of gravitational lensing measured in CMB data is consistent with the cosmology preferred by the lensing-marginalized CMB observables is yet another illustration of the internal consistency within and among the three CMB data sets treated here.
We revisit this scenario adding BAO data in \cref{sec:alens_BAO}.

\begin{figure}
	\includegraphics[width=\columnwidth]{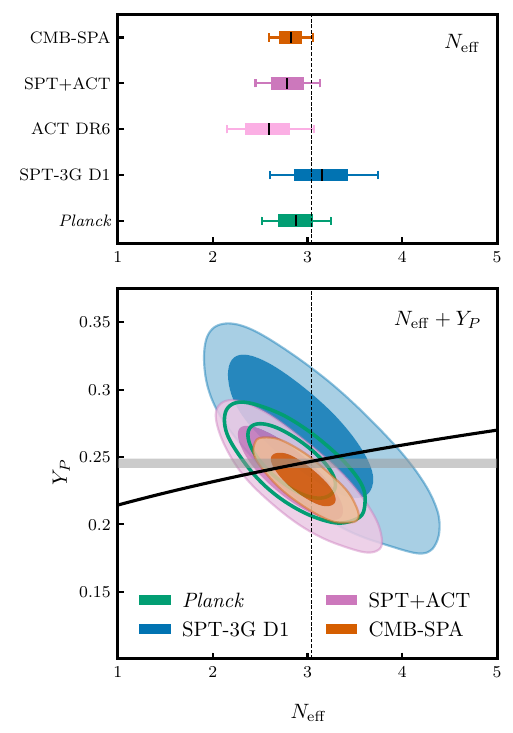}
	\caption{Constraints on light relics from CMB data: \sptlr (blue), \ground{} (purple), \planck{} (green), and \cmball{} (red).
	\textit{Top panel:} Marginalized one-dimensional posterior on $\Neff{}$, when only varying this parameter.
	The \ground{} constraint is tighter than the \planck{} one, highlighting the exquisite constraining power of ground-based CMB data.
	All CMB constraints are within $2 \, \sigma$ of the standard model prediction of $\Neff{}=3.044$.
	\textit{Bottom panel:}
	Constraints on $\neff{}$ and the primordial helium abundance $\Yp$, when the two parameters are varied simultaneously. The black line represents BBN consistency, while the gray band corresponds to the $1\,\sigma$ region of the $\Yp$ constraint from observations of metal-poor galaxies~\citep{aver20}.
	We find no evidence for new light particles in the early universe from CMB data.}
	\label{fig:neff_constraints}
\end{figure}

\subsubsection{New light particles}
\label{sec:neff}

\begin{table*}[t!]
	\centering
	\setlength{\tabcolsep}{12pt}
	\begin{tabular}{c c c c c}
		\hline
		& \sptlr & \ground{} & \cmball{} & \sptlr{} + DESI\\
		\hline \hline
		\neff & $3.1\replace{7}{8}^{+0.29}_{-0.33}$ & $2.7\replace{7}{8}\pm 0.17$ & $2.8\replace{1}{2}\pm 0.12$ & $3.5\replace{2}{3}\pm 0.23$\\
		\hline
		\Yp & $0.264\pm 0.022$ & $0.226^{+0.015}_{-0.013}$ & $\replace{0.2285\pm 0.0085}{0.2294\pm 0.0083}$ & $0.279\pm 0.022$\\
		\hline
		\neff & $2.9\replace{7}{8}^{+0.40}_{-0.64}$ & $2.8\replace{5}{6}^{+0.32}_{-0.40}$ & $\replace{2.99^{+0.22}_{-0.26}}{2.96^{+0.21}_{-0.24}}$ & $3.6\replace{3}{4}^{+0.39}_{-0.44}$\\
		\Yp & $0.269^{+0.040}_{-0.030}$ & $0.236^{+0.025}_{-0.021}$ & $\replace{0.231\pm 0.014}{0.233\pm 0.013}$ & $0.241\pm 0.034$\\
		\hline
	\end{tabular}
	\caption{Constraints on the effective number of relativistic species \neff{} and the primordial helium abundance $\Yp$ from CMB data alone and in combination with DESI.
	We report constraints on \neff{} and $\Yp$ when varying each parameter individually (first and second row) and when they are varied independently of each other at the same time (last two rows).
	The results are consistent with the standard model prediction of $\neff{}=3.044$, and the $\Yp$ constraint is consistent with the \lcdm{} BBN consistency prediction \citep{parthenope17, pitrou18} and the measurement of~\citep{aver20}.
	For the above models, we do not report constraints for \ground{} and \cmball{} in combination with DESI as these combinations of CMB and BAO data do not meet our consistency requirements for joint analyses (see \cref{sec:CMB+BAO}).
	}
	\label{tab:neff_yp}
\end{table*}

We now search for the signature of new light particles in the early universe by using our data to estimate the effective number of relativistic species, \neff{}. 
This parameter quantifies the density of relativistic particles other than photons. It is defined 
in terms of the density expected from an equivalent number of relativistic species under the assumption they have negligible chemical potential 
and none of the entropy of the electrons and positrons gets transferred to the neutrinos, so that:
\begin{equation}
	\rho_{\rm rad}
 	= \rho_\gamma\left[1+(7/8) \times (4/11)^{4/3}\neff{}\right],
\end{equation}
where $\rho_{\rm rad}$ is the radiation density and $\rho_\gamma$ is the photon density.
In the standard model, a small fraction of the entropy of the electrons and protons does get transferred to the neutrinos so that 
$\neff = 3.044$
\citep{froustey20, bennett20,akita_precision_2020,drewes_towards_2024}.

The impact of light relics on cosmological observables, and the CMB in particular, has been studied in many papers~\citep{jungman96b, hu96,bashinsky04,hou13,follin15,baumann16,cyr22,Ge:2022qws}. 
One of the more recent treatments~\citep{Ge:2022qws} emphasized the connection between changes to various rate ratios and dimensionless observables (such as those of the CMB and BAO), as was done by~\citep{zahn03} in the context of a time-varying gravitational constant $G$. 
The relevant rates here are free-fall rates for each component, $\sqrt{G \rho(z)}$, the Hubble rate $H(z)$, and the Thomson scattering rate $\sigma_{\rm T} n_{\rm e} (z)$.
Extending \lcdm{} to include additional light relics opens up a scaling transformation that preserves important rate ratios and, therefore, leads to an approximate degeneracy in the parameter space. 
BAO observables, $\rd H(z)$ and $\rd /D_{\rm M}(z)$, where $\rd$ is the comoving size of the sound horizon at the end of the baryon drag epoch, $H(z)$ the expansion rate, and $D_{\rm M}(z)$ the transverse comoving distance, are particularly unaffected by this transformation so we refer to this as ``BAO scaling.''
Note that important angular scales to which CMB power spectra are highly sensitive such as the angular scale of the sound horizon and the angular scale of the comoving Hubble length at matter-radiation equality, both projected from the last-scattering surface, are also approximately preserved under this transformation.\footnote{These are exactly preserved if the redshift of last-scattering does not change.}

The degeneracy given CMB data is only approximate since other important rate ratios inevitably change. 
Chief among these is the ratio of the Thomson scattering rate to the Hubble rate, $\sigma_{\rm T}  n_{\rm e}(z)/H(z)$,\footnote{
	where $\sigma_{\rm T}$ is the Thomson scattering cross-section and $n_{\rm e}(z)$ the electron density.
	} 
which impacts Silk damping and polarization generation~\citep{zahn03}. 
Other important effects, recently reviewed in~\citep{Ge:2022qws}, further contribute to lift the degeneracies.

Due to these collective effects, the approximate degeneracy direction given CMB data is not along the BAO scaling direction, 
but instead along a parameter direction in which the matter density, $\rho_{\rm m}$, 
scales up more slowly than the radiation density, $\rho_{\rm rad}$, and the dark energy density, $\rho_\Lambda$. 
While BAO scaling leaves $\Omega_{\rm m}$ unaltered, the CMB degeneracy direction has $\Omega_{\rm m}$ decreasing with increasing $N_{\rm eff}$, as we discuss in \cref{sec:neff_BAO}. 

Since changes to $\Neff$ lead to changes in diffusion damping, polarization generation, and acoustic peak locations, the weighting of SPT data toward small angular scales and polarization spectra make them particularly interesting for this search.
Varying \neff{} in our analysis, we report
\begin{align}
	\Neff{} &= 3.1\replace{7}{8}^{+0.29}_{-0.33} \ \text{for} \  \sptlr, \\
	\Neff{} &= 2.7\replace{7}{8}\pm 0.17 \ \text{for} \  \ground, \\
	\Neff{} &= 2.8\replace{1}{2}\pm 0.12 \ \text{for} \  \cmball.
\end{align}
The SPT constraint is in excellent agreement with the standard model prediction of $\neff = 3.044$ and the data allow for a $\sim 10\%$ determination of \Neff{}\footnote{
	A previous constraint on \neff{} was derived in \citep{ge24} using unlensed \EE{} and lensing \pp{} band powers, yielding $2.70^{+0.70}_{-0.91}$. 
	The determination of \neff{} using lensed \sptlr{} \EE{} and the same \pp{} band powers yields a posterior centered at $\Neff{} = 3.36^{+0.77}_{-0.88}$, which is almost $1\,\sigma$ higher. 
	As mentioned in footnote 27, the pipelines employ different methodologies and use unequal amounts of data.
	End-to-end simulations are required to fully evaluate the consistency of the measurements.
}.
This is weaker than the result computed from the ACT DR6 likelihood introduced in \cref{tab:dataset}, which yields $N_\mathrm{eff} = 2.58^{+0.22}_{-0.25}$.
Constraints on \Neff{} from different combinations of data sets are explored in \citep{calabrese25}.
However, the SPT data carry considerable weight in the \ground{} combination, which yields a posterior of slightly tighter width to the \planck{} constraint, $\Neff{} = 2.8\replace{6}{7}\pm 0.19$~\citep{planck18-5}.
We report no evidence for a deviation from the standard model prediction of $\neff = 3.044$ for the \ground{} and \cmball{} combinations.
Marginalized \Neff{} posteriors for these cases are shown in the top panel of \cref{fig:neff_constraints}.

The constraints above enforce consistency between \Neff{} and the helium abundance $\Yp$ in the framework of Big Bang Nucleosynthesis (BBN).\footnote{We use the BBN tables based on~\citep{parthenope17}, which are the default in the \class version used in our analyses. 
To be precise, we define $\Yp$ as the helium \emph{mass} fraction.}
$\Yp$ is relevant here, because the abundance of light relics affects the production of helium during BBN, and also because changes to $\Yp$ lead to changes to the Thomson scattering rate through the electron density $n_{\rm e} (z)$.
For a fixed baryon density, increasing $\Yp$ reduces $n_{\rm e}(z)$. This is because more neutrons are captured in helium atoms instead of decaying to protons and electrons during BBN, and because helium recombines earlier than hydrogen, thus decreasing the electron density at the time of hydrogen recombination.
Freely varying $\Yp$ means that the ratio of the Thomson and Hubble rates can be preserved, even while $H(z)$ is scaling up with \neff{}.
This leads to a partial degeneracy between $\Yp$ and \neff{} for CMB data.

We now vary \Neff{} and $\Yp$ simultaneously in our analysis and show the resulting constraints in \cref{fig:neff_constraints}.
We report 68\% confidence intervals for the two parameters:
\begin{align}
	&
	\left.
	\begin{array}{l}
		\Neff{} = 2.9\replace{7}{8}^{+0.40}_{-0.64} \\
		\Yp = 0.269^{+0.040}_{-0.030}
	\end{array}\right\} \quad \text{for} \  \sptlr, \\ 
	&
	\left.
	\begin{array}{l}
		\Neff{} = 2.8\replace{5}{6}^{+0.32}_{-0.40} \\
		\Yp = 0.236^{+0.025}_{-0.021}
	\end{array}\right\} \quad \text{for} \  \ground, \\ 
	&
	\left.
	\begin{array}{l}
		\Neff{} = \replace{2.99^{+0.22}_{-0.26}}{2.96^{+0.21}_{-0.24}} \\
		\Yp = \replace{0.231\pm 0.014}{0.233\pm 0.013}
	\end{array}\right\} \quad \text{for} \  \cmball.
\end{align}
All of the \Neff{} results match the standard model prediction of $3.044$.
The helium fraction constraints are consistent with the predictions for each data sets under BBN (using the calculations of~\citep{parthenope17} or of~\citep{pitrou18}) as well as the $\Yp = 0.2453 \pm 0.0034$ determination of~\citep{aver20}, which is based on observations of metal-poor galaxies.
Lastly, varying only $\Yp$ while keeping \Neff{} fixed to the \lcdm{} value also shows no deviation from the standard model.
Therefore, we conclude that we see no evidence for additional light relics in the early universe in CMB data.
All results of this section are given in \cref{tab:neff_yp}.
We revisit these models adding BAO data in \cref{sec:neff_BAO}.

\subsubsection{Modified recombination}
\label{sec:modrec}

In the final extension to \lcdm{} explored with CMB data alone, we reconstruct the free electron fraction, $X_{\rm e}(z) = n_{\rm e}(z) / (n_{\rm HI}(z) + n_{\rm HII}(z))$, during the epoch of cosmological recombination, approximately spanning redshifts $500 < z < 1600$.
In modified recombination scenarios, a new degeneracy between $\Hubble$ and $X_{\rm e}(z)$ emerges which preserves the angular scale of the sound horizon at last scattering, $\thetastar = r_{\rm s}^\star / D_{\rm A}^\star$.
As such, modifications to the standard recombination scenario have enjoyed recent attention due to their potential role in resolving the Hubble tension~\citep[e.g.][]{hart19,sekiguchi20, lee22,lynch24a,mirpoorian24}.
To explore these models in a non-parametric way, we use the \texttt{ModRec} model and emulator introduced in~\citep{lynch24a} and used in \citep[e.g.,][]{calabrese25,lynch25}, which uses seven free parameters to define the deviation of $X_{\rm e}(z)$ from the standard model prescription.\footnote{Non-parametric here means that no particular functional form is imposed on $X_{\rm e}(z)$. However, we are still using a parametric cosmological model, which the seven $X_{\rm e}(z)$ parameters are now part of.}
For MCMC analyses of this model, we use the same priors as~\citep{lynch24a}.
We restrict ACT data to $\ell < 4000$, as the accuracy of the \texttt{ModRec} emulator has not been validated at higher multipoles.
We do not include lensing data due to compatibility issues with the \texttt{ModRec} emulator; since this model does not substantially change lensing predictions, we expect this to be a small effect (see Fig. 12 in~\citep{prabhu24}).
For this model, we consider chains to be sufficiently converged when the Gelman-Rubin statistic is $R-1<0.03$.

\begin{figure}[t]
    \centering
    \includegraphics{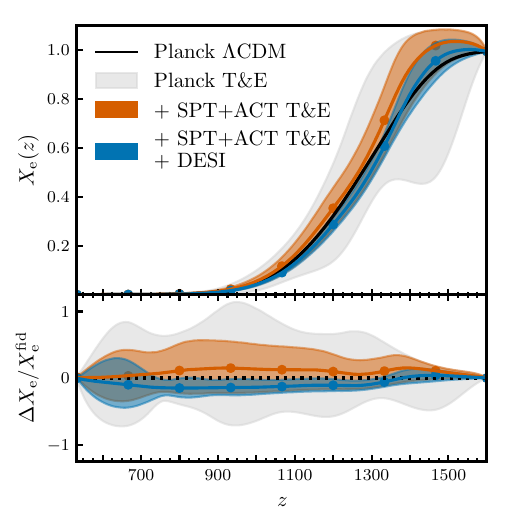}
    \caption{\textit{Top:} Reconstructions of the ionization fraction $X_{\rm e}(z)$ during the epoch of recombination, using the \texttt{ModRec} model.
	Solid lines show the mean reconstructed ionization fraction, with points indicating the placement of the seven control points.
	Bands indicate the 95\% confidence intervals.
	The black line is the fiducial ionization fraction for standard recombination with \planck{} cosmology~\citep{planck18-6}.
	\textit{Bottom:} The fractional change relative to a fiducial ionization fraction calculated using the \planck{} T\&E mean parameters~\citep{planck18-6}.
	Note that nodes can be significantly correlated and hence the significance of deviations from zero is difficult to judge by eye.
	The reconstructions based on CMB data are in excellent agreement with the standard model prediction; adding DESI BAO data leads to a mild $1.8\,\sigma$ preference for earlier recombination as discussed in \cref{sec:modrec_BAO}.}
    \label{fig:xe_reconstructions}
\end{figure}

Whereas we have previously compared the width of individual parameter posteriors across CMB data sets to assess constraining power, this is not informative in this model as we have seven considerably correlated extension parameters.
Therefore, we use the volume reduction (VR) statistic to quantify the information data sets hold in this model space introduced in \cref{sec:lcdm}.
Here, we report natural logarithms of the square-root of determinant ratios and report values relative to the \planck{} T\&E case, i.e.
\begin{equation}
    \log {\rm VR} = \log \left( \det(\Sigma_{\text \planck}) / \det(\Sigma)\right).
\end{equation}
For Gaussian posteriors, this quantity gives the number of e-folds by which the posterior volume is reduced compared to the \planck{} posterior.
Additionally, to assess the significance of the deviation from \lcdm{}, we translate the goodness-of-fit improvement compared to the standard model to the equivalent significance for a one dimensional Gaussian distribution, accounting for the seven additional degrees of freedom introduced.

Results from CMB data are in excellent agreement with the standard model prediction:
for SPT+\planck{} data, the mean reconstructed ionization fraction matches the standard scenario within $0.18\,\sigma$.
The addition of SPT to \planck{} data significantly reduces uncertainties in the reconstruction of $X_{\rm e}(z)$:
\begin{equation}
    \log {\rm VR} = 5.9 \text{ for \planck{} T\&E + \sptlr T\&E}
\end{equation}
This reduction in posterior volume of almost 6 e-foldings is due to the inclusion of low-noise measurements of the EE and TE damping tail at $\ell>1800$, which contain information on $X_{\rm e}(z)$ at early times~\citep{lynch24a}.
For comparison, we report the reduction of $\log \mathrm{VR=7.1}$ for the case when ACT data are added to \planck{} instead.
Adding both SPT and ACT data to \planck{} reduces the reference posterior volume by about an additional e-fold:
\begin{equation}
    \log {\rm VR} = \replace{8.3}{7.6} \text{ for \cmball{} T\&E}.
\end{equation}
The significance of the deviation from the standard recombination scenario is $1.1\,\sigma$ in this case and we show this reconstruction in \cref{fig:xe_reconstructions}.
Overall, we find no evidence for modified recombination from CMB data alone; given the volume reduction from \planck{}, this is a non-trivial validation of the standard recombination model on the new ground-based data.
We add BAO data to this analysis in \cref{sec:modrec_BAO}.

\subsection{Evaluating the consistency of CMB and BAO data in \lcdm{}}
\label{sec:CMB+BAO_lcdm}

We now discuss the consistency of CMB and BAO measurements in the \lcdm{} model.
The state-of-the-art BAO data set is the DESI Data Release 2 (DR2)~\citep{desi25}.
Under \lcdm{}, BAO data constrain $\omm$ and $\hrd$ and the DESI data prefer a lower matter density and a higher $\hrd$ than CMB data~\citep{efstathiou25b,garciaquintero25}.\footnote{\citet{tang25} show that if mock SDSS BAO data are generated assuming a dynamical dark energy model, a naive analysis in \lcdm{} can lead to non-negligible biases in $\omm$.
As DESI data greatly improve on SDSS data, more work is needed to understand to what extent this effect may apply to DESI data as well.}
In the discussion below, we translate the differences in the $\omm$ and $\hrd$ constraints from CMB and DESI data to equivalent statistical significances for a one-dimensional Gaussian distribution; we consider data sets consistent if they agree to better than $3\,\sigma$ according to this metric and allow ourselves to combine them.

\begin{figure}
	\includegraphics[width=\columnwidth]{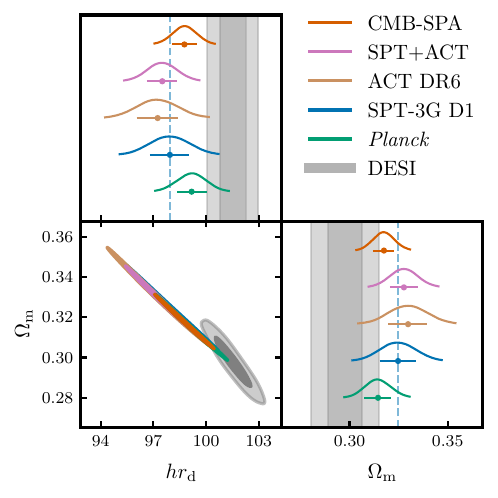}
 	\begin{tabular}{l c c c c}
		\toprule
		                 & $100\,\omm$       & \hrd{}\,\rdragunit & Distance to \DESI{} \\
		\midrule
		\cmball{}        & $\replace{31.66\pm 0.50}{31.75\pm0.55}$   & $\replace{98.89\pm 0.63}{98.77\pm 0.69}$    & $2.8\,\sigma$       \\
		\ground{}        & $32.77\pm 0.72$   & $97.51\pm 0.87$    & $3.7\,\sigma$       \\
		\SPT{}+\planck{} & $31.89\pm 0.54$   & $98.63\pm 0.67$    & $3.0\,\sigma$       \\
		\ACTDR{}         & $33.0\pm 1.0$     & $97.2\pm 1.2$      & $3.1\,\sigma$       \\
		\sptlr           & $32.47\pm 0.91$   & $97.9\pm 1.1$      & $2.5\,\sigma$       \\
		\planck{}        & $31.45\pm 0.67$   & $99.18\pm 0.84$    & $2.0\,\sigma$       \\
		\DESI{}          & $29.76 \pm 0.87 $ & $101.52\pm 0.73$   &                     \\
		\bottomrule
	\end{tabular}
	\caption{\emph{Top}: Comparison of \lcdm{} constraints from DESI and CMB data in the $\omm$-$\hrd$ plane. \emph{Bottom}: Mean and 68\% confidence intervals, as well as significance of the discrepancy, ranging between $2.0$-$3.7\,\sigma$ for CMB data sets and particularly $2.8\,\sigma$ for the \cmball{} combination.}
	\label{fig:DESI_hord}
\end{figure}

We compare \lcdm{} constraints from DESI and CMB data in \cref{fig:DESI_hord}.
Using SPT data we find $\omm= 0.3247\pm0.0091$ and $\hrd{}= 97.9\pm1.1\,\rdragunittxt$.
Accounting for the correlations between the two parameters, the difference with DESI translates to a one-dimensional Gaussian fluctuation of $2.5\,\sigma$.
This mild discrepancy becomes stronger when more CMB data are added.
Among CMB data sets, ACT data prefer the highest $\omm$, and the tension between ACT and DESI is at the $3.1\,\sigma$ level.\footnote{
	We show the role that the $\tau$ prior plays in the consistency between ACT and DESI in \cref{app:BAO_consistency,fig:hord_act}.
}
We note that the ACT results used here include primary T\&E CMB data as well as CMB lensing from \ACTDR{} alone. This is different than what was done in~\citep{garciaquintero25}, where the ``ACT'' data combination also contained \planck{} lensing, and also used a different prior on $\taureio$, reducing the difference with DESI to the $2.7\,\sigma$ level.\footnote{
	We reproduce this result when using the same data combination, see \cref{app:BAO_consistency}.
}
A combination of ground-based experiments into \ground{} yields tighter error bars.
This increases the distance with DESI to $3.7\,\sigma$.
Similarly, the joint SPT+\planck{} constraints are different than DESI at $3.0\,\sigma$ due to the constraining power of the combined CMB data sets, even though the \sptlr{} and \planck{} constraints individually lie below the $3\,\sigma$ threshold.
However, the \planck{} large angular scale data that are added going from \ground{} to \cmball{} favor a lower $\omm$~\citep{planck18-6}, and we report a difference with DESI of $2.8\,\sigma$ for \cmball{}.

\begin{table}
    \begin{tabular}{l | l l} \hline
        \textbf{Parameter} & \bf \makecell[l]{{\sptlr}                                             \\ {+ \DESI{}}} \quad & \bf \makecell[l]{\cmball{} \\ + \DESI{}} \quad \\ \hline
        \hline
        \emph{Sampled}                                                                         \\
        $10^4$\thetastar     & $ 104.227\pm 0.056$ & $ \replace{104.180\pm 0.022}{104.179\pm 0.022}$ \\
        100\,\ombh           & $ 2.218\pm 0.022$   & $ \replace{2.2452\pm 0.0089}{2.2478\pm 0.0091}$ \\
        100\,\omch           & $ 11.749\pm 0.079$  & $ \replace{11.813\pm 0.058}{11.809\pm 0.060}$  \\
        \ns                & $ 0.949\pm 0.012$ & $ \replace{0.9728\pm 0.0027}{0.9726\pm 0.0028}$ \\
        \logA              & $ 3.066\pm 0.014$   & $ \replace{3.0574\pm 0.0094}{3.0586\pm 0.0094}$ \\
        \taureio           & $ 0.0559\pm 0.0056$ & $ \replace{0.0625\pm 0.0050}{0.0613\pm 0.0051}$ \\
        \hline
        \emph{Derived}                                                                         \\
        \Hubble\,[\kmsmpc] & $ 68.21\pm 0.31$    & $ \replace{68.06\pm 0.24}{68.10\pm 0.24}$    \\
        \Age\,\Ageunit     & $ 13.795\pm 0.025$  & $ \replace{13.783\pm 0.012}{13.781\pm 0.012}$  \\
        \clamp             & $ 1.920\pm 0.021$ & $ \replace{1.8773\pm 0.0055}{1.8841\pm 0.0079}$ \\
        \omegal            & $ 0.6983\pm 0.0039$ & $ \replace{0.6950\pm 0.0033}{0.6954\pm 0.0033}$ \\
        \omm               & $ 0.3017\pm 0.0039$ & $ \replace{0.3049\pm 0.0033}{0.3045\pm 0.0033}$ \\
        \rdrag\,\rdragunit & $ 147.99\pm 0.33$   & $ \replace{147.51\pm 0.17}{147.50\pm 0.18}$   \\
        \sigmaeight        & $ 0.8079\pm 0.0059$ & $ \replace{0.8120\pm 0.0038}{0.8122\pm 0.0039}$ \\
        \hline
    \end{tabular}
    \caption{Joint \lcdm{} parameters constraints from \sptlr{} and \cmball{} with \DESI{}. We report mean values and 68\% confidence intervals.}
    \label{tab:cmb_desi_lcdm}
\end{table}

We report joint constraints on \lcdm{} parameters from \sptlr{} and \cmball{} with DESI data in \cref{tab:cmb_desi_lcdm}.
The addition of the BAO data tightens constraints on some parameters, yielding notably $\Hubble = 68.21\,\pm\,0.31\,\kmsmpc$ and $100\,\omch=11.749\,\pm\,0.079$ for SPT+DESI.
We forego reporting the combination of DESI with \ground{} as it does not meet our $3\,\sigma$ consistency requirement;\footnote{Though \ground{} is a subset of \cmball{}, given the consistency of CMB data we have demonstrated in \cref{sec:lcdm}, we expect that the addition of large scale \planck{} data pushes the joint CMB constraints closer to the underlying mean.
As the difference between \cmball{} and DESI is below our $3\,\sigma$ threshold, we report the joint results.}
if we were to do so, differences in the favored $\omm$ and $\hrd$ values would also lead to sizeable shifts in other cosmological parameters in the joint constraints compared to the CMB-preferred values.
Due to the degeneracies of the model space, $\Hubble$, $\sigmaeight$, and $\ns$ are particularly vulnerable to this effect.
While special attention is often given to the first two parameters as they can be accurately determined by different cosmological probes (see \cref{sec:lcdm}), the precise value of $\ns$ has consequences for the allowed model space of inflationary theories~\citep{yi25, maity25, byrnes25}; this compels caution in the interpretation of CMB+DESI results in \lcdm{}.

As one would expect, the previous data release from DESI finds $\omm$ and $\hrd$ values similar to DESI DR2, though with larger error bars which reduces the significance of the discrepancy with CMB data~\citep{desi24-6, desi24-7, desi25}.
DESI DR2 is also consistent with SDSS BAO data~\citep{beutler11,ross15,alam17,alam21, desi25}.
While the SDSS data support values of $\omm$ and $\hrd$ that are more consistent with CMB data in \lcdm{}, the data are also far less constraining than the DESI data, which prevents a meaningful comparison.

\begin{figure*}
	\includegraphics[width=\textwidth]{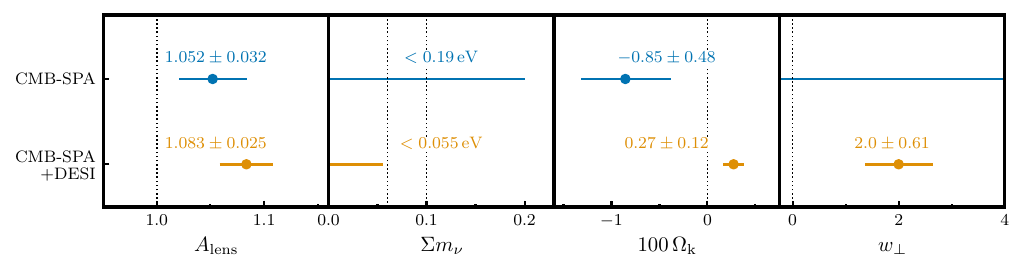}
	\caption{Constraints on extension parameters from \cmball{} (top, blue) and \cmball{}+DESI (orange, bottom); from left to right: \Alens{}, \mnu{}, \curv{}, and \wperp{} (which summarizes $\wo$ and $\wa$ constraints as defined in \cref{eq:wperp}).
	We give the mean and $68\%$ confidence interval above each constraint; for \mnu{} we indicate the $95\%$ upper limit with arrows.
	We indicate the value of the extension parameter to which the model reverts in \lcdm{} as the dotted vertical line; for \mnu{} we indicate the minimum values for the normal and inverted hierarchies ($0.06\,\mathrm{eV}$ and $0.1\,\mathrm{eV}$, respectively).
	We indicate the insensitivity of CMB data to the late-time evolution of dark energy by a horizontal blue line in the final panel.
	We find that differences between CMB and DESI data lead to sizeable shifts from the CMB-preferred values in joint constraints and moderate departures from \lcdm{}.
	We see a similar behavior for modified recombination scenarios (\cref{sec:modrec,sec:modrec_BAO}) (not shown in this figure).
	}
	\label{fig:cmball_DESI_summary}
\end{figure*}

\begin{table*}[th!]
	\centering
	\setlength{\tabcolsep}{12pt}
	\begin{tabular}{c c c c c c}
		\hline
		                    & \bf \cmball{}            & \bf \DESI{}             & \multicolumn{3}{c}{\bf \cmball{}+\DESI{}}                                                     \\
		\cline{4-6}
		Model               & $\chi^2_{\mathrm{CMB}}$  & $\chi^2_{\DESI{}}$      & $\chi^2_{\mathrm{CMB}}$                   & $\chi^2_{\DESI{}}$ & $\chi^2_{\mathrm{CMB+DESI}}$ \\
		\hline
		\hline
		\lcdm{}             & 603.7                    & 10.3                    & 606.7                                     & 14.6               & 621.3                        \\
		$\Alens{}$          & 601.5 ($2.1, 1.5\sigma$) & -                       & 603.4                                     & 10.9               & 614.3 ($6.9, 2.6\sigma$)     \\
		$\texttt{ModRec}$   & 517.1 ($8.5, 1.1\sigma$) & -                       & ($11.0$)                                  & ($2.2$)            & ($13.2, 1.8\sigma$)          \\
		$\curv$             & 600.8 ($2.8, 1.7\sigma$) & 10.0 ($0.3, 0.6\sigma$) & 606.0                                     & 11.1               & 617.0 ($4.2, 2.1\sigma$)     \\
		$\curv + m_{\rm e}$ & -                        & -                       & 604.0                                     & 10.8               & 614.7 ($6.5, 2.1\sigma$)     \\
		$\mnu{}$            & 603.0 ($0.7, 0.8\sigma$) & -                       & 604.0                                     & 12.2               & 616.2 ($5.1, 2.3\sigma$)     \\
		$\wo \wa$           & -                        & 5.6 ($4.7, 1.7\sigma$)  & 602.3                                     & 7.4                & 609.7 ($11.6, 3.0\sigma$)    \\
		\hline
	\end{tabular}
	\caption{$\chi^2$ values of \cmball{} and DESI for different cosmological models (first column)  at their individual best-fit points (second and third columns, respectively) and for a joint analysis.
		For the joint fit, we list the total $\chi^2$ value (last column) and the contributions from the CMB and DESI likelihoods (fourth and fifth columns, respectively).
		In parentheses, we report the $\chi^2$ improvement with respect to the relevent \lcdm{} reference case (first row) and translate this number to an equivalent frequentist significance for a one-dimensional Gaussian distribution.
		As we do not use lensing data when constraining modified recombination scenarios (see \cref{sec:modrec}), the absolute $\chi^2$ values are not directly comparable with the rest of the table; we instead only report differences with a corresponding \lcdm{} analysis.}
	\label{tab:CMB_DESI_chi2}
\end{table*}

It is worth highlighting that the \lcdm{} parameter constraints from CMB data presented in this work hinge on a determination of the optical depth to reionization \taureio{} from \planck{} large scale polarization data~\citep{planck18-5}.
Crucially \taureio{} and $\omm$ are anticorrelated for CMB constraints, and it has been noted that
raising $\taureio{} \approx 0.09$ would bring \lcdm{} predictions from CMB data and DESI into better agreement and regularize neutrino mass constraints~\citep[see \cref{sec:mnu}, ][]{craig24, green25, sailer25, jhaveri25}.
However, such a high value of \taureio{} is not supported by the \planck{} E-mode data and there is no known significant systematic contamination in the measurement at this level. 
The prior-free estimation of $\taureio$ reported at the end of \cref{{sec:lcdm}} is consistent both with our \planck{}-based prior, but also within $1\,\sigma$ of $\taureio=0.09$.
Together with the fact that the \planck{} large-scale polarization data are not sample variance limited, this motivates revisiting $\taureio$ using E-mode measurements from, e.g., the CLASS telescope~\citep{essinger14, watts18, li25} or the LiteBIRD mission~\citep{hazumi20, sakamoto22}.

We conclude that while \lcdm{} provides an excellent fit to CMB and BAO data separately, there are certain combinations of CMB and BAO data that differ at a level that is borderline statistically significant. 
This motivates the collection of more precise data as well as the search for a cosmological model that better fits both types of data simultaneously.

\subsection{Constraints from CMB and BAO data on extended cosmological models}
\label{sec:CMB+BAO}

In this section, we combine DESI and CMB data to constrain different extended (beyond-\lcdm) cosmological models.
The combination of BAO and CMB data has the potential to constrain extensions beyond just the combined statistical power of the two probes. BAO data are sensitive to the expansion rate and angular-diameter distances at low redshifts, breaking geometric degeneracies that are otherwise present when analyzing CMB data alone.
Additionally, we are motivated by the aforementioned possibility of identifying a model that improves on the joint description of CMB and BAO data provided by \lcdm{}.
To justify a joint analysis, we require that the individual constraints from CMB and DESI data are consistent in a given model space.
We use the same metric and threshold we used to assess consistency in \lcdm{} in \cref{sec:CMB+BAO_lcdm}; the differences in constraints in the $\omm{}$-$\hrd{}$ plane between the data sets translate to a one-dimensional Gaussian fluctuation of less than $3\,\sigma$.
For models where DESI data are sensitive to the extension parameters, we also include these in the comparison.

\RR{In the discussion below, we focus on model extensions probed by CMB data in \cref{sec:cosmo_cmb}, as well as other models of specific interest in combination with BAO data.}
We revisit constraints on the lensing amplitude (\cref{sec:alens_BAO}), light relics (\cref{sec:neff_BAO}), and modified recombination (\cref{sec:modrec_BAO}).
We also report constraints on spatial curvature, alone (\cref{sec:curv}) and in addition to varying the electron mass (\cref{sec:curv+me}), the sum of neutrino masses (\cref{sec:mnu}), and time-evolving dark energy (\cref{sec:w0wa}).
\RR{\citet{khalife2025spt3gd1axionearly} also explore constraints on an axion-like early dark energy model using CMB and DESI data.}
We find shifts away from the CMB-preferred values and at times moderate fluctuations away from the standard model.
We report best-fit $\chi^2$ values for \cmball{} and DESI in \cref{tab:CMB_DESI_chi2} and show constraints for select models in \cref{fig:cmball_DESI_summary}.
We highlight similarities among the different constraints throughout the text.

\begin{figure*}
	\includegraphics[width=\textwidth]{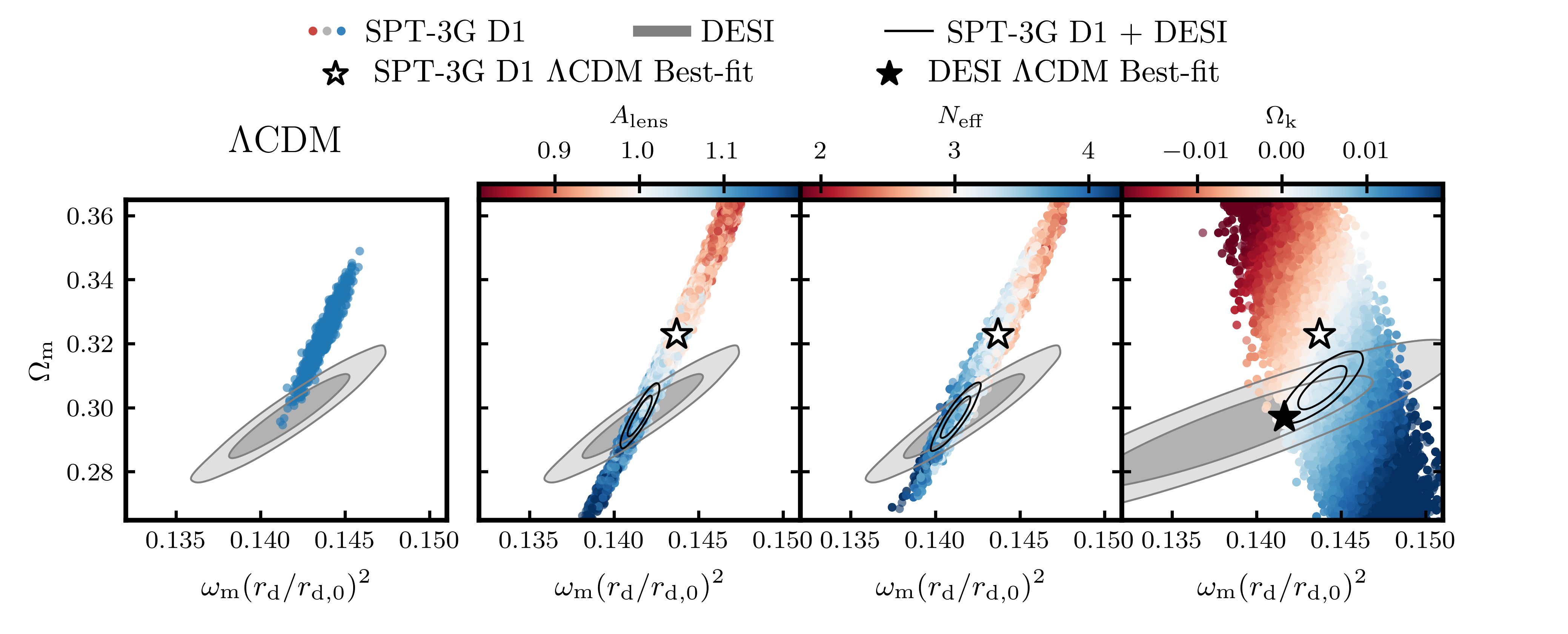}
	\caption{Differences between CMB and DESI data project onto extension parameters in joint analyses.
	We show constraints in the $\omm$-$\ommh (\rd/r_{\mathrm{d,0}})^2$ plane, where $r_{\mathrm{d,0}} = 147\,\mathrm{Mpc}$, first in \lcdm{}, then for extensions with free $A_{\rm{lens}}$, $\neff$, and $\curv$.
	In all panels, colored points represent samples from \sptlr{} chains, filled gray contours indicate DESI posteriors, and solid black line contours indicate the joint \sptlr{} + DESI posteriors.
	The black and white stars indicate the \lcdm{} best-fit points of either probe individually, when these do not coincide with the centers of the contours or samples shown (\sptlr{} in white, DESI in black).\\
	\textit{Far-left panel:} \sptlr{} MCMC samples (blue dots) and DESI posteriors (gray contours) in \lcdm{}.
	The SPT data favor higher values of $\omm$ and $\ommh \rd^2$ than the DESI data.\\
	\textit{Center-left panel:} same as far left but for a model with free CMB lensing amplitude (dots colored according to \Alens{}).
	In this model space, a degeneracy between $\omm$ and $\ommh \rd^2$ at constant $\thetastar$ extends the SPT posterior from \lcdm{} into the parameter region supported by DESI data; with values \Alens{} varying along the band.
	Since the DESI constraints intersect the SPT band at comparatively low $\omm$ and $\ommh \rd^2$ values, the joint analysis favors $\Alens>1$.\\
	\textit{Center-right panel:} same as far left but for a model with free effective number of relativistic species (dots colored according to $\Neff{}$).
	For CMB data, lower $\omm$ and $\ommh \rd^2$ can, to a certain degree, be accommodated by raising $\Neff{}$, leading to a moderate positive shift in the best-fit \Neff{} value in a joint analysis with DESI data.\\
	\textit{Far-right panel:} same as far left but for a model with free mean spatial curvature (points colored according to $\curv$).
	CMB data constrain a band in the $\omm$-$\ommh \rd^2$ plane; lines of constant $\curv$ dissect the band diagonally, along the \lcdm{} degeneracy direction.
	As in the other extension models, the DESI contour intersects the SPT band away from the \lcdm{} value of the extension parameter, in this case leading to a mild preference for an open universe in the joint constraints.\\
	}
	\label{fig:DESI_ext_proj}
\end{figure*}

\subsubsection{Lensing amplitude revisited}
\label{sec:alens_BAO}

We begin by revisiting constraints on the consistency of the signature of gravitational lensing across cosmological probes.
We repeat the analysis of \cref{sec:alens}, but now add DESI data.
We stress again that the models considered here are non-physical.

We first consider the case of $\Alens$, i.e. the coherent modification of the amplitudes of lensing in the CMB primary power spectrum and the CMB lensing reconstruction, as defined in \cref{sec:alens}.
Adding DESI to CMB data changes the $\Alens$ constraint from \cref{eq:alens_withlensing} to
\begin{align}
	\Alens &= 1.084\pm 0.035 \, \text{for} \  \sptlr+\DESI{}, \\
	\Alens &= 1.092\pm 0.026 \, \text{for} \  \ground{}+\DESI{}, \\
	\Alens &= \replace{1.084\pm 0.024}{1.083\pm 0.025} \, \text{for} \  \cmball{}+\DESI{}.
\end{align}
which are deviations from the standard model prediction of $2.4\,\sigma$, $3.5\,\sigma$, and $\replace{3.5}{3.3}\,\sigma$, respectively.
We note that in moving from \ground{} to \cmball{} we only include \planck{} data that by themselves do not prefer an anomalous $\Alens$ (in particular by limiting \planck{} \TT{} data to $\ell<1000$).
We find similar results when allowing $\Atwopt$ and $\Arecon$ to vary independently from one another, e.g. for \ground{}+\DESI{} we report a deviation from \lcdm{} of $3.1\,\sigma$ (see also~\citep{ge24}).

When considering only CMB data, allowing $\Alens$ to vary does not improve the goodness-of-fit for ground based data, as explored in \cref{sec:alens}.
The joint \cmball{}+DESI fit improves on the \lcdm{} minimum $\chi^2$ value by $\replace{9.5}{6.9}$ points; the fit to CMB data is improved by $\replace{5.6}{3.3}$ points and the fit to BAO data by $3.\replace{9}{7}$ points (see \cref{tab:CMB_DESI_chi2}).
This is a non-negligible improvement given the introduction of one additional parameter.
Translating the quality-of-fit improvement to an equivalent one-dimensional Gaussian significance, this corresponds to a $\replace{3.1}{2.6}\,\sigma$ event.

This result is a projection of differences in the CMB and DESI data that can be understood in the $\omm$-$\ommrdsq$ plane, where $\omega_{\rm m} \equiv \omm h^2$ (for a more complete discussion, see~\citep{loverde24}).
The precise determination of $\thetastar$ by CMB data translates to a thin contour in this plane.
When $\Alens$ is allowed to vary, the contour is extended into a narrow band, along which the extension parameter varies.
The extended contour meets the DESI constraint at comparatively low values of $\omm$ and $\ommrdsq$, which necessitates $\Alens>1$ in order to not degrade the fit to the CMB data.
We illustrate this effect in \cref{fig:DESI_ext_proj}.
Physically, the lower matter density preferred by DESI would imply less gravitational lensing.
This is at odds with the amplitude of the effect in CMB data (see \cref{sec:alens}) and hence $\Alens$ is raised to compensate.
As such, the deviation from $\Alens=1$ in joint constraints with DESI is a rephrasing of the marginal agreement of the two probes in \lcdm{}.

\subsubsection{New light particles revisited}
\label{sec:neff_BAO}

Next, we revisit constraints on light relics, now adding DESI to the CMB data.
As discussed in \cref{sec:neff}, varying \neff{} from its standard model prediction opens up the BAO scaling transform, under which BAO observables are effectively invariant~\citep{Ge:2022qws}.\footnote{Assuming a fixed redshift for the end of the baryon drag epoch, $z_{\rm d}$, BAO observables remain exactly invariant.
Even including the associated changes to $z_{\rm d}$, the locations of the predicted BAO correlation peaks change by $<0.2\%$ for the extreme case of $\Neff=5$, which, as we saw in \cref{sec:neff}, is ruled out by CMB data.}
In contrast, CMB observables change as mentioned in \cref{sec:neff} and as discussed in~\citep{Ge:2022qws}, with the result that CMB data can be best fit if the matter density scales up more slowly than the radiation density.
The different responses of the two probes to BAO scaling make it interesting to study \Neff{} using the combination of the two.

In this model space, the constraints from \ground{} and \cmball{} on \omm{} and \hrd{} are discrepant with those from DESI at more than $3\,\sigma$ and hence do not meet our requirement for joint analyses (see the start of \cref{sec:CMB+BAO}).
This is also true for the cases of \yp{} and \neff{}+\yp{}.
Hence, we only report results for \sptlr{} in combination with DESI data.

We first allow for $\Neff{}$ to vary in our analysis while maintaining BBN consistency.
Adding DESI to \sptlr{}, we report:
\begin{align}
	\Neff{} &= 3.5\replace{2}{3}\pm 0.23 \, \ \text{for} \  \sptlr+\DESI{}.
\end{align}
The inclusion of DESI data shifts the central value of the posterior up by $1\,\sigma$ and we report a mild preference for $\Neff{}>3.044$ at $2.1\,\sigma$.
The \neff{} posterior tightens by $20\%$ compared to the SPT CMB-only constraint.

Varying \neff{} and \yp{} simulatenously, we report:
\begin{align}
	&
	\left.
	\begin{array}{l}
		\Neff{} = 3.6\replace{3}{4}^{+0.39}_{-0.44} \\
		\Yp = 0.241\pm 0.034
	\end{array}\right\} \quad \text{for} \  \sptlr+\DESI{}.
\end{align}
Again, the inclusion of DESI data tightens the $\Neff{}$ posterior and leads to mild $\sim\,1\,\sigma$ shifts up in $\Neff{}$ and down in $\yp{}$ from the CMB-preferred values as these two extension parameters are anticorrelated.
The \Neff{} value is within $1.4\,\sigma$ of the \lcdm{} value and the \yp{} constraint matches the BBN predictions of~\citep{parthenope17} and~\citep{pitrou18} and the measurement of~\citep{aver20} at $\leq 0.2\,\sigma$.
We find similar results when allowing only $\Yp$ to vary while fixing $\Neff{}=3.044$ (see \cref{tab:neff_yp}).

As we illustrate in \cref{fig:DESI_ext_proj} for the case of only varying \neff{}, at the level of parameter degeneracies this result can be understood in a similar way to the $\Alens$ case above.
Again, the $\omm$-$\ommrdsq$ posterior of \sptlr{} is extended along its existing degeneracy direction compared to the \lcdm{} case; \Neff{} varies along this direction, with values higher than the standard model prediction being supported at lower $\omm$ and $\ommrdsq$ values, as explained in \cref{sec:neff}.
This is where the DESI posterior lies in the $\omm$-$\ommrdsq$ plane and hence the inclusion of the BAO data pulls the joint constraints towards higher $\Neff{}$ compared to the values prefered by \sptlr{}.
The picture is the same when opening up the helium abundance; we find a shift down in $\Yp$ as this parameter is anticorrelated with \neff{}, but positively correlated with $\omm$ and $\ommrdsq$.

\subsubsection{Modified recombination revisited}
\label{sec:modrec_BAO}

We examine the impact of DESI data on the previously presented reconstructions of the ionization fraction $X_{\rm e}$ (see \cref{sec:modrec}). We find that the inclusion of DESI further significantly improves the reconstruction, with:
\begin{equation}
    \log {\rm VR} = \replace{12.3}{11.8} \text{ for \cmball{} T\&E + DESI}.
\end{equation}
This is a reduction by about four e-foldings compared to the \cmball{} reconstruction and is due to the fact that the low-redshift measurements of the expansion history provided by BAO break the $\Hubble$-$X_{\rm e}$ degeneracy that is present when only using CMB data.

Along with sharpening the posterior, we find that the inclusion of DESI data leads to a mild $1.8\,\sigma$ preference for earlier recombination as shown in \cref{fig:xe_reconstructions}.
This goes hand-in-hand with higher values of $\Hubble$:
\begin{align}
    \Hubble &= 69.34\pm 0.70 \text{ for \planck{} + SPT T\&E + DESI} \\
    \Hubble &= 69.\replace{48}{55}\pm 0.65 \text{ for \cmball{} T\&E + DESI}
\end{align}
This result is in line with previous work that has found that DESI data lead to a preference for higher $\Hubble$ values in modified recombination scenarios when analyzed alongside CMB data~\citep{pogosian24, lynch24b, mirpoorian24, calabrese25, mirpoorian25}.
However, differences between the above \Hubble{} constraints and local measurements from SH0ES remain $>3\,\sigma$.
There is no statistically significant preference for this model over \lcdm{} for \cmball{}+DESI.
Though the best-fit $\chi^2$ value is reduced by $14.2$ points, when accounting for the seven additional degrees of freedom this only translates to a mild $2\,\sigma$ preference for modified recombination (see \cref{tab:CMB_DESI_chi2}).
Still, a lower-dimensional model able to reproduce the essential features of \texttt{ModRec} could in principle lead to a comparable goodness-of-fit improvement and an increased significance.\footnote{Common parametrizations of primordial magnetic fields are of interest in this context~\citep[see][]{jedamzik20, thiele21, rashkovetskyi21, galli22, jedamzik23}.}

Our findings above are again a result of features in the DESI data that project onto $\omm$ and $h \rd$.
Through earlier recombination, the \texttt{ModRec} model improves consistency between BAO and CMB data;
this allows the model to adjust to higher $\Hubble$ values~\citep[see Fig. 2 in][]{lynch24b}, while keeping the ratio $\thetastar = r_{\rm s}^\star/D_{\rm A}^\star$ consistent with CMB allowed values.

\subsubsection{Spatial curvature}
\label{sec:curv}

Inflation is expected to reduce primordial spatial curvature to levels well below current experimental sensitivity.
A deviation from this prediction would constitute a major challenge for the standard model of cosmology.
Though CMB data are sensitive to spatial curvature by themselves, they suffer from geometric degeneracies that make constraints rather weak.

For CMB alone, we find
\begin{align}
	100\curv & = 0.2^{+1.5}_{-1.2} \ \text{for} \  \sptlr,          \\
	100\curv & = -0.06^{+0.81}_{-0.70} \ \text{for} \  \ground{}, \\
	100\curv & = -\replace{0.88 \pm 0.48}{0.85\pm 0.48} \ \text{for} \  \cmball{}.
\end{align}
Though loose, these constraints are all compatible with a flat universe at $<2\,\sigma$.

The addition of BAO data breaks the limiting degeneracies and tightens the posteriors by about a factor of four:
\begin{align}
	100\curv & = 0.40 \pm 0.20 \ \text{for} \  \sptlr+\DESI{},     \\
	100\curv & = 0.51 \pm 0.17 \ \text{for} \  \ground{}+\DESI{},     \\
	100\curv & = \replace{0.26 \pm 0.11}{0.27\pm 0.12} \ \text{for} \  \cmball{}+\DESI{}.
\end{align}
The \sptlr{}+DESI, \ground{}+DESI, and \cmball{}+DESI constraints are $2.0\,\sigma$, $3.0\,\sigma$, and $2.\replace{4}{5}\,\sigma$ away from spatial flatness, respectively.
The above deviations are reflected in moderately improved best-fit $\chi^2$ values for joint CMB+DESI analyses as listed in \cref{tab:CMB_DESI_chi2}.
For \cmball{}+\DESI{} we report a reduction by $\replace{6.3}{4.2}$ points compared to \lcdm{}, which corresponds to a $2.\replace{5}{1}\,\sigma$ preference.

At the parameter level, this deviation from \lcdm{} can be understood similarly to the $\Alens$ and \Neff{} cases above, though when allowing for non-zero spatial curvature, the degeneracy direction between $\omm$ and $\ommrdsq$ flips compared to \lcdm{} and the parameters are now anticorrelated.
However, the data constrain a relatively thick band in this plane across which lines of constant $\curv$ trace the original degeneracy direction of $\omm$ and $\ommrdsq$ present in \lcdm{} (as well as in $\Alens$ and \Neff{}).
Hence, CMB data can support lower $\omm$ without substantially modifying $\ommrdsq$ if $\curv>0$.
This is illustrated in the right-most panel of \cref{fig:DESI_ext_proj}.
Note that a positive $\curv$ is at odds with the excess lensing preferred by the high-$\ell$ \planck{} TT spectrum, which tends to push $\curv$ to negative values and $\omm$ to higher values (the opposite of the low $\omm$ preferred by DESI). 
Thus, it is possible that replacing the \ACTDR\ high-$\ell$ TT used in the \cmball{}+\DESI{} combination with the one from \planck\ could weaken the preference for a $\curv>0$.
For a more detailed discussion of CMB and BAO constraints in this model space, we point the reader to~\citep{chen25}.

\subsubsection{Spatial curvature and a varying electron mass}
\label{sec:curv+me}

We now consider constraints when allowing for a varying electron mass in a non-flat geometry, $m_{\rm e} + \curv$.
In the review of potential Hubble tension solutions by~\citep{Review_H0}, this model was one of the most promising, and we re-evaluate it in the context of the most up-to-date and stringent CMB and BAO data here.
We take the result of~\citep{breuval_small_2024} from the SH0ES collaboration as the reference $\Hubble$ measurement for this section.

From a theoretical standpoint, scalar fields predicted by fundamental theories could couple to elementary particles, specifically electrons~\citep{VarMe1,Planck_VarMe}, 
which could result in electrons having an effective mass at high redshift different than that measured in laboratory experiments today.
Increasing the electron mass causes recombination to occur earlier, which in turn decreases the size of the sound horizon
at that epoch.
Moreover, allowing the spatial curvature to vary accommodates this change in a way that fits late-universe measurements, specifically BAO data.
Together, these two effects allow for an increase in $\Hubble$~\citep{H0_Olympics,Review_H0,Tristan_Nils}.
We model the transition of the electron mass as a step function at $z\sim 50$~\citep{VarMe2015,hart19}; this is the simplest implementation currently available in \class and is sufficient for the precision of current cosmological data.
We allow for the value at high redshift, $m_{e}$, to deviate from today's value, $m_{{\rm e},0}$, such that $m_{\rm e}/m_{{\rm e},0} \in (0.1,2)$.
Due to the heavy numerical cost of this model, we loosen our convergence requirement to $R-1<0.04$.

\begin{figure}
	\includegraphics[width=\columnwidth]{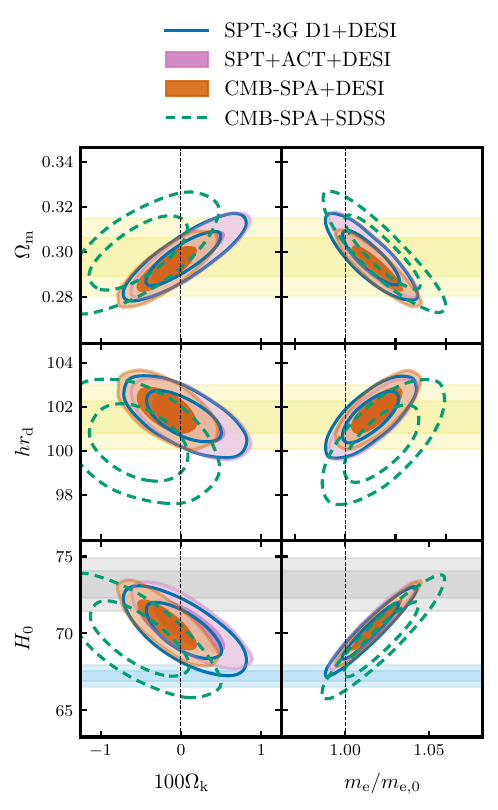}
	\caption{Constraints on $m_{\rm e} + \curv$ from \sptlr + DESI (blue, empty), \ground{} + DESI (purple, filled), and \cmball{} + DESI (orange, filled). In the two top rows, the yellow bands correspond to the 68\% and 95\% confidence levels of \DESI{} assuming \lcdm{}. In the bottom row, the light blue bands correspond to the 68\% and 95\% confidence levels for \cmball{} in \lcdm{}, while the gray bands indicate the SH0ES measurement~\citep{breuval_small_2024}.
	This model is able to considerably reduce the Hubble tension (from $Q_{\rm MPCL} = 5.9\,\sigma$ to $\replace{1.9}{2.1}\,\sigma$, see text for details).
	We show the constraints when replacing DESI with SDSS BAO data as dashed green line contours.}
	\label{fig:Me_Omk}
\end{figure}

In \cref{fig:Me_Omk}, we show joint constraints from CMB and BAO data on key parameters.\footnote{We do not report CMB-only constraints as CMB data alone yield large parameter uncertainties in this model, rendering results inconclusive~\citep{Review_H0, Nils_Me2024}.}
In this model space, current CMB data and DESI constrain  $\Hubble$ to be:
\begin{align}
	\Hubble &= 70.1\pm 1.2 \, \kmsmpc \text{ for} \  \sptlr+\DESI{}, \\
	\Hubble &= 70.5\pm 1.2 \, \kmsmpc \text{ for} \  \ground+\DESI{}, \\
	\Hubble &= 70.\replace{6}{8}\pm 1.1 \, \kmsmpc \text{ for} \  \cmball+\DESI{}.
\end{align}
Due to the degeneracies that $\left(m_{\rm e}/m_{{\rm e},0},\curv\right)$ exhibit with $\left(\omm,\hrd\right)$, raising $\Hubble$ brings $\omm$ and $\hrd$ in line with the DESI-preferred values in \lcdm{}.
Comparing the best-fit $\chi^2$ value of \cmball{}+DESI in this model to \lcdm{} yields an improvement by $\replace{6.8}{6.5}$ points, which given the introduction of two new free parameters, corresponds to a mild $2.1\,\sigma$ preference (see \cref{tab:CMB_DESI_chi2}).
Full parameter results can be found in \cref{tab:Me_Omk}.
We note that when replacing DESI data with SDSS BAO measurements, similarly high values of $\Hubble$ can be achieved, as shown in \cref{fig:Me_Omk}.

\begin{table}
	\begin{tabular}{lllll}
		\toprule
		{ Parameters} & \sptlr{} & \ground{} & \cmball \\
		& +\DESI{} & +\DESI{} & +\DESI{} \\
		\midrule
		\midrule
		$\Hubble {\scriptstyle\left[{\kmsmpc}\right]}$ & $70.1\pm 1.2$ & $70.5\pm 1.2$ & $70.\replace{6}{8}\pm 1.1$ \\
		$100\,\omm$ & $29.74\pm 0.79$ & $29.83\pm 0.78$ & $\replace{29.20\pm 0.68}{29.26\pm0.69}$ \\
		$\hrd{\scriptstyle\left[\text{Mpc}\right]}$ & $102.5 \pm 0.8$ & $101.4 \pm 0.7$ & $101.\replace{9}{8} \pm 0.7$ \\
		$100\left(\frac{m_\mathrm{e}}{m_\mathrm{e,0}}-1\right)$ & $1.5\pm 1.1$ & $1.6\pm 1.1$ & $\replace{1.97\pm 0.98}{1.9\pm1.0}$  \\
		$100\,\curv$ & $0.04\pm 0.31$ & $0.10\pm 0.30$ & $\replace{-0.21\pm 0.25}{-0.18\pm0.25}$ \\
		$Q^{\shoes{}}_{\rm MPCL}$  $\left[\sigma\right]$ & 2.3 & 2.0 & \replace{1.9}{2.1} \\
		\bottomrule
	\end{tabular}
	\caption{Cosmological parameters for the $m_{\rm e} + \curv$ model for \sptlr, \ground{}, and \cmball{} data sets combined with DESI BAO data.
		We also report the Marginalized Posterior Compatibility Level, $Q^{\shoes{}}_{\rm MPCL}$, which quantifies the tension with SH0ES data~\citep{breuval_small_2024}.}
	\label{tab:Me_Omk}
\end{table}

To assess to what degree this model reduces the tension between CMB+BAO and \shoes\ data, we use two metrics.
First, we calculate the Marginalized Posterior Compatibility Level, $Q_\text{MPCL}$,\footnote{This statistic quantifies the agreement between two data sets, generalized to the case where the posterior distributions are not necessarily Gaussian. It computes consistency directly from the MCMC chains, instead of using parameter covariance matrices (see Section 4.3 of~\citep{Review_H0} for more details).} introduced by~\citep{Review_H0}~\citep[see also][]{Raveri_Doux,leizerovich2023tensions}.
We report a reduction from $5.9\,\sigma$ in \lcdm{} to $\replace{1.9}{2.1}\,\sigma$ in this model for \cmball{}+DESI compared to \shoes{}.
Second, we report the Difference of the Maximum a Posteriori (DMAP) criterion, defined as:
\begin{equation}
	Q_{\text{DMAP}} = \sqrt{\chi^2_{\mathcal{D}+\text{SH0ES}} - \chi^2_{\mathcal{D}}},
	\label{eq:QDMAP}
\end{equation}
where the best-fit $\chi^2$ of the model is evaluated with a given data set $\mathcal{D}$ with and without the \shoes{} likelihood.
By comparing the two $\chi^2$ values, the statistic indicates whether the model fit to the data worsens when including \shoes{} information.
For \cmball{}+\DESI{}, we find a reduction of the tension with \shoes{} data from $Q_{\text{DMAP}} = 5.6\,\sigma$ in \lcdm{} to $\replace{2.6}{2.3}\,\sigma$ for $\curv + m_{\rm e}$.
The model is able to reduce the tension to below the $3\,\sigma$ threshold, which was not the case in~\citep{Review_H0} using the previous SPT-3G data from~\citep{balkenhol23}, \planck, Pantheon SNIa, and part of the SDSS BAO measurements.
We conclude that while at current sensitivity the CMB and BAO data show no statistically significant preference for this model over \lcdm{}, the model passes key tests in the comparison with the \shoes{} $\Hubble$ measurement and may therefore still be considered as a possible solution to the Hubble tension.

\subsubsection{Neutrino cosmology}
\label{sec:mnu}

From neutrino oscillation experiments, we know that at least two neutrinos have a non-zero mass, which implies that the sum of neutrino masses is either larger than $0.06\,\mathrm{eV}$ or $0.1\,\mathrm{eV}$ assuming a normal or inverted mass hierarchy, respectively~\citep{fukuda98, ahmad02}.
As shown in many works, massive neutrinos hinder the formation of structure, leading to many observable effects, including a reduction in the amplitude of deflections due to gravitational lensing~\citep{lesgourgues06b}.
As \citep{lynch25} illustrate, CMB and BAO data have opposing degeneracies in a parameter space in which neutrino mass is a free parameter, which motivates a joint analysis.
In this section, we fit jointly to CMB and DESI data allowing for arbitrary, positive $\mnu{}$.

\begin{figure}[!h]
	\includegraphics[width=\columnwidth]{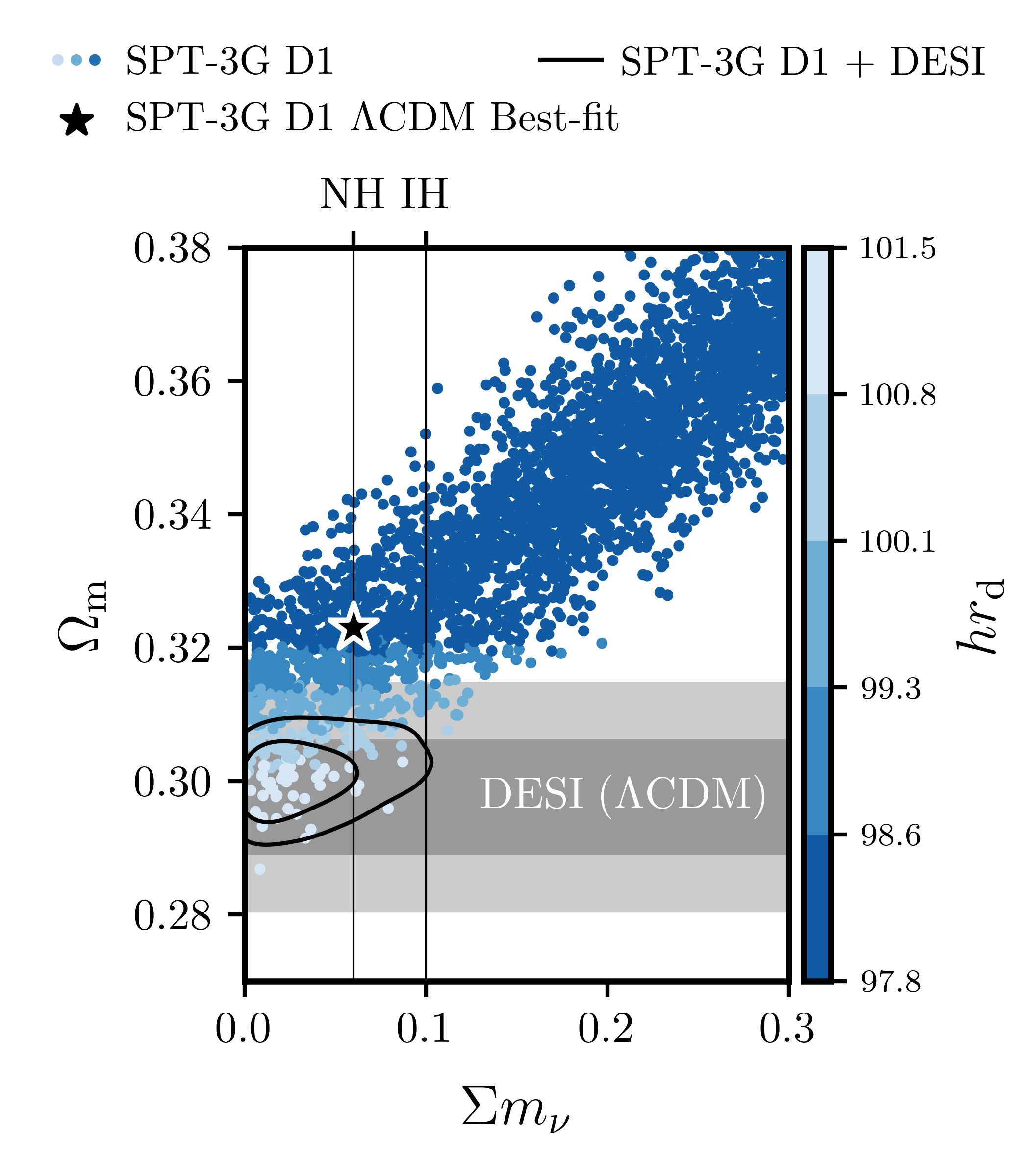}
	\caption{
	Samples from an MCMC analysis of \sptlr{} constraining $\Lambda\mathrm{CDM}+\mnu$ (blue dots).
	The dot color corresponds to the $\hrd$ value, where light blue is the mean for DESI data in \lcdm{} ($\hrd=101.5\,\rdragunittxt$) and dark blue is $\geq 4\,\sigma$ low ($\hrd=98.6\,\rdragunittxt$).
	The horizontal gray bands indicate the \lcdm{} DESI $\omm$ constraint.
	As a reference, the black star shows the best-fit point for SPT in \lcdm{}.
	We mark the lower limits for a normal (NH) and inverted (IH) neutrino hierarchy with black vertical lines.
	The differences between DESI and CMB data projected into $\omm$ and $\hrd$ lead to a preference for as small a value of $\mnu$ as possible in joint constraints (black line contours).}
	\label{fig:mnu_bao}
\end{figure}

We remind the reader that the \cmball{} combination removes \planck{} data at $\ell>1000$ to avoid correlations with ACT data.
Because of the known preference of \planck{} \TT{} data for a higher lensing amplitude (see \cref{sec:alens}), if we were to use the \planck{} data at those multipoles instead of ACT in that combination, we would likely find an even tighter upper limit on \mnu{}.
This effect is discussed in~\citep{calabrese25}.
Hence, for this model, our construction of the \cmball{} data set, following the P-ACT prescription introduced in \citep{louis25}, is conservative.
As a point of comparison, we first report 95\% confidence upper limits for the CMB data alone of:
\begin{align}
	\mnu{} &< 0.77 \, \mathrm{eV} \ \text{for} \  \sptlr, \\
	\mnu{} &< 0.58 \, \mathrm{eV} \ \text{for} \  \ground, \\
	\mnu{} &< \replace{0.17}{0.19} \, \mathrm{eV} \ \text{for} \  \cmball{}.
\end{align}
Though these upper limits remain away from the minimum \mnu{} values allowed based on neutrino oscillation experiments, we note that the \cmball{} constraint improves on the \planck{} one of $<0.25\, \mathrm{eV}$ by about $30\%$.

We now add DESI data.
In this model space, the constraints from \ground{} and DESI on \omm{} and \hrd{} are discrepant at more than $3\,\sigma$, exceeding our requirement for joint analyses, and we do not report results for this combination.
As in the \lcdm{} case (\cref{sec:CMB+BAO_lcdm}), the addition of \planck{} data to \ground{} regularizes the CMB constraint such that the differences between \cmball{} and DESI are below our threshold.
When adding DESI to \sptlr{} and \cmball{}, we find at the 95\% confidence level:
\begin{align}
	\mnu{} &< 0.081 \, \mathrm{eV} \ \text{for} \  \sptlr+\DESI{}, \\
	\mnu{} &< 0.055 \, \mathrm{eV} \ \text{for} \  \cmball+\DESI.
\end{align}
As expected, adding BAO data tightens the constraint substantially.
While the upper limit derived from SPT data alone is consistent with neutrino oscillation data, with a posterior that peaks slightly away from zero, the \cmball{}+\DESI{} combination appears to rule out the normal and inverted hierarchies at 96.6\% and 99.9\% confidence, respectively.
Moving \mnu{} close to zero reduces the best-fit $\chi^2$ value by $\replace{7.8}{5.1}$ points for joint CMB and BAO analyses compared to the minimal value for the normal hierarchy, which for one additional parameter corresponds to a $2.\replace{8}{3}\,\sigma$ significance (see \cref{tab:CMB_DESI_chi2}).

The drive toward as low of a value for $\mnu{}$ as allowed in joint constraints (even negative $\mnu{}$ values if the model is phenomenologically extended into this regime) is a known effect, and there exists a growing literature dissecting and contextualizing cosmological neutrino mass constraints~\citep[see e.g.][]{craig24, loverde24, green25, lynch25}.
Increasing \mnu{} raises $H(z)$ during the matter-dominated epoch; to keep $\thetastar$ consistent with CMB data, the cosmological constant decreases and \omm{} increases.
Though $\rd$ does not change, lowering $\Lambda$ decreases $\Hubble$ and hence \hrd{} and \mnu{} are anti-correlated~\citep{pan15, loverde24}.
As DESI data prefer a low $\omm$ and a high $\hrd$ compared to CMB data, this forces the joint posterior against the $\mnu{}=0$ boundary.
This gives rise to the tight upper limits seen above and is illustrated in \cref{fig:mnu_bao}.
These results are in growing discord with neutrino oscillation experiments;
improved cosmological data sets will allow for more scrutiny.

\subsubsection{Time-evolving dark energy}
\label{sec:w0wa}

Lastly, we turn our attention to time-evolving dark energy.
Instead of assuming a cosmological constant model, we allow for the equation of state of dark energy to vary~\citep{linder02, chevallier01} according to
\begin{equation}
	\label{eq:wowa}
	w(z) = \wo + \wa \frac{z}{1+z}.
\end{equation}
This model has recently received attention as the combination of DESI data with CMB and uncalibrated SNe Ia data show a $\gtrsim 3\,\sigma$ preference for a deviation from \lcdm{}~\citep{desi24-6, desi25, sabogal25}.

Combining \cmball{} and DESI data, we report 
\begin{align}
	\wo &= -0.4\replace{1}{0}\pm 0.20, \\
	\wa &= -1.\replace{78}{82}\pm 0.55.
\end{align}
In the $\wo$-$\wa$ plane, this is a $2.9\,\sigma$ deviation from $(\wo, \wa) = (-1,0)$.
Adding Pantheon+ uncalibrated SNe Ia data moves the constraint to $\wo = \replace{-0.831\pm 0.054}{-0.830\pm 0.054}$ and $\wa = \replace{-0.66\pm 0.19}{-0.66\pm 0.20}$ and the $\wo$-$\wa$ central values remain $3\,\sigma$ from the \lcdm{} prediction.\footnote{Using Union3~\citep{rubin_union_2024} or DES-SN5YR~\citep{abbott_dark_2024} SNe Ia data instead of Pantheon+, the deviation is expected to be even larger~\citep{desi25}.}

This deviation is reflected by improved best-fit $\chi^2$ values (see \cref{tab:CMB_DESI_chi2}).
The goodness-of-fit improves by $\replace{13.5}{11.6}$ points for \cmball{}+DESI in this model compared to \lcdm{}.
This is close to the value reported by the DESI collaboration when using only \planck{} CMB data ($12.5$)~\citep{desi25} and equals to a $3.\replace{2}{0}\,\sigma$ event for a one-dimensional Gaussian distribution.
The contributions from CMB and DESI data to the total $\chi^2$ improve by $\replace{6.0}{4.4}$ and $7.\replace{5}{2}$ points compared to the standard model, respectively.
We note that restricting the allowed parameter space or otherwise imposing priors in the \wowa{} plane to ensure the evolution of dark energy is physical tends to weaken the significance of departures from a cosmological constant~\citep{peirone17, raveri17}.
More generally, one can argue that our ignorance of the nature of dark energy makes it difficult to apply theoretically motivated priors on this model and hence particularly strong evidence from the data in favor of this model is needed~\citep[for a recent discussion in the context of DESI results, see][]{efstathiou25b}.
Still, the sensitivity of BAO data to the late-time evolution of dark energy makes this model interesting and further data will help assess the robustness of the trends we are seeing.

\begin{figure}[!h]
	\includegraphics[width=\columnwidth]{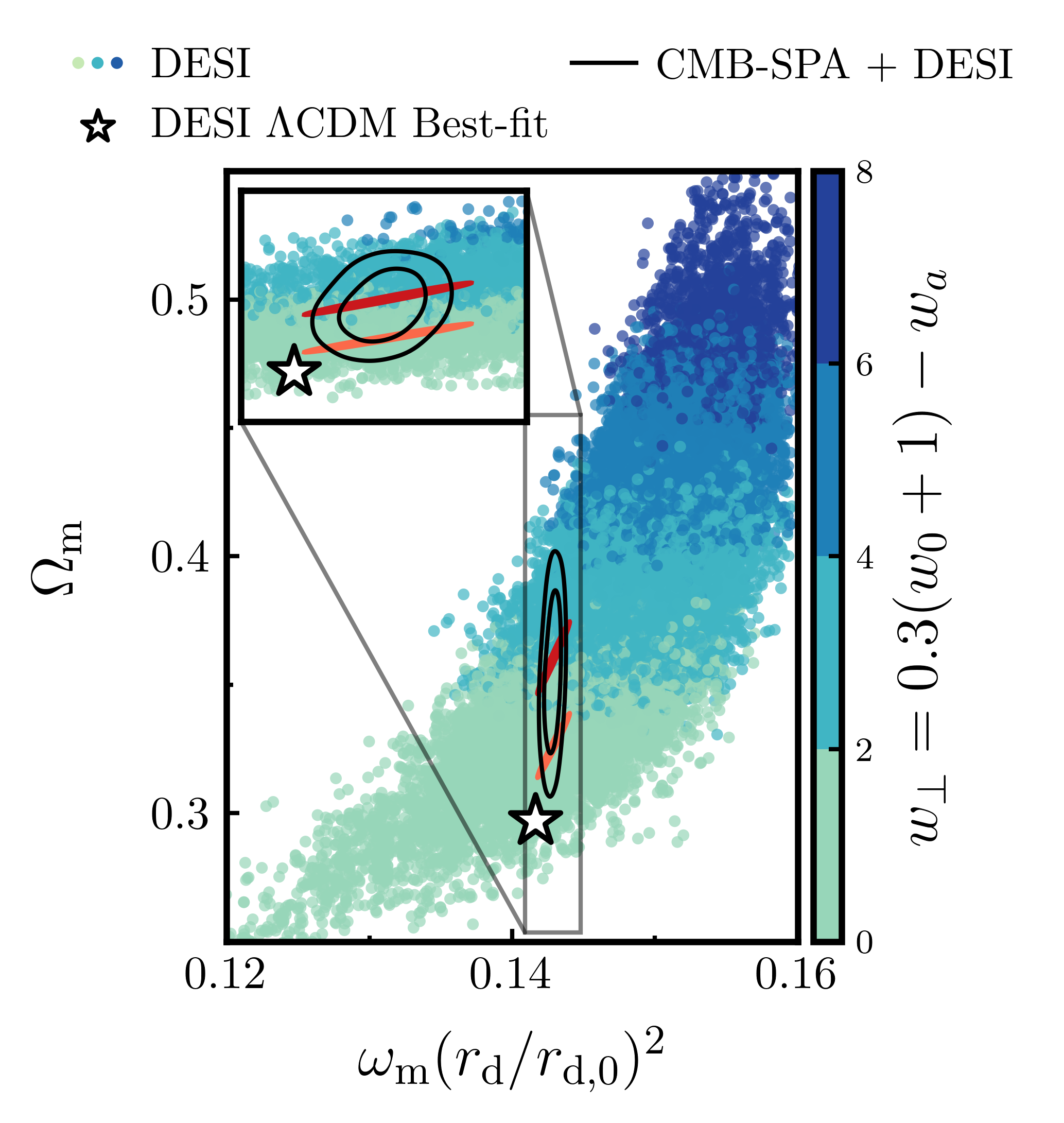}
	\caption{DESI MCMC samples in the $\omm$-$\ommrdsq{}$ plane for $\Lambda\mathrm{CDM}+\wo \wa$ (colored dots).
	The DESI constraints in this plane are substantially widened compared to the \lcdm{} case.
	Samples are colored according to $\wperp$ (see \cref{eq:wperp}), which is highly correlated with $\omm$.
	The white star marks the DESI \lcdm{} best-fit point for reference.
	We indicate the CMB information by showing the \lcdm{} \cmball{} $2\,\sigma$ contour (orange ellipse) and the \cmball{} $2\,\sigma$ contour fixing $\wperp$ to the DESI mean value of $\wperp=3.\replace{1}{2}$ (red ellipse).
	By constraining $\ommrdsq$, the CMB data break the degeneracies of the DESI data, which substantially tightens the joint $\wperp$ posterior (\cmball{}+DESI shown as black line contours).}
	\label{fig:w0wa}
\end{figure}

To better understand the above result we introduce the summary parameter
\begin{equation}
	\label{eq:wperp}
	\wperp \equiv \replace{\wa + 3.5(\wo+1)}{0.3(\wo+1)-\wa},
\end{equation}
which maximally varies along the $\wo$-$\wa$ degeneracy direction of DESI data, such that $\wperp=0$ corresponds to $\wo, \wa = (-1,0)$ and $\wperp>0$ to $\wo>-1, \wa<0$.
As shown in \cref{fig:w0wa}, allowing for dynamical dark energy greatly relaxes the DESI constraints in the $\omm$-$\ommrdsq$ plane (compare to the left-most panel of \cref{fig:DESI_ext_proj}); in this model, $\wperp$ is highly correlated with $\omm$.
For DESI data, we report $\wperp = \replace{3.1 \pm 1.6}{3.2 \pm 1.7}$.
CMB data, on the other hand, do not constrain $\wperp$: they are not sensitive to the detailed time evolution of the dark energy equation of state at late times, but only to the integral effect on the angular diameter distance to the last scattering surface.
However, CMB data still provide a tight constraint on $\ommrdsq$ in this model and, therefore, break the degeneracies of the DESI data.
The error bar on $\wperp$ shrinks substantially in joint constraints and we report $\wperp = \replace{1.91 \pm 0.57}{2.0 \pm 0.61}$ for \cmball{}+DESI.


\section{Conclusions}
\label{sec:conclusion}

In this work, we have presented CMB temperature and E-mode polarization power spectrum measurements based on SPT-3G data
and the constraints on cosmological parameters enabled by these measurements, both individually and in combination with other data sets.
We used observations of the \mainfield{} field at $95$, $150$, and $220\,\ghz$ collected in the austral winter seasons of 2019 and 2020.

The temperature and polarization maps made from these data,
and the series of null tests used to demonstrate that these maps
are free from significant systematic effects,
are described in detail in Q25.
These maps are the deepest high-resolution CMB data
to be used for a measurement of the \TTTEEE{} power spectra,
with a coadded white noise level of $3.3\,\ukarcmin$ in temperature and $5.1\,\ukarcmin$ in polarization.

This unprecedented sensitivity motivated a series of pipeline modernizations which are presented in this analysis.
They represent a major step forward compared to previous SPT \TTTEEE{} analyses such as~\citep{balkenhol23}.
These improvements were summarized in the introduction \cref{sec:intro}
and described in detail in \cref{sec:maps,sec:power_spec,sec:likelihood}.

The power spectra estimated from these maps cover angular scales
$400 \leq \ell \leq 3000$ in \TT\ and $400 \leq \ell \leq 4000$ in \TE\ and \EE.
An extensive campaign of consistency tests, at the power spectrum and cosmological parameter level, was performed while blind to final results to ensure robustness, as described in \cref{sec:validation}.
While we respected all pre-established criteria to unblind the results, we still found that substantial changes to the pipeline were required after unblinding,
in particular the modeling of higher-order temperature-to-polarization leakage and polarized beams, as described in detail in \cref{sec:quadrupolar,sec:polbeams}, and \cref{app:post-unblind}. The evidence for these effects is established in cosmology-independent tests.
Constraining the polarized beams from \TTTEEE{} data alone degrades sensitivity to cosmological parameters, and 
a precise independent characterization of the polarized beams would increase the constraining power of the \EE{} band powers in particular.

Overall, the model for astrophysical contaminants and systematic effects is
accurate enough to ensure the consistency between power spectra
across frequencies to better than $0.01\%$ of sample variance.
The resulting binned \TT, \TE, and \EE\ power spectra, or band powers,
constructed using data in pairs of SPT-3G frequency bands, are shown in \cref{fig:bandpowers}.
The minimum-variance combinations of all frequency pairs
in \TT, \TE, and \EE{} are shown in \cref{fig:experiments}.
The minimum-variance band powers in \EE\ and \TE\ are the most
constraining to date at $\ell=1800$-$4000$ and $\ell= 2200$-$4000$, respectively, as shown by the signal-to-noise ratio in \cref{fig:snr}.

We use these data to infer constraints on cosmological parameters which are
summarized in \cref{sec:summary} and reported in detail in \cref{sec:pars}.
The main results are:
\begin{enumerate}

    \item The \lcdm\  standard model of cosmology provides an excellent
          description of our data, as shown in the residual plots in \cref{fig:bandpowers}.
          We find excellent agreement with the results of \planck\ and \ACT{}
          as shown in \cref{fig:experiments}, \cref{fig:lcdm_main}, and \cref{tab:lcdm}. 
          
    \item The \sptnew{} data provide constraints on some cosmological parameters, such as the Hubble constant, which are comparable to those from \planck.
          In particular, we find $\Hubble = 66.66 \pm 0.60\,\kmsmpc$, a $6.2\,\sigma$ tension between \sptlr data alone and the latest SH0ES results~\citep{breuval_small_2024}, as also shown in \cref{fig:Hubble_compilation}.

    \item For the first time, a combination of data from ground-based experiments, namely \ground, reaches the constraining power of the \planck\ satellite data on some cosmological parameters, such as $\Hubble$ and $\sigmaeight$.
          This is a milestone for modern cosmology and the beginning of a new era for CMB experiments.
          The combination of these three CMB experiments in \cmball\ provides the tightest CMB constraints to date.
          The Hubble tension with \shoes{} grows to $6.4\,\sigma$ with the constraint of $\Hubble{} = \replace{67.24 \pm 0.35}{67.19\pm0.38}\,\kmsmpc$ derived from \cmball.
          While the \sptlr data alone have large uncertainties on the spectral index $\ns$,  we highlight that \cmball\ provides $\ns=\replace{0.9684\pm 0.0030}{0.9679\pm 0.0033}$ in \lcdm, a $\replace{10.5}{9.8}\,\sigma$ difference from a scale invariant spectrum with $\ns=1$.

\item The values of $\sigmaeight$ and $\omm$ from SPT-3G are in excellent agreement with the findings of other CMB experiments. We report $\sigmaeight=0.8158\pm0.0058$ and $\omm=0.3246\pm0.0091$. We confirm that some recent results from large scale structure probes, such as the weak lensing measurements from DES-Y3~\citep{abbott22a} and KiDS~\citep{wright25}, align well with the CMB data, to better than $2\,\sigma$, as shown in \cref{fig:S8}.

    \item The \ground\ primary anisotropy data prefer a lensing amplitude $\Alens$ in CMB spectra in agreement with the \lcdm\ expectations, contrary to the 
    slight excess found in \planck\ data, as illustrated in \cref{fig:alens_triangle}.

    \item There are borderline statistically significant differences between the CMB and BAO measurements from DESI in \LCDM\ in the $\omm$-$\hrd$ parameter space, as quantified in
    \cref{fig:DESI_hord}.

    \item The above discrepancy is relaxed in extended models of cosmology. While the CMB alone does not prefer any deviations from \lcdm{} at greater than $1.5\,\sigma$, the combination with DESI shifts the extension parameters away from the CMB best-fit values.
        The most-preferred departures from the standard model are $\Alens > 1$, $\mnu < 0.06\,\mathrm{eV}$, $\Omegak > 0$, and $\wo{} > -1, \wa < 0$; after accounting for the introduction of additional parameters, the corresponding models are preferred over \lcdm{} at $2$-$3\,\sigma$ each, as summarized in \cref{fig:cmball_DESI_summary} and \cref{tab:CMB_DESI_chi2}.
\end{enumerate}

At this point, we do not interpret the joint CMB plus DESI results as definitive evidence for a breakdown of the standard model;
the goodness-of-fit improvements extended models offer over \lcdm{} are moderate.
The hints of new physics are driven by the combination of the two probes and are not detected by either of them independently.
We conclude that while the differences between CMB and DESI data are an interesting avenue in the search for new physics, the possibility that these are due to statistical or systematic effects is not ruled out.
Hopefully, improved data from these two probes and others will provide new insight.

This analysis is the third, after~\citep{ge24,zebrowski25}, in a series of CMB power spectrum papers based on the SPT-3G observations carried out in 2019 and 2020.
We are currently improving our measurement of the distortion of the CMB due to weak gravitational lensing by including temperature information, expanding on the results from polarization alone used in this paper and presented in~\citep{ge24}.

One of the main limitations to the statistical power of this analysis is the sample variance resulting from the small sky fraction observed.
This will be improved by the measurement of the primary CMB spectra on an additional $2\,600\,\degsq$ observed during the austral summer in 2019-2020 and 2020-2021 (the SPT-3G Summer fields, see \cref{fig:footprints}).
The analysis of these fields is ongoing and is in an advanced stage.
Furthermore, the analysis of additional $6\,000\,\sqdeg$ observed in 2024 is also ongoing.
Together with the \mainfield{} and Summer fields, this will allow us to infer cosmological constraints from $10\,000\,\degsq$, approximately $25\%$ of the sky.
Finally, the full SPT-3G survey will include a total of at least seven years of observations of the Main field, reaching a coadded sensitivity of no worse than $1.6\,\ukarcmin$ in temperature.
This will allow us to probe even smaller angular scales in polarization with high precision.
We forecast that these upcoming data sets will provide substantial improvements compared to the results presented in this paper~\citep{prabhu24}.
The analysis of this data will benefit from techniques introduced here which serve as a blueprint for future work.
The full SPT-3G data set will enable precise new tests of the \LCDM\ cosmological model and searches for physics beyond it.

\section{Acknowledgements}
\label{sec:acknowledgements}
The South Pole Telescope program is supported by the National Science Foundation (NSF) through awards OPP-1852617 and OPP-2332483. Partial support is also provided by the Kavli Institute of Cosmological Physics at the University of Chicago. 
Argonne National Laboratory’s work was supported by the U.S. Department of Energy, Office of High Energy Physics, under contract DE-AC02-06CH11357. 
The UC Davis group acknowledges support from Michael and Ester Vaida. 
Work at the Fermi National Accelerator Laboratory (Fermilab), a U.S. Department of Energy, Office of Science, Office of High Energy Physics HEP User Facility, is managed by Fermi Forward Discovery Group, LLC, acting under Contract No. 89243024CSC000002.
The Melbourne authors acknowledge support from the Australian Research Council’s Discovery Project scheme (No. DP210102386). 
The Paris group has received funding from the European Research Council (ERC) under the European Union’s Horizon 2020 research and innovation program (grant agreement No 101001897), and funding from the Centre National d’Etudes Spatiales. 
The SLAC group is supported in part by the Department of Energy at SLAC National Accelerator Laboratory, under contract DE-AC02-76SF00515.

We gratefully acknowledge the computing resources provided on Crossover, a high-performance computing cluster operated by the Laboratory Computing Resource Center at Argonne National Laboratory. This work has made use of the Infinity Cluster hosted by Institut d'Astrophysique de Paris. We thank Stephane Rouberol for smoothly running this cluster for us.
The CAPS authors are supported by the Center for AstroPhysical Surveys (CAPS) at the National Center for Supercomputing Applications (NCSA), University of Illinois Urbana-Champaign. 
This work made use of the Illinois Campus Cluster, a computing resource that is operated by the Illinois Campus Cluster Program (ICCP) in conjunction with the National Center for Supercomputing Applications (NCSA) and which is supported by funds from the University of Illinois at Urbana-Champaign. 
This work relied on the \texttt{NumPy} library for numerical computations~\citep{numpy}, the \texttt{SciPy} library for scientific computing~\citep{scipy}, the \texttt{JAX} library for automatic differentiation and GPU/TPU acceleration~\citep{jax18}, and the \texttt{Matplotlib} library for plotting~\citep{matplotlib}.
Posterior sampling analysis and plotting were performed using the \texttt{GetDist} package~\citep{Lewis_getdist}.

\section{Data Availability}
\label{sec:data_availability}
The CMB band powers and associated likelihood are made public~\citep{sptwebsite}.
The MCMC chains used to obtain those results are made public~\cite{chainslambda}.

\clearpage

\appendix
\crefalias{section}{appendix}
\crefalias{subsection}{appendix}
\crefalias{subsubsection}{appendix}

\section{Filtering and transfer function}
\label{app:transferfunction}
In \cref{sec:tf}, we detailed the computation of the filtering artifacts and the transfer function. In this appendix, we showcase a map-level example of filtering artifacts. We also detail how we propagate the error on the estimation of the transfer function to parameters, and we describe how we evaluate the consistency of the simulation pipelines.

\subsection{Filtering artifacts}
\label{app:filtering-artifacts-map}
\begin{figure}[h]
	\centering
	\includegraphics[width=\columnwidth]{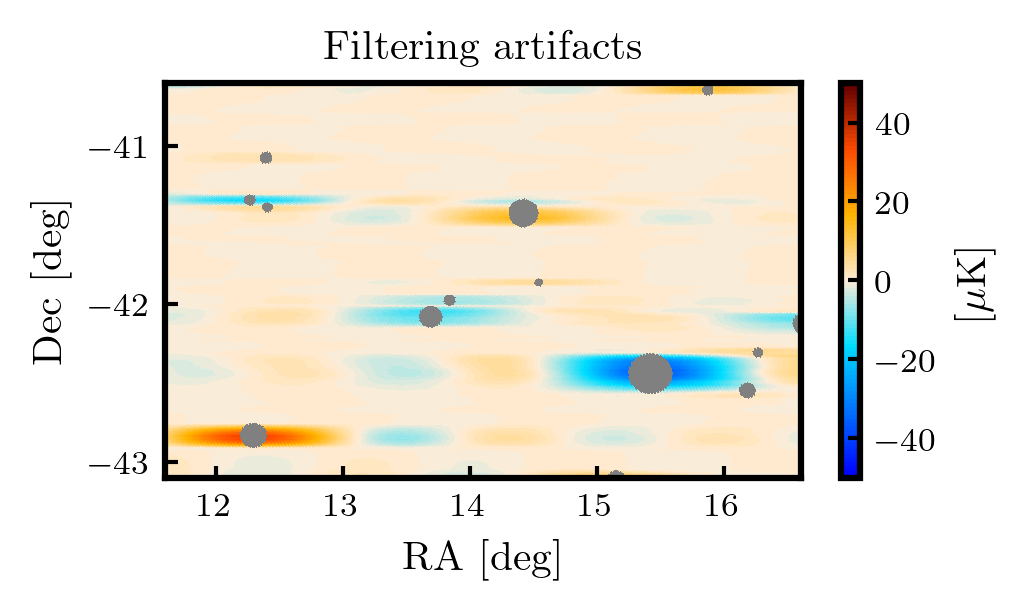}
	\caption{Figure of filtering artifacts in the \T{} map. The gray disks indicate the mask applied to remove point sources. Colored regions highlight filtering artifacts, which appear as elongated stripes aligned with the scan direction (horizontal axis). These artifacts result from the masking of point sources during timestream filtering in the map-making process.}
	\label{fig:artifacts_map}
\end{figure}

As discussed in \cref{sec:filtering-artifacts}, the filtering of timestream data, particularly the masking of point sources, introduces artifacts in the maps. 
These artifacts manifest as elongated stripes aligned with the scan direction in the \T{} map, as shown on a simulation in \cref{fig:artifacts_map}. 
Following the notations of \cref{sec:filtering-artifacts}, we display a map computed as 
\begin{equation}
	T^{\rm artifacts} \equiv T^{\ff;\mmmask} - T^{\ff;\nommmask}.
\end{equation}
The stripes follow lines of constant elevation, making them highly localized in Fourier space. This localization enables effective post-processing using an additional \alm{} weighting, as described in \cref{eq:wieneralm}.

\subsection{Error on the estimation of the transfer function}
When estimating the transfer function, we rely on $\Nmc{}=2000$ MC simulations of the data as described in \cref{sec:tf}. We expect a residual error on the estimation of the transfer function that we propagate to cosmological parameters. Since $F_\ell$ is the ratio of averages of correlated variables, the standard deviation of the error made on its estimate $\Delta F_\ell$ is related to the standard deviation of the individual power spectra $\Delta C_\ell\equiv \VEV{\left(\delta C_\ell\right)^2}^{1/2}$ by
\begin{align}
	\frac{\Delta F_{\ell}}{F_{\ell}}
	 & =
	\left(\frac{1+F^2_\ell/H_\ell-2 F_\ell E_\ell/H_\ell}{\Nmc-1} \right)^{1/2}
	\frac{\Delta C_{\ell}^{\ff}}{C_{\ell}^{\ff}},
	\\
	 & =
	\left(\frac{1+H_\ell/F^2_\ell-2 E_\ell/F_\ell}{\Nmc-1} \right)^{1/2}
	\frac{\Delta C_{\ell}^{\uu}}{C_{\ell}^{\uu}},
\end{align}
where $H_\ell$ is defined in \cref{eq:Hl_def} and we introduce
\begin{align}
	E_{\ell} & \equiv \frac{\VEV{\delta C_{\ell}^{\XY,\ff}\ \delta C_{\ell}^{\X'\Y',\uu} }} %
	{\VEV{ \delta C_{\ell}^{\XY,\uu}\ \delta C_{\ell}^{\X'\Y',\uu}}}.
	\label{eq:El_def}
\end{align}
Assuming $F_\ell \simeq H_\ell \simeq E_\ell$, as confirmed by simulations, we can simplify the above expressions to
\begin{align}
	\frac{\Delta F_{\ell}}{F_{\ell}}
	 & \simeq
	\left(\frac{1-F_\ell}{\Nmc-1} \right)^{1/2}
	\frac{\Delta C_{\ell}^{\ff}}{C_{\ell}^{\ff}},
	\\
	 & \simeq
	\left(\frac{1/F_\ell-1}{\Nmc-1} \right)^{1/2}
	\frac{\Delta C_{\ell}^{\uu}}{C_{\ell}^{\uu}}.
\end{align}
Setting aside the \Nmc{} factor, it shows that, because the ratio of \cref{eq:Fl_def} cancels some of the cosmic variance, for all exploitable multipoles (where $0 \le F_\ell \le 1$), the relative error on $F_\ell$ is smaller than the one on $C_\ell^{\ff}$, and also smaller than the one on $C_\ell^{\uu}$ for scales such that $F_\ell\ge 1/2$.

\subsection{Consistency of the simulation pipelines}

\quickmock and \fullmock simulations do not agree perfectly on the output $C(\ell)$ they produce for the same input maps. We found that the relative discrepancy between the respective transfer functions can be recast as
\begin{align}
	\frac{F^{\QM;\XY;\mu\nu}_{\ell}}{F^{\FM;\XY;\mu\nu}_{\ell}} -1 & = \rho_\ell + \varepsilon_\ell^{\XY;\mu\nu}
\end{align}
where
\begin{align}
	\rho_\ell & \equiv \left(1.4\times 10^{-3} + 2000/\ell^2\right)\left(1 - (350/\ell)^2\right)
\end{align}
is the same for all spectra and frequency combinations, having a peak value of $\rho=4.8116\times 10^{-3}$ at $\ell=517.66$ while the residual error
\begin{align}
	\left|\varepsilon_\ell^{\TT;\mu\nu}\right| & \lessapprox  10^{-3} < \frac{1}{10}\frac{\Delta C_\ell^{\TT;\mu\nu}}{C_\ell^{\TT;\mu\nu}} \ \rm{for} \ \ell \le 5000,    \\
	\left|\varepsilon_\ell^{\EE;\mu\nu}\right| & \lessapprox 2\times 10^{-3} < \frac{1}{10}\frac{\Delta C_\ell^{\EE;\mu\nu}}{C_\ell^{\EE;\mu\nu}} \ \rm{for} \ \ell \le 3000,
\end{align}
where $\Delta C_\ell$ is the power spectrum standard deviation for a bin size of $\Delta\ell = 50$.
The final transfer functions are then defined as
\begin{align}
   F^{\XY;\mu\nu} &\equiv \frac{F^{\QM;\XY;\mu\nu}}{1+\rho_\ell}.
\end{align}

\section{Post-unblinding changes}
\label{app:post-unblind}

As discussed in the description of the blinding procedure in \cref{sec:blinding}, 
after unblinding we discovered and corrected two previously untreated systematic effects:
(1) quadrupolar \TtoP leakage and (2) partial depolarization of the beam sidelobes. 
The quadrupolar \TtoP leakage is seen most clearly in the coadded \Q{} and \U{} maps of bright point sources and was determined to be the cause of the failure of \TE{} and \EE{} band-power difference tests above $\ell = 3000$, which initially led to the exclusion of these data from the analysis.
This effect had a substantial impact on the consistency between the \TT{} and \TE{} likelihoods. 
We addressed this by implementing models based on leakage beams measured from bright point sources and propagated the correction to the band powers.
After applying this correction, all band-power difference tests passed within the baseline $\ell$ range, 
and improved agreement under the \lcdm{} model was achieved between the \TT{} and \TE{} likelihoods.

Evidence for the depolarization of beam sidelobes—or, equivalently, for differences between the temperature and polarization beams—arises primarily from discrepancies in the \EE{} power measured on the same sky at 95 and 150\,GHz. 
While correcting for this effect clearly improves the pre-unblinding frequency consistency tests, we note that these tests formally passed prior to the correction. 
The observed sidelobe depolarization also led to differences in the cosmological parameter values favored by the \TT{} and \EE{} likelihoods, most notably in the \ns{} and \ombh{} plane. 
Although this initial discrepancy motivated further investigation of residual systematics in our data, we emphasize that we have compelling, cosmology-independent evidence supporting this model.

The models we use to correct for both of these effects were described briefly 
in \cref{sec:beams} and \cref{sec:data_model}; 
in this appendix we provide more detailed explanations of these 
effects and discuss their impact on inferred cosmological parameters.

\subsection{Quadrupolar beam leakage}
\label{app:t2p}
\begin{figure*}
	\centering
	\includegraphics[width=\columnwidth]{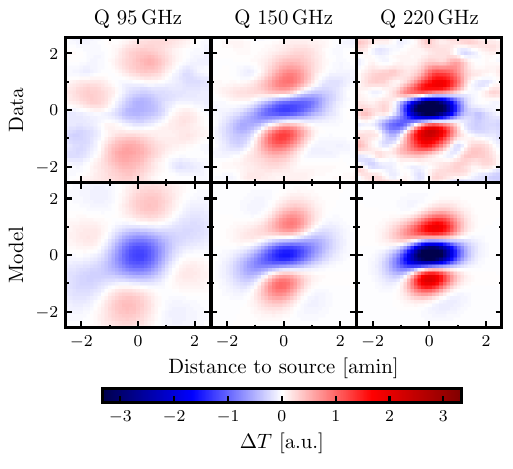}
	\includegraphics[width=\columnwidth]{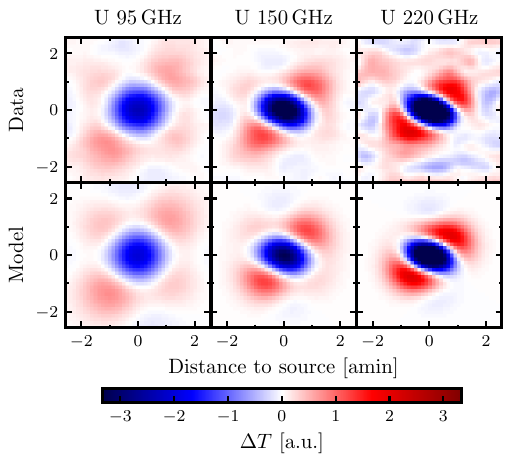}
	\caption{Residual point source \TtoP leakage in the Q and U maps in arbitrary units. \emph{Top}: Coaddition of 418 bright point-like sources in the Q and U maps after monopole subtraction. \emph{Bottom}: Template obtained from the fit of the leakage beams, using the model of \cref{eq:t2p_beams}.}
	\label{fig:t2p}
\end{figure*}

After unblinding, we found that we were not accounting for a substantial quadrupolar \TtoP leakage.
This leakage was responsible for an initial failure of the band power-level consistency tests in \TE{} and \EE{} spectra for $\ell \in [3000, 4000]$, which led us to abandon these data before unblinding.
Additionally, it resulted in a bias in the \TE{} band powers and incompatible results between temperature and polarization data in \lcdm{}.
As illustrated in \cref{fig:leak}, this leakage exhibits coherence across frequencies at large angular scales and is largest, relative to the uncertainty on the power spectrum, at small scales. 
This explains why the leakage was not identified in the interfrequency consistency tests, except in the previously excluded $\ell \in [3000,4000]$ range, and underscores a limitation of our blinding procedure, which proved insufficient for identifying and treating all systematic effects present in the data. 
This leakage was not corrected for in the MUSE-based analysis of~\citep{ge24}. 
That work was based on \EE{}-only measurements and did not detect significant quadrupolar leakage. 

A quadrupolar \TtoP leakage arises because the linearly polarized beams from individual detectors are slightly elliptical along the axis of polarization. The difference in the response to an unpolarized source for two orthogonally polarized detectors, with this polarization-direction-dependent ellipticity, results in a leakage signal~\citep{hivon_quickpol_2017,hu03,shimon_cmb_2008}.
This leakage is a quadrupolar pattern in the Q and U maps, which is particularly visible around point sources, as shown in \cref{fig:t2p}. In this figure, we show the residual point source \TtoP leakage in the Q and U maps after monopole subtraction, as well as the template for the quadrupolar signal obtained from the model fit described below.

Although the amplitude of this leakage is expected to be small, it can have a significant impact on power spectrum estimation if not accounted for.
To properly capture this effect, we assume that the \TtoP leakage is a convolution of the underlying temperature map with leakage beams specific to the Q and U maps, which we label $B^{\rm T\to Q}$ and $B^{\rm T\to U}$, respectively.
These functions correspond to the Q and U response from an unresolved pure temperature signal.
We model the leakage beams using an expansion of orthogonal Hermite polynomials:
\begin{align}
	B^{\rm T\to Q;\mu}(x, y) & = B_{\sigma_\mu}(x, y) \sum_{m+n} a_{m,n}^{\rm T\to Q;\mu} H_{m,n}\left(\frac{x}{\sigma_\mu}, \frac{y}{\sigma_\mu}\right), \nonumber \\
	B^{\rm T\to U;\mu}(x, y) & = B_{\sigma_\mu}(x, y) \sum_{m+n} a_{m,n}^{\rm T\to U;\mu} H_{m,n}\left(\frac{x}{\sigma_\mu}, \frac{y}{\sigma_\mu}\right)
	\label{eq:t2p_beams}
\end{align}
with:
\begin{equation}
	H_{m,n}\left(\frac{x}{\sigma_\mu}, \frac{y}{\sigma_\mu}\right)=H_{m}\left(\frac{x}{\sigma_\mu}\right)H_{n}\left(\frac{y}{\sigma_\mu}\right)
\end{equation}
where, for each frequency $\mu$, $B_{\sigma_\mu}$ is a Gaussian of width $\sigma_\mu$, $a_{m,n}^{\rm T\to Q;\mu}$ and $a_{m,n}^{\rm T\to U;\mu}$ are the coefficients of the expansion, and $H_{m}\left(x/\sigma_\mu\right)$ are Hermite polynomials orthogonal with respect to the measure $e^{-x^2/\sigma_\mu^2}$.
Hermite polynomials depend on the Gaussian width $\sigma_\mu$ to guarantee their orthogonality. 

\begin{figure}
	\includegraphics[width=\columnwidth]{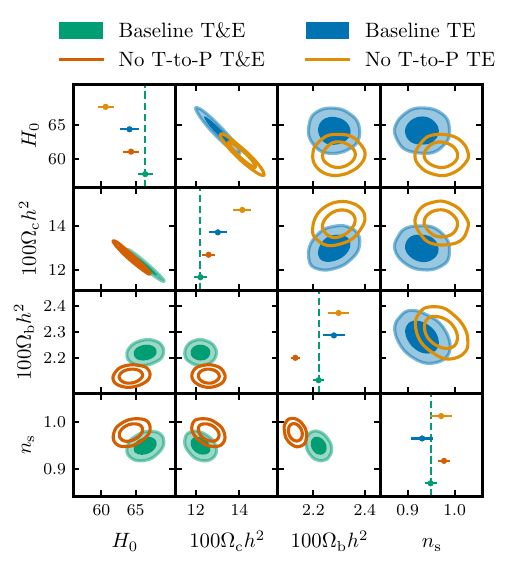}
	\caption{Comparison between the baseline likelihood and a likelihood without the quadrupolar \TtoP leakage correction. 
	Contours indicate the $68\%$ and $95\%$ confidence levels. 
	The dashed vertical line marks the mean value of the results of the baseline likelihood.
	The upper right panel shows \TE{} constraints, and the lower left shows \TTTEEE{} constraints. 
	}
	\label{fig:shift_t2p}
\end{figure}

We combine $418$ thumbnail maps centered at the location of bright sources into $25$ coadded maps of the leakage in Q and U.
The coaddition allows us to reach a significant detection over the CMB signal and noise.
Each of the coadded maps contains a variable number of sources in order to reach comparable signal to noise. 
The coaddition is straightforward thanks to the telescope's scanning strategy, which makes the detectors always oriented in the same direction with respect to the sky;
the measured beams are thus independent of the sky location.
The upper row of \cref{fig:t2p} displays the coadded maps from the location of all $418$ bright sources used in this analysis.
We find no significant detection of monopole leakage, as expected given the monopole deprojection described in \cref{sec:calibration}, but detect a significant quadrupolar and a mildly significant hexadecapolar leakage. 
We measure the mean and the standard deviation of the beam leakage parameters from fits to the $25$ coadded maps.
The lower row of \cref{fig:t2p} shows the mean template obtained from this procedure.
We see a  pattern of decreasing angular scale for the leakage with increasing frequency as expected from the temperature beam shapes.
The pattern in U is similar to the one in Q but rotated by an angle of $\pi/4$.

We propagate the measured model coefficients to the band powers. 
This is done based on an analytical model of the leakage described in \cref{eq:t2p} and confirmed through simulations to be the convolution of the temperature map by the leakage beam.
The amplitude of the quadrupolar leakage is quantified by 
\begin{equation}
	\epsilon^{\mu}_2 = \left(a^{\rm T\to Q;\mu}_{2,0} - a^{\rm T\to Q;\mu}_{0,2} + a^{\rm T\to U;\mu}_{1,1}\right)/2.
\end{equation}
As expected, the \TE{} and \EE{} contribution is only connected to the spin-2 component of the leakage at the map level, without the pure spin-0 radial contribution $(a^{\rm T\to Q}_{2,0}+a^{\rm T\to Q}_{0,2})$. 
We report these values in \cref{sec:quadrupolar}.
We show the expected contribution to the band powers from this leakage in \cref{fig:leak}.

After accounting for this effect in the data model, all consistency tests between bands and spectra passed across the full angular multipole range.
We tested variations of this model where the leakage was assumed to contain additional orders (dipole, octopole, hexadecapole).
Although hexadecapolar leakage is detected with mild significance in the map-level analysis, we verified that including it in the model  does not impact cosmological parameter estimation.
In this analysis, we incorporate the leakage through forward modeling in the theoretical prediction. 
As a cross-check, we also implemented an alternative approach in which the leakage is subtracted directly from the data band powers using the measured \TT{} and \TE{} band powers rather than the \lcdm{} model band powers.
Both methods yield consistent results.
Given no clear preference for variations on the quadrupolar leakage model presented here, it was adopted in the baseline analysis described in \cref{sec:quadrupolar}.

In \cref{fig:shift_t2p}, we show the impact of the \TtoP leakage correction on cosmological parameters. 
We compare \lcdm{} constraints obtained from the baseline \TTTEEE{} and \TE{} likelihoods with those obtained without the \TtoP leakage correction. 
The impact on the \TE{} constraints is significant, with close to $3\,\sigma$ shifts in \Hubble{}, \ns{}, and \omegac{}.
We also show the effect of \TtoP leakage correction on the full \TTTEEE{} likelihood, but we caution that these results are more difficult to interpret, because without this correction the \TE{} \lcdm{} constraints are not sufficiently compatible with those from \TT{} and \EE{} to be properly combined. 
With this caveat, we note that the biggest shift for the \TTTEEE{} likelihood is on \ombh{} due to the degeneracy of this parameter with both \Hubble{} and \omegac{}. 
Including the \TtoP leakage correction in the \TTTEEE{} likelihood yields a $\Delta \chi^2 = 94$ improvement in the fit to the data. 
This improvement is highly significant, considering that the model introduces only three additional degrees of freedom.

\subsection{Polarized beams}
\label{app:polbeams}
Unlike for the temperature beam, there are no sufficiently bright and polarized detected sources that can be used to directly map the polarized beam with the required signal-to-noise. 
As has been done for previous CMB analyses, prior to unblinding we assumed that the polarized beam shape was identical to that of the measured temperature beam. 
The measured temperature beam has significant sidelobe power, arising from sources such as diffraction from primary mirror panel gaps and multiple reflections from optical elements, that is not captured in simulations and for which we have no knowledge of the optical path. 
In retrospect, it is unrealistic to assume that this sidelobe power is polarized identically to the main beam. 

Similarly to the \muse{} analysis~\citep{ge24}, after unblinding it became apparent that our assumptions about the polarized beam were unjustified.
As stated in \cref{sec:beams} and \cref{sec:data_model}, prior to unblinding we assumed beam sidelobes polarized equally to the main beam, i.e. $\betapol=1$.
In the baseline model adapted for  this work and for the \muse{}~\citep{ge24} analysis of the same data, we allow the degree of sidelobe polarization for each of the bands relative to the main beam, $\betapol^{\nu}$, to vary.

There is clear evidence from the data supporting this model.
First, consistency of the polarization data in the three observation bands (particularly for $\ell < 1200$) requires fractional depolarization of the beam sidelobes.
Under the original, rigid, assumption of identical temperature and polarized beams, the PTE of the $150\times150\ghz$ vs $95\times150\ghz$ difference test was 0.004.
Though borderline passing our blinding threshold of $0.28\%$, this is low and an inconsistency was visible in the difference spectrum (see \cref{fig:EEdifference}), particularly in all the comparisons between 95 and 150\ghz\ channels.
When allowing for a varying sidelobe polarization fraction, the PTE of the difference test is $51\%$, a substantial improvement.
When considering only the agreement at multipoles below $\ell=1000$, the PTE for the $(95\times95\ghz, 95\times150\ghz)$, $(95\times95\ghz, 150\times150\ghz)$, and $(95\times150\ghz, 150\times150\ghz)$  \EE{} difference tests improve from $4.4\%$, $1.1\%$, and $0.27\%$ to $60\%$, $43\%$, and $24\%$, respectively.

Second, a cosmological-model-free reconstruction of the binned CMB power spectra strictly prefers depolarization of the beam sidelobes. 
We perform this reconstruction without the need to assume a cosmological dependence using the \emph{lite} framework introduced in \cref{sec:lite_likelihood}.
In order to marginalize over nuisance and foreground contributions, the framework first estimates the best fit parameters of the likelihood, replacing in \cref{eq:datamodel} the cosmological set of parameters $\psi$ by binned CMB spectra~\citep{balkenhol25}.
Within the \emph{lite} framework, we call this procedure the estimation of the reconstruction likelihood. 
This relies on the assumption that up to nuisance and foregrounds parameters, the power of the CMB signal is the same in bins covering the same multipole range of the same spectrum.\footnote{E.g. the CMB power is the same in the \EE{} bin covering $1000 < \ell < 1050$ at $95\times150\,\ghz$ and $220\times220\,\ghz$. This assumption does not strictly hold as the window functions are not identical across frequencies, though deviations are less than 0.2\% and hence negligible for this purpose.}
In other words, this allows us to estimate the best foreground and nuisance model parameters that minimize the discrepancy 
between cross-frequency spectra, without assuming any cosmological model. 
Note that minimizing the foreground and nuisance-induced discrepancy between cross-spectra is not immune from any coherent effect across frequency channels. 
With this limitation in mind, the \emph{lite} framework is an efficient tool to test for different models of the systematics, 
in a cosmology-independent way and compare the best-fit $\chi^2$ values.

We use this approach to investigate the preference of the data for different models of the instrumental beam. 
We test two cases: 
(1) the baseline (post-unblinding) model that includes a fractional depolarization of the beam sidelobes and 
(2) the pre-unblinding model with identical beam shapes in temperature and polarization, i.e. fixing $\beta^\nu_{\mathrm{pol}}=1$ in \cref{eq:datamodel}. 
We minimize the reconstruction likelihoods and calculate the associated $\chi^2$ values. 
We report $\chi^2$ (PTE) values of $1198.64$ ($31.05\%$) for model (1) and $1229.46$ ($14.57\%$) for model (2), respectively. 
While both models provide an acceptable fit to the data, the difference in $\chi^2$ is $30.8$.
The $\chi^2$ improvement is concentrated on large angular scales with $22.7$ points below $\ell=1000$ and $26.9$ points below $\ell=1500$.
This is significant and, when adjusting for the additional degrees of freedom between the models, the $\chi^2$ improvement on the full $\ell$ range translates to a difference in the Akaike Information Criterion of $24.8$.
The data strongly prefer the  beam sidelobe depolarization model, 
when assuming no cosmological model for the CMB signal in the data, 
and reject the alternative hypothesis that the temperature and polarized beam shapes are identical.

Third, we note that, similar to the \sptlite{} reconstruction discussed above, the \muse{} analysis~\citep{ge24} also found strong evidence in favor of the polarized beam model, independent of any cosmological model assumptions. 
In that pipeline, which is close to simulation-based inference, the binned \EE{} and $\phi\phi$ spectra are first reconstructed—together with systematics—without imposing a cosmological model.
At this stage, the baseline polarized beam model presented here was preferred over the rigid one.

Finally, we can evaluate this preference when assuming a \lcdm{} model. 
In this case, the polarized beam model is preferred over the rigid one at $\Delta \chi^2 = 30$.
This number is consistent with the $\Delta \chi^2$ obtained from the \emph{lite} reconstruction likelihood, highlighting that the preference for the polarized beam model does not require assuming any particular cosmological model, but is rather required by the differences between frequency bands.
One expects the largest signature of polarized beams in the \EE{} data, which in fact does contribute $\Delta \chi^2 = 24$.

\begin{figure*}
	\includegraphics[width=\columnwidth]{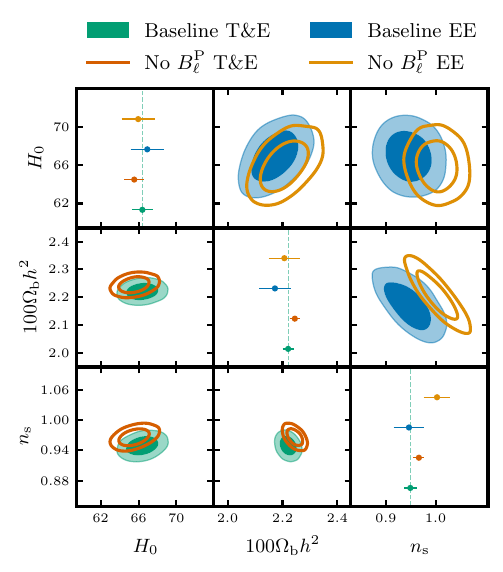}
	\includegraphics[width=\columnwidth]{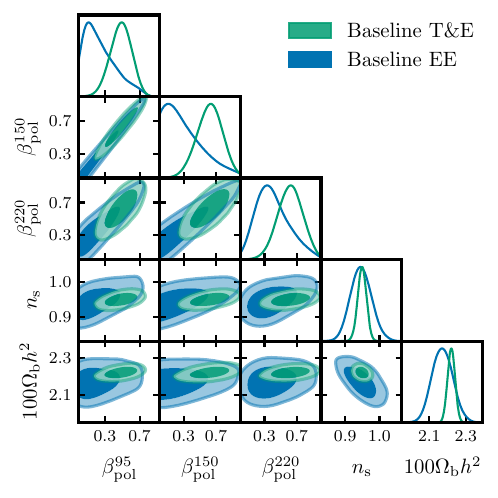}
	\caption{
		\emph{Left}: 
		Comparison between the baseline likelihoods with and without the polarized beam correction for \TTTEEE{} and \EE{} alone.
		The upper right panel shows \EE{} constraints, and the lower left shows \TTTEEE{} constraints. 
		\emph{Right}: Posterior distributions of the polarized beam parameters from the baseline \TTTEEE{} and \EE{} likelihoods, 
		shown alongside parameter constraints in the $\ns$-$\ombh$ plane.
		The distributions are non-Gaussian and unbounded, so we do not show whiskers in the diagonal plots.
		The figure highlights the correlation between beam parameters. 
		The \TTTEEE{} and \EE{} contours show good agreement. 
		This pronounced preference for sidelobe depolarization is driven by the mismatch between the 
		95 and 150\ghz\ data at $\ell < 1200$, resulting in a clear exclusion of the point 
		$(\betapol^{95},\betapol^{150}) = (1, 1)$ in the upper left two-dimensional panel.
	}
	\label{fig:polbeampars}
\end{figure*}

In the left panel of \cref{fig:polbeampars}, we show the impact of the polarized beam correction on the cosmological parameters. 
We compare the \lcdm{} constraints obtained from the baseline \TTTEEE{} and \EE{} likelihoods with those obtained without the polarized beam correction. 
The polarized beam model mostly affects the \ns{} and \ombh{} plane, with a $2\,\sigma$ shift in \ns{} for \EE{} alone. 
In the right panel, we show that fitting the beam parameters within the \lcdm{} model yields consistent values for $\betapol$ when using either the \EE{}-only or the full \T\&\E{} data set. 
Both likelihoods yield a $5\,\sigma$ detection of $(\betapol^{95},\betapol^{150},\betapol^{220})\neq(1,1,1)$ in the associated 3d parameter space.
The small statistical shift observed between the two cases arises from common modes in the polarized beam corrections that, within the uncertainties,  are degenerate with cosmological parameters.
This degeneracy is reduced when all spectra are included, leading to tighter constraints. 
The cosmological parameters most sensitive to the beam depolarization effect shift by at most $1\,\sigma$. 
As demonstrated in \cref{sec:cosmoparametertests} and \cref{fig:parameter-difference-test}, accounting for the correlation between the \T\&\E{} and \EE{}-only data sets, both likelihoods yield fully consistent results.

It is important to highlight that this model is not specifically designed to treat small angular scales, where beam  differences propagate to relative slopes (see~\citep{louis25}), but has the largest impact on large angular scales where the data are poorly described without it.
Excising the large scale data naturally weakens the detection of $(\betapol^{95},\betapol^{150},\betapol^{220})\neq(1,1,1)$, though it does not shift cosmological constraints by more than the expected amount, as demonstrated in \cref{sec:lcdm_integrity}.
In particular, constraints on \ns{} and \ombh{} are stable when removing $\ell<1200$, $\ell<1500$, or $\ell<2000$ data.
In fact, forcing $(\betapol^{95},\betapol^{150},\betapol^{220}) = (1,1,1)$ does not impact constraints on \lcdm{} parameters when fitting data from $\ell>2000$ alone.

\begin{figure}
	\includegraphics[width=\columnwidth]{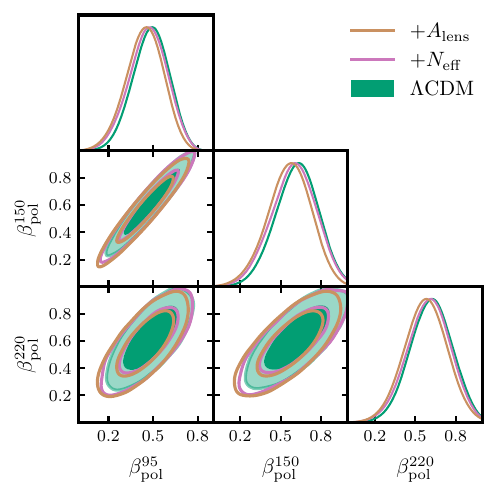}
	\caption{Posterior distributions of the polarized beam parameters from the baseline \TTTEEE{} likelihood in \lcdm{} + \Alens{} and \lcdm{} + \neff{} cosmologies.}
	\label{fig:polbeam_ext}
\end{figure}

In addition, we performed several tests of the underlying assumptions of the polarized beam model. 
First, we investigated the possibility of a systematic effect mimicking reduced sidelobe efficiency in the \TT{} spectra by applying the same polarized beam model to the temperature data. 
This serves as a check of the analysis pipeline, as there is no physical motivation for such an effect. 
We find no evidence for reduced sidelobe efficiency in the temperature data; the measured sidelobe efficiencies are consistent with unity within $1.5\,\sigma$. 
The absence of evidence for reduced sidelobe efficiency is further supported by the \TT{} difference tests presented in \cref{sec:difference}.
Second, we tested a variation of the beam model which allowed for a sidelobe polarization fraction that varies as a function of scale. 
Introducing this freedom did not further improve the consistency between the band powers from different frequencies using the \emph{lite} framework or the fits to \lcdm{} cosmology. 
Third, we tested another variation where the shape of the main beam, which is calculated analytically and informs the shape of the sidelobes, is replaced by its best-fit Gaussian approximation for each frequency band, instead of using the calculated beams.
Despite this extreme change, the preference for sidelobe depolarization 
and the resulting cosmological parameters were nearly identical. 
We conclude that the polarized beam model is insensitive to reasonable variations of the 
main beam shape and the assumption of a uniform depolarization of the beam sidelobes. 
The data are well described by the simple model we present, with a 
highly polarized main beam and a sidelobe 
that is uniformly fractionally polarized compared to the main beam and described by a single polarization parameter for each band.
\RR{Finally, we verified that the preference for the polarized beam model is also found in simple extension of \lcdm{}, such as \lcdm{} + \Alens{} and \lcdm{} + \neff{}. We find consistent $\betapol$ values when including the polarized beam model in these extended cosmologies, and the preference for the polarized beam model remains strong. This is shown in \cref{fig:polbeam_ext}.}

Altogether, there is strong evidence in favor of the polarized beam model, which we adopt as the baseline.
We stress that this is a conservative choice and that there is no evidence in our data to support the rigid assumption of identical temperature and polarized beams that has been commonly adopted for CMB analyses.
We choose to parametrize and marginalize over the uncertainty in the sidelobe polarization fraction with the goal of eliminating bias at the cost of some constraining power on cosmological parameters.
Freeing the sidelobe polarization fractions from their best fit posterior values degrades the constraint on \ns{} by $20\%$, as can be seen in \cref{fig:polbeampars}.
Direct measurements of the polarized beams have the potential to recover this lost constraining power. However, there are no sufficiently bright and highly polarized detected sources in the southern sky that would allow us to measure the polarized beam with the required precision.
Because of the large primary mirror of the SPT, it is prohibitive to place a ground-based polarized source both in the far-field and at an elevation sufficiently high to prevent detector saturation from atmospheric loading.
Satellite based polarized sources have the potential to enable a direct high signal-to-noise measurement of polarized beams for a large aperture telescope and reduce this source of uncertainty~\citep{Ritacco_2024,s21103361}.

\begin{figure}
	\includegraphics[width=\columnwidth]{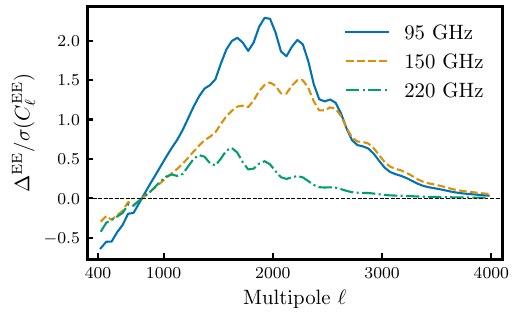}
	\caption{Impact of the polarized beam model on the \EE{} band powers.
	The quantity plotted is defined as in \cref{eq:beam_pol_variation} and is shown in units of the statistical uncertainty of the \EE{} band powers.}
	\label{fig:beam_pol_variation}
\end{figure}

\RR{\cref{fig:beam_pol_variation} shows the impact of the polarized beam model on the band powers.
We display the change in the \EE{} band powers induced by differences between the temperature and polarization beams as a function of multipole.
Specifically, we define the plotted quantity as
\begin{equation}
	\Delta^{\EE;\mu\mu} \equiv C_\ell^{\EE;\mu\mu} \left(\frac{B_\ell^{\P;\mu}}{B_\ell^{\T;\mu}}-1\right)^2.
	\label{eq:beam_pol_variation}
\end{equation}
The quantity is shown in units of the statistical uncertainty of the \EE{} band powers to emphasize the multipole ranges where the effect is significant.
As discussed above, the effect is most pronounced at large and intermediate angular scales, where the sidelobes contribute substantially to the overall beam shape.}

\subsection{Miscellaneous post-unblinding changes}
\label{sec:misc}

In this subsection, we present various minor updates that have been made to the pipeline after unblinding.
These updates did not have a significant impact on the results; they are listed for completeness.

First, we realized that the covariance used for unblinding was erroneously missing the lensing contribution described in \cref{sec:covariance}, as well as the regularization factor (\cref{eq:fudge}) even though the decision had been made to include both of them already. Adding the lensing contribution increases the error bars on cosmological parameters by 10\%. The covariance conditioning is required to correctly interpret the $\chi^2$ of the data, but has a negligible impact on cosmological parameters. We also updated CMB and foreground template spectra used to compute the covariance for the best-fit model obtained from the \TTTEEE{} analysis to ensure consistency between the model and the data.

We corrected a mistake in the implementation of the tSZ-CIB cross-correlation model in the likelihood software. We also fixed the amplitude of \EE{} Poisson power to zero based on expectations from the source masking threshold and the low polarization of radio sources, see~\citep{chou25}.

Finally, minor improvements were introduced in the beam pipeline, resulting in changes to the beam $B_\ell$ smaller than $0.2\%$.
We also removed unnecessary priors on polarization calibration for the \TTTEEE{} likelihood, letting the data drive the calibration.
We further removed priors on the relative inter-frequency temperature calibration of the 95 and 220\ghz{} channels.
However, we restore priors on temperature and polarization calibration when analyzing subsets of the data (e.g. \EE{}  or \TE{} alone) to break degeneracies.
We also increased the uncertainty on the global temperature calibration prior calculated from the comparison with \planck{} data to account for systematic changes depending on the choice of the \planck{} map.
The temperature calibration prior changed from $\sigma(A_{\rm ext}^{\rm cal})=0.0019$ to $\sigma(A_{\rm ext}^{\rm cal})=0.0036$.
This conservative change has an overall negligible impact on the cosmological parameters, though it widens the \logA{} posterior by $4\%$.

\subsection{Initial cosmological results}

\begin{figure}
	\includegraphics[width=\columnwidth]{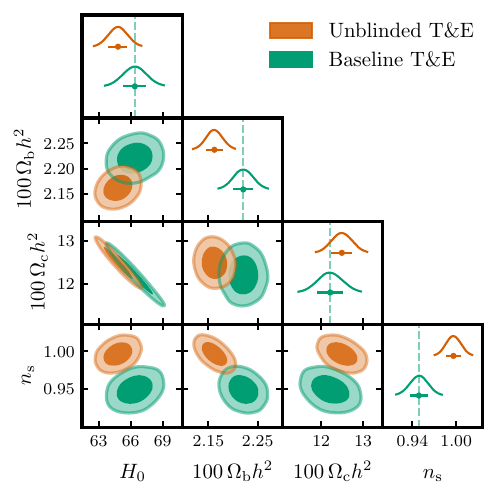}
	\caption{Initial cosmological results from the \TTTEEE{} likelihood at unblinding, compared to final results. Contours indicate the $68\%$ and $95\%$ confidence levels. The unblinded results revealed strong inconsistencies among the individual \TT{}, \TE{}, and \EE{} likelihoods, motivating the investigations detailed in \cref{app:t2p,app:polbeams}. The unblinded contours plotted here rely on the combination of inconsistent subsets and are not formally meaningful.}
	\label{fig:unblinding}
\end{figure}

In accordance with our blinding procedure, we did not examine the cosmological results prior to unblinding. 
The initial results obtained from the \TTTEEE{} likelihood at the time of unblinding are shown in \cref{fig:unblinding}. 
The observed shifts in the \lcdm{} parameters are substantial and can be attributed to specific systematic effects. 
First, the primary source of the shift in \ombh{} is the quadrupolar \TtoP leakage, which biases the \TE{} band powers, as illustrated in \cref{fig:leak,fig:shift_t2p}. 
The \TE{}-derived parameters were highly inconsistent with those from \TT{} and \EE{}, with none of the PTE tests passing. 
Enforcing agreement among the channels within the \lcdm{} model resulted in the significant discrepancies shown in the figure. 
Then, the shift in \ns{} is driven by both the \TtoP leakage and the introduction of the polarized beam sidelobe model. 

Our experience highlights that the blinding procedure implemented in this analysis was insufficient to identify all systematic effects present in the data, as evidenced by the substantial shifts in cosmological parameters following unblinding. 
The procedure was primarily designed to flag systematics that manifest as incoherent features across frequencies. 
Consequently, the quadrupolar \TtoP leakage went undetected at $\ell<3000$, and the depolarized sidelobes were similarly missed due to their near coherence across frequencies. 
Notably, a low PTE in the \EE{} band-power difference test at $\ell<1200$ did provide an early indication of the latter effect.

\section{Covariance matrix}
\label{app:cov}
The covariance matrix computation is detailed in \cref{sec:covariance}. In this appendix, we justify the additional noise term in \cref{eq:covnoisecontribution} and we showcase the mixing matrix for deeper understanding of the data set.

\subsection{Cross-bundle covariance matrix}
\label{app:covcross}

In this work, we estimate the power spectrum by taking cross-products of maps from different bundles to avoid a noise bias.
Another approach is to estimate the band powers directly from all the data compiled together in a single coadd.
The covariance matrix of the mean cross-bundle power spectrum is different than the covariance matrix of the coadd power spectrum.
Assuming that the noise $N_\ell$ is Gaussian and uncorrelated between $n_{\rm b}$ bundles and that maps cover the full sky, the covariance matrix of the mean cross-bundle power spectrum is
\begin{align}
	\label{eq:lueker_cov}
	\Sigma_{\ell\ell}^{\rm bundles} \propto \left(2C_\ell^2 + 2C_\ell N_\ell + \frac{2n_{\rm b}}{n_{\rm b}-1}N^2_\ell\right),
\end{align}
where $C_\ell$ is the fiducial signal.
This equation is derived from the framework introduced in~\citep{lueker10}.
The noise variance is boosted by a factor of $n_{\rm b}/(n_{\rm b}-1)$ compared to the coadd covariance.
This is due to the fact that we avoid auto-bundle spectra in our framework, again to avoid a noise bias.
For our case of $30$ bundles, the noise variance is increased by $3\%$.
In the infinite bundle limit, the covariance matrix of the mean cross-bundle power spectrum is the same as the covariance matrix of the coadd power spectrum. 

\subsection{Mixing matrix}
\label{app:mixingmatrix}
From the covariance computed in \cref{sec:covariance}, we can compute the mixing matrix, which describes the contribution of each of the cross-frequencies to the minimum variance estimator of the band powers. 
The mixing matrix is defined as 
\begin{equation}
	\label{eq:mixingmatrix}
	M^{\rm mix} = \left(X^\top \Sigma^{-1}X\right)^{-1}X^\top \Sigma^{-1},
\end{equation}
where $X$ is the design matrix that dictates which cross-frequency spectra are used to estimate the band powers and $\Sigma$ is the covariance matrix. It makes it possible to compute the minimum variance estimator of the band powers as used in \cref{fig:experiments,fig:snr} (for details on the procedure see ~\citep{planck15-11, mocanu19, balkenhol25}).
\begin{figure}
	\includegraphics[width=\columnwidth]{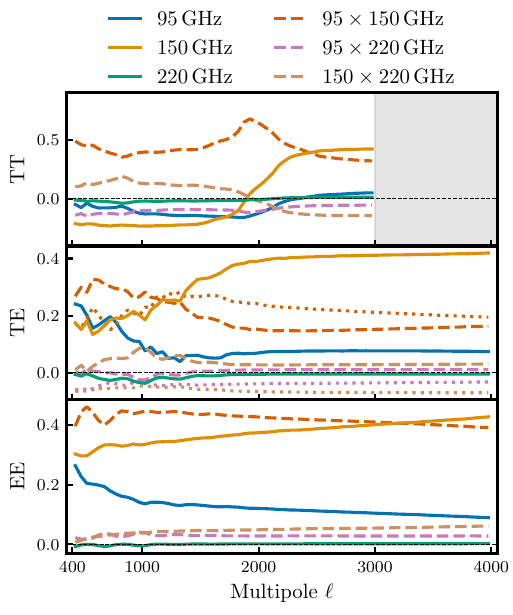}
	\caption{Diagonal elements of the mixing matrix $M^{\rm mix}$ for the \TT{} (top), \TE{} (middle), and \EE{} (bottom) spectra.
	The dashed lines indicate the contributions from the cross-frequency spectra, $95\times150\ghz$, $95\times220\ghz$, and $150\times220\ghz$.
	The dotted lines in the \TE{} plot indicate the contributions from the inverted cross frequency spectra, $150\times95\ghz$, $220\times95\ghz$, and $220\times150\ghz$.
	The mixing matrix is defined in \cref{eq:mixingmatrix}. 
	The diagonal elements of the mixing matrix are the weights of each cross-frequency spectrum in the minimum variance estimator of the band powers. 
	We normalize the mixing matrix here such that the absolute value of all elements at a given $\ell$ sum to one for  \TT{}, \TE{}, and \EE{} separately.}
	\label{fig:mixing_matrix}
\end{figure}

As shown in \cref{fig:mixing_matrix}, the mixing matrix is straightforward for \EE{}, highlighting that the deepest channels, 95 and 150\ghz{}, are contributing the most to the measurement of the band powers. This also explains why we are particularly sensitive to discrepancies between the two channels. In \TE{}, the mixing matrix is still dominated by the 95 and 150\ghz{} channels, although the contributions from $95\times150\ghz$ is different than $150\times95\ghz$, due to the noise structure. Negative terms in the mix matrix can arise when channels are correlated. Finally, the \TT{} mixing matrix displays two regimes.
At large scales, the signal and the noise are correlated;
while the $95\times150\ghz$ spectrum contributes the most weight here, the information is more equally distributed between the cross-frequency spectra compared to the \TE{} and \EE{} cases.
For $\ell\gtrsim 2000$ the noise correlation falls (see \cref{fig:noise}) and the mixing matrix is then dominated by the $150\times150\ghz$ and $95\times150\ghz$ spectra.

\section{Likelihood nuisance parameters}
\label{app:nuisance_pars}
A set of nuisance parameters is included in the likelihood to model uncertainties in calibration, beam characterization, and residual systematics, as described in \cref{sec:data_model}. 
\cref{tab:nuisance-priors} summarizes these parameters, providing their descriptions, priors, and default values used in simulation fits. 
For calibration and polarization efficiency parameters, we also specify the priors adopted when analyzing data subsets; these are derived from comparisons with \textit{Planck} data over the same sky region (see \cref{sec:calibration}). 
Beam eigenmode priors are standard normal distributions, reflecting the amplitude uncertainties of the associated modes (see \cref{eq:beammodes}). 
Priors on the quadrupolar beam leakage amplitudes are determined from map-level analyses (see \cref{app:t2p}). 
Uniform priors are assigned to the sidelobe polarization fractions (see \cref{app:polbeams}). 
For each parameter, we report the best-fit value and the mean with 68\% confidence intervals from the \lcdm{} analysis. 

We find that the posterior values of the subset calibration parameters are consistent with the priors derived from the calibration pipeline, which are not applied in the baseline analysis.

To assess the robustness of the \RR{beam} systematic model, we broaden the priors on the beam eigenmodes by factors of $[2, 5, 10, 20, 50, 100]$. 
This procedure increases the flexibility of the beam model. 
\RR{For each case, we find at most a $1.5\,\sigma$ Gaussian distance between the best-fit beam parameters and the prior means, indicating that the data are consistent with the beam model.} 
\RR{The cosmological parameters remain relatively stable as the beam priors are broadened, with shifts of less than $0.5\,\sigma$ for \ns{} and $0.2\,\sigma$ for all parameters in the worst case of a factor of $100$ broadening, and less than $0.1\,\sigma$ for all parameters for a factor of $5$ broadening.} 
Since $\ns$ characterizes the tilt of the primordial power spectrum, it is naturally correlated with the beam parameters \RR{and is more sensitive to changes in the beam priors.}
\replace{The observed shift is not statistically significant, supporting the robustness of the beam modeling. }
Next, we allowed the quadrupolar beam leakage amplitudes to vary freely and found less than a $3\,\sigma$ Gaussian distance between the likelihood and the prior set by the map-based analysis. 
Thus, the data are consistent with the quadrupolar beam leakage model.

\begin{table*}
	\begin{tabular}{l l l c c c c}
		\hline
		\hline
		\bf Parameter                         & \bf Description                                      & \bf Prior                                    & \bf Subset Prior           & \bf \lcdm{} Best-fit         & \bf 68\% CL limit                     & \bf Section                          \\
		\hline
		\hline
		\boldmath $A_{\rm cal}^{\rm ext}$     & External calibration                                 & ${\mathcal{N}(1.0,0.0036)[1.0]}$                &                            & $1.00002                   $ & ${1.0002\pm 0.0036}          $ & \cref{sec:calibration}                  \\		\hline
		\boldmath $A_{\rm cal}^{\rm rel;95}$  & \multirow{2}{*}{\shortstack[l]{Relative calibration                                                                                                                                                                                                      \\ factors}}             & \multirow{2}{*}{$\mathcal{U}(0.8,1.2)[1.0]$}         & $\mathcal{N}(1.0, 0.0024)$  & $1.000196                  $ & $1.00020\pm 0.00039        $          & \multirow{2}{*}{\cref{sec:calibration}}      \\
		\boldmath $A_{\rm cal}^{\rm rel;220}$ &                                                      &                                                        & $\mathcal{N}(1.0, 0.010)$  & $1.00853                   $ & $1.0087\pm 0.0012          $          &                                      \\
		\hline
		\boldmath $E_{\rm cal}^{\rm ext}$     & Polarization efficiency                              & $\mathcal{U}(0.8,1.2)[1.0]$                            & $\mathcal{N}(1.0, 0.0095)$ & $1.0085                    $ & $1.0095\pm 0.0051          $          & \cref{sec:calibration}                  \\
		\hline
		\boldmath $E_{\rm cal}^{\rm rel;95}$  & \multirow{2}{*}{\shortstack[l]{Relative polarization                                                                                                                                                                                                     \\ efficiencies}} & \multirow{2}{*}{$\mathcal{U}(0.8,1.2)[1.0]$}     & $\mathcal{N}(1.0, 0.0022)$     & $0.99869                   $ & $0.9986\pm 0.0011          $          & \multirow{2}{*}{\cref{sec:calibration}}      \\
		\boldmath$E_{\rm cal}^{\rm rel;220}$  &                                                      &                                                        & $\mathcal{N}(1.0, 0.0067)$ & $0.99579                   $ & $0.9957\pm 0.0030          $          &                                      \\
		\hline
		{\boldmath$\beta_1        $}          & \multirow{9}{*}{Beam eigenmodes}                     & \multirow{9}{*}{${\mathcal{N}(0.0, 1.0)[1.0]}$} &                            & $-0.47                     $ & ${-0.45\pm 0.95}             $ & \multirow{9}{*}{\shortstack[l]{\cref{sec:beams} \\ \cref{sec:sys}}} \\ 

		{\boldmath$\beta_2        $}          &                                                      &                                                        &                            & $-0.62                     $ & ${-0.61\pm 0.88  }           $ &                                      \\

		{\boldmath$\beta_3        $}          &                                                      &                                                        &                            & $0.47                      $ & ${0.51\pm 0.92 }             $ &                                      \\

		{\boldmath$\beta_4        $}          &                                                      &                                                        &                            & $-0.76                     $ & ${-0.87\pm 0.70 }            $        &                                      \\

		{\boldmath$\beta_5        $}          &                                                      &                                                        &                            & $-0.06                     $ & ${-0.03\pm 0.93 }            $ &                                      \\

		{\boldmath$\beta_6        $}          &                                                      &                                                        &                            & $-1.11                     $ & ${-1.10\pm 0.90 }            $ &                                      \\

		{\boldmath$\beta_7        $}          &                                                      &                                                        &                            & $0.15                      $ & ${0.13\pm 0.72  }            $        &                                      \\

		{\boldmath$\beta_8        $}          &                                                      &                                                        &                            & $-0.36                     $ & ${-0.36\pm 0.99 }            $ &                                      \\

		{\boldmath$\beta_9        $}          &                                                      &                                                        &                            & $0.20                      $ & ${0.20\pm 0.98 }             $ &                                      \\
		\hline
		\boldmath $\epsilon_2^{95}$           & \multirow{3}{*}{\shortstack[l]{2nd order                                                                                                                                                                                                                 \\ T$\rightarrow$P leakage \\ amplitudes}}        & $\mathcal{N}(-0.0065, 0.0011)[0.0]$                    & &$-0.00687                  $ & $-0.00690\pm 0.00078       $          & \multirow{3}{*}{\shortstack[l]{\cref{sec:quadrupolar} \\ \cref{sec:sys}}} \\
		\boldmath $\epsilon_2^{150}$          &                                                      & $\mathcal{N}(-0.012, 0.0021)[0.0]$                     &                            & $-0.01458                  $ & $-0.0146\pm 0.0015         $          &                                      \\
		\boldmath $\epsilon_2^{220}$          &                                                      & $\mathcal{N}(-0.023, 0.0066)[0.0]$                     &                            & $-0.02844                  $ & $-0.0285\pm 0.0036         $          &                                      \\
		\hline
		\boldmath $\betapol^{95}$      & \multirow{3}{*}{\shortstack[l]{Sidelobe                                                                                                                                                                                                                  \\ polarization\\ fractions}}          & \multirow{3}{*}{$\mathcal{U}(0.0, 1.0)[1.0]$}          & &$0.555                     $ & $0.48^{+0.13}_{-0.12}      $          & \multirow{3}{*}{\shortstack[l]{\cref{sec:polbeams} \\ \cref{sec:sys}}}       \\
		\boldmath $\betapol^{150}$     &                                                      &                                                        &                            & $0.709                     $ & $0.62^{+0.17}_{-0.15}      $          &                                      \\
		\boldmath $\betapol^{220}$     &                                                      &                                                        &                            & $0.687                     $ & $0.62\pm 0.15              $          &                                      \\
		\hline
	\end{tabular}
	\caption{Summary of systematic nuisance parameters in the \SPT{} \T\&\E{} likelihood. 
	The third column lists the priors used in the \lcdm{} analysis; we fix the nuisance parameters to the values given in brackets when fitting the simulations because they are not included in the simulation generation. 
	Calibration and polarization efficiency priors are derived from comparison with \planck{} maps; 
	only the external temperature calibration prior is used for the full likelihood, while ``subset priors'' are applied when analyzing data subsets. 
	For each nuisance parameter, we report the best-fit value from the \lcdm{} analysis, along with the mean and 68\% confidence interval. 
	The final column references the section where each parameter is discussed.}
	\label{tab:nuisance-priors}
\end{table*}

\section{Foreground model}
\label{app:foreground_functions}

\RR{\subsection{Foreground model functional forms}}

In this section, we present the explicit functional forms of the foreground model components used in the analysis (see \cref{sec:foreground} for additional discussion). The model includes the following terms.

First, unresolved sources are modeled as a Poisson component with a power-law dependence on multipole $\ell$:
\begin{equation}
	D^{\mathrm{Poisson}}_{\ell, \mu\nu} = A_{\mu\nu}^{\mathrm{Poisson}} \left(\frac{\ell}{3000}\right)^2,
\end{equation}
where $A_{\mu\nu}^{\mathrm{Poisson}}$ is the amplitude at $\ell=3000$ for each cross-frequency pair $\nu\times\mu$.
Based on expectations from our source masking threshold and the results of \citep{chou25} we set the Poisson amplitude to zero for all \TE{} and \EE{} cross-frequency pairs.

Second, the clustered CIB component is modeled as a power law in $\ell$:
\begin{equation}
	D^{\mathrm{CIB}}_{\ell, \mu\nu} = A_{\mu\nu}^{\mathrm{CIB}} \left(\frac{\ell}{3000}\right)^{\alpha^{\mathrm{CIB}}},
\end{equation}
where $A_{\mu\nu}^{\mathrm{CIB}}$ is the amplitude at $\ell=3000$ for each cross-frequency and $\alpha^{\mathrm{CIB}}$ is the spectral index. The CIB clustering term is fit only for the $150\times150\ghz$, $150\times220\ghz$, and $220\times220\ghz$ spectra; for other cross-frequencies, it is marginalized over in the covariance, as it is not significantly detected.

Third, the tSZ contribution is modeled using a fixed template with a frequency-dependent scaling:
\begin{equation}
	D^{\mathrm{tSZ}}_{\ell, \mu\nu} = A^{\mathrm{tSZ}}\, f_{\nu_0}^{\mathrm{tSZ}}(\nu,\mu)\, D_\ell^{\mathrm{tSZ,\,template}},
\end{equation}
where $A^{\mathrm{tSZ}}$ is the amplitude at $\ell=3000$, $D_\ell^{\mathrm{tSZ,\,template}}$ is the template power spectrum, and $f_{\nu_0}^{\mathrm{tSZ}}(\nu,\mu)$ encodes the SED of the standard tSZ frequency dependence relative to primary CMB fluctuations~\citep{shaw10}. The template power spectrum is obtained from \agora{} simulations~\citep{omori22} and is fixed in the analysis.

Fourth, the kSZ contribution is constant in CMB units and is modeled with a fixed template:
\begin{equation}
	D^{\mathrm{kSZ}}_{\ell, \mu\nu} = A^{\mathrm{kSZ}} D_\ell^{\mathrm{kSZ,\,template}},
\end{equation}
where $A^{\mathrm{kSZ}}$ is the amplitude at $\ell=3000$ and $D_\ell^{\mathrm{kSZ,\,template}}$ is the template power spectrum. The kSZ template is obtained from \agora{} simulations~\citep{omori22} and is fixed in the analysis.

Finally, the Galactic dust contribution is modeled as a modified black body with a power-law dependence on $\ell$:
\begin{equation}
	D^{\mathrm{dust}}_{\ell, \mu\nu} = A^{\mathrm{dust}} g_{\nu_0}^{\mathrm{dust}}(\mu,\nu,\beta^{\mathrm{dust}}) \left(\frac{\ell}{80}\right)^{\alpha^{\mathrm{dust}}+2},
\end{equation}
where $A_{80}^{\mathrm{dust}}$ is the amplitude at $\ell=80$, $g_{\nu_0}^{\mathrm{dust}}(\mu,\nu,\beta^{\mathrm{dust}})$ describes the frequency dependence of the dust emission and $\alpha^{\mathrm{dust}}$ is the power-law index. 
The frequency dependence is given by:
\begin{align}
	g_{\nu_0}(\mu,\nu,\beta) &= \frac{g(\mu)g(\nu)}{g(\nu_0)^2} \left(\frac{\mu\nu}{(\nu_0)^2}\right)^{\beta},
\end{align}
where $g(\nu)$ is the modified black body function.
We use this model in \TT{}, \TE{}, and \EE{} spectra, with independent amplitudes and priors.

In an extended version of the likelihood, we included a model for the tSZ–CIB correlation:
\begin{equation}
	D_{\ell,\mu\nu}^{\rm{tSZ}\times\rm{CIB}} = -\xi_{\rm tSZ\times CIB} \left( \sqrt{
		D_{\ell,\mu\mu}^{\rm tSZ} D_{\ell,\nu\nu}^{\rm CIB}
	} + \mu \leftrightarrow \nu \right),
	\label{eq:tszcibcorr}
\end{equation}
where the tSZ–CIB cross-correlation coefficient, $\xi_{\rm tSZ-CIB}$, was treated as a free parameter with a Gaussian prior $\mathcal{N}(0.18, 0.33)$. 
We found, however, that the data do not strongly constrain this parameter, and allowing it to vary substantially increased the computational cost of the likelihood evaluation. 
Therefore, in the baseline analysis, we fix $\xi_{\rm tSZ-CIB}$ to its best-fit value of $0.26$ and marginalize over its uncertainty in the covariance matrix computation.
We verified that fixing $\xi_{\rm tSZ-CIB}$ and marginalizing over the uncertainty has a negligible impact on cosmological parameters, see \cref{sec:foreground_variation}.

In \cref{tab:foreground-priors}, we summarize the priors on the foreground parameters used in the analysis. 
The priors originate from our previous \TTTEEE{} analysis~\citep{balkenhol23}, except on the Poisson and CIB amplitudes, which are set to be uniform in the range $[0, 200]\,\uk^2$. 
We also updated the prior on the CIB clustering power law index to $\mathcal{N}(0.53, 0.1)$, to reflect the results of~\citep{mak_measurement_2017}.

\begin{figure}
	\includegraphics[width=\columnwidth]{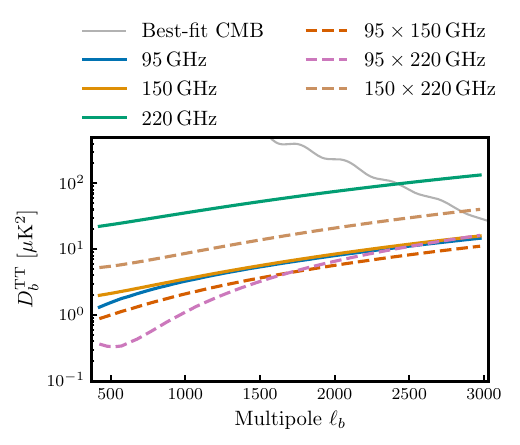}
	\caption{
	Total foreground model contributions to the \TT{} band powers at each frequency, assuming \lcdm{} cosmology.
	We show the auto-frequency channels as solid lines and the cross-frequency channels as dashed lines.
	The total foreground model is the sum of the Poisson, CIB clustering, tSZ, kSZ, and Galactic dust contributions.
	The best-fit \lcdm{} CMB prediction for the \sptbp{} is shown as a gray line.
	}
	\label{fig:foregrounds}
\end{figure}

In \cref{fig:foregrounds}, we present the best-fit total foreground contributions as determined from the data. 
At small angular scales, the foregrounds are dominated by Poisson sources, while clustered CIB becomes increasingly important at higher frequencies. 
The feature near $\ell \sim 500$ in the $95\times220\,\mathrm{GHz}$ spectrum arises from the tSZ–CIB cross-correlation, which is negative; this effect is only mildly significant given the uncertainties.
We note that the foreground model provides a good fit to the data.
In particular, the successful passing of the \TT{} power spectrum difference and conditional tests (see \cref{sec:powspenulltests}) demonstrates that the foreground model is sufficiently flexible to account for the observed differences in the band powers across frequencies.

\begin{table*}[ht]
	\centering
	\begin{tabular}{lllcc}
		\hline
		\hline
		\bf Parameter                                 & \bf Description                             & \bf Prior                                              & \bf \lcdm{} Best-fit         & \bf Posterior                 \\
		\hline
		$A^{\rm tSZ}\,[\uk^2]$                        & Thermal SZ amplitude                        & $\mathcal{N}(3.23, 2.4) \ \& \ \mathcal{U}(0, \infty)$ & $1.03                      $ & $0.93^{+0.41}_{-0.60}      $  \\
		$A^{\rm kSZ}\,[\uk^2]$                        & Kinetic SZ amplitude                        & $\mathcal{N}(3.7, 4.6) \ \& \ \mathcal{U}(0, \infty)$  & $0.09                      $ & $< 2.90                    $  \\
		\hline
		$A^{\rm CIB}_{150\times 150}\,[\uk^2]$        & \multirow{3}{*}{CIB clustering amplitude}   & $\mathcal{U}(0, 200)$                                  & $2.17                      $ & $1.88\pm 0.80              $  \\
		$A^{\rm CIB}_{150\times 220}\,[\uk^2]$        &                                             & $\mathcal{U}(0, 200)$                                  & $7.24                      $ & $7.5\pm 1.7                $  \\
		$A^{\rm CIB}_{220\times220}\,[\uk^2]$         &                                             & $\mathcal{U}(0, 200)$                                  & $34.0                      $ & $35\pm 5                   $  \\
		\hline
		$\alpha_{\rm CIB}$                            & CIB clustering power law index      \qquad  & $\mathcal{N}(0.53, 0.1)$                               & $0.464                     $ & $0.513\pm 0.092            $  \\
		\hline
		$A^{\rm Poisson}_{95\times95}\,[\uk^2]$       & \multirow{6}{*}{Poisson amplitude}          & $\mathcal{U}(0, 200)$                                  & $12.44                     $ & $10.7^{+2.2}_{-1.7}        $  \\
		$A^{\rm Poisson}_{95\times 150}\,[\uk^2]$     &                                             & $\mathcal{U}(0, 200)$                                  & $10.22                     $ & $8.6^{+2.1}_{-1.6}         $  \\
		$A^{\rm Poisson}_{95\times 220}\,[\uk^2]$     &                                             & $\mathcal{U}(0, 200)$                                  & $18.47                     $ & $16.7^{+2.5}_{-2.3}        $  \\
		$A^{\rm Poisson}_{150\times 150}\,[\uk^2]$    &                                             & $\mathcal{U}(0, 200)$                                  & $13.42                     $ & $11.9^{+2.1}_{-1.7}        $  \\
		$A^{\rm Poisson}_{150\times 220}\,[\uk^2]$    &                                             & $\mathcal{U}(0, 200)$                                  & $34.19                     $ & $32.1\pm 2.8               $  \\
		$A^{\rm Poisson}_{220\times 220}\,[\uk^2]$    &                                             & $\mathcal{U}(0, 200)$                                  & $98.9                      $ & $95.1^{+6.5}_{-5.5}        $  \\
		\hline
		$\kappa$                                      & Super-sample lensing                        & $\mathcal{N}(0., 0.00045)$                             & $0.3\cdot 10^{-5}          $ & $0.00000\pm 0.00045        $  \\
		\hline
		$A^{\rm Dust}\,[\uk^2]$                       & Galactic dust amplitude                            & $\mathcal{N}(1.88, 0.96)$                              & $1.80                      $ & $2.00\pm 0.82              $  \\
		$\alpha^{\rm Dust}$                           & Galactic dust power law index                      & $\mathcal{N}(-2.53, 0.05)$                             & $-2.5312                   $ & $-2.530\pm 0.049           $  \\
		$\beta^{\rm Dust}$                            & Galactic dust spectral index                       & $\mathcal{N}(1.48, 0.02)$                              & $1.4801                    $ & $1.480\pm 0.020            $  \\
		\hline
		$A^{\rm PolGalDust}_{\TE }\,[\uk^2]$          & \TE{} polarized dust amplitude              & $\mathcal{N}(0.12, 0.051)$                             & $0.1040                    $ & $0.104\pm 0.034            $  \\
		$\alpha^{\rm PolGalDust}_{\TE }$              & \TE{} polarized dust power law index        & $\mathcal{N}(-2.42, 0.04)$                             & $-2.4307                   $ & $-2.434\pm 0.039           $  \\
		$\beta^{\rm PolGalDust}_{\TE }$               & \TE{} polarized dust spectral index         & $\mathcal{N}(1.51, 0.04)$                              & $1.5139                    $ & $1.511\pm 0.040            $  \\
		\hline
		$A^{\rm PolGalDust}_{\EE }\,[\uk^2]$          & \EE{} polarized dust amplitude              & $\mathcal{N}(0.05, 0.022)$                             & $0.0566                    $ & $0.057\pm 0.013            $  \\
		$\alpha^{\rm PolGalDust}_{\EE }$              & \EE{} polarized dust power law index  \quad & $\mathcal{N}(-2.42, 0.04)$                             & $-2.4140                   $ & $-2.417\pm 0.039           $\ \\
		$\beta^{\rm PolGalDust}_{\EE }$               & \EE{} polarized dust spectral index         & $\mathcal{N}(1.51, 0.04)$                              & $1.5089                    $ & $1.508\pm 0.040            $  \\
		\hline
		\hline
		\multicolumn{3}{l}{\bf \replace{Marginalized}{Fixed} in baseline likelihood}                                                                                                                                                         \\
		\hline
		\hline
		$A^{\rm CIB-cl.}_{\rm 95\times 95}\,[\uk^2]$  & \multirow{3}{*}{CIB clustering amplitude}   & $\mathcal{N}(0.26, 0.15)$                                                                                             \\
		$A^{\rm CIB-cl.}_{\rm 95\times 150}\,[\uk^2]$ &                                             & $\mathcal{N}(0.04, 0.53)$                                                                                             \\
		$A^{\rm CIB-cl.}_{\rm 95\times 220}\,[\uk^2]$ &                                             & $\mathcal{U}(0, 200)$                                                                                                 \\
		\hline
		$\xi_{\rm tSZ-CIB}$                           & tSZ-CIB correlation                         & $\mathcal{N}(0.18, 0.33) \ \& \ \mathcal{U}(-1, 1)$                                                                   \\
		\hline
	\end{tabular}
	\caption{Overview of the foreground model parameters for the \SPT{} \T\&\E{} likelihood. 
	The first part of the table lists the parameters that are varied in the baseline analysis, while the second part lists the parameters that are \RR{fixed to the best-fit value of the extended likelihood, and associated uncertainty is }marginalized over by adding a suitable constant contribution to the band power covariance matrix.
	The best-fit values and 68\% confidence limits are shown for the \lcdm{} model. 
	The priors are either Gaussian $\mathcal{N}(\mu, \sigma)$ or uniform $\mathcal{U}(a, b)$ distributions. }
	\label{tab:foreground-priors}
\end{table*}

\subsection{Robustness of the foreground model}
\label{sec:foreground_variation}

\begin{figure}
	\includegraphics[width=\columnwidth]{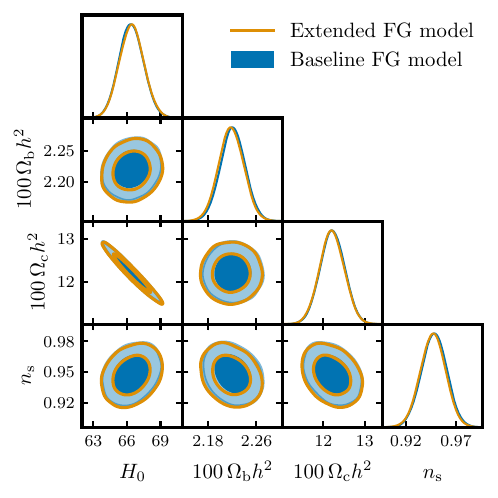}
	\caption{
		We show the shifts in \lcdm{} parameters when using the extended foreground model relative to the baseline analysis. 
		Posterior distributions are consistent between the two analyses.
	}
	\label{fig:extended_vs_baseline}
\end{figure}

We compare the ``extended'' likelihood that treats the tSZ–CIB correlation as a free parameter and includes CIB clustering amplitudes for the $95\times95\ghz$, $95\times150\ghz$, and $95\times220\ghz$ spectra, rather than marginalizing over them in the covariance matrix computation, to the baseline likelihood described in \cref{sec:foreground}.
This extended model adds four foreground degrees of freedom relative to the baseline and was not adopted for the final analysis because it increases computational cost while providing weak constraints on the extra parameters.
We compare the cosmological parameters obtained by MCMC exploration of the extended likelihood to those from the baseline analysis in \cref{fig:extended_vs_baseline}. 
We find no shifts between the posterior distributions, indicating that the baseline foreground model is sufficient to capture the relevant astrophysical contributions in the data.

We evaluated the robustness of the foreground model by varying foreground-related analysis choices and measuring the impact on cosmological parameters.
Specifically, we considered several alternative foreground treatments and compared their resulting parameter shifts to the statistical uncertainties.
For this exploration, we approximate the parameter posteriors using the Fisher matrix at the best-fit point from the baseline analysis, which allows for rapid evaluation of multiple variants, but may slightly overestimate the parameter shifts compared to full MCMC sampling.

\begin{figure}
	\includegraphics[width=\columnwidth]{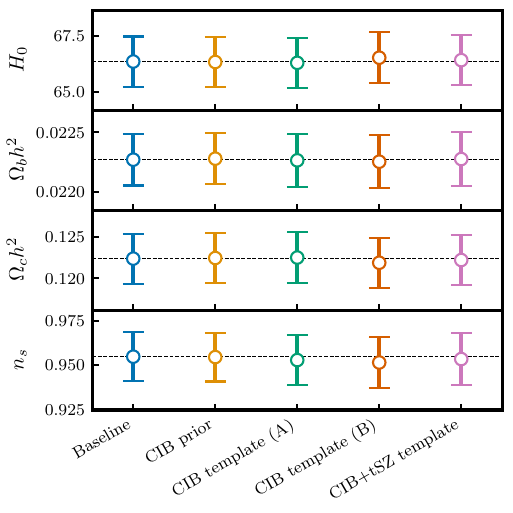}
	\caption{\RR{Robustness of cosmological parameters to variations in the foreground model.
		We show the shifts in \lcdm{} parameters when varying the foreground treatment relative to the baseline analysis, using the Fisher matrix approximation.
		All shifts are small compared to statistical uncertainties, indicating that the cosmological results are robust to foreground modeling choices.}}
	\label{fig:foreground_robustness}
\end{figure}

We varied the prior on the CIB clustering index $\alpha_{\rm CIB}$ from the baseline $\mathcal{N}(0.53,0.1)$ to $\mathcal{N}(0.8,0.1)$, motivated by \citep{reichardt21}.
This variant is labeled ``CIB prior'' in \cref{fig:foreground_robustness}.

We replaced the CIB power-law description with a template derived from the \agora{} simulations \citep{omori22} and varied a single common amplitude for all cross-frequency spectra that include the CIB.
We refer to this test as ``CIB template (A)'' and neglect tSZ–CIB correlation for implementation simplicity.

We also repeated the template test while allowing independent amplitudes for each cross-frequency combination where the CIB is included.
This case is labeled ``CIB template (B)'' and likewise neglects tSZ–CIB correlation.

Finally, we replaced the tSZ, kSZ, and CIB clustering power laws by their respective \agora{} templates and varied separate amplitudes for each component, common across cross-frequency spectra where the component applies.
This configuration is labeled ``CIB+tSZ template''.

All tests are summarized in \cref{fig:foreground_robustness}.
In every case the cosmological parameter shifts were small compared to statistical errors.
The maximum observed shifts were below $0.2\,\sigma$ for all parameters and tests.
These results indicate that the baseline foreground model yields stable cosmological inferences while offering a reasonable balance between model complexity and computational cost.
A comprehensive study of foreground parameter constraints and astrophysical implications will be reported in a future work.

\section{Construction and performance of the CMB-only likelihood}
\label{app:cmblite}

We provide additional information on the construction and performance of the \emph{lite} likelihood, \sptlite{}.
The framework for the construction of the CMB-only likelihood introduced by~\citep{balkenhol25} and used here cannot translate the information of parameter boundaries for nuisance parameters to the covariance of the CMB-only band powers~\citep{raveri19} (unless one resorts to MCMC sampling the reconstruction likelihood~\citep{dunkley13}, which we would like to avoid due to the numerical cost).
However, the multi-frequency likelihood and the chosen nuisance parameter priors are fairly constraining so this is typically not an issue.
The only exceptions are the \TT{} Poisson power parameters and the beam sidelobe depolarization parameters, as the \emph{lite} framework is by design only sensitive to frequency-differences.
In response, we condition \sptlite{} on the results of the multi-frequency likelihood in \lcdm{} by (1) imposing regularization priors on the \TT{} Poisson parameters centered on the best-fit values and 10 times wider than their corresponding constraints and (2) setting the beam sidelobe depolarization parameters to their best-fit values.
This breaks the otherwise complete degeneracy of the Poisson parameters and ensures the bulk of the posterior mass is in the physical range for all parameters; this allows the reconstruction procedure to better capture the uncertainty due to the nuisance parameters in the covariance of the CMB-only band powers.
Though this procedure is no longer strictly independent of cosmology, we prefer to condition the \emph{lite} likelihood this way to improve its performance: indeed, as we show below this leads to a good match between the parameter constraints inferred from the multi-frequency likelihood and \sptlite{} in \lcdm{} and beyond.
We have verified that changing the width of the Poisson priors or offsetting the $\betapol$ parameters in a frequency-coherent way has a small impact on cosmological parameters.

\begin{figure*}
	\includegraphics[width=\textwidth]{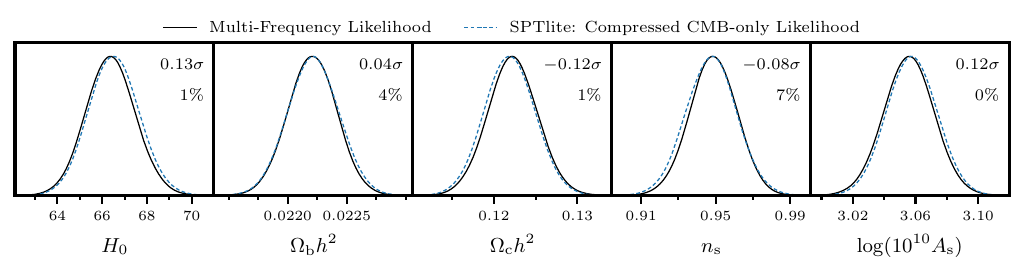}
	\caption{Comparison of \lcdm{} parameter constraints from the multi-frequency likelihood (black lines) and the compressed CMB-only \sptlite{} likelihood (blue dashed lines).
	The means of the one-dimensional marginalized posterior distributions of the \emph{lite} likelihood shift by $\lesssim 0.1\,\sigma$ compared to the multi-frequency likelihood, where $\sigma$ is the width of the posterior when using \sptlite{}.
	Similarly, the width of the posteriors match to $\leq 7\%$.
	Details on the construction of \sptlite{} can be found in \cref{sec:lite_likelihood}.}
	\label{fig:mf_vs_lite}
\end{figure*}

We compare the one-dimensional marginalized posterior distributions assuming \lcdm{} for $\Hubble$, $\ombh$, $\omch$, $\ns$, and $\logA$ obtained from the multi-frequency likelihood and \sptlite{} in \cref{fig:mf_vs_lite}.
The means of the marginalized posterior distributions shift by $\lesssim 0.1\,\sigma$, where $\sigma$ is the width of the \emph{lite} posteriors.
Error bars match to $\leq 6\%$.
We calculate the size of the mean shift $\Delta p$ in the full $N$-dimensional parameter space respecting the correlation of the parameters as given by the \sptlite{} parameter covariance $C$ via: $\sqrt{(\Delta p^T C^{-1} \Delta p)/N}$.
This metric calculates the Euclidean distance in the parameter space transformed according to $C$ and adjusts it for the dimensionality; in the one-parameter case, it reduces to the familiar $\Delta p / \sigma$.
Calculating this metric yields an average offset of $0.09$ per parameter, which is negligible; we conclude that the \sptlite{} likelihood performs well in \lcdm{}.

The good performance also holds up in extended model spaces.
We further explore the comparison of \sptlite{} and the multi-frequency likelihood by extending \lcdm{} by $A_{\rm lens}$, $\Neff{}$, $\Omega_k$, and $\mnu{}$ separately.
For the latter two cases we also add DESI data, as primary CMB data by themselves suffer from strong degeneracies in these models.
We compare the same parameters as for \lcdm{} plus the relevant extension parameters.
This yields a total of $24$ parameters to compare; the means differ by $<0.3\,\sigma$ and error bars match to $<10\%$ in all cases.
Though we record slightly larger fluctuations than in \lcdm{}, given the increased size of tests this is not surprising.
As before, we calculate the size of the mean shift in the full parameter space respecting the correlation of parameters, finding all offsets to be $<0.2$.
For $\mnu{}$, the $95\%$ confidence limit inferred from the \emph{lite} likelihood is $15\%$ higher.
In general, the \emph{lite} likelihood leads to a minor widening of posteriors due to the treatment of the beams detailed above; this makes it a conservative choice.
We conclude that the \emph{lite} likelihood performs well for all model spaces considered and that the \lcdm{} conditioning performed has a negligible impact on other cosmological models.

\section{Power spectrum consistency tests between frequencies}
\label{app:consistency}

During the validation of the analysis pipeline detailed in \cref{sec:validation}, we perform a series of tests to check the consistency of the data across different frequency combinations. 
We only show an extract of those in \cref{fig:errorbars}. In this section, we show the complete results of these tests.

\subsection{Probability to exceed}
\label{app:pte}
We outline the procedure for computing the PTE for each test presented in this work. First, we construct the vector $\Delta$ of observed values and the corresponding predicted covariance matrix $\Sigma^\Delta$. The associated $\chi^2$ statistic is then calculated as:
\begin{equation}
	\chi^2=\Delta^T \Sigma^{\Delta} \Delta.
\end{equation}
We then compute the PTE as the probability of observing a $\chi'^2$ value greater than or equal to the observed $\chi^2$ value, given the degrees of freedom $\Ndof$ of the test. The PTE is defined as:
\begin{align}
	\label{eq:difftestpte}
	p = & 1- P(\chi'^2 \leq \chi^2,\Ndof)\\
	= & 1-\int_0^{\chi^2}d\chi'^2 P(\chi'^2,\Ndof)
\end{align}
where $P(\chi'^2 \leq \chi^2,\mathrm{dof})$ is the cumulative distribution function of the $\chi^2$ probability function $P(\chi'^2,\Ndof)$ for $\Ndof$ degrees of freedom.

\subsection{Difference tests}
\label{app:frequencydiff}
To validate the pipeline, we conduct a series of frequency-difference tests, comparing spectra derived from different frequency combinations. 
The objective is to assess the consistency among the various frequency channels and to identify any potential issues in the data or analysis pipeline. 
The methodology is described in \cref{sec:difference} and a subset of the results is presented in \cref{fig:errorbars}. 
Here, we provide the full results of these tests. 
The figures in this section display the frequency-difference tests for \TT, \TE, and \EE, along with the probability to exceed (PTE) for each spectral combination.
\begin{figure*}[p]
	\includegraphics[width=\textwidth]{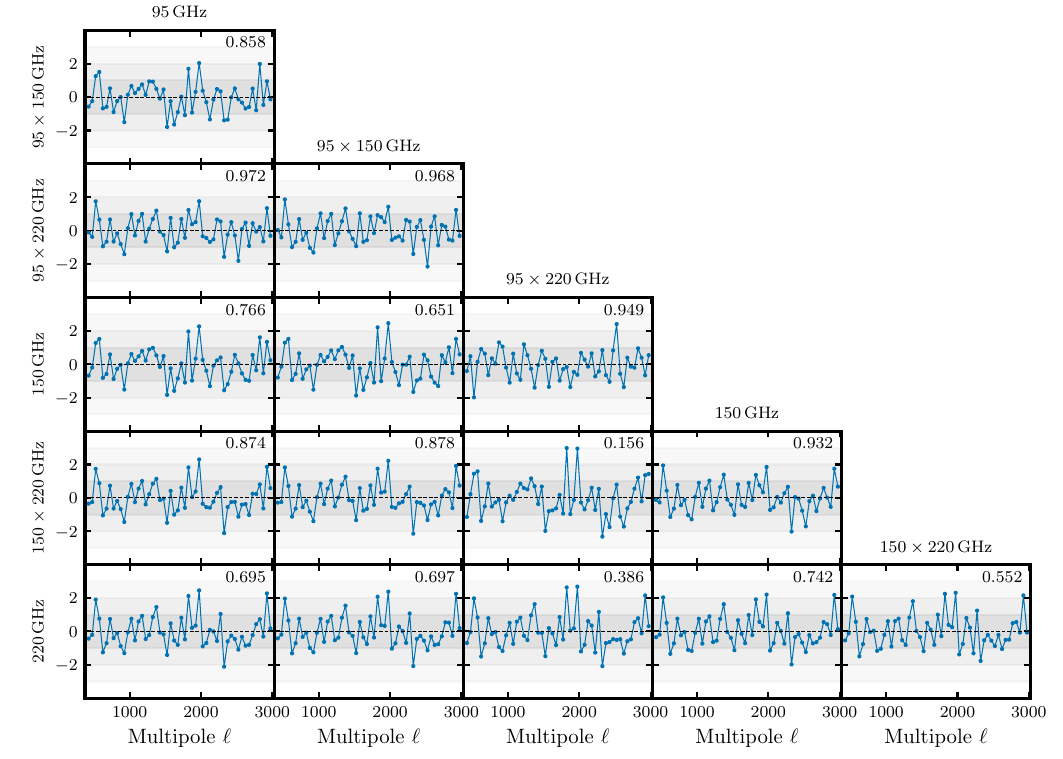}
	\caption{Difference tests for \TT{} spectra. Each block shows the difference between the two cross-frequency spectra $\mu\times\nu$ (row labels) and $\alpha\times\beta$ (column labels), e.g. the upper left panel is the difference spectrum $C_\ell^{95\times150} - C_\ell^{95\times95}$. We show quantities in units of the expected error bars, as calculated from the covariance matrix in \cref{eq:difftestspec,eq:difftestcov}. The PTE values quoted on the upper right corners are calculated from \cref{eq:difftestpte}, with associated number of degrees of freedom per tests being $\Ndof=52$. Combining \TT, \TE, and \EE{}, there are 18 independent difference tests, thus the PTE threshold is $0.05/18 = 0.0028$. The gray shaded regions indicate the $1,2,{\rm and}\, 3\,\sigma$ regions of the expected distribution.}
	\label{fig:TTdifference}
\end{figure*}
\suppressfloats
\begin{figure*}[p]
	\includegraphics[width=\textwidth]{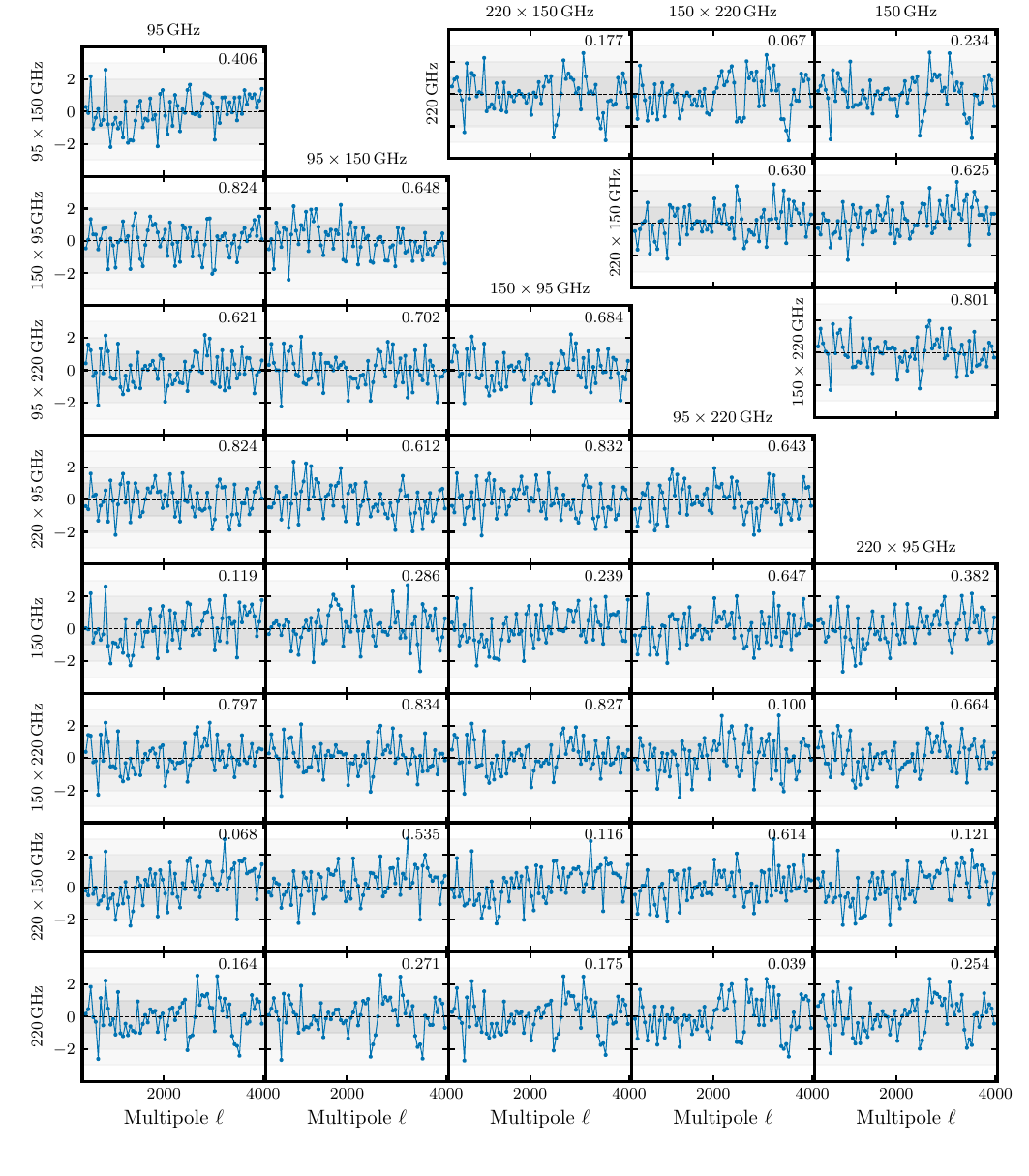}
	\caption{Difference tests for \TE{} spectra. Each block shows the difference between the two cross-frequency spectra $\mu\times\nu$ (row labels) and $\alpha\times\beta$ (column labels), e.g. the upper left panel is the difference spectrum $C_\ell^{95\times150} - C_\ell^{95\times95}$. We show quantities in units of the expected error bars, as calculated from the covariance matrix in \cref{eq:difftestspec,eq:difftestcov}. The PTE values quoted on the upper right corners are calculated from \cref{eq:difftestpte}, with associated number of degrees of freedom per tests being $\Ndof=72$. Combining \TT, \TE, and \EE{}, there are 18 independent difference tests, thus the PTE threshold is $0.05/18 = 0.0028$. The gray shaded regions indicate the $1,2,{\rm and}\, 3\,\sigma$ regions of the expected distribution.}
	\label{fig:TEdifference}
\end{figure*}
\suppressfloats
\begin{figure*}[p]
	\includegraphics[width=\textwidth]{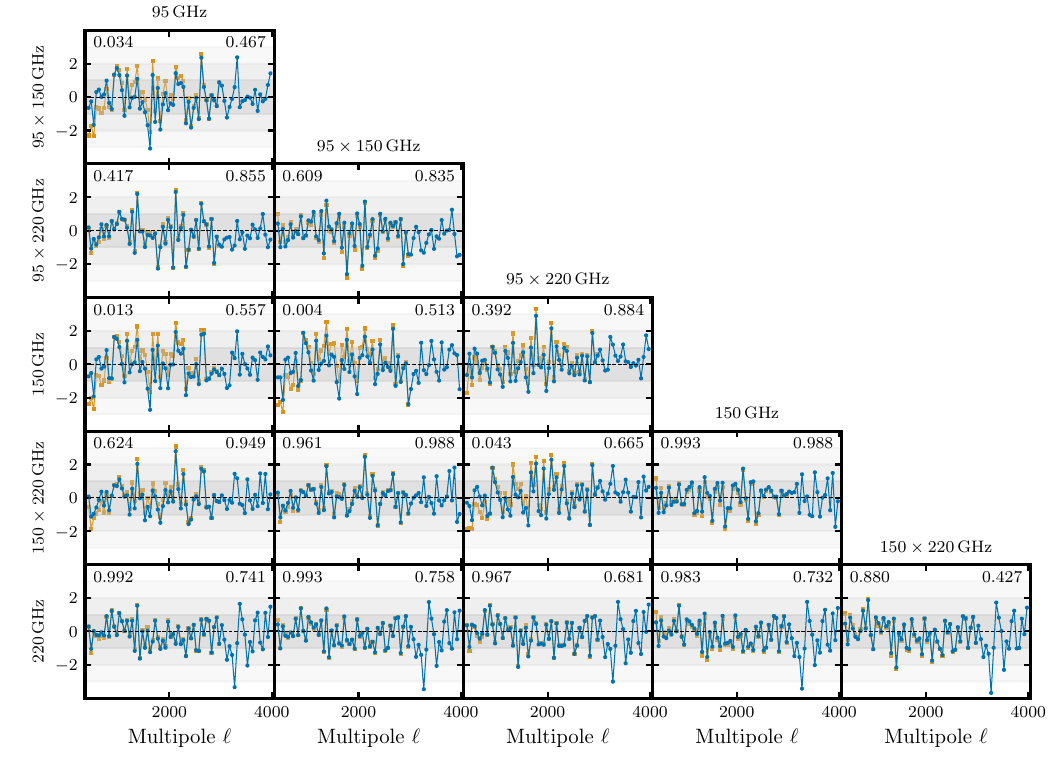}
	\caption{Difference tests for \EE{} spectra. Each block shows the difference between the two cross-frequency spectra $\mu\times\nu$ (row labels) and $\alpha\times\beta$ (column labels), e.g. the upper left panel is the difference spectrum $C_\ell^{95\times150} - C_\ell^{95\times95}$. We show quantities in units of the expected error bars, as calculated from the covariance matrix in \cref{eq:difftestspec,eq:difftestcov}. The PTE values quoted on the upper right corners are calculated from \cref{eq:difftestpte}, with associated number of degrees of freedom per tests being $\Ndof=72$. The orange line and the PTEs quoted on the upper left are those obtained prior to unblinding, exhibiting a close-to-threshold value in 95-150\ghz, which is improved by the polarized beam modeling. Combining \TT, \TE, and \EE{}, there are 18 independent difference tests, thus the PTE threshold is $0.05/18 = 0.0028$. The gray shaded regions indicate the $1,2,{\rm and}\, 3\,\sigma$ regions of the expected distribution.}
	\label{fig:EEdifference}
\end{figure*}
\suppressfloats

\subsection{Conditional tests}
\label{app:conditionals}
In addition to the frequency-difference tests, we conduct conditional tests to evaluate the consistency of the data across different frequency combinations. 
Unlike the frequency-difference tests, which compare pairs of spectra, the conditional tests assess a given spectrum against its prediction based on all other spectra. 
This approach yields more stringent constraints on data consistency, as it leverages the information from the full set of remaining spectra to predict the spectrum in question.
The methodology is outlined in \cref{sec:conditional}, and selected results are shown in \cref{fig:errorbars}.
Here, we present the complete results of these tests.
In this section, \cref{fig:TTdifference,fig:TEdifference,fig:EEdifference} display the conditional tests for \TT, \TE, and \EE, along with the PTE for each spectral combination.

\begin{figure*}[p]
	\includegraphics[width=\textwidth]{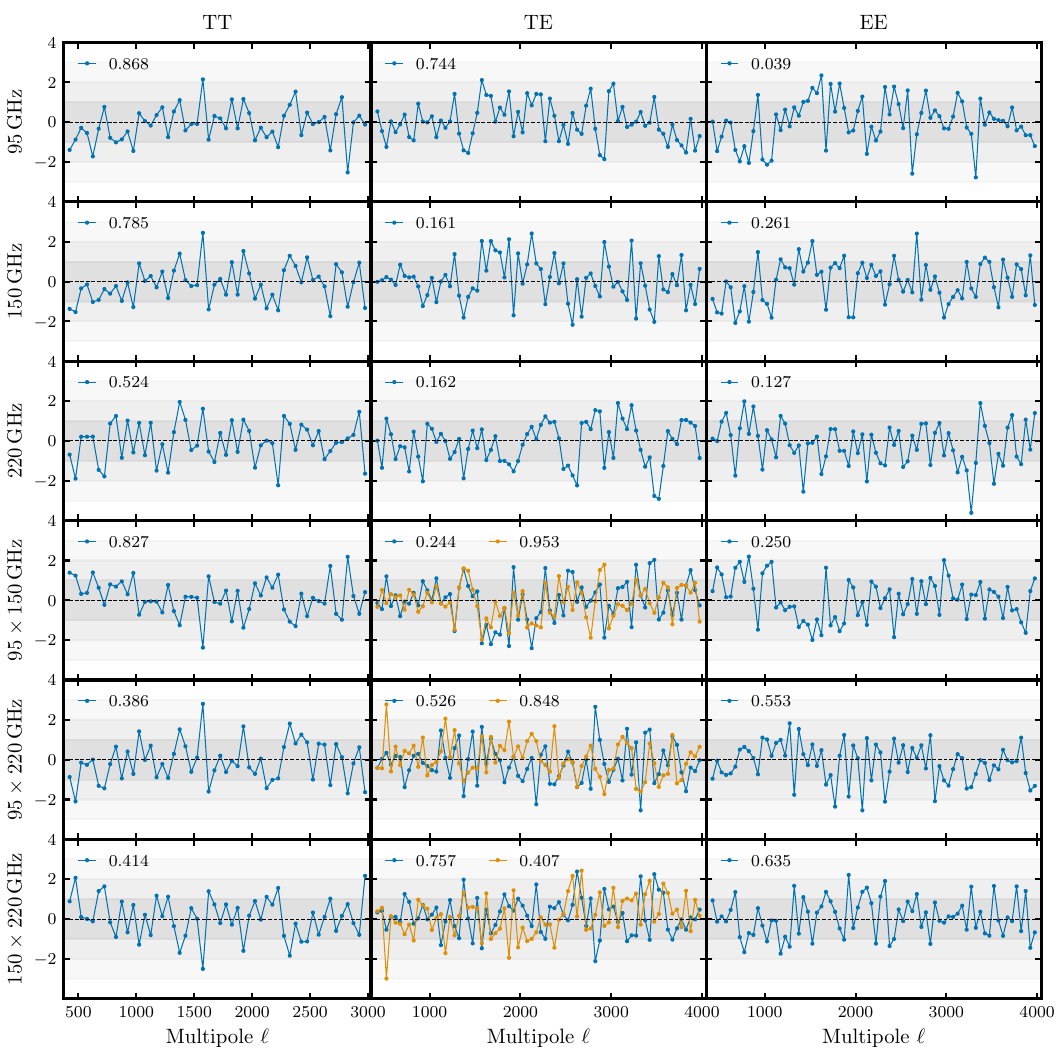}
	\caption{Conditional tests. For each \TT{}, \TE{}, and \EE{} cross frequency spectrum, we show the difference between the data and the prediction conditioned on the other cross-frequency spectra of the same channel. We show the difference in units of the expected error bars, as calculated from the covariance matrix in \cref{eq:difftestspec,eq:difftestcov}. The PTE values quoted on the upper right corners are calculated from \cref{eq:difftestpte}, with associated number of degrees of freedom per tests being $\Ndof=52$ for \TT{} and $72$ for the others. There are 18 independent tests, thus the PTE threshold is $0.05/18 = 0.0028$. The gray shaded regions indicate the $1,2,{\rm and}\, 3\,\sigma$ regions of the expected distribution. For the \TE{} asymmetrical cross-frequencies case, we show on the same plot the $\mu\times\nu$ and $\nu\times\mu$ combinations in blue and orange, respectively.
	}
	\label{fig:freq_cond}
\end{figure*}
\suppressfloats

\section{Comparison with data recorded in 2018}
\label{sec:compare_2018}

We compare these results against those derived from data recorded during the 2018 observing season reported in~\citep{balkenhol23}.
While these data are also subject to quadrupolar leakage and 
the depolarization of beam sidelobes,
these effects went unmodeled as they were unknown at the time.
Though this may bias the results of~\citep{balkenhol23}, the uncertainties of the 2018 data set were much larger; they therefore did not necessitate the sophisticated methods employed here to be modeled accurately and any relative bias is expected to be smaller than what it would be for the new data.
We stress that the 2018 data and results are superseded by this work.

In principle, a comparison at the band power level would necessitate a detailed understanding of the correlation between the two data sets, which is difficult to model for various reasons.\footnote{
	Different analysis choices were made regarding, for example, the filtering strategy, the source flux cut threshold, whether to analyze maps in a flat- or curved-sky framework, whether or not to inpaint sources, or how to bin the power spectrum measurement into band powers.
	Moreover, the focal plane was replaced after the 2018 observing season; this leads to small beam and bandpass changes.
	Together, these aspects lead to differences in the covariance matrices and band power window functions that are non-trivial to model.
}
Instead, we perform a parameter-level comparison assuming the \lcdm{} model.
Since this model provides a good description of both data sets it allows for a qualitative check.
We restrict the two data sets to the common multipole moment range and perform MCMC analyses jointly of the full set of \TTTEEE{} spectra, as well as of each spectrum type individually.
The parameter set we choose to compare constraints across includes the parameters $\thetastar$, $\ombh$, $\omch$, $\ns$, and $\clamp$.
We quantify parameter differences in two ways, either by assuming the data sets are independent or by assuming the 2018 data is a subset of the new data, which mathematically corresponds to either adding or subtracting the parameter covariance matrices, respectively.
Strictly speaking both of these tests are inappropriate; while in the first case, we ignore the shared sample variance fluctuations, in the second case we assume shared sample variance but also some common noise fluctuations.
Still, by having an overly conservative and an overly optimistic test, we can gain a qualitative understanding of the consistency.

We first carry out the optimistic test, comparing constraints assuming the data sets are independent.
We assume a total of three independent tests for each case and regard PTEs above $2.5\%/3=0.83\%$ to be passing.
For the full set of \TTTEEE{} spectra we obtain a PTE of $76.96\%$, signaling good agreement.
For \TT{}, \TE{}, and \EE{} fits individually, we obtain PTEs of $85.98\%$, $71.54\%$, and $1.07\%$.
All of these lie above the PTE threshold.
Though differences in \EE{} constraints may be related to the updated beam and leakage modeling, the associated PTE is statistically normal.
Therefore, we conclude that the results presented here and the ones of~\citep{balkenhol23} are broadly consistent.

The conservative test exhibits numerical difficulties as the resulting parameter difference covariance matrices are generally not positive definite.
This is particularly the case for the combined \TTTEEE{} constraints, where changes in the degeneracy directions between parameters from~\citep{balkenhol23} compared to this work lead to instabilities.
Still, using the diagonal of the covariance calculated this way, all individual parameter shifts are $<3\,\sigma$ for all cases.
For \TTTEEE{} specifically, four out of the five parameters are offset by $<1.5\,\sigma$.
For individual spectrum fits, we are able to numerically stabilize the test by restricting ourselves to the two parameters that are the most discrepant in the one-dimensional marginalized posteriors.
For \TT{}-only fits, these are $\ns$ and $\clamp$, for which we obtain a PTE of $1.5\%$.
Since the spectral tilt profits from the lower-noise measurement of the CMB damping tail that the new data set offers, it is not surprising to find a fluctuation in this parameter plane.
We note that debiasing the new temperature band powers using the 2018 beam leads to a negligible shift in the \TT{}-derived parameter constraints.
Moreover, we have verified that the shift in the $\ns - \clamp$ plane produced by multiplying or dividing the \TT{} band powers by up to eight powers of the pixel window function leads to a comparatively small shift that does not fully align with the direction of differences between the old and new data.
Given that the other three parameters agree to $\leq1.5\,\sigma$ we do not regard this PTE as problematic.
For \TE{}, we perform the test over $(\ns, \thetastar)$ and report a PTE of $4.83\%$, whereas for \EE{} we restrict ourselves to the $(\ns, \omch)$ plane and report a PTE of $2.25\%$.
We conclude that even under overly conservative assumptions, the results of this analysis agree with the predecessor work of~\citep{balkenhol23}.

\section{Consistency of ACT DR6 and DESI data}
\label{app:BAO_consistency}

\begin{figure}[H]
	\includegraphics[width=\columnwidth]{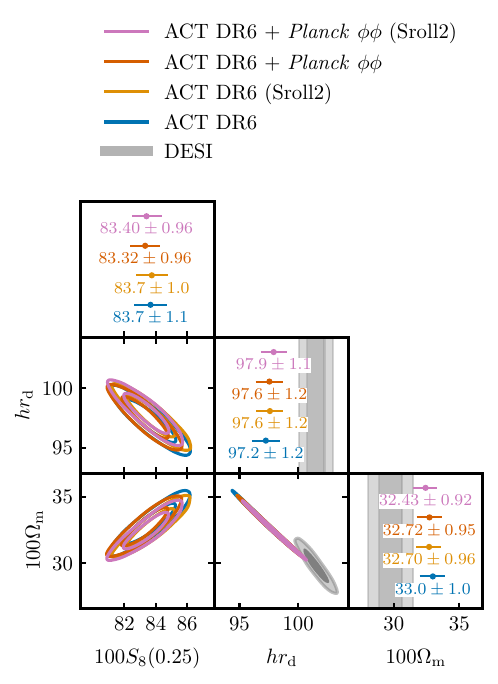}
	\centering
	\begin{tabular}{l c}
		\toprule
		\ACTDR{} & $3.1\,\sigma$ \\
		\ACTDR{} (Sroll2) & $2.9\,\sigma$ \\
		\ACTDR{} + \planck{} $\phi\phi$ & $2.9\,\sigma$ \\
		\ACTDR{} + \planck{} $\phi\phi$ (Sroll2) & $2.7\,\sigma$ \\
		\bottomrule
	\end{tabular}
	\caption{
		Comparison of the $\omm$-$\hrd$ parameters between the \ACTDR{} and DESI data sets.
		We varied the $\tau$ prior, using Sroll2 likelihood~\citep{pagano2020reionization} as used in~\citep{louis25}.
		The lensing reconstruction combined with the \T\&\E{} data is either the \PP{} data from \ACTDR{} alone~\citep{qu24} or incorporates the \planck{} lensing data~\citep{carron22}.
		Reported values in the table indicate the Gaussian distance between the data sets and the \DESI{} constraints in the $\omm$-$\hrd$ parameter plane.
	}
	\label{fig:hord_act}
\end{figure}
The \ACTDR{} \lcdm{} cosmological parameters show a mild tension with the \DESI{} data set, corresponding to a $3.1\,\sigma$ separation in the $\omm$-$\hrd$ plane.
When substituting our $\tau$ prior with the Sroll2 likelihood~\citep{pagano2020reionization}, which is the baseline choice in~\citep{louis25}, we find a $2.9\,\sigma$ distance.
We note that we are using the \texttt{ACT-lite} likelihood, whereas the baseline \ACTDR{} results are based on multi-frequency likelihoods, and this may lead to a small difference in the reported distance.
In addition, a recent study by the DESI collaboration~\citep{garciaquintero25} found that the \ACTDR{}+\planck{} $\phi\phi$ \lcdm{} cosmological parameters are $2.7\,\sigma$ distant from the DESI data set
The results reproduce those of~\citep{garciaquintero25}.
In \cref{fig:DESI_hord}, we also examine how this result changes when replacing our $\tau$ prior with the Sroll2 likelihood.
We conclude that the significance of the \ACTDR{}–\DESI{} discrepancy is sensitive to analysis choices. 
Throughout this work, we reported values with consistent priors, based on the NPIPE $\tau$ prior, see \cref{tab:dataset}.

\section{Glossary of cosmological parameters}
\label{app:par_defs}

We provide an overview of cosmological parameters used in this manuscript in \cref{tab:par_def}.

\begin{table*}[h]
    \centering
    \begin{tabular}{c l}
    \hline
    \textbf{Parameter} & \multicolumn{1}{c}{\textbf{Definition}}\\
    \hline
    \hline
    \thetastar & Angle of the sound horizon at recombination\\
    \thetaMC & Approximate angle of the sound horizon at recombination\\
    \Hubble & Hubble constant, expansion rate today in $\kmsmpc$\\
    $h$ & $\Hubble/100\,\kmsmpc$\\
    \ombh & Physical baryon density\\
    \omch & Physical cold dark matter density\\
    \omm & Fractional matter density\\
    \As & Amplitude of the power spectrum of initial scalar fluctuations\\
    \ns & Tilt of the power spectrum of initial scalar fluctuations\\
    \taureio & Optical depth to reionization\\
    \wowa & Dark energy equation of state parameters (see \cref{eq:wowa})\\
    \wperp & Deviation from a cosmological constant along the BAO degeneracy direction, $\replace{\wa + 3.5(\wo+1)}{0.3(\wo+1)-\wa}$\\
    \curv & Mean spatial curvature\\
    \sigmaeight & Root mean square of matter fluctuations today in linear theory in a sphere of comoving radius of $8\ h^{-1}$ Mpc\\
    \neff & Effective number of neutrino species\\
    \yp & Primordial helium abundance (mass fraction)\\
    \mnu & Sum of neutrino masses\\
    \rd & Comoving size of the sound horizon at the end of the baryon drag epoch\\
    \Atwopt & Amplitude of the signature of gravitational lensing in the primary CMB power spectra\\
    \Arecon & Amplitude of the CMB gravitational lensing power spectrum\\
    \Alens & Coherent variation of the amplitude of gravitational lensing, $\Atwopt=\Arecon$\\
    $m_{\rm e}$ & Electron mass\\
    $m_{\rm e, 0}$ & Electron mass today\\
    $X_{\rm e}$ & Ionization fraction of the universe at recombination\\
    \hline
    \end{tabular}
    \caption{Definition of all cosmological parameters used.}
    \label{tab:par_def}
\end{table*}

\clearpage
\bibliography{spt,TTTEEE1920}

\end{document}